\documentclass{aa}
\usepackage[utf8]{inputenc}

\usepackage{graphicx}
\usepackage[varg]{txfonts}

\usepackage{amsmath}
\usepackage{amssymb}
\usepackage{booktabs}
\usepackage{xtab}
\usepackage{textcomp}
\usepackage{microtype}
\usepackage{bm}
\usepackage{listings}
\usepackage{siunitx}
\usepackage{pdflscape}
\usepackage{float}
\usepackage{longtable}
\usepackage{xcolor}
\usepackage{pgf}
\usepackage{multirow}
\usepackage{url}
\usepackage{collcell}
\usepackage{microtype}
\usepackage{todonotes}

\usepackage{paralist}
\usepackage{soul}

\usepackage{pdflscape}

\usepackage{natbib,twoopt}
\usepackage[breaklinks=true]{hyperref}

\bibpunct{(}{)}{;}{a}{}{,} 

\makeatletter
\newcommandtwoopt{\citeads}[3][][]{\href{http://adsabs.harvard.edu/abs/#3}{\def\hyper@linkstart##1##2{}\let\hyper@linkend\@empty\citealp[#1][#2]{#3}}}
\newcommandtwoopt{\citepads}[3][][]{\href{http://adsabs.harvard.edu/abs/#3}{\def\hyper@linkstart##1##2{}\let\hyper@linkend\@empty\citep[#1][#2]{#3}}}
\newcommandtwoopt{\citetads}[3][][]{\href{http://adsabs.harvard.edu/abs/#3}{\def\hyper@linkstart##1##2{}\let\hyper@linkend\@empty\citet[#1][#2]{#3}}}
\newcommandtwoopt{\citeyearads}[3][][]{\href{http://adsabs.harvard.edu/abs/#3}
{\def\hyper@linkstart##1##2{}\let\hyper@linkend\@empty\citeyear[#1][#2]{#3}}}
\makeatother

\usepackage{color}
\definecolor{mygreen}{RGB}{0,128,0}

\hypersetup{colorlinks=true,linkcolor=blue,citecolor=blue,urlcolor=blue}

\usepackage{etoolbox}
\makeatletter

\makeatletter
\ifaa@referee

\else

\fi
\makeatother

\newcommand{\muas}{\ensuremath{\,{\mu}\text{as}}}

\newcommand{\gaia}{\textit{Gaia}}

\newcommand*{\nHz}{\text{nHz}}

\newcommand*{\aGW}{\alpha_{\mathrm{gw}}}
\newcommand*{\dGW}{\delta_{\mathrm{gw}}}
\newcommand*{\hPlusSin}{h_{+}^{\mathrm{s}}}
\newcommand*{\hPlusCos}{h_{+}^{\mathrm{c}}}
\newcommand*{\hTimesSin}{h_{\times}^{\mathrm{s}}}
\newcommand*{\hTimesCos}{h_{\times}^{\mathrm{c}}}
\newcommand*{\Tobs}{T}
\newcommand*{\Pgw}{\ensuremath{P_{\rm gw}}}

\newcommand*{\lmax}{\ell_\mathrm{max}}

\newcommand*{\numtwo}[1]{\num[round-mode=places,round-precision=2]{#1}}
\newcommand*{\numthree}[1]{\num[round-mode=places,round-precision=3]{#1}}
\newcommand*{\numfour}[1]{\num[round-mode=places,round-precision=4]{#1}}

\newcommand*{\hps}{\hPlusSin}
\newcommand*{\hpc}{\hPlusCos}
\newcommand*{\hts}{\hTimesSin}
\newcommand*{\htc}{\hTimesCos}

\newcommand{\nusl}{\nu_\mathrm{SL}}
\newcommand{\nuone}{\nu_\mathrm{1yr}}

\newcommand{\orcit}[1]{\protect\href{https://orcid.org/#1}{\protect\includegraphics[width=8pt]{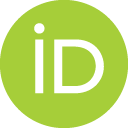}}}

\definecolor{LightGray}		{gray}{0.9}
\definecolor{Gray}		{gray}{0.5}
\definecolor{DarkGray}     	{gray}{0.2}
\definecolor{listinggray} 	{gray}{0.96}
\definecolor{DarkGreen}     	{rgb}{0.0,0.4,0.0}
\definecolor{DarkRed}     	{rgb}{0.6,0.0,0.0}
\definecolor{DarkBlue}     	{rgb}{0.0,0.0,0.6}
\definecolor{DarkCyan}     	{rgb}{0.7,0.7,0.2}
\definecolor{DarkDarkGreen}	{rgb}{0.0,0.4,0.0}

\begin{document}
   \title{Influence of a continuous plane gravitational wave on \gaia-like astrometry}

   \subtitle{}
\author{R.~Geyer \orcit{0000-0001-6967-8707}
          \inst{1}
          \and
	  S.A.~Klioner \orcit{0000-0003-4682-7831}
          \inst{1}
          \and
	  L.~Lindegren \orcit{0000-0002-5443-3026}
          \inst{2}
          \and
	  U.~Lammers \orcit{0000-0001-8309-3801}
          \inst{3}
          }
          
\authorrunning{R.~Geyer et al.}

   \institute{Lohrmann Observatory, Technische Universität Dresden,
              Mommsenstraße 13, 01062 Dresden, Germany\\
              \email{robin.geyer@tu-dresden.de}
              \and
              Lund Observatory, Division of Astrophysics, Department of
              Physics, Lund University, Box 43, 22100 Lund, Sweden
%              \email{lennart.lindegren@fysik.lu.se}
              \and
              European Space Agency (ESA), European Space Astronomy Centre
              (ESAC), Camino bajo del Castillo, s/n, Urbanización Villafranca del
              Castillo, Villanueva de la Cañada, 28692 Madrid, Spain
%              \email{uwe.lammers@esa.int}
             }

\date{Received / Accepted}

  \abstract
{A gravitational wave (GW) passing through an astrometric
     observer causes periodic shifts of the apparent star positions
     measured by the observer. For a GW of sufficient amplitude 
     and duration, and of suitable frequency, these shifts
     might be detected with a \gaia-like astrometric telescope.}
{This paper aims to analyse in detail the effects of GWs on an astrometric 
     solution based on \gaia-like observations, which are one-dimensional,
     strictly differential between two widely separated fields of view, and
     following a prescribed scanning law.}
{We present a simple geometric model for the astrometric 
     effects of a plane GW in terms of the time-dependent positional 
     shifts. Using this model, the general interaction between the 
     GW and a \gaia-like observation is discussed. Numerous
     \gaia-like astrometric solutions are made, taking as input 
     simulated observations that include the effects of a continuous 
     plain GW with constant parameters and periods ranging from
     $\sim\,$50~days to 100~years.  The resulting solutions are 
     analysed in terms of the systematic errors on astrometric and 
     attitude parameters, as well as the observational residuals.}
{It is found that a significant part of the GW signal is absorbed 
   	 by the astrometric parameters, leading to astrometric errors of 
     a magnitude (in radians) comparable to the strain parameters. 
     These astrometric errors are in general not possible to detect, 
     because the true (unperturbed) astrometric parameters are not
     known to corresponding accuracy. The astrometric errors are 
     especially large for specific GW frequencies that are linear 
     combinations of two characteristic frequencies of the scanning law.  
     Nevertheless, for all GW periods smaller than the time span covered 
     by the observations, significant parts of the GW signal also go into 
     the astrometric residuals. This fosters the hope for a GW
     detection algorithm based on the residuals of standard
     \gaia-like astrometric solutions.}
{}

   \maketitle
   
\newpage

\section{Introduction}
\label{sec_intro}

The fact that a gravitational wave (GW) causes astrometric effects has
been known and analysed for quite some time 
(\citeads{1990NCimB.105.1141B}; \citeads{1996ApJ...465..566P};
\citeads{1997ApJ...484..545K}; \citeads{1997ApJ...485...87G};
\citeads{1999PhRvD..59h4023K}; \citeads{2011PhRvD..83b4024B}).  Since
the high-accuracy astrometry of \gaia\ has become a reality
(e.g.\ \citeads{2023A&A...674A...1G}), there has been a
growing interest in investigating GW effects specifically for
these (or similar) astrometric observations (e.g.\
\citeads{2018CQGra..35d5005K}; \citeads{2018PhRvD..98b4020O};
\citeads{2018PhRvD..97l4058M}; \citeads{2017PhRvL.119z1102M};
\citeads{2018PhRvD..98l4036B}; \citeads{2018ApJ...861..113D};
\citeads{2024JCAP...05..030C}).

Indeed, \gaia\ astrometry, being not only of microarcsecond accuracy
but also global, offers interesting possibilities to investigate
GWs. Here it is, however, very important that the specific
observational principles of astrometric scanning satellites
(\citeads{2010EAS....45..109L}; \citeads{2016A&A...595A...1G}) 
are properly taken into account. One example is the fact that the 
instantaneous pointing (attitude) of the satellite must be
determined from its own astrometric observations, and that the
observations of a \gaia-like astrometric instrument are effectively
one-dimensional. Another aspect is the prescribed observational 
schedule known as the scanning law. Although 
\citetads{2018CQGra..35d5005K} take these specifics into account,
that is often not the case in other publications.

It is known from the literature cited above that a GW passing through an
astrometric observer causes time-dependent apparent shifts of the
positions of astrometric sources. The direction and magnitude of the
shift depends on time, the celestial position of the astrometric source, 
and the parameters of the GW (e.g.\ \citeads{2011PhRvD..83b4024B};
\citeads{2018CQGra..35d5005K}). The GW effect in astrometry is global
in the sense that it can be detected over a large fraction of the sky, if it
is at all detectable. This is so because the amplitude of the shift is proportional 
to $\sin\theta$, where $\theta$ is the angular distance between the 
source and the direction of propagation of the GW. Moreover, the effect
does not depend on the distance to the astrometric source, as long as it 
is much greater that the wavelength of the GW 
(e.g.\ \citeads{2011PhRvD..83b4024B}). These properties make global 
astrometry especially interesting for the study of possible GW signals.

This paper is the first in a series of publications discussing the
interaction of GW signals with astrometric observations made from a 
\gaia-like scanning telescope. In particular, this paper is devoted to a
detailed discussion of the effects that a single plane continuous GW 
has on a \gaia-like astrometric solution.

For several reasons the effects of GWs cannot be part of the standard
relativistic model for high-accuracy astrometry (see, e.g.,  
\citeads{2003AJ....125.1580K}; \citeyearads{2004PhRvD..69l4001K}
describing the model used for \gaia). The standard relativistic model
is intended to correct for all accurately known relativistic
effects like those coming from the gravitational field of the Solar
System.  However, no GW sources producing relevant
astrometric effects are currently known. On the other hand, without
good a~priori estimates of the GW parameters, it is in practice not
feasible to include the GW parameters as unknowns in the global 
astrometric solution, which is essentially a linear (or weakly non-linear)
least-squares estimation. Presently, the only practicable approach is 
therefore to apply a dedicated detection algorithm to the residuals of
a standard astrometric solution. This is the topic of a separate
publication in this series.

A detailed introduction to the standard astrometric solution used 
for \gaia\ can be found in \citetads{2012A&A...538A..78L}. Given 
that this solution is a simultaneous determination of the astrometric 
(source) parameters, the spacecraft attitude, and the calibration 
parameters, it is not a~priori clear to what extent an unmodelled 
GW signal in the data is absorbed by these solution unknowns, and 
therefore merely produces a biased astrometric solution; and to what 
extent it instead goes into the residuals, where it could be detected 
by a dedicated post-processing algorithm.

The goal of this first paper is precisely to give a detailed qualitative 
and quantitative description of the effects that an individual GW 
will have on a \gaia-like global astrometric solution and its residuals. 
We present both theoretical and numerical results to illustrate the effects. 
The investigation is limited to the case of a single plane continuous GW 
with constant parameters, but considering a wide range of the parameters.  

The structure of the paper is as follows.  Section~\ref{sec__gw_model}
presents the key parameters of the GW model and a brief overview 
of the astrometric effects of a GW. In Sect.~\ref{sec__general_influence} 
we discuss the mechanism of interaction between a GW signal and 
a \gaia-like astrometric solution.  Section~\ref{sec__methodology}
describes the numerical simulations used in this study. Simulation 
results are presented in Sect.~\ref{sec__source_errors} and
summarised in Sect.~\ref{sec__conclusion}.

Additional details are given in appendices. 
Appendix~\ref{section-GW-model} gives a simple geometric description of 
the apparent positional shifts from a GW, providing new insight into the nature 
of the astrometric effects.  
Appendix~\ref{sec__GW-signal} gives a statistical overview of the expected
astrometric GW signal in observational data.
Appendix~\ref{apx__freqRange} justifies our choice of GW parameters.
In Appendix~\ref{section-examples} we give examples of
the GW-induced errors of the \gaia-like astrometric solutions for some
given parameters of the GW. Finally, Appendix~ \ref{section-shvsh}
contains the results of the analysis of the GW-induced error patterns
in astrometric parameters, using expansions in scalar and vector
spherical harmonics.

Throughout the paper, the term `source' means the astronomical
object (typically a star or a quasar) subject to the apparent shifts from
GWs that are of interest here; only in a few places do we mean the source 
of the gravitational radiation itself, in which case we explicitly write 
`GW source'. Moreover, `error' is always used in its strict sense as the
difference between a computed (or perturbed) quantity and its true
value, never to mean the uncertainty of the quantity.
 
\section{Astrometric effects of a plane GW}
\label{sec__gw_model}

The theory of light deflection by a plane gravitational wave and its
influence on astrometric measurements has been worked out by
\citetads{1990NCimB.105.1141B}, \citetads{1996ApJ...465..566P},
and \citetads{1997ApJ...485...87G}, and further refined by
\citetads{2011PhRvD..83b4024B} and \citetads{2018CQGra..35d5005K}. The
latter publication formulates a model suitable for astrometric data reduction,
which is summarised and reformulated in an improved way in
Appendix~\ref{section-GW-model}. Here, only the most important
features of the model are recalled.

A plane GW is described by seven parameters. Three of the parameters are 
non-linear, namely the frequency $\nu$ of the GW and the angles 
$(\aGW, \dGW)$ specifying the direction in which the GW propagates.
The remaining four parameters -- the strain and phase parameters $\hPlusCos$,
$\hPlusSin$, $\hTimesCos$, and $\hTimesSin$ -- are linear and describe 
the magnitude and phase of the two polarisation components of the GW
in General Relativity. 
For astrometric measurements, the effect of the GW is a time-dependent 
shift of the apparent positions of sources over the whole sky. The magnitude 
and direction of the shift depends on the GW parameters as well as on time
and the position of the source. In this study we consider a continuous GW 
with constant parameters, modelled as described in 
Appendix~\ref{section-GW-model} and \ref{apx__freqRange}.

\begin{figure}[tb]
\centering
 \includegraphics[keepaspectratio,width=0.99\hsize]{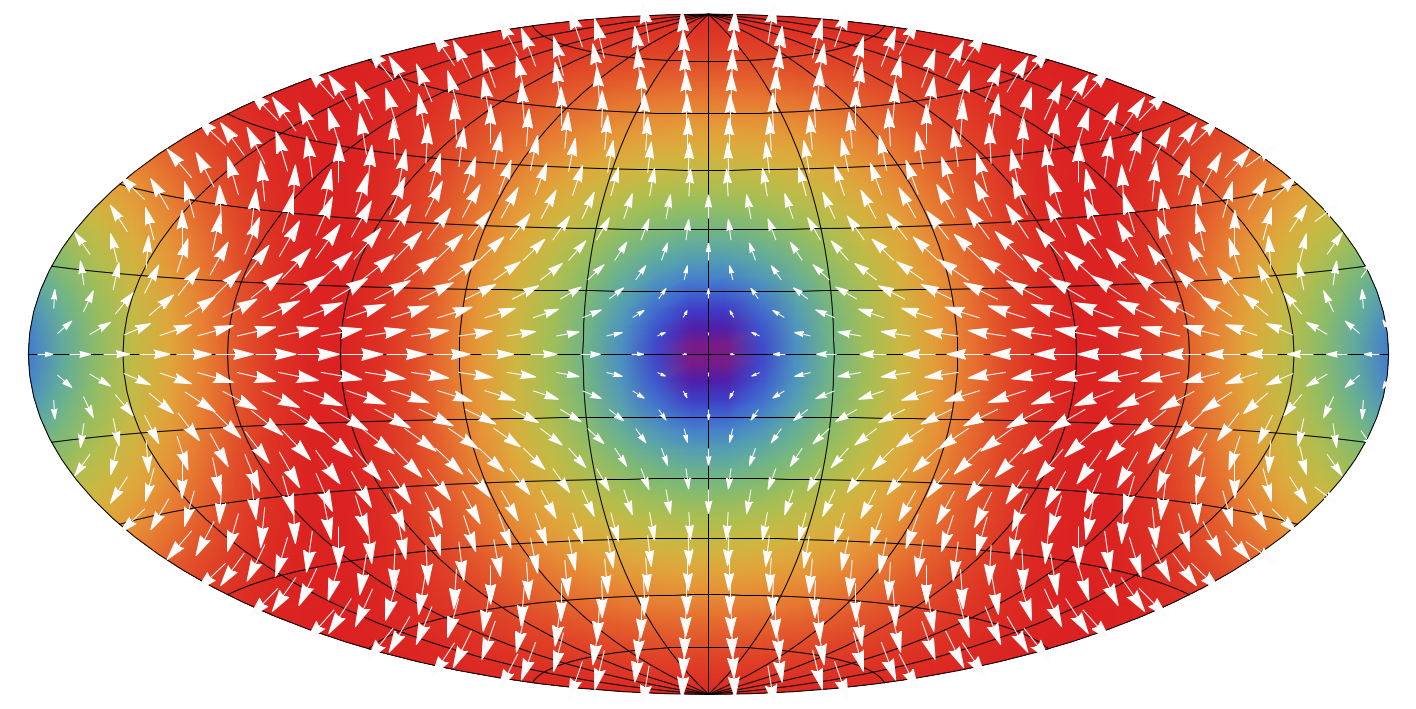}
 \hspace*{-0.65mm}\includegraphics[keepaspectratio,width=0.99\hsize]{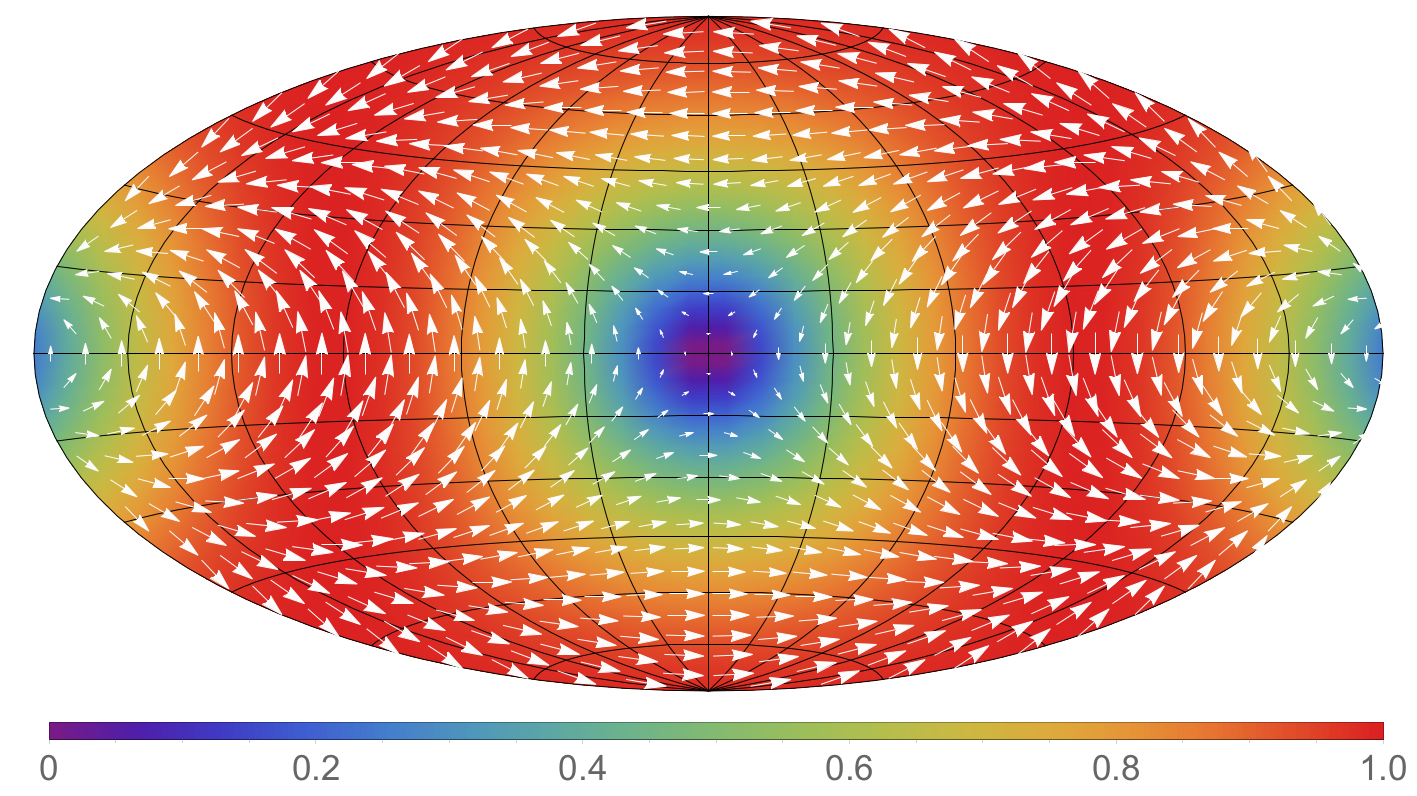}
 \vspace*{-1mm}
 \hspace*{4mm}$|\delta\vec{u}|/\Delta_{\rm max}$
 \smallskip
 \caption{Apparent shifts $\delta\vec{u}$ in source positions caused 
   by the two polarisation components of a plane GW propagating 
   towards the centre of the maps, i.e.\ the `$+$' component (\emph{top})
   and `$\times$' component (\textit{bottom}). The background colour 
   encodes the magnitude of the shift, $|\delta\vec{u}|$, in units of the 
   maximum shift $\Delta_{\rm max}$. This and all other full-sky maps 
   use the Hammer–Aitoff projection in equatorial coordinates, with 
   $\alpha=\delta=0$ at the centre, north up, and $\alpha$ increasing 
   from right to left.\label{fig__PGW_Vector_Field_Plots}}
\end{figure}

Figure~\ref{fig__PGW_Vector_Field_Plots} shows, for some fixed 
moment of time, the two vector fields $\delta\vec{u}_{+}$ and 
$\delta\vec{u}_{\times}$ describing the apparent positional 
shifts at different points on the sky, created by the two 
polarizations of the GW. Here, $\vec{u}$ is the (unperturbed)
source position given by Eq.~\eqref{vec-u}. The plots are all-sky maps in
Hammer--Aitoff projection, with white arrows representing the vector 
field at each location, and the background colour encoding  
the magnitude of the effect $|\delta\vec{u}|$. In
both plots the GW propagates towards the centre of the map. The
moment of time is chosen for maximum magnitude of the effect; 
this happens simultaneously over the whole sky, at which time
$|\delta\vec{u}|=\Delta$ as given by Eq.~\eqref{Delta-amplitude}.
In the plots one can clearly see the distribution of the overall
strength of the signal, as well as the differences between the two
polarisations.  In agreement with Eq.~\eqref{Delta-amplitude}, the
magnitude $\Delta$ of the shift is zero in the direction of propagation 
and in the opposite direction, where $\sin\theta=0$. Sources in 
a direction normal to the propagation ($\sin\theta=1$) attain  
the maximal shift $\Delta_{\rm max}$ given by 
Eq.~\eqref{Delta-amplitude-max}.

\section{Imprint of the GW signal on a \gaia-like astrometric solution}
\label{sec__general_influence}

We shall now discuss how the astrometric effects of a GW described in the previous 
section might be seen in a \gaia-like astrometric solution. For this it is necessary 
to consider the specific way observations are made with a \gaia-like instrument, 
as well as their interplay with the model parameters in the astrometric solution.
These interactions are in general quite complicated, and a full characterisation 
can only be obtained by means of numerical simulations, as described in the 
next sections. Under simplifying assumptions, theoretical predictions can
nevertheless be derived for some of the interactions, which is the subject of 
this section.

\subsection{Nature of \gaia-like observations}
\label{sec___gaialike_observations}

We consider astrometric observations from a scanning space observatory
based on the principles of the ESA missions \textsc{Hipparcos} and \gaia,
using simultaneous observations of sources in two fields of view
(FoVs) separated by a large angle $\Gamma$ (basic angle).
The nominal basic angle is $106.5^\circ$ for \gaia\ and $58^\circ$ for
\textsc{Hipparcos}.  The instrument rotates in space according to a
prescribed scanning law, with a rotation period of hours.

The two FoVs are called preceding and following according to the
direction of the instrument's rotation. The position of a source in
each FoV at a certain time can be specified by two field angles measured
from the centre of the field: the along-scan (AL) angle $g$, and the 
across-scan (AC) angle $h$. We denote the field angles in the preceding 
FoV as $g_{\rm p}$, $h_{\rm p}$, and in the following FoV as 
$g_{\rm f}$, $h_{\rm f}$ (cf.\ Fig.~1 and Sects.~2.1 and 2.2 of 
\citeads{2017A&A...603A..45B}). 

The size of each FoV is of the order of 1~deg, so at the times 
of observation the field angles are always $\lesssim 10^{-2}$~rad 
in absolute value. To first order one can then neglect the finite size 
of the FoV and all the details of what happens when the source transits
the FoV, effectively regarding the observations as instantaneous
(cf.\ Sect.~4.4 of \citeads{2017A&A...603A..45B}).
Any time-dependent variation (`signal') in the positions of sources 
on the sky then translates into corresponding changes 
$(\delta g_{\rm p},\,\delta h_{\rm p})$ or 
$(\delta g_{\rm f},\,\delta h_{\rm f})$ of their measured field angles 
at the respective times of observation. Neglecting the geometrical
calibration of the detectors within each FoV, the accurately measured 
AL and AC field angles constitute the elementary astrometric observations
of a \gaia-like instrument.

\subsection{Global astrometric solution}
\label{sec___agis}

Given the basic angle and the accurate attitude of the instrument at the 
moment of observation, the observed field angles can be transformed 
into an instantaneous position of the source on the sky. The astrometric 
parameters of the source can in principle be determined from a sequence 
of such positions. However, the accuracy with which the actual attitude 
and calibration of the instrument is known a~priori is not sufficient for the 
astrometric accuracy to be achieved. Therefore, the attitude and calibration 
parameters must also be determined from the astrometric observations 
(self-calibration). 
Consistency among the various source, attitude, and calibration 
parameters is ensured by solving all of them together in a global 
least-squares solution, as described for \gaia\ in 
\citetads{2012A&A...538A..78L}. The residuals of the solution
(the difference between the observed and fitted field angles)
may then contain imprints of any unmodelled signal in the data, 
such as the astrometric effects of the GW. 

Naively, one might think that the GW effect would remain 
practically unchanged in the residuals of the astrometric solution. 
However, as discussed in the literature (\citealp{Klioner2014};
\citeads{2017A&A...603A..45B}; \citeads{2018CQGra..35d5005K}), this is
not the case.  The two main reasons are (i) that part of any
unmodelled signal is absorbed by the attitude parameters, and (ii)
that another part of the signal may be absorbed by the source 
parameters. A significant part of the GW signal does however remain 
in the residuals, provided that the period of gravitational wave $\Pgw$ 
is not considerably larger than the time span $T$ of the astrometric data 
(see also Sect.~2 of \citeads{2018CQGra..35d5005K}).

Part of the GW signal could also be absorbed by the calibration parameters, 
which in turn could modify the source and attitude parameters in the 
global solution. This interaction cannot easily be described in 
general terms, because the calibration model tends to be very specific 
to a given instrument and tailored to its behaviour. One exception is 
the calibration of a possible variation of the basic angle, which is 
briefly addressed below.

\subsection{Interaction with attitude and basic angle determination}
\label{sec___attitude_absorption}

Consider an arbitrary perturbation of the observed stellar positions,
that is a smooth function $\delta\vec{u}(t,\alpha,\delta)$ of both time 
and position on the sky. Examples of such perturbations are: 
(i) a global shift of all parallaxes; 
(ii) a large-scale distortion of the positions and proper motions; 
(iii) the astrometric effects of a low-frequency GW.

Generalising the pioneering work of \citet{lindegren77},
\citet{Klioner2014} and \citetads{2017A&A...603A..45B} described an
important problem here. Thanks to the scanning law, which is also a 
smooth function of time, any such perturbation will produce smooth, 
time-dependent variations $\delta g_{\rm p}$, $\delta h_{\rm p}$, 
$\delta g_{\rm f}$, $\delta h_{\rm f}$ of the observed field angles. 
These variations are observationally indistinguishable from a certain 
time-dependent variation of the attitude parameters and of the basic 
angle. Specifically, Eq.~(7) of \citetads{2017A&A...603A..45B} shows 
that the AC components of a signal, described by 
$\delta h_{\rm f}$ and $\delta h_{\rm p}$, as well as the average 
of the AL components, $(\delta g_{\rm f}+\delta g_{\rm p})/2$, 
are equivalent to a certain change of the attitude. 
Because the attitude is determined as a free function of time, with a 
time resolution (seconds) much higher than the rotation period (hours),
these components of $\delta\vec{u}$ are completely absorbed by the 
attitude parameters. The remaining AL part of the signal is effectively 
equivalent to a variation of the basic angle, 
$\delta\Gamma=\delta g_{\rm f}-\delta g_{\rm p}$. This signal 
can influence the astrometric source parameters, be partially absorbed
by the basic-angle calibration, or remain in the residuals of the 
astrometric solution, depending on the calibration model used 
in the astrometric solution. As mentioned before, the interaction
with the calibration model cannot be discussed in general terms 
and will be ignored in this paper.

This general mechanism applies in particular to the case of the
astrometric signal produced by a GW. We therefore conclude that the AC
components of the GW signal and the mean of the AL components in the
two FoVs are absorbed, to first order, by the attitude as determined
in the astrometric solution (see \citeads{2012A&A...538A..78L} for 
the standard attitude model of \gaia.) Only the differential AL effect 
$\delta g_{\rm f}-\delta g_{\rm p}$ carries information about the
GW signal, but even part of that may be absorbed by the source
parameters and/or the model of basic-angle variations.

These theoretical expectations have been confirmed in dedicated numerical
simulations (similar to those described in Sect.~\ref{sec__methodology}), 
in which only certain
parts of a GW signal were used to compute the simulated data: only the
AC part, only the AL part, only the mean AL signal in both FoVs, etc.

\subsection{Interaction with astrometric source parameters}
\label{sec__subsec_interaction_with_sourceparams}

In the standard astrometric model of stellar motion (e.g.\ Sect.~3.2 of
\citeads{2012A&A...538A..78L}), the motions of sources on the sky are
described by five free parameters per source: two components of the 
position ($\alpha$, $\delta$), two components of the proper motion 
($\mu_{\alpha*}$, $\mu_{\delta}$), and a parallax ($\varpi$). 
As explained above, the component of the GW signal that is not absorbed 
by the attitude parameters will partly alter the source parameters and 
partly remain in the residuals of the solution. In full generality, this 
interaction will be investigated below by means of the numerical 
simulations described in Sect.~\ref{sec__methodology}.

However, a simple model can elucidate the effect on the source
parameters at least in the case of a GW with period $\Pgw$
sufficiently longer than the time span $\Tobs$ covered by observations. 
In this case we assume that the effect on the parallax is
negligible. A change of parallax would cause a periodic shift of the
apparent position on the sky with a period equal to the orbital period
of the observer (satellite) in its motion around the barycentre of the
Solar System. For \gaia\ this period is about one year. From
Appendix~\ref{section-GW-model} one can see that a GW signal causes
elliptic variations of the apparent positions with a period equal to
that of the gravitational wave $\Pgw$. Therefore, if $\Pgw \gg T \gg
1$ year, it is a plausible assumption that the effect of such a GW on
parallax is negligible. This is explicitly justified by the 
simulations below.

Therefore, we only consider variations of positions
$\Delta\alpha^*=(\,\widetilde\alpha-\alpha)\cos\delta$ and
$\Delta\delta=\widetilde\delta-\delta$, and proper motions
$\Delta\mu_{\alpha*}=\widetilde\mu_{\alpha*}-\mu_{\alpha*}$ and
$\Delta\mu_{\delta}=\widetilde\mu_{\delta}-\mu_{\delta}$. Here, 
$\widetilde\alpha$, $\widetilde\delta$, $\widetilde\mu_{\alpha*}$, 
and $\widetilde\mu_{\delta}$ are the parameters computed
under the influences of a GW signal, while $\alpha$, $\delta$,
$\mu_{\alpha*}$, and $\mu_{\delta}$ are the true ones or those
computed with no GW signal in the observations.  Here and below, the
asterisk in $\alpha^*$ indicates that the $\cos\delta$ factor is
implicit, as in $\mu_{\alpha*}=\dot\alpha\,\cos\delta$.

\begin{figure}[htb]
 \centering
\includegraphics[keepaspectratio,width=0.98\hsize]{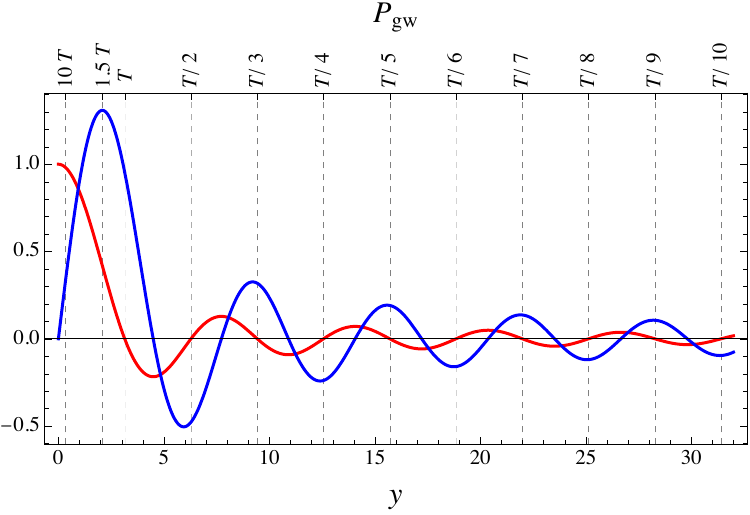}
\caption{The averaging functions $f(y)$ and $g(y)$ shown in red and
  blue, respectively. The equivalent GW period $\Pgw=\pi\Tobs/y$ is given
  on the top axis in terms of $\Tobs$, the time span of the observations.
  For fixed $\Tobs=5$~yr, $y$ is equivalent to the GW frequency 
  $\nu = y/(\pi\Tobs)\simeq y\times 2.017$~nHz.
  \label{figure-fg}}
\end{figure}

We further assume that there is an infinite number of observations
uniformly distributed over the time interval of observations
$t-t_0\in[-T/2,T/2]$, where $t_0$ is the middle of the observational
period and also the reference epoch of the catalogue (that is, the source
positions are defined at $t_0$). Now, we fit the GW effect 
$\delta\alpha^*$, $\delta\delta$ as given by Eq.~\eqref{delta=D.b} 
in Appendix~\ref{section-GW-model} by the linear functions
$\Delta\alpha^* + \Delta\mu_{\alpha*}\,(t-t_0)$ and $\Delta\delta +
\Delta\mu_{\delta}\,(t-t_0)$. The small R\o{}mer correction 
$-c^{-1}\,\vec{p}\cdot\vec{x}_{\rm obs}(t)$ in Eq.~\eqref{Phase}
is ignored, so that the meaning of the strain and phase parameters 
$\hPlusCos$, $\hPlusSin$, $\hTimesCos$, and $\hTimesSin$
is defined by the reference epoch $t_{\rm ref}$ in Eq.~\eqref{Phase}.
Without loss of generality we set $t_{\rm ref} = t_0$.
We also neglect that a part of the GW signal is absorbed by the
attitude parameters. Considering all the assumptions, standard
linear regression then gives
\begin{eqnarray}
\label{lin-pos}
  \begin{pmatrix}
    \Delta\alpha^*\\
    \Delta\delta
  \end{pmatrix}
  &=&{1\over 2}\sin\theta\ \vec{R}\,\vec{h}_{\rm c}\,f(y)\,,\\
\label{lin-pm}
  \begin{pmatrix}
    \Delta\mu_{\alpha*}\\
    \Delta\mu_\delta
  \end{pmatrix}
  &=&{1\over T}\,\sin\theta\ \vec{R}\,\vec{h}_{\rm s}\,g(y)\,,
\end{eqnarray}
where
\begin{eqnarray}
\label{hc}
  \vec{h}_{\rm c}&=&
    \begin{pmatrix}
      \hPlusCos\\
      \hTimesCos
    \end{pmatrix}\,,\\
\label{hs}    
\vec{h}_{\rm s}&=&
    \begin{pmatrix}
      \hPlusSin\\
      \hTimesSin
    \end{pmatrix}\,,  
\end{eqnarray}
and
\begin{eqnarray}
\label{lin-f}
    f(y)&=&{\sin y\over y}=1-{1\over 6}\,y^2+{\cal O}(\,y^4\,),
    \\
\label{lin-g}
    g(y)&=&-3{df\over dy}={3\over y^2}\,\left(\sin y-y\,\cos y\right)=y-{1\over 10}\,y^3+{\cal O}(\,y^4\,) 
\end{eqnarray} 
are the the functions illustrated in Fig.~\ref{figure-fg} with argument
\begin{equation}
\label{lin-y}
    y=\pi\nu T=\pi T/\Pgw\,.
\end{equation}
Here, $\theta$ is the angular distance between the direction of
observation and the direction of propagation of the GW, given by
Eq.~\eqref{angular-distance-theta}, and $\vec{R}$ is the rotation
matrix given by Eq.~\eqref{matrix-R}. 

The residuals of this linear fit,
\begin{equation}
\label{residuals-definition}
\vec{r}=
\begin{pmatrix}
 \delta\alpha^*\\
 \delta\delta
\end{pmatrix}
-
\begin{pmatrix}
 \Delta\alpha^*\\
 \Delta\delta
\end{pmatrix}
-  
\begin{pmatrix}
 \Delta\mu_{\alpha*}\\
 \Delta\mu_\delta
\end{pmatrix}\,(t-t_0)\,,
\end{equation}
have zero mean. One can derive also the following result for the standard deviations 
of the residuals in AL and AC directions 
($\sigma_{r,{\rm AL}}$ and $\sigma_{r,{\rm AC}}$):
\begin{multline}
\label{sigma-r}
\sigma^2_{r,{\rm AL}}+\sigma^2_{r,{\rm AC}}={1\over 4}\sin^2\theta\\
  \times\,\left({1\over 2}\,h^2+{1\over 2}\,(h_{\rm c}^2-h_{\rm s}^2)\,f(2y)-h_{\rm c}^2 f^2(y)-{1\over 3}\,h_{\rm s}^2\,g^2(y)\right)\,,
\end{multline}
where $h_{\rm c}=|\vec{h}_{\rm c}|$, $h_{\rm s}=|\vec{h}_{\rm s}|$, and $h^2=h^2_{\rm c}+h^2_{\rm s}$ 
in agreement with Eq.~(\ref{h}).

It is seen that the effective change in position
only depends on the cosine parameters $\hPlusCos$ and $\hTimesCos$,
while the proper motion is only affected by the sine parameters
$\hPlusSin$ and $\hTimesSin$.  This is easily understood from the
parity properties of the sine and cosine terms with respect to the
reference epoch $t_0 = t_{\rm ref}$. The functions $f(y)$ and $g(y)$ 
depicted in Fig.~\ref{figure-fg} represent the averaging over the 
observations: a part of the elliptic motion described in 
Appendix~\ref{gw-model-ellipses} is approximated by a straight 
line corresponding to a constant proper motion. The argument 
$y$ is proportional to the ratio $\Tobs/\Pgw$ of the time span of 
the observations to the GW period. In the limit when 
$\Pgw\gg\Tobs$ (small $y$) one has $f(y)\simeq 1$ and $g(y)\simeq y$, 
in which case, for a given set of strain parameters, the GW has maximal 
effect on the position, while the proper motion effect is proportional to 
the GW frequency $\nu =1/\Pgw=y/(\pi\Tobs)$. The function $g(y)$ 
reaches a maximum value of $1.3086$ for $y\simeq 2.0816$, which 
corresponds to $\Pgw\simeq 1.5092\,T$. For a given amplitude of
the strain parameters, the proper motion is therefore maximally 
sensitive to GW periods of about 1.5 times the observation interval.

The linear regression \eqref{lin-pos}--\eqref{lin-y} can be reasonably
applied for $\Pgw\gtrsim T$, provided that $T$ is considerably greater
than the orbital period of the observer with respect to the barycentre
of the Solar System. The latter orbital period drives the parallactic
signal in the astrometric data. In this regime one can expect that
virtually the whole GW signal goes into the corresponding changes of
the positions and proper motion so that both parallax and the
residuals are only minimally affected. On the other hand, for $y\gg1$,
one sees that the averaging functions $f(y)$ and $g(y)$ tend to zero
so that under the assumptions of this model, the effects of a GW of a
higher frequency on positions and proper motions gradually
diminish. From the numerical simulations presented below, we find
that significant changes in the positions and proper motions occur 
also for higher GW frequencies. This can be attributed to various
complications neglected in the present analysis: the actual scanning 
law, the finite number of observations, the partial absorption of the 
GW signal by attitude parameters, and the interaction with parallax. 
For $\Pgw\gg\Tobs$, on the other hand, Fig.~\ref{fig_simulated20yr} 
shows that Eqs.~\eqref{lin-pos}--\eqref{lin-y} accurately describe 
the simulated variations of the astrometric parameters.
 
\section{Numerical simulations}
\label{sec__methodology}

The goal of the work presented here and in the next section is to show 
typical characteristics of the influence of a GW on a \gaia-like astrometric
solution for various GW parameters. To assess these effects we use
simulated data and compute the corresponding astrometric
solution with realistic modelling of sources and attitude. 

We use the AGISLab software \citepads{2012A&A...543A..15H} for
the simulations. This software offers various options for
simulating observations, computing the astrometric solution and
exporting all kinds of parameters as well as the residuals of the
solution.  In particular, the software allows one to add arbitrary
signals to the model used when generating the simulated observations,
and to omit or include random noise (photon noise) in the observations.

Using AGISLab, we simulate \gaia-like observations including a signal
of a GW with some specific parameters, but without random observational noise. 
These observations are processed inside AGISLab, using standard models 
for the sources and the satellite attitude as described e.g.\ in
\citetads{2012A&A...538A..78L}. Because the standard models do not 
take into account any GW effects, the resulting astrometric solution 
directly shows the effect of the GW on the source parameters, spacecraft 
attitude, and solution residuals. (Without the GW signal, such a solution
gives zero residuals and reproduces the assumed model parameters to
within the numerical precision of the computations, $\sim 10^{-3}~\mu$as.) 

It is clear that the number of simulations required to exhaustively
explore the seven-dimensional parameter space of a plane continuous GW
is prohibitively large. (For example, to cover possible GW source
directions, with $\simeq 10^\circ$ spacing, requires about
four hundred simulations for just a single GW frequency and a specific set
of strain parameters.) Instead, we select GW parameters that are
representative for a wide region of the parameter space to
demonstrate the most important effects, both qualitatively and
quantitatively.

The simulations cover GW periods ranging from $\Pgw=30$\,yr
($\nu_{\min}\simeq 1.05627$\,nHz) to approximately 50.2\,d with a
fixed spacing in frequency of 1.5\,nHz. This gives 154 frequency
values with $\nu_{\max}=\nu_{\min}+153\times 1.5\,{\rm nHz}\simeq
230.55627$\,nHz.  Additionally, we incorporate five GW frequencies
with periods between 100\,yr and 5\,yr to achieve a denser coverage in
the low-frequency range, and the 11 special frequencies affecting
positions and/or parallax listed Table~\ref{tab__list_peaks} that are
combinations of the fundamental frequencies of the scanning law (see
Eq.~\ref{eq___peak_freqs}).  Consequently, in total 170 GW frequencies
are used.  The rationale behind this choice of frequency interval is
given in Appendix~\ref{apx__freqRange}.

For each frequency, five different sets of strain parameters are
chosen as follows:
\begin{enumerate}
\item We set all four parameters to the same value. This gives
  GWs with eccentricity $e=1$ (see Eq.~\ref{eccentricity-e}).
\item We set $\hPlusSin = \hTimesCos \neq 0$ and $\hPlusCos =
  \hTimesSin = 0$ to get GWs with $e = 0$.
\item We randomly select each strain parameter from the same uniform
  distribution. The resulting eccentricity $e$ can have any value from 
  0 to 1, but higher values are considerably more probable.
\item We randomly choose the strain parameters in such a way that
  the eccentricity is approximately uniformly distributed from 0 to
  1. This was done numerically, by first generating a very large 
  number of combinations of random amplitudes, binning them by 
  eccentricity, and then randomly selecting first an eccentricity bin,
  and then one of the amplitude combinations within that bin.
\item The fifth set was generated exactly as the fourth, only using
a different random seed. This set was added to reduce sampling
noise, after it was found that the uniform distribution of eccentricity
gives the largest variation in attitude errors and residual statistics.
\end{enumerate}
These recipes define the strain parameters up to some normalising
factor. Because the astrometric GW signal is linear in the strain 
parameters, the choice of normalisation factor is in principle arbitrary,
as long as the resulting astrometric effects are considerably greater 
than the numerical noise in the astrometric solutions. We choose the 
normalisation that gives $\Delta_{\rm max}=1$\,mas in every simulation. 
Although this implies unrealistically large values for the strain, it does 
not matter thanks to the linearity of the effect and because all numerical 
results are reported relative to $\Delta_{\rm max}$.

For each set of frequency and strain parameters we randomly select the
GW propagation direction from a uniform distribution on the sky. This
yields $170 \times 5 = 850$ independent simulations that form the 
basis for the following investigations.

All simulations used (1) one million sources randomly distributed on
the sky; (2) no observational noise; (3) the nominal \gaia\ instrument
configuration; (4) the nominal \gaia\ scanning law; and (5) a mission
duration of $\Tobs=5$\,yr. The reference epoch $t_{\rm ref}$ for the
GW parameters (as in Eq.~\ref{Phase}) and that for the source
parameters $t_0$ were both chosen to be exactly in the middle of the
simulated mission. The number of sources was selected in order to
ensure a stable astrometric iterative solutions on the one side, and a
sufficiently fine sampling of the astrometric GW effect in the
solution on the other. We stress that the statistical characteristics
of the GW effect in the solution as discussed in
Sect.~\ref{sec__source_errors} are independent of the number of
sources in the simulations.

In addition to these simulations, we made some additional simulations for
specific GW parameters. These are used, for instance, to illustrate
the distribution of astrometric errors on the sky 
(Fig.~\ref{fig___error_sky_maps}), to give additional statistics 
for selected frequencies (Table~\ref{tab_statstable2}), and to discuss 
the effects of the GW phase for a very long period 
(Table~\ref{tab___longperiod_effects_by_phase}). 
The technical setup for these simulations was the same as above,
except for the specific GW parameters described in each case.

\begin{figure}[htb]
	\centering
    \includegraphics[width=0.95\hsize,keepaspectratio]{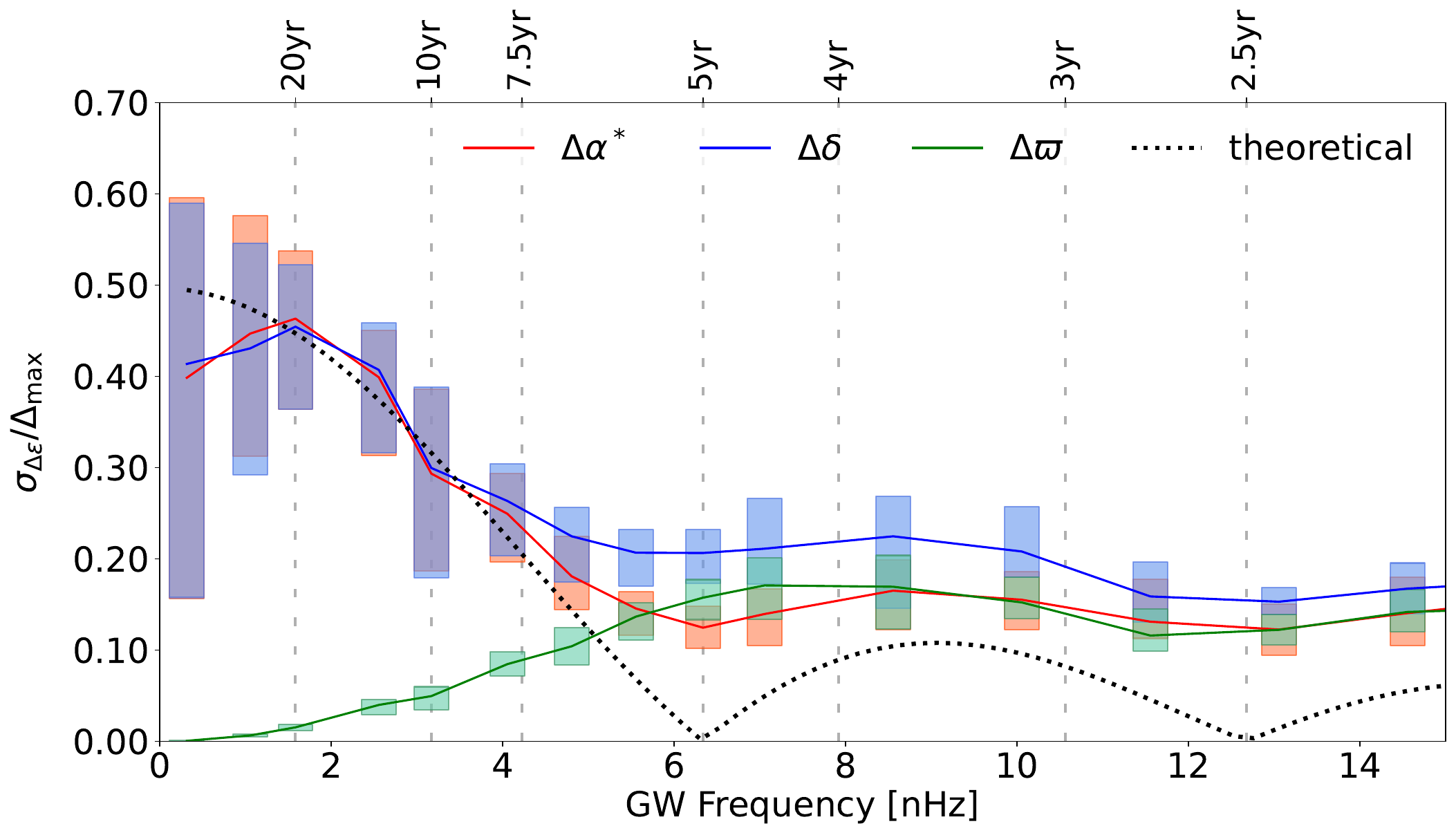}
    \includegraphics[width=0.95\hsize,keepaspectratio]{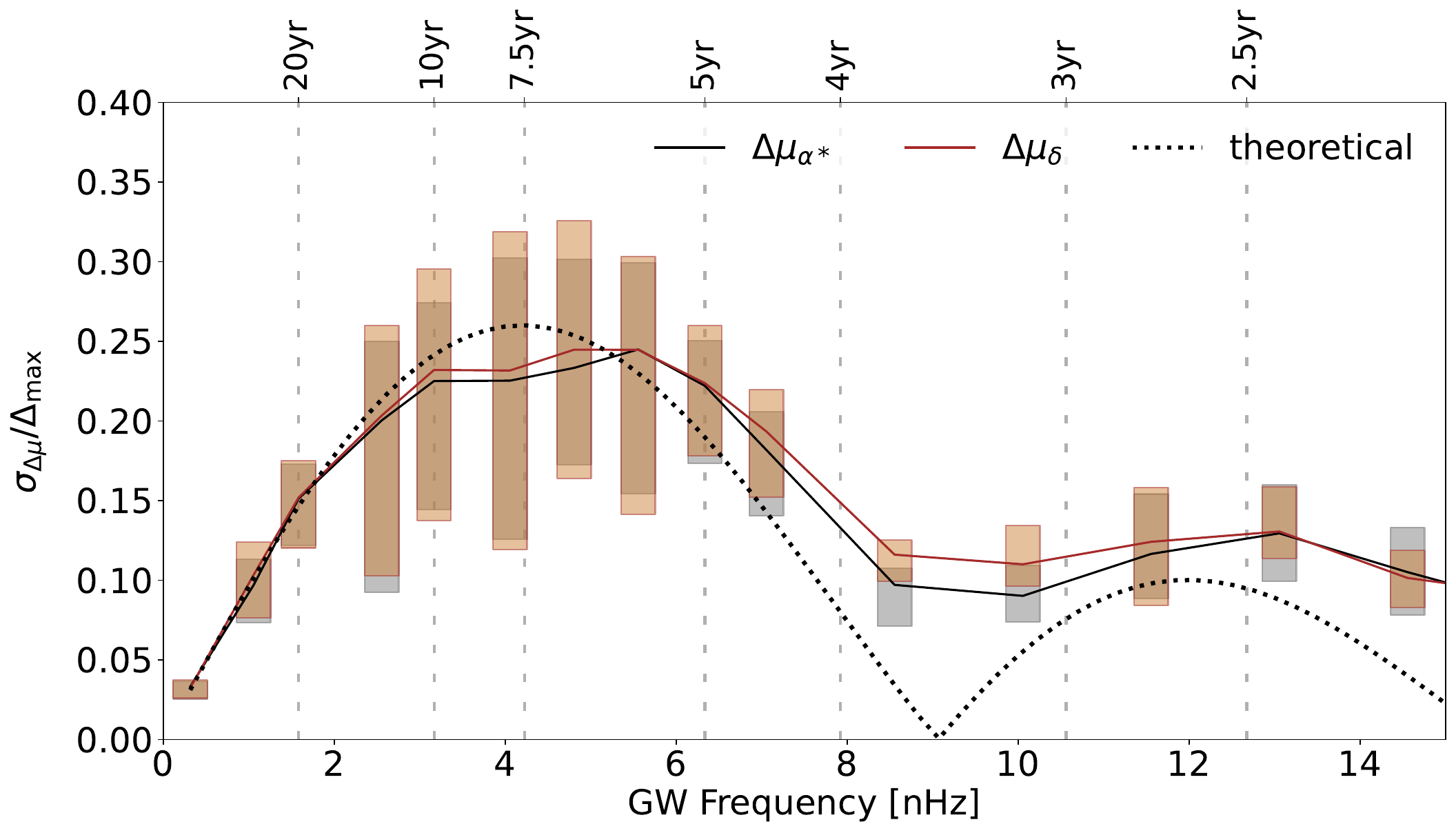}	
    \includegraphics[width=0.95\hsize,keepaspectratio]{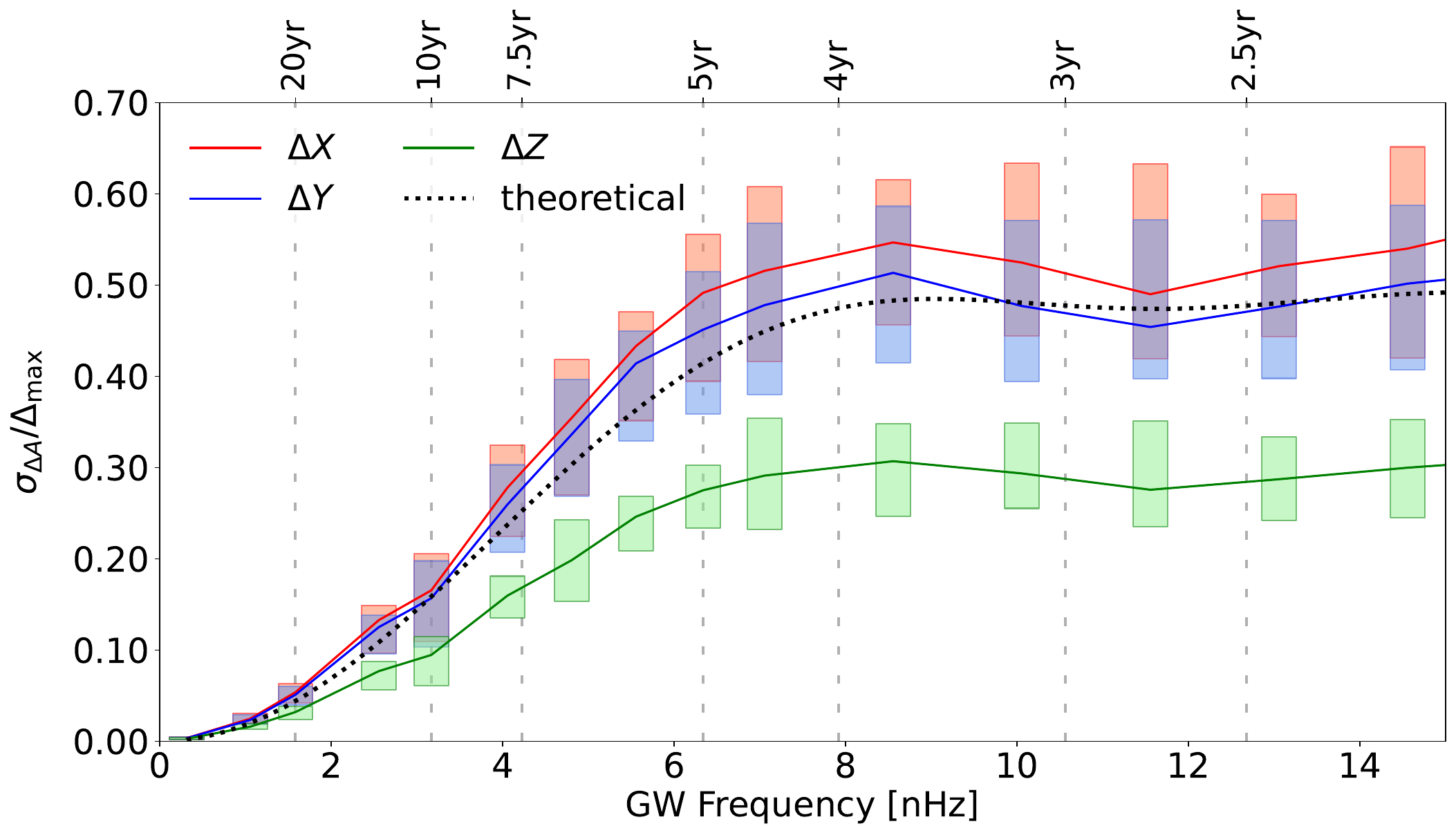}
	\includegraphics[width=0.95\hsize,keepaspectratio]{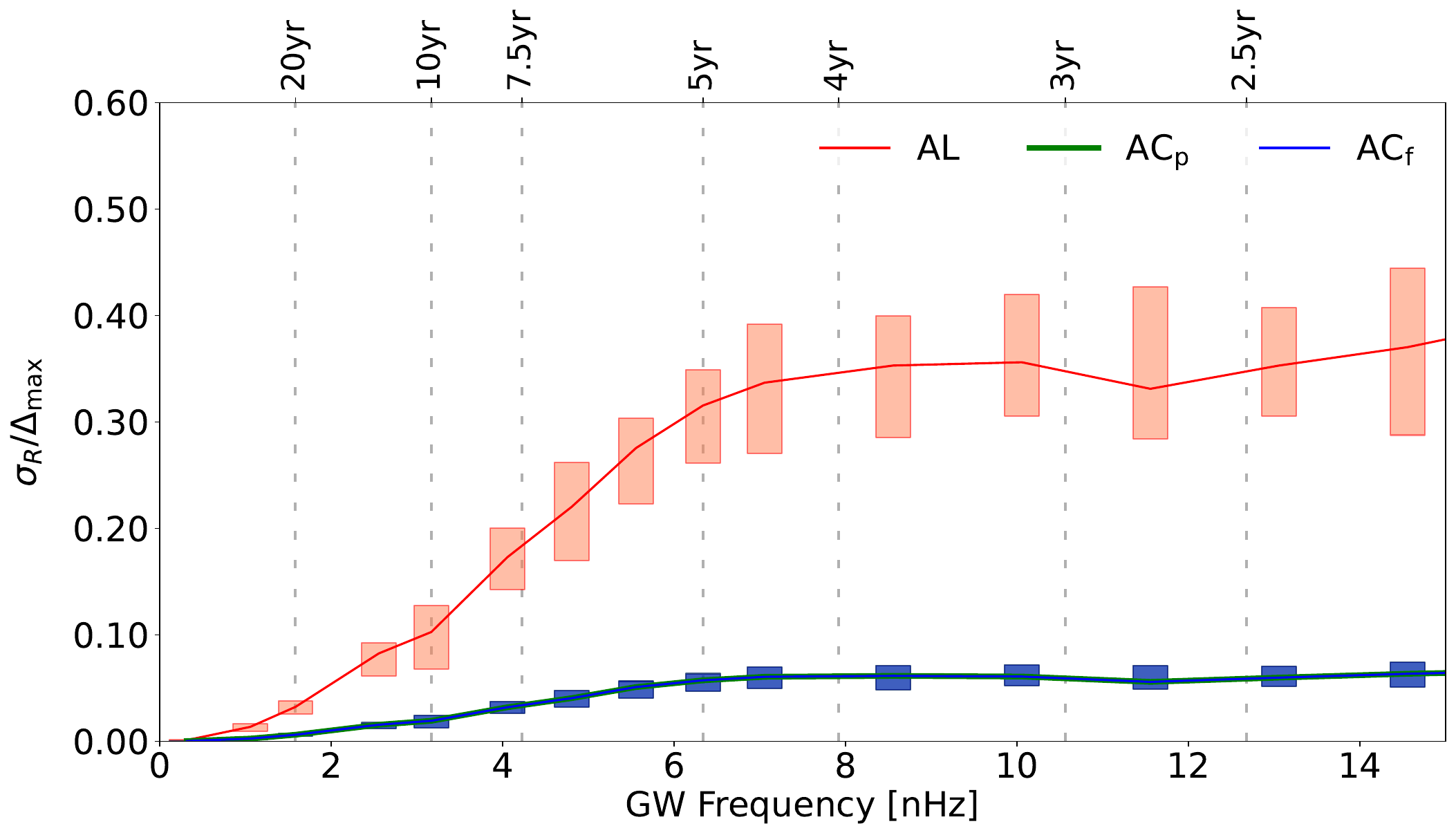}
	\caption{Standard deviations of the errors in source parameters, attitude, 
	      and residuals, for GW periods longer than
          2.1\,yr.  These plots are zoomed versions of the ones in
          Figs.~\ref{fig___freq_overview-source}--\ref{fig___freq_overview-att-res},
          but with a few more points added to improve the coverage at the lowest
          frequencies. Theoretical curves shown on the three upper plots are given by
          Eqs.~(\ref{normalized-sigma-pos}), (\ref{normalized-sigma-pm}), and (\ref{normalized-sigma-res}),
          respectively (see text for further explanations).
          \label{fig___freq__overviewslow}}
\end{figure}

\makeatletter
\ifaa@referee
\def\scalethree{0.8}
\else
\def\scalethree{0.78}
\fi
\makeatother

\begin{figure}[htb]
  \centerline{\includegraphics[keepaspectratio,width=\scalethree\hsize]{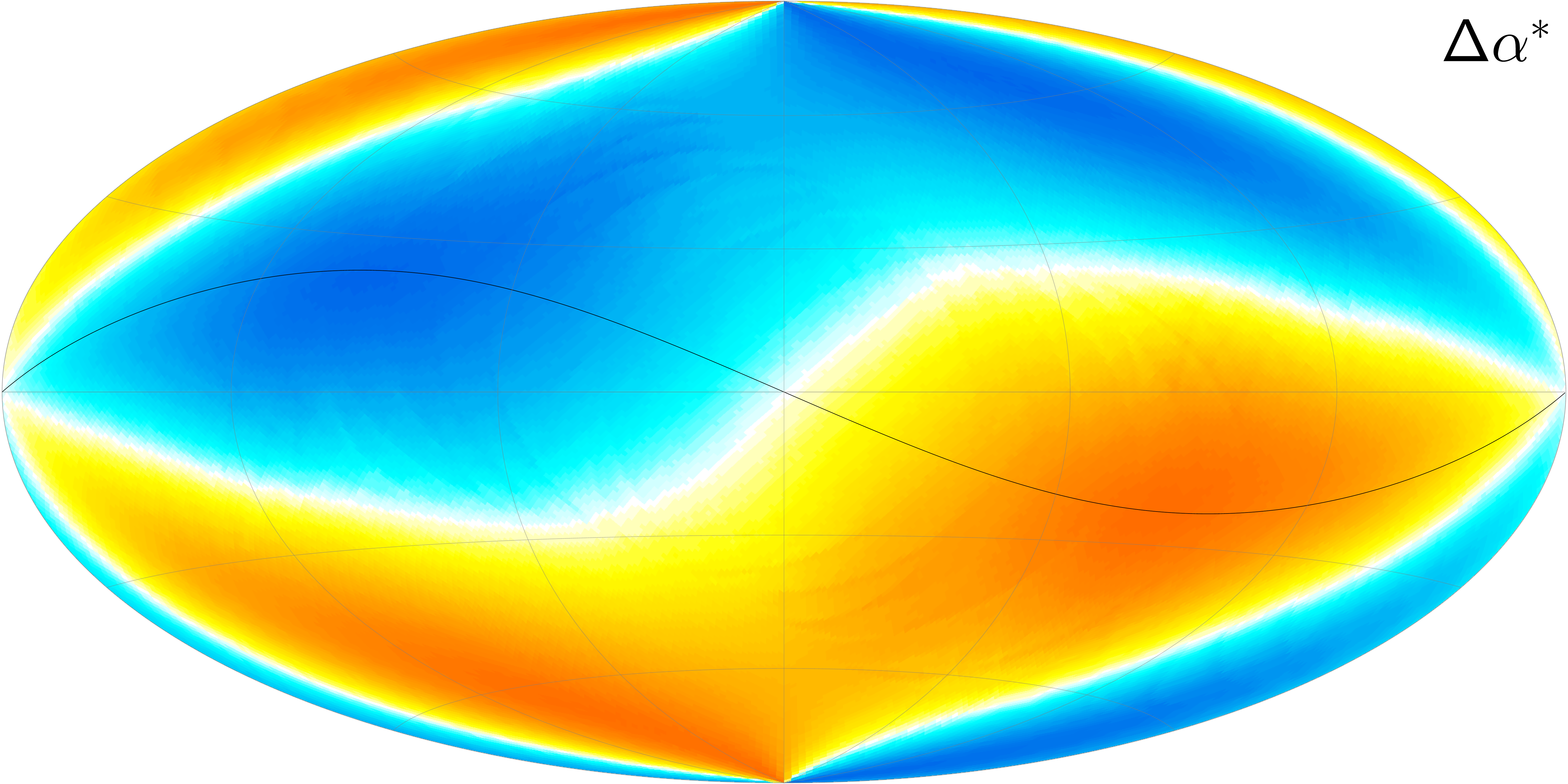}}
\vspace*{1mm}
\centerline{\includegraphics[keepaspectratio,width=\scalethree\hsize]{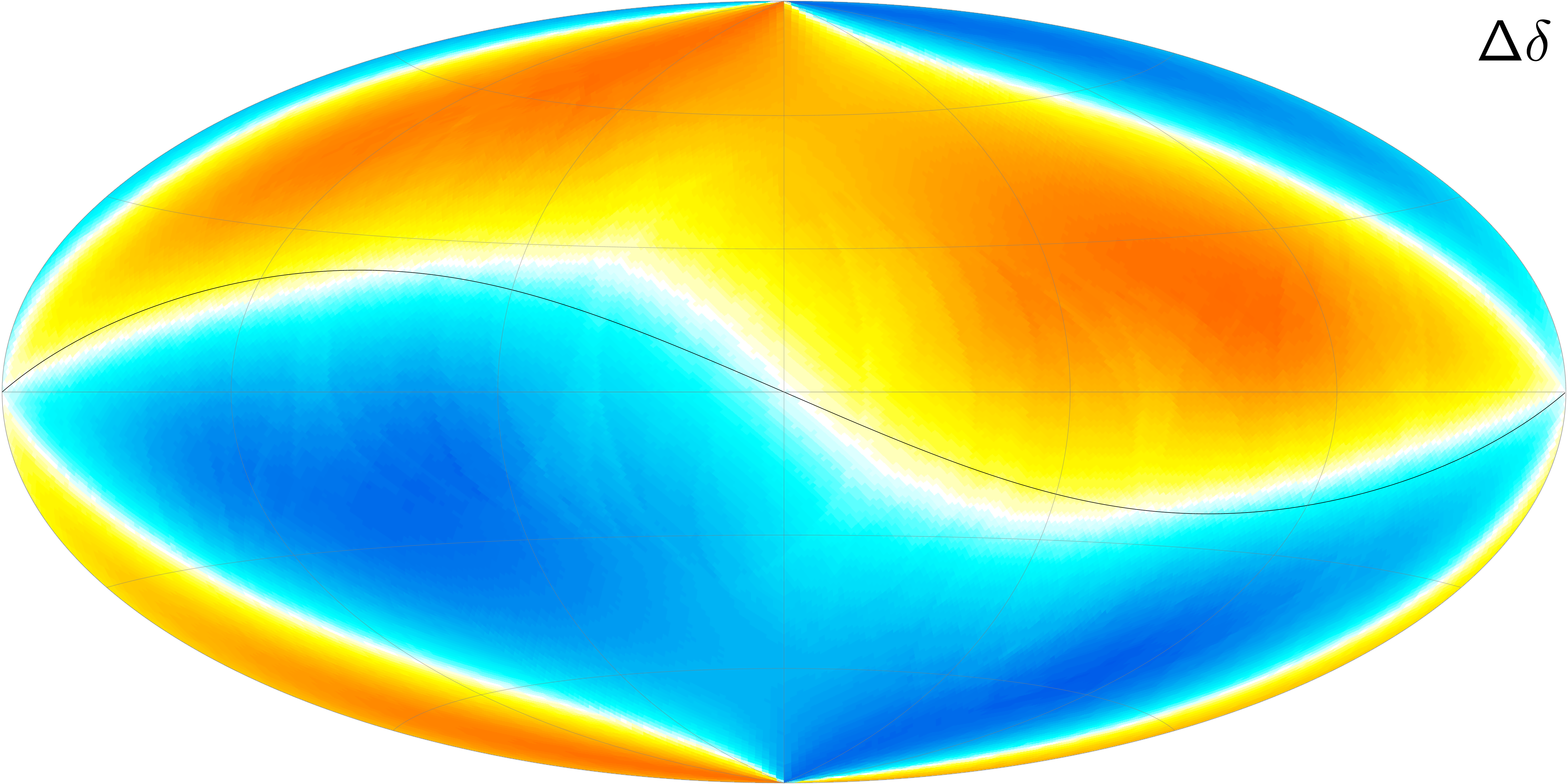}}
\vspace*{1mm}
\centerline{\includegraphics[keepaspectratio,width=\scalethree\hsize]{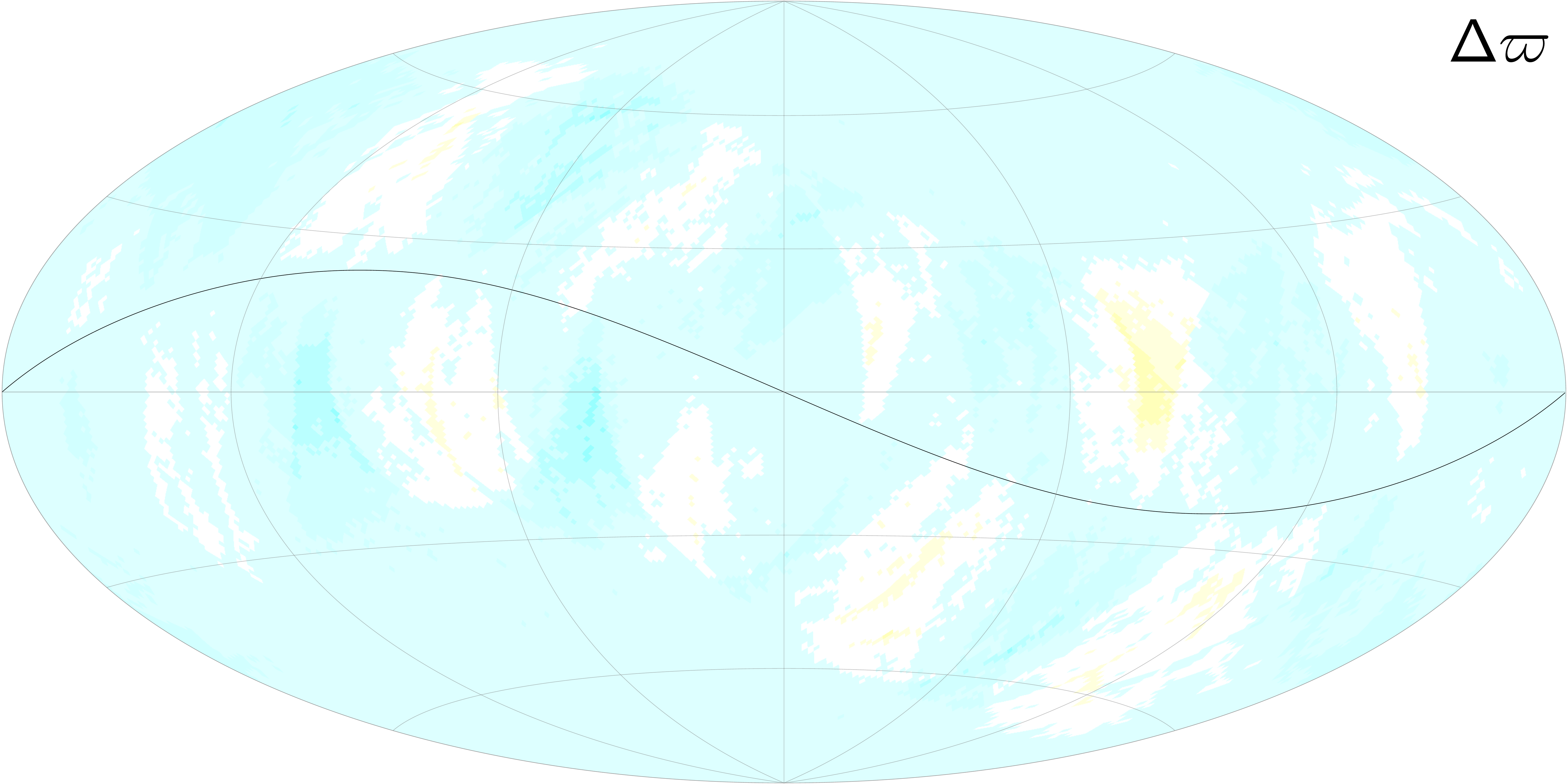}}
\vspace*{1mm}
\centerline{\includegraphics[keepaspectratio,width=\scalethree\hsize]{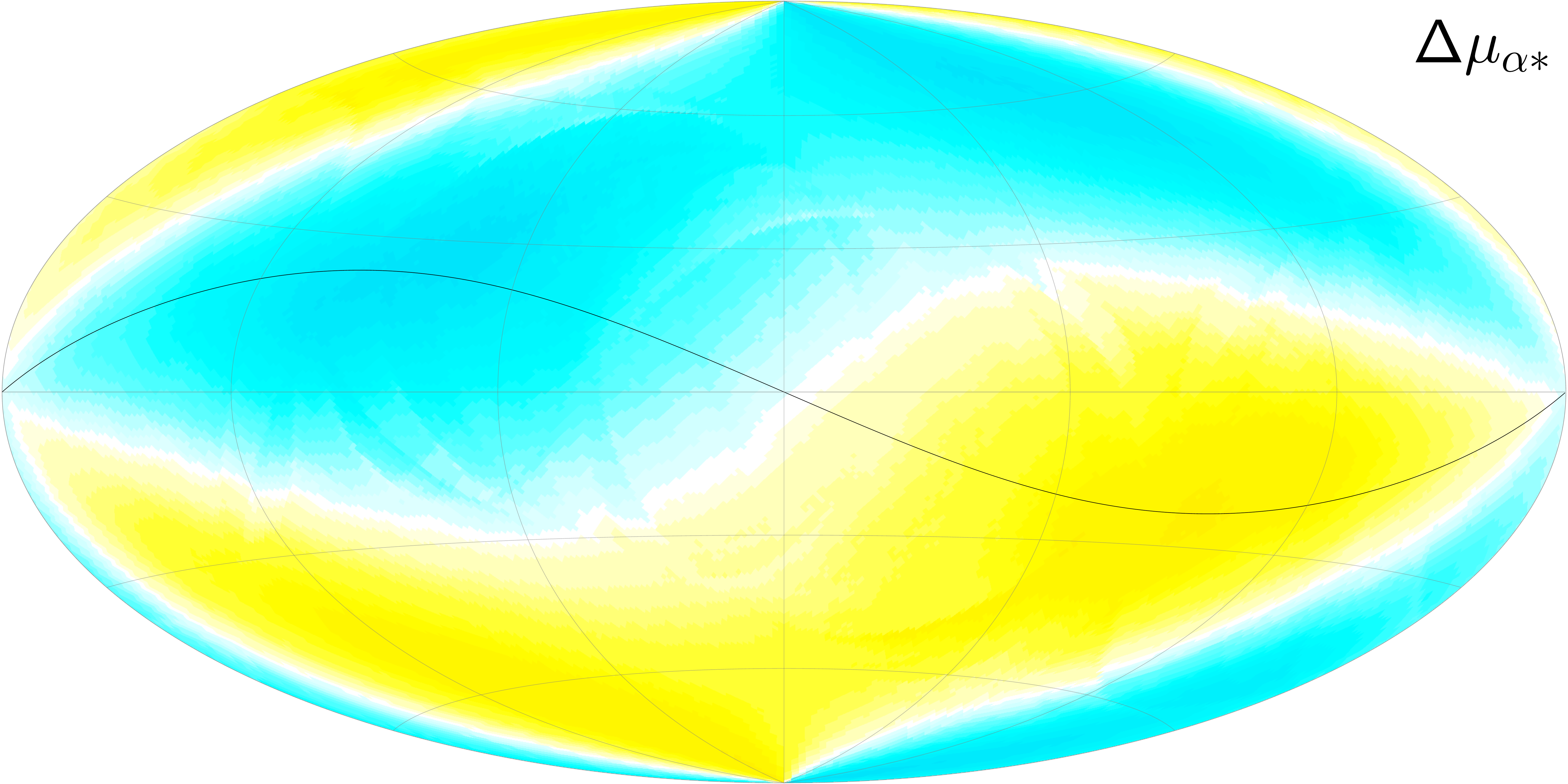}}
\vspace*{1mm}
\centerline{\includegraphics[keepaspectratio,width=\scalethree\hsize]{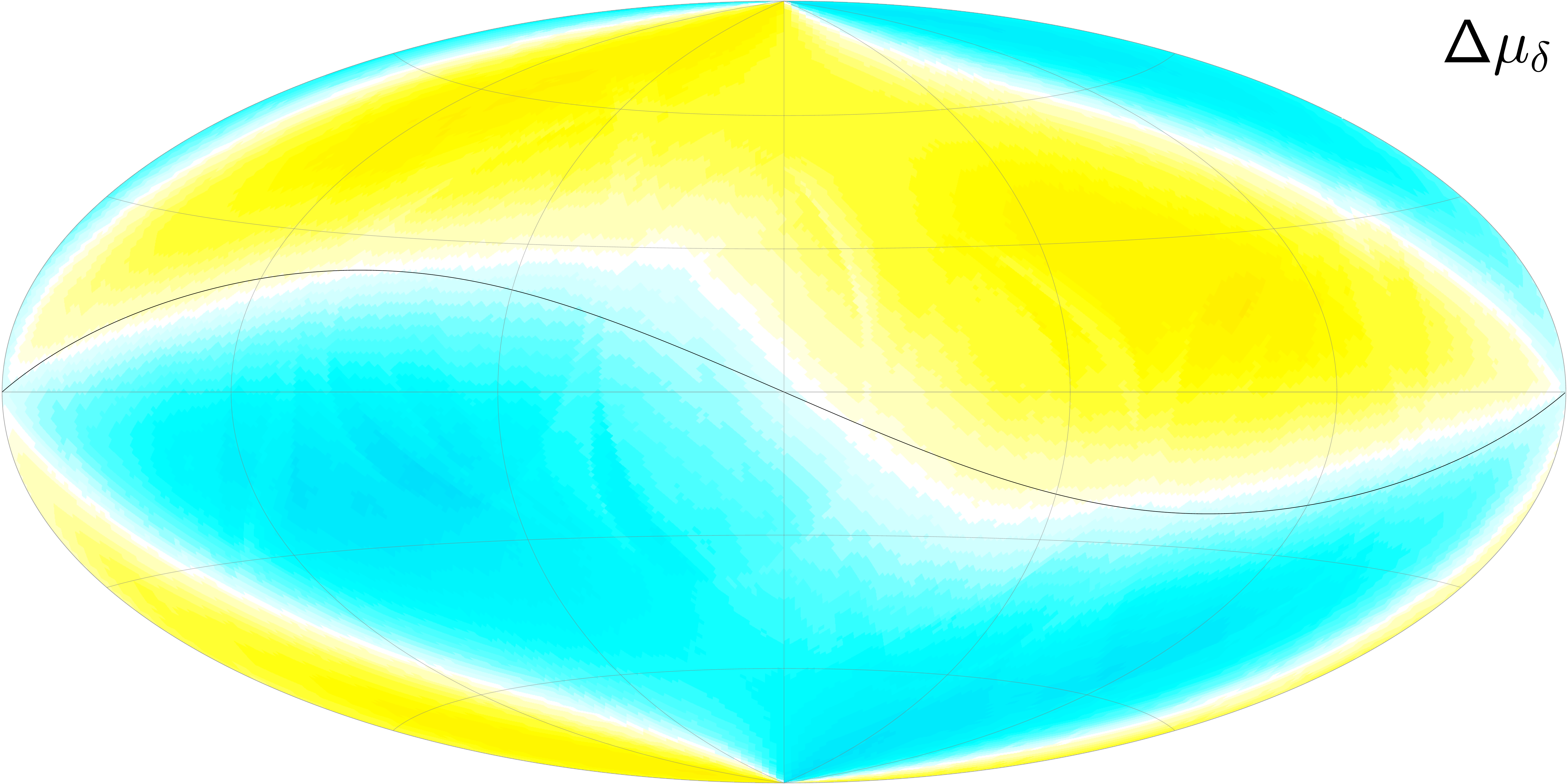}}
\vspace*{1mm}
%\includegraphics[keepaspectratio,width=1\hsize]{20yrImgs/colorlegend.pdf}
%\rotatebox[origin=c]{90}{\includegraphics[keepaspectratio,width=0.8\hsize,clip]{20yrImgs/colorlegend3.pdf}}
\centerline{\includegraphics[keepaspectratio,width=0.7\hsize]{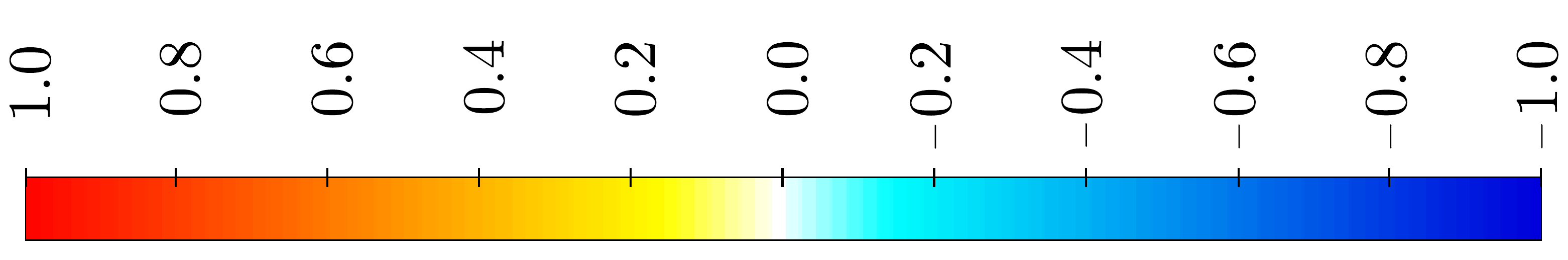}}
%  \vspace*{1mm}
%  \hspace*{8mm} 
\centerline{${\rm med}(\Delta\varepsilon)/\Delta_{\rm max}$}
  \caption{Astrometric errors caused by a GW with $\Pgw=20$\,yr
    ($\nu\simeq 1.58$\,nHz) propagating towards the centre of the maps
    with strain parameters $\hPlusCos=\hPlusSin=\hTimesCos=\hTimesSin$. 
    In the legend, $\varepsilon$ is a placeholder for any of the five astrometric 
    parameters. The maps show the median error, normalised by $\Delta_{\rm max}$, 
    using a pixel size of $\simeq 0.84$\,deg$^2$ (HEALPix level~6). 
    The projection is the same as in Fig.~\ref{fig__PGW_Vector_Field_Plots}.
    The black curve marks the ecliptic. \vspace*{-2mm} 
    \label{fig_simulated20yr}}
\end{figure}

\section{GW-induced errors in the astrometric solution}
\label{sec__source_errors}

\subsection{Overview of the numerical results}
\label{sec__results_overview}

A main result of our simulations is that any injected GW signal,
regardless of its frequency and other parameters, generates errors
in the astrometric source parameters and in the spacecraft attitude, 
as well as in the residuals of the solution. How the errors are distributed
in the various parts of the solution mostly depends on the GW frequency 
$\nu$, or period $\Pgw=1/\nu$. For periods comparable to or greater 
than the duration of the data, $\Pgw\gtrsim\Tobs$ (subsequently 
referred to as the low-frequency regime), the positions and 
proper motions are altered most, while the residuals are barely affected. 
By contrast, in the high-frequency regime ($\Pgw\lesssim\Tobs$) 
all parts of astrometric solutions are affected. For some specific GW
frequencies -- those related to the fundamental frequencies of the
scanning law -- the astrometric errors are substantially larger than
for nearby frequencies. At the same time, 
the errors in the attitude are significantly increased while the effect on 
the residuals is smaller. This shows that substantial parts of the GW 
signal are absorbed by the astrometric and attitude parameters for 
these specific GW frequencies. We also note that the errors
in the astrometric parameters are not randomly distributed on the
sky but display a variety of complex patterns.

In the following subsections these findings are discussed in greater detail.
Figures~\ref{fig___freq_overview-source} and \ref{fig___freq_overview-att-res},
summarising basic error statistics versus GW frequency, provide 
a useful reference for the discussion.

\subsection{Low-frequency regime: $\Pgw\gtrsim\Tobs$}
\label{sec__errors_slowGWs}

\begin{figure*}[htb]
\centering
\includegraphics[width=0.49\textwidth,keepaspectratio]{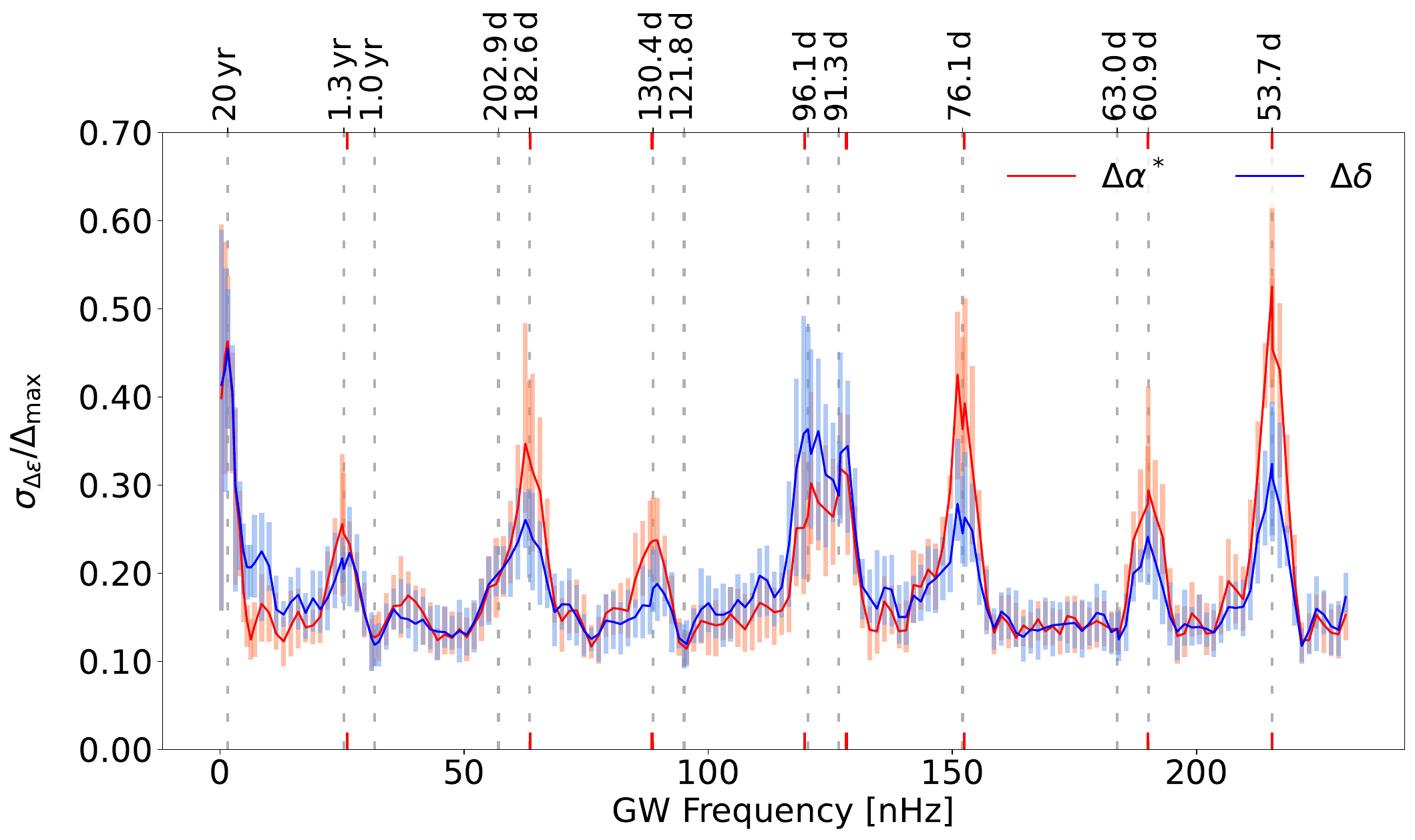}
\includegraphics[width=0.49\textwidth,keepaspectratio]{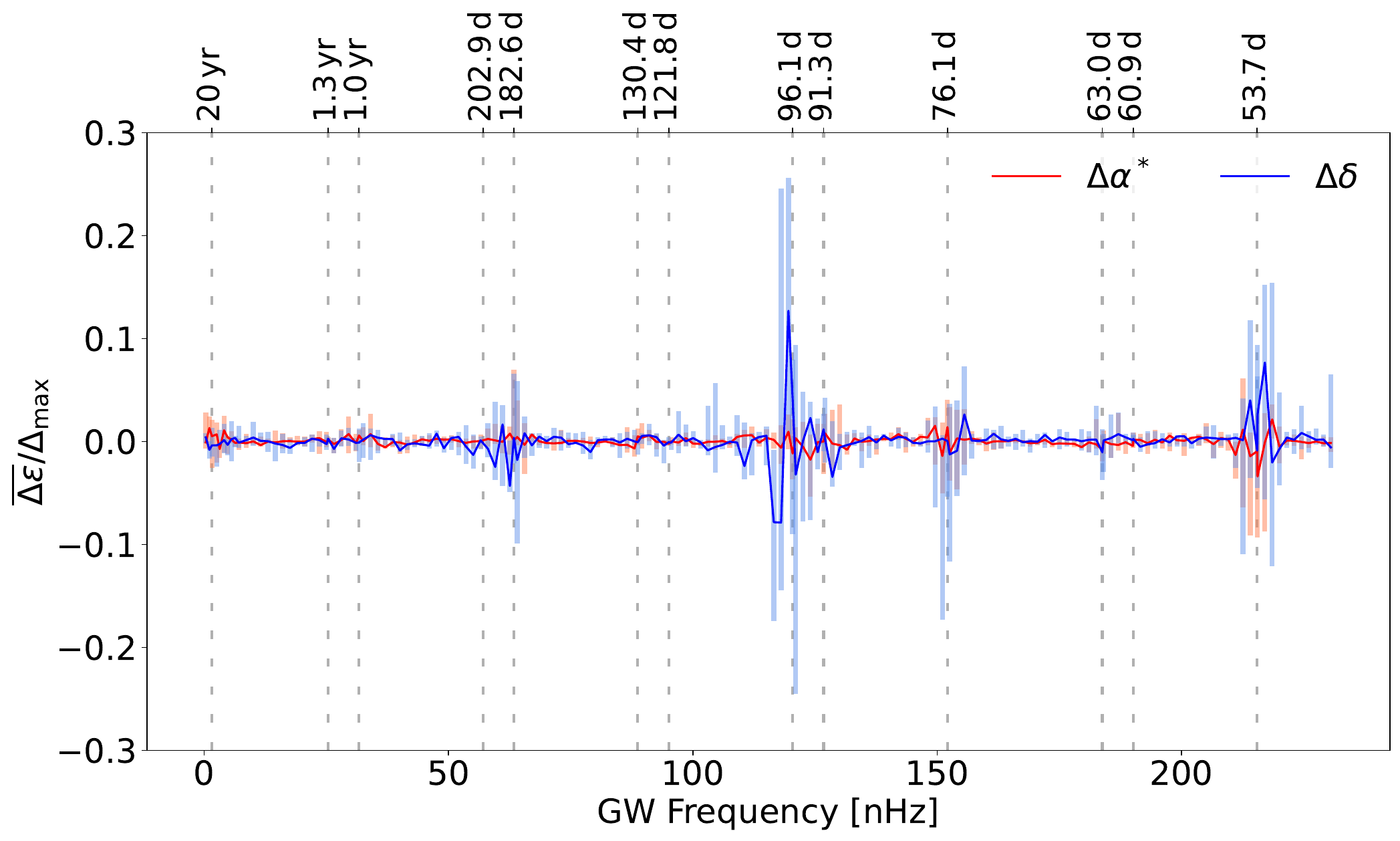}
\includegraphics[width=0.49\textwidth,keepaspectratio]{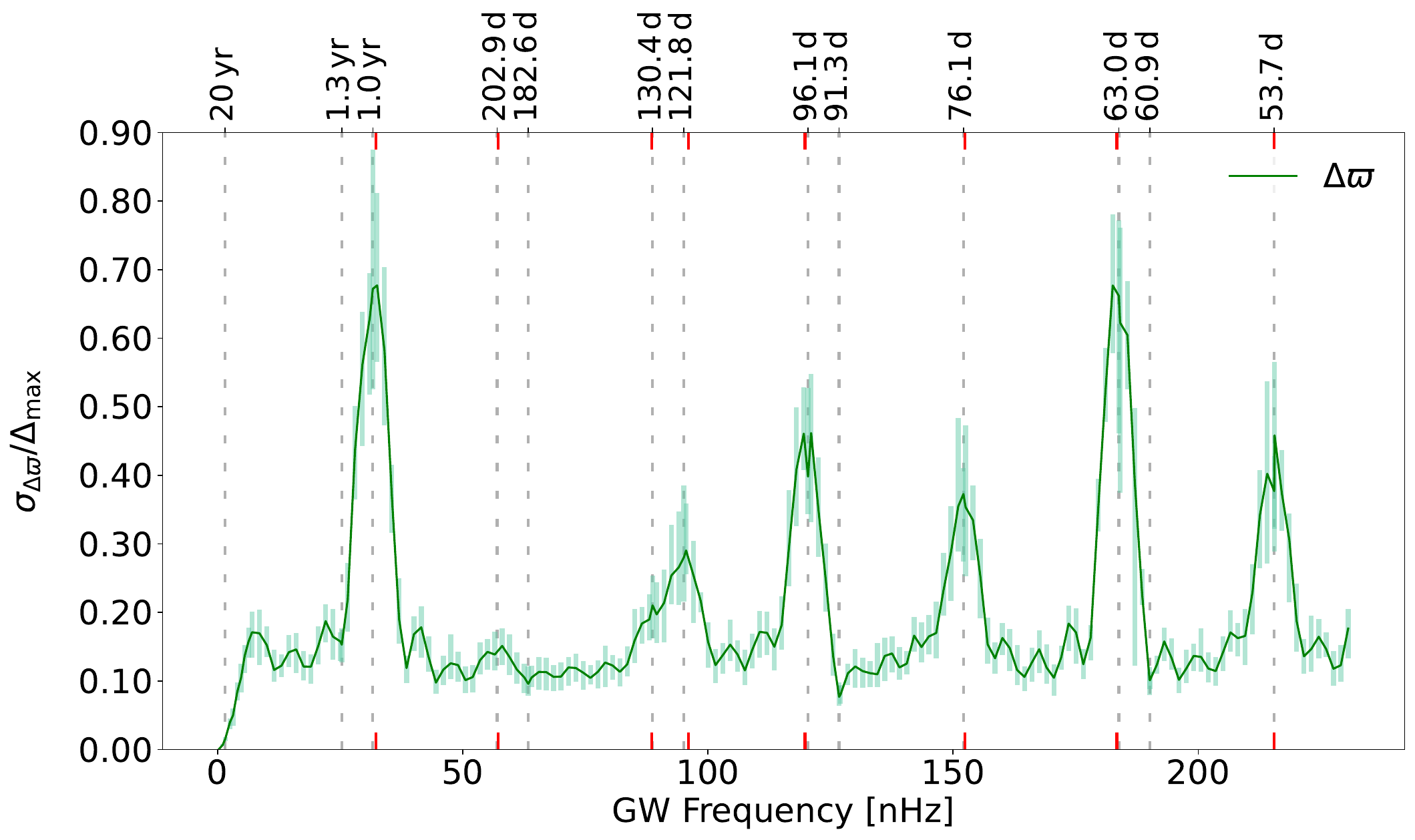}
\includegraphics[width=0.49\textwidth,keepaspectratio]{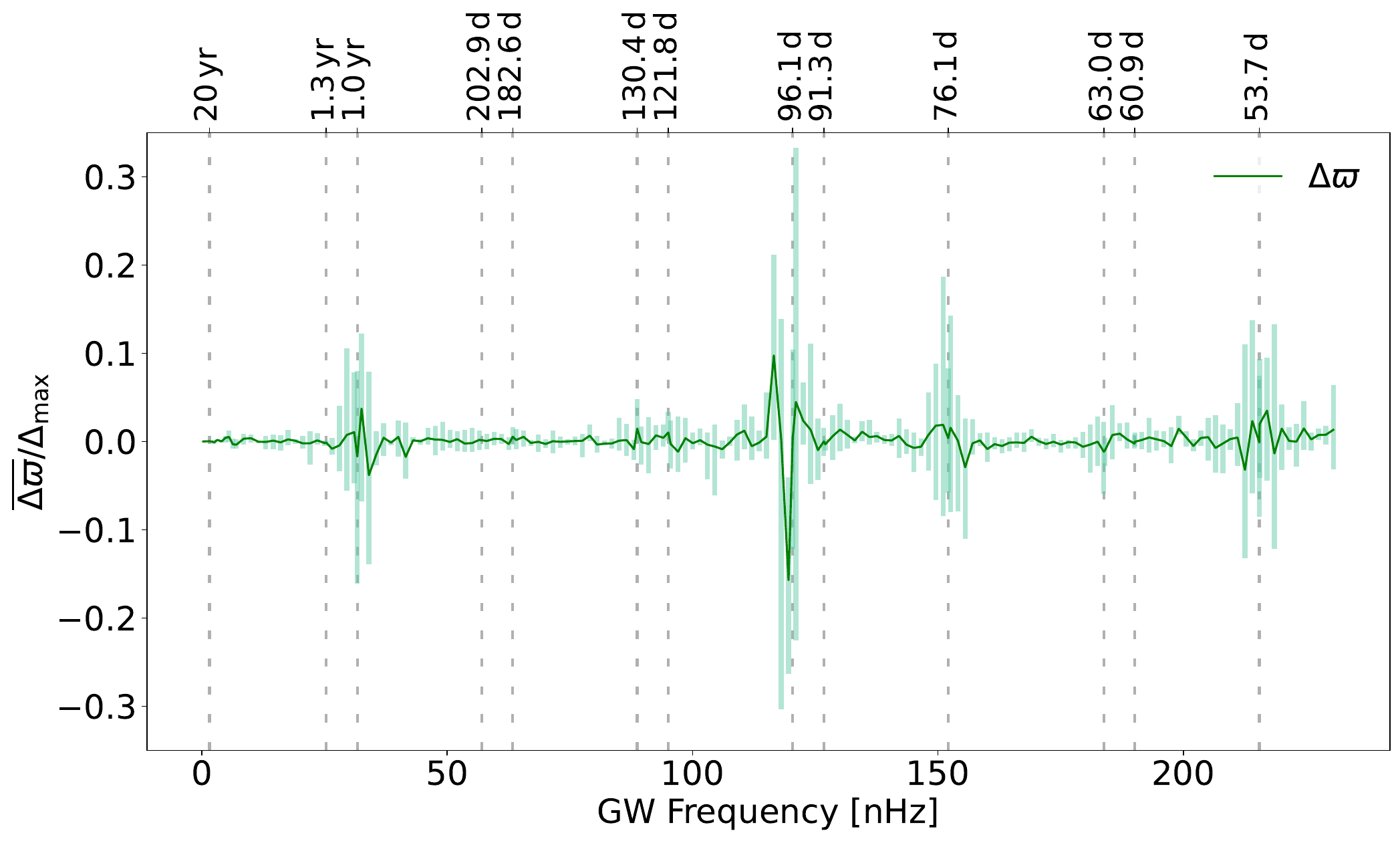}
\includegraphics[width=0.49\textwidth,keepaspectratio]{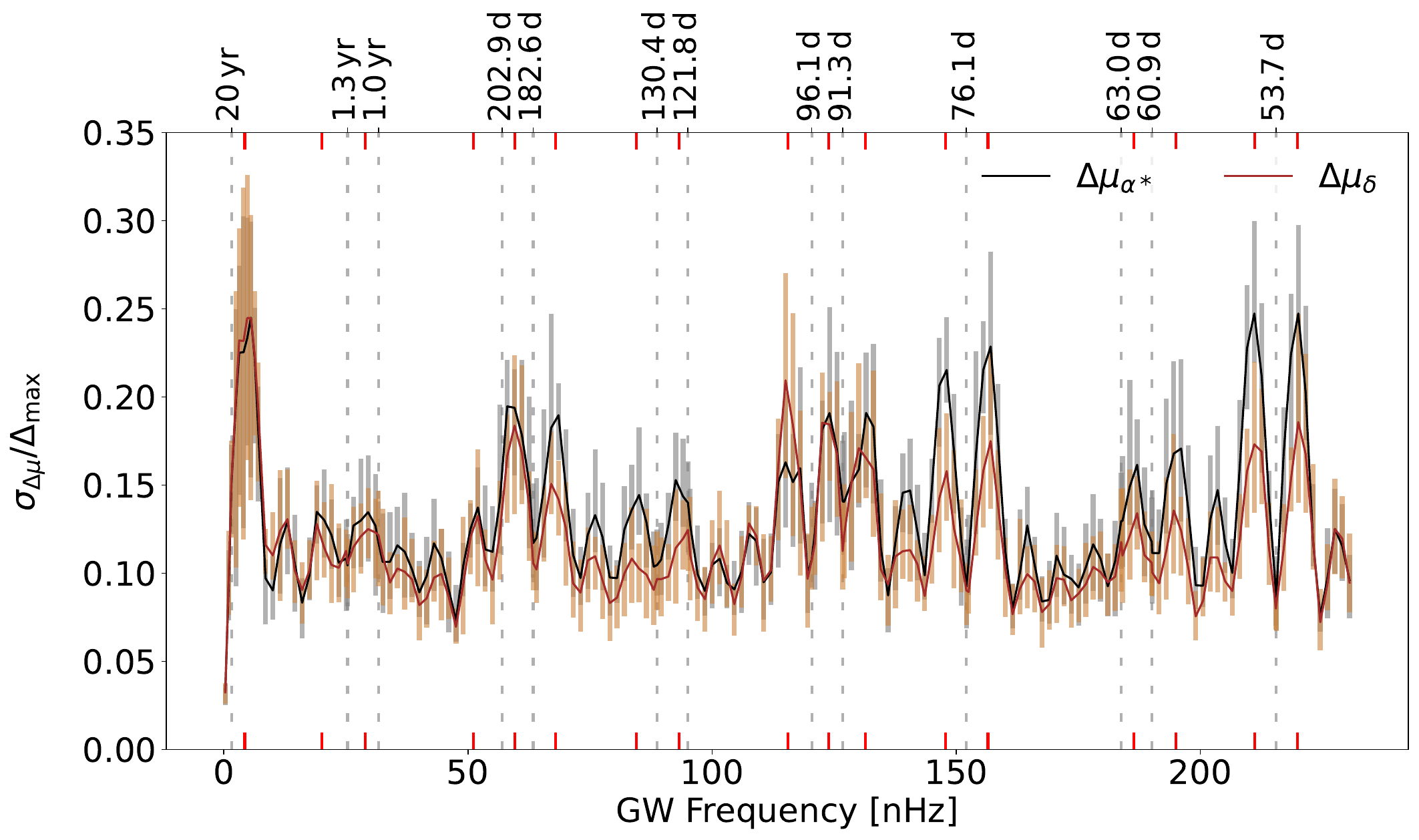}
\includegraphics[width=0.49\textwidth,keepaspectratio]{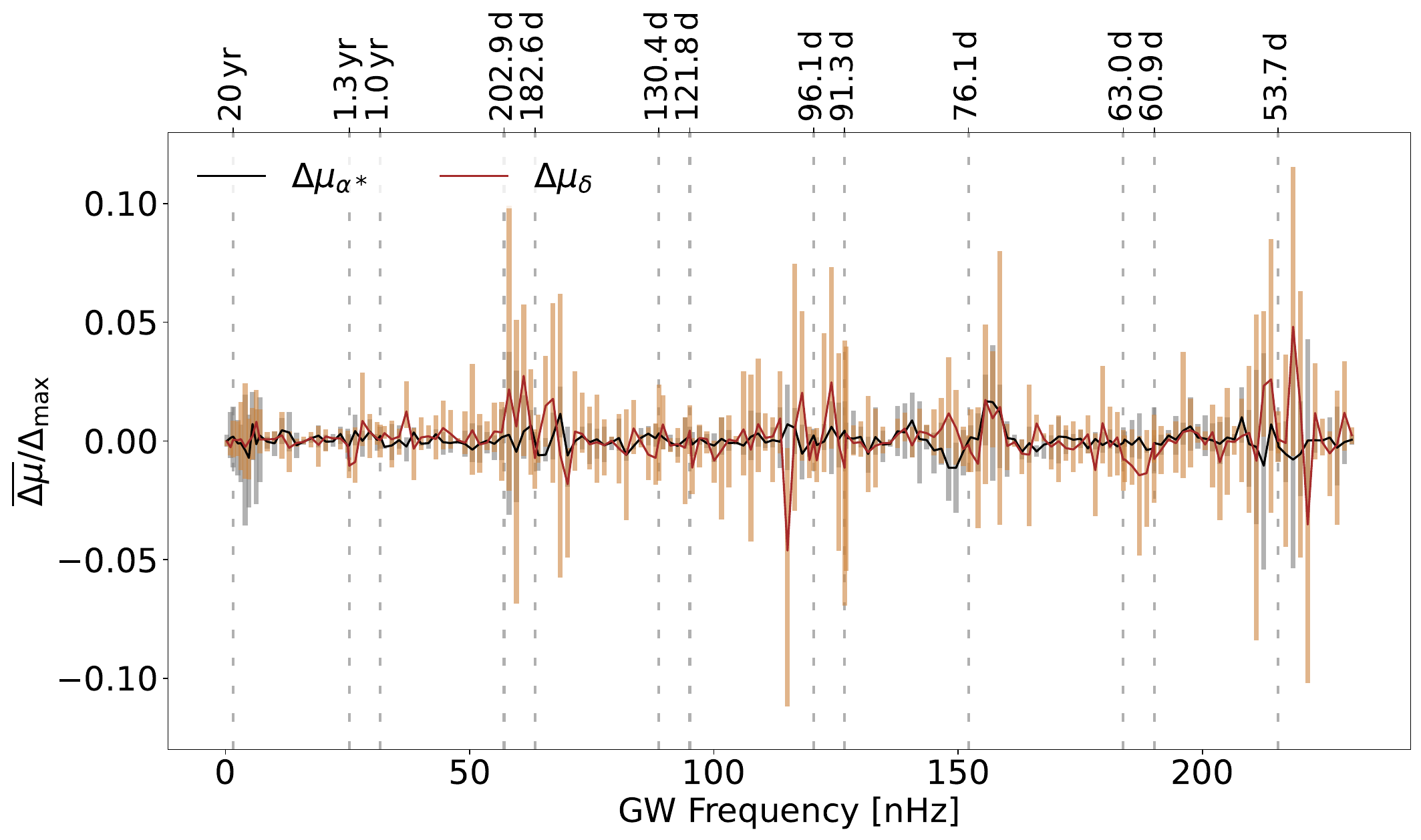}
\caption{Statistics of the errors of the astrometric source parameters versus
  the frequency of the injected GW. 
  \textit{Left}: standard deviations of the errors.
  \textit{Right}: averages of the errors.
  \textit{Top} to \textit{bottom}: errors in position ($\alpha$ and $\delta$), 
  parallax ($\varpi$), and proper motion ($\mu_{\alpha *}$ and $\mu_\delta$).
  All statistics are normalised to the maximum astrometric amplitude of
  the GW effect, $\Delta_{\rm max}$. The coloured bars show the range 
  of statistics obtained in the five simulations for each frequency 
  (see Sect.~\ref{sec__methodology}); for improved visibility the coloured lines 
  connect the average statistics in the corresponding bars. The red tick marks
  on both lower and upper horizontal axes of the left pictures show the positions
  of the GW frequencies and periods shown in Table~\ref{tab__list_peaks}.
  \label{fig___freq_overview-source}}
\end{figure*}

\begin{figure*}[htb]
	\centering
        \includegraphics[width=0.49\textwidth,keepaspectratio]{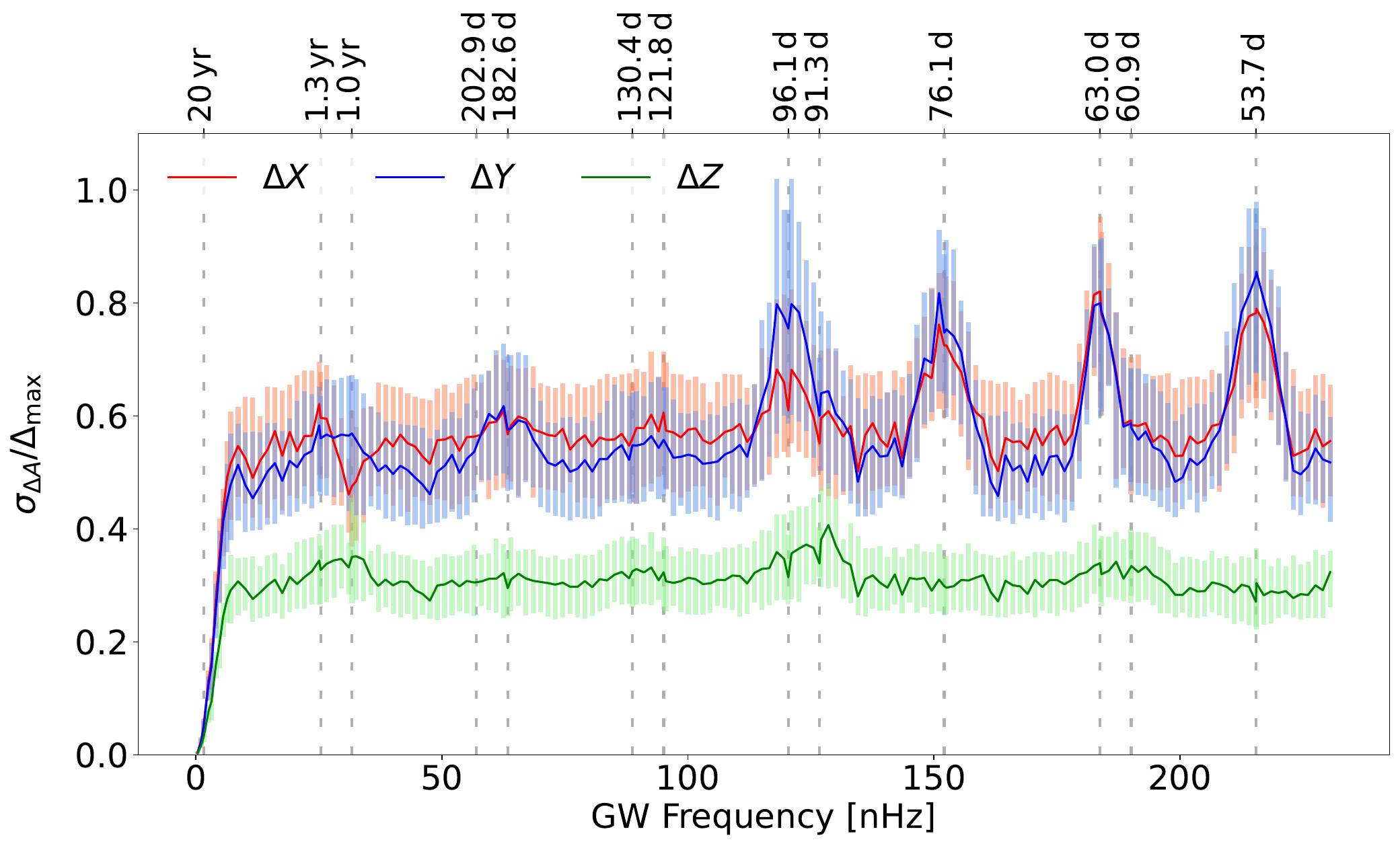}
        \includegraphics[width=0.49\textwidth,keepaspectratio]{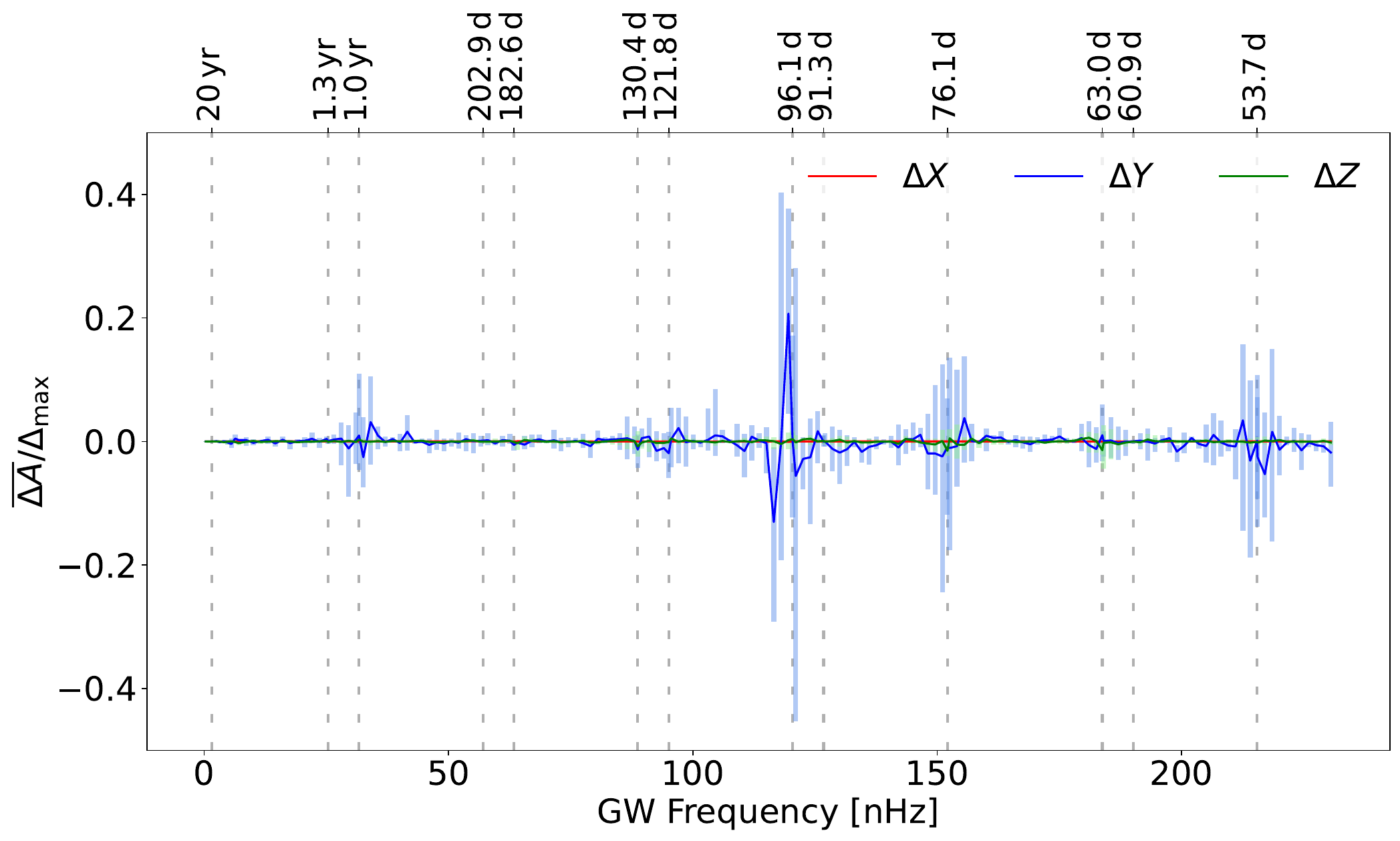}        
        \includegraphics[width=0.49\textwidth,keepaspectratio]{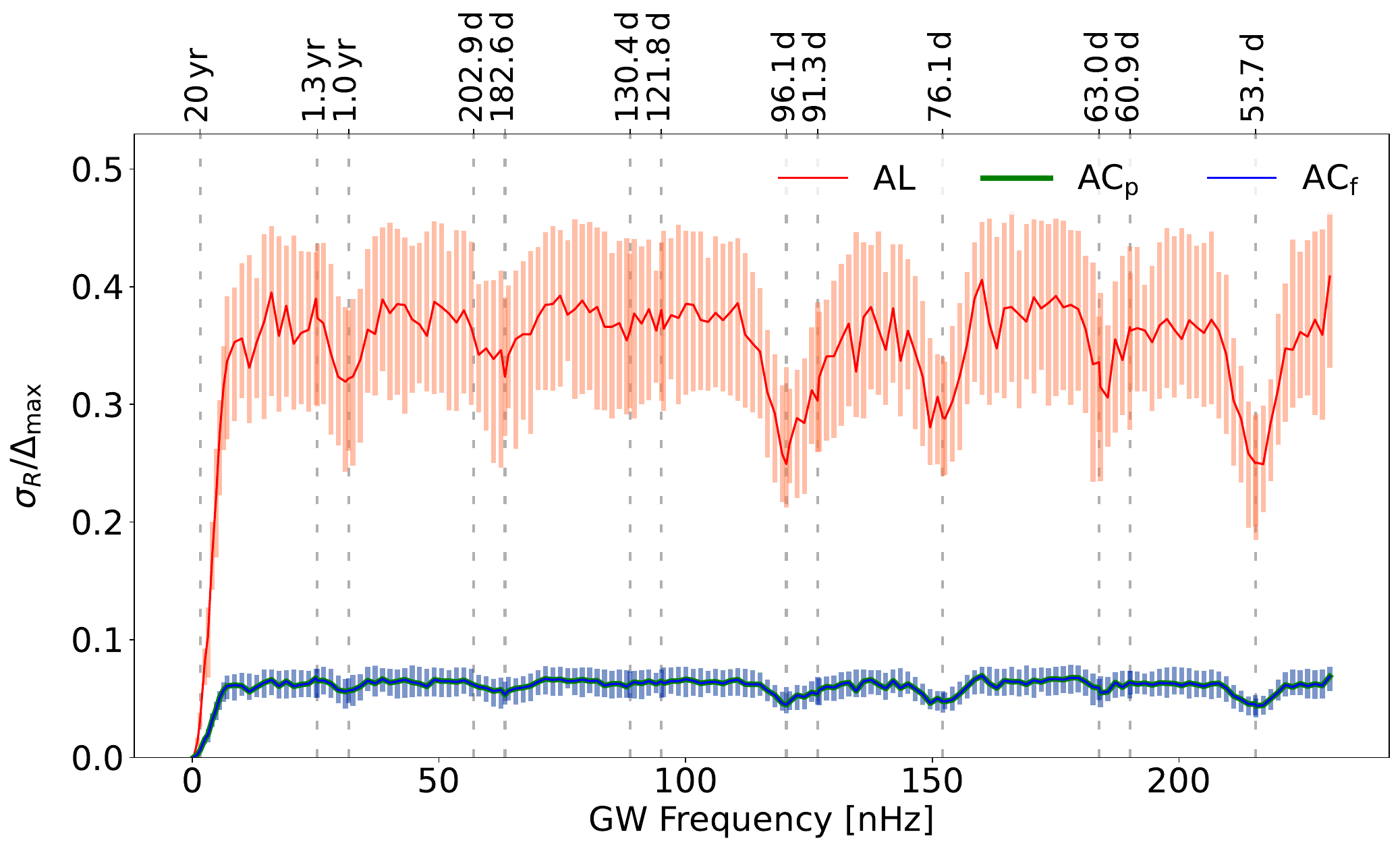}
        \includegraphics[width=0.49\textwidth,keepaspectratio]{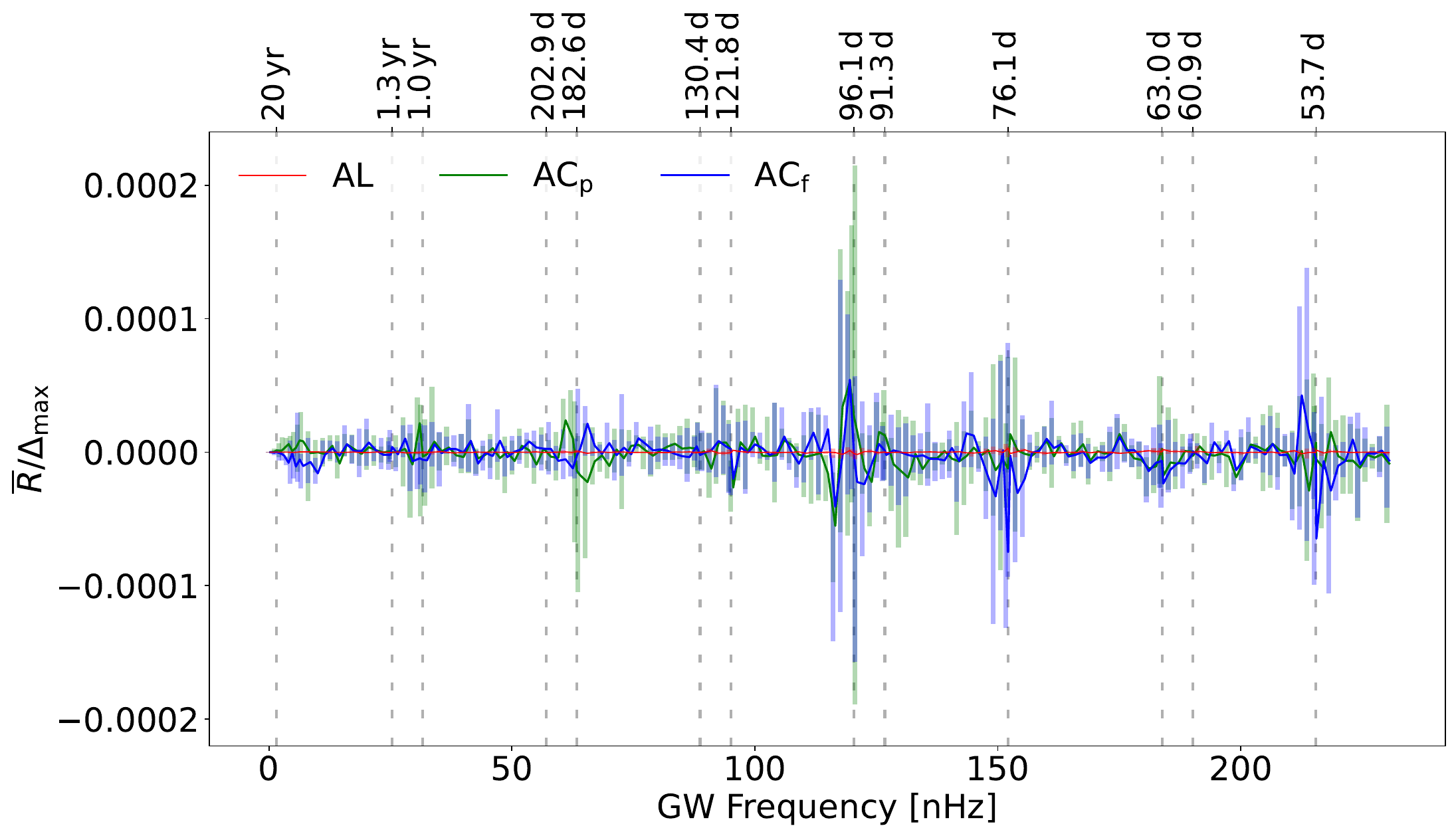}
\caption{Statistics of the attitude errors (\textit{top}) and astrometric residuals
  (\textit{bottom}) versus the frequency of the injected GW. 
  \textit{Left}: standard deviations of the errors or residuals.
  \textit{Right}: averages of the errors or residuals.
  In the upper panels, $A$ in $\Delta A$ is a placeholder for $X$, $Y$, or $Z$, 
  the three axes of the Scanning Reference System (Sect.~3.1 of
  \citeads{2012A&A...538A..78L}). Similarly, in the lower panels, $R$ is a placeholder
  for the residual along-scan (AL), across-scan in the
  preceding FoV (${\rm AC}_\mathrm{p}$), or across-scan in the following FoV
  (${\rm AC}_\mathrm{f}$). The coloured bars show the range 
  of statistics obtained in the five simulations for each frequency 
  (see Sect.~\ref{sec__methodology}); for improved visibility the coloured lines 
  connect the average statistics in the corresponding bars. 
  The standard deviations of the residuals in ${\rm AC}_\mathrm{p}$ and 
  ${\rm AC}_\mathrm{f}$ virtually coincide.
  \label{fig___freq_overview-att-res}}
\end{figure*}

The simulations for GWs with periods greater than the duration of
observations (corresponding to $\nu\lesssim 6.3$\,nHz for $\Tobs=5$\,yr)
largely confirm the predictions of the simplified theoretical model in
Sect.~\ref{sec__subsec_interaction_with_sourceparams}, namely that
the source parameters in this frequency regime absorb most of the 
astrometric effects of the GW. This is most clearly seen in 
Fig.~\ref{fig___freq__overviewslow}, which zooms in on the 
low-frequency end ($\nu<15$\,nHz, or $\Pgw>2.1$\,yr) of the 
standard deviations in Figs.~\ref{fig___freq_overview-source} and 
\ref{fig___freq_overview-att-res}. The errors in position and proper 
motion from the simulations behave qualitatively as expected, with
maxima at roughly the frequencies where the theoretical curves in 
Fig.~\ref{figure-fg} have extrema, and minima where the theoretical
curves go through zero. For example, the maximal errors in proper 
motions were predicted for $\Pgw \simeq 1.5\Tobs=7.5$\,yr for
the simulations, which agrees very well with the maxima in the
second panel of Fig.~\ref{fig___freq__overviewslow} (cf.\  
Table~\ref{tab__list_peaks}, which puts the maximum at 
$\Pgw\simeq 7.2$\,yr as measured from the simulation results).

Moreover, the model (\ref{lin-pos})--(\ref{sigma-r}) valid for any
particular source allows one to compute a theoretical prediction of
the normalized standard deviations of the GW effects in the position,
proper motions and residuals of an astrometric solution involving many
sources homogeneously distributed over the sky. Indeed, assuming that
both components of positions ($\Delta\alpha^*$ and $\Delta\delta$) and
proper motions ($\Delta\mu_{\alpha*}$ and $\Delta\mu_\delta$) have the
same standard deviations as well as considering that statistically in our
simulations $h_{\rm c}\simeq h_{\rm s}\simeq h/\sqrt{2}$ one gets
\begin{eqnarray}
  \label{normalized-sigma-pos}
  &&{\sigma_{\Delta\alpha^*}\over\Delta_{\rm max}}={\sigma_{\Delta\delta}\over\Delta_{\rm max}}=
  \sqrt{{2-\langle e^2\rangle\over 6}}\ \left|f(y)\right|\,,\\
  \label{normalized-sigma-pm}
  &&{\sigma_{\Delta\mu_{\alpha*}}\over\Delta_{\rm max}}={\sigma_{\Delta\mu_\delta}\over\Delta_{\rm max}}={2\over \Tobs}\,
  \sqrt{{2-\langle e^2\rangle\over6}}\ \left|g(y)\right|\,,
\end{eqnarray}
\noindent
where $\langle e^2\rangle\approx{8\over 15}$ is the averaged value of $e^2$ in our simulations. 
These two theoretical curves are shown on two upper plots of Fig.~\ref{fig___freq__overviewslow}.
Then assuming that the residuals in AL and AC have the same standard deviations (so that
$\sigma_{r,{\rm AL}}\simeq\sigma_{r,{\rm AC}}$) one gets from Eq.~(\ref{sigma-r})
\begin{equation}
  \label{normalized-sigma-res}
{\sigma_{r,{\rm AL}}\over\Delta_{\rm max}}={\sigma_{r,{\rm AC}}\over\Delta_{\rm max}}=
\sqrt{{2-\langle e^2\rangle \over 6}}\ {\left(1-f^2(y)-{1\over 3}\,g^2(y)\right)}^{1/2}\,.
\end{equation}
\noindent
We remind that Eq.~(\ref{normalized-sigma-res}) is derived ignoring
interaction of the GW signal with the attitude.
Section~\ref{sec___attitude_absorption} explains that the AC effects
of the GW signal are fully absorbed by the AC attitude, while the AL
effects are partially absorbed by the AL attitude and partially
equivalent to a variation of the basic angle. Since we do not model any
variation of the basic angle in our simulation, that second part
remains in the AL residuals. The theoretical curve
(\ref{normalized-sigma-res}) is shown on the attitude plot on
Fig.~\ref{fig___freq__overviewslow} where it reasonably agrees with
the normalized standard deviation of the variations in AC attitude
$\Delta X$ and $\Delta Y$. For the AL effects, one has $\sigma_{r,{\rm
    AL}}^2\simeq\sigma_{\Delta Z}^2+\sigma_{\rm AL}^2$, where
$\sigma_{\Delta Z}$ is the standard deviation of the AL attitude
variations and $\sigma_{\rm AL}$ is the standard deviation of the AL
residuals (shown on the two lowest plots of
Fig.~\ref{fig___freq__overviewslow}).

The theoretical model was derived with many simplifying assumptions,
including that there would be no effect of the GW on the parallaxes
and attitude.  Figure~\ref{fig___freq__overviewslow} shows that there
is some such effect at all frequencies, but that it becomes
progressively smaller towards zero frequency.
Figure~\ref{fig_simulated20yr} shows the errors of the astrometric
parameters induced by a GW with a period of 20\,yr.  At this low
frequency the errors in parallax are indeed negligible compared to the
errors in the other parameters. The large-scale variations of the
errors in position and proper motion shown here closely follow the
predictions using Eqs.~\eqref{lin-pos}--\eqref{lin-y}.

\begin{table}[htb]
	\caption{Errors induced by the different strain parameters in the low-frequency regime.\label{tab___longperiod_effects_by_phase}}
	\centering
%	\resizebox{\columnwidth}{!}{%
	{\small
	\begin{tabular}{lccccc}
	\hline\hline\noalign{\smallskip}
	\shortstack[l]{~}      & $\sigma_{\alpha*}/\Delta_{\rm max}$ & $\sigma_\delta/\Delta_{\rm max}$ & $\sigma_\varpi/\Delta_{\rm max}$ & $\sigma_{\mu_{\alpha*}}/\Delta_{\rm max}$ & $\sigma_{\mu_{\delta}}/\Delta_{\rm max}$ \\
	\noalign{\smallskip}\hline\noalign{\smallskip}
        $\hPlusCos$  & \numthree{0.5190}     & \numthree{0.5949}     & \numthree{0.0057284} & \numthree{0.0049768} & \numthree{0.0042499}\\
	$\hPlusSin$  & \numthree{0.00042712} & \numthree{0.00058728} & \numthree{0.00058422} & \numthree{0.096530}   & \numthree{0.1106}\\
        $\hTimesCos$ & \numthree{0.5947}     & \numthree{0.5188}     & \numthree{0.0061520} & \numthree{0.0048166} & \numthree{0.0036897}\\
        $\hTimesSin$ & \numthree{0.00039225} & \numthree{0.00057881} & \numthree{0.00057403} & \numthree{0.1106}     & \numthree{0.096530}\\
        \noalign{\smallskip}\hline
	\end{tabular}
}
        \tablefoot{The table gives standard deviations of the errors in source
          parameters, normalised by $\Delta_{\rm max}$, for a GW signal
          of frequency $\numthree{0.925}$\,nHz  ($\Pgw\simeq\numtwo{34.223}$\,yr).
          Each line represents a separate astrometric solution for simulated data with only
          one non-zero strain parameter, as indicated in the first column. The data cover
          an interval of $\Tobs=5$\,yr, with the reference epoch in the middle of the interval.
          The values for the proper motion components are given per year.}
\end{table}

Equations~\eqref{lin-pos}--\eqref{lin-y} show that the effect of the GW on
positions and proper motions in the low-frequency regime depends on the phase
of the GW: only the cosine-related strain parameters $\hPlusCos$ and
$\hTimesCos$ influence the positions, while only the sine-related
parameters $\hPlusSin$ and $\hTimesSin$ produce an effect in the proper
motions. Also this aspect of the theoretical model in 
Sect.~\ref{sec__subsec_interaction_with_sourceparams} has been
confirmed by the dedicated simulations reported in 
Table~\ref{tab___longperiod_effects_by_phase}. 
The table also shows the much smaller effect on parallax for $\Pgw\gg\Tobs$.

\subsection{High-frequency regime: $\Pgw\lesssim\Tobs$}
\label{sec__errors_fastGWs}

From Fig.~\ref{fig___freq_overview-source} it is seen that GWs with
periods shorter than the duration of observations $\Tobs$ produce
errors in the astrometric parameters with a typical standard
deviation of 0.1--0.2$\Delta_{\rm max}$. At certain frequencies 
the errors do however reach much higher values, as discussed
in Sect.~\ref{sec__freq_bands_high_errors}. Averaged over all sources, 
the errors are typically small (less than $\pm 0.01\Delta_{\rm max}$), 
but again at specific frequencies they can be much larger. For the 
components of proper motion and for $\alpha$ the average error 
barely reaches $\pm 0.1\Delta_{\rm max}$, but for $\delta$ and 
$\varpi$ it can reach $\pm 0.3\Delta_{\rm max}$. Generally, the 
average errors in right ascension are considerably smaller than in 
$\delta$ and $\varpi$. The reason for this anisotropy is not fully 
understood.  
Appendix~\ref{section-examples} contains additional results 
from the simulations, illustrating the characteristics of 
the astrometric errors produced by GWs.

Figure~\ref{fig___freq_overview-att-res} shows the standard deviations
and average values of the attitude errors and residuals of the
solution. The standard deviation of the attitude errors is typically 
0.25--0.35$\Delta_{\rm max}$ for the rotation around the nominal
spin axis $Z$, corresponding to the AL attitude (see Sect.~3.1 of 
\citeads{2012A&A...538A..78L} for the definition of the Scanning 
Reference System of \gaia), and 0.4--0.7$\Delta_{\rm max}$ around
the other two axes $X$ and $Y$, corresponding for the AC attitude.
For some specific GW frequencies the standard deviation can reach 
0.5$\Delta_{\rm max}$ in $Z$ (AL) and $\simeq 1.0\Delta_{\rm max}$ 
for $X$ and $Y$ (AC). The average errors are typically close to zero, 
but can reach $\pm 0.4\Delta_{\rm max}$ for the rotation around the 
$Y$ axis.

Concerning the residuals shown in the lower panels of 
Fig.~\ref{fig___freq_overview-att-res}, it can be noted that the 
average values (bottom right) are always very small, within 
$\pm 0.0002\Delta_{\rm max}$. 
Corresponding to the peaks in the astrometric and/or attitude 
errors, the residuals (bottom left) show decreased standard 
deviations at the same frequencies. The most prominent example 
is seen around a GW period of approximately 96.1\,d. 

Based on the theoretical model in Sect.~\ref{sec___attitude_absorption} 
we expect the AC attitude (around the $X$ and $Y$ axes) to absorb 
the entire AC signal of a GW, while the AL attitude (around $Z$) only 
absorbs the mean of the AL signals in the two FoVs. This behaviour is
largely confirmed by the standard deviations of the residuals shown 
in the bottom left panel of Fig.~\ref{fig___freq_overview-att-res}:
whereas the standard deviation of the AL residuals is significant 
(0.25--0.45$\Delta_{\rm max}$), that of the AC residuals is 
much smaller (0.04--0.07$\Delta_{\rm max}$).
The standard deviation of the AL residuals can be compared to
the standard deviations given of the full GW signal given in
Table~\ref{tab__gw_sig_stats} as 0.4--0.6$\Delta_{\rm max}$. One can see that
the AL residuals only contain a part of the total GW signal.
That the standard 
deviation of the AC residuals is not completely negligible can be 
attributed to the simplifying assumptions adopted in the theoretical 
model. One such assumption was that second-order (differential) 
effects within the FoV, caused by the finite FoV size, can be neglected. 
But this can only explain a minor part of the AC residuals, of the 
order of $0.01\Delta_{\rm max}$. Instead, the dominating contribution 
to the AC residuals comes from the interaction between source and 
attitude parameters that was also neglected in the theoretical model.

To cross-check our understanding of these interactions, a series of 
dedicated simulations were made in which only specific components 
of the GW signal were added to the observations. For example, if the 
AL components ($\delta g_{\rm f}$, $\delta g_{\rm p}$) of the GW 
signal were included, but not the AC components ($\delta h_{\rm f}$, 
$\delta h_{\rm p}$), it was found that the source parameters did not 
change from a reference simulation including all four AL and AC 
components of the GW signal. If, on the other hand, the AC components 
were included together with the mean value of the AL components (so 
the signal $\frac{1}{2}(\delta g_{\rm f}+\delta g_{\rm p})$ was applied in 
both FoVs), the attitude errors were found to be the same as in the 
reference simulation, while the source parameters were virtually 
unaffected by the applied signal, resulting in very small AL and AC 
residuals. Two important results from these experiments are 
(i) that errors in the source parameters are caused almost exclusively 
by the differential AL component of the GW signal, 
$\delta g_{\rm f}-\delta g_{\rm p}$; 
(ii) that these source errors in turn produce residuals in both 
coordinates, although they are much smaller (by a factor $\sim$5) 
AC than AL. The first result is expected based on the simplified model
of Sect.~\ref{sec___attitude_absorption}. The second result can only 
be understood by considering the way the attitude and source 
parameters are determined in the global astrometric solution by 
minimising the sum of squared residuals (SSR). As detailed e.g.\ 
in Eq.~(24) of \citetads{2012A&A...538A..78L}, both AL and AC 
observations are used, so 
$\text{SSR}=\text{SSR}_\text{AL}+\text{SSR}_\text{AC}$. 
Although there is an attitude that fit the AC observations 
perfectly also in the presence of the GW signal (corresponding 
to $\text{SSR}_\text{AC}=0$ in the present noise-free simulations), 
a somewhat different attitude may be preferred in terms of the 
total SSR, if it allows a slight reduction of the AL residuals.

\begin{figure}[htb]
	\centering
	\includegraphics[width=0.99\hsize,keepaspectratio]{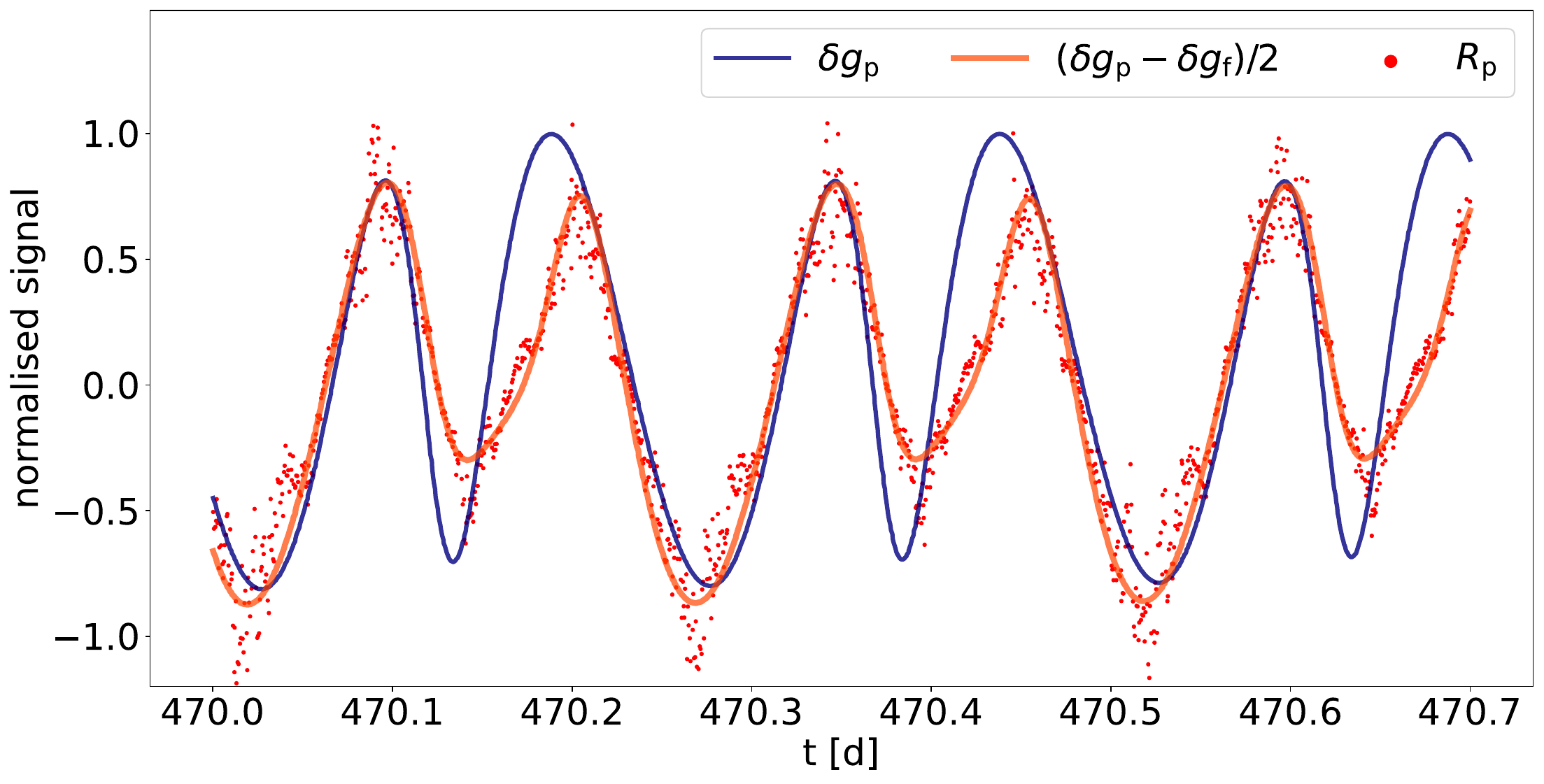}
	\caption{Example of the GW signal from one of the simulations and how 
	it appears in the astrometric solution. 
	The black curve ($\delta g_\mathrm{p}$) is the AL component of the GW signal 
	in the preceding FoV. The red dots ($R_\text{p}$) show the AL residuals of the 
	solution in the same FoV. For comparison, the red curve shows half the
	differential GW effect, $(\delta g_{\rm p}-\delta g_{\rm f})/2$.
	All data are normalised by $\Delta_{\rm max}$. The frequency of the GW signal
	was $\nu = \numthree{101.16}$\,\nHz, equivalent to a period of 
	$\Pgw \simeq \numtwo{114.42}$\,d.
	\label{fig___residuals_vs_signal}}
\end{figure}

Even though parts of the GW signal are absorbed by the source and 
attitude models, the AL residuals generally contain a large fraction of 
the AL component of a high-frequency GW signal. This is illustrated 
in Fig.~\ref{fig___residuals_vs_signal}, showing an arbitrary short 
segment of the data from one of the simulations. It is seen that the 
residuals $R_\text{p}$ in the preceding FoV approximately 
follow the curve giving half the differential GW signal,
$(\delta g_{\rm p}-\delta g_{\rm f})/2$. For the full dataset 
the correlation coefficient is 0.92 in this example. The residuals in 
the following field (not shown in the plot) similarly follow half the
differential GW signal, but with the opposite sign. It should be noted 
that the actual GW signal in that FoV ($\delta g_{\rm p}$), shown 
by the black curve, is not well reproduced by the residuals
(correlation coefficient 0.72). The simulations also show that
the AC residuals have no correlation with the AC component of
the GW signal.

Another way to look at the astrometric errors is to consider the
errors in position, and separately the ones in proper motion as a
vector field on the sphere. For each star, the errors in positions or
proper motions have a direction and a magnitude, given by either the
combination of errors $\Delta\alpha^*$ and $\Delta\delta$, or
$\Delta\mu_{\alpha *}$ and $\Delta\mu_\delta$, respectively. The
errors in parallax $\Delta\varpi$ can be considered as a scalar field
on the sphere. Then one can consider the RMS of those vector or scalar
fields. However, this way would not give any new information compared
to what we have above. Indeed, for any $n$-dimensional vector $x$ the
components of which are to be investigated one has a simple relation
between the RMS value ($r_x$), the standard deviation ($\sigma_x$) and 
the average values (${\overline x}$): 
$r_x^2={\overline x}^2+\frac{n-1}{n}\sigma_x^2$.
Since we have already considered the standard deviations and
averages of the astrometric errors, there is no reason to consider also
the RMS of the corresponding scalar and vector fields.

\subsection{Frequency bands of special significance}
\label{sec__freq_bands_high_errors}

A striking feature of Figs.~\ref{fig___freq_overview-source} and 
\ref{fig___freq_overview-att-res} is the multitude of peaks in the standard 
deviations of the astrometric and attitude errors, and sometimes also in the
average errors. As Fig.~\ref{fig___freq_overview-att-res} shows, there is an 
increase in the AC attitude errors at some of these frequencies, accompanied 
by a decrease mainly in the AL residuals. We have verified that the peaks 
occur at the same frequencies independent of the direction of propagation
of the GW.

Most of the peaks are approximately centred on GW frequencies 
that can be expressed as
\begin{equation}
\nu_{k,l,m} = k\nuone + l\nusl + m\nu_T
= k\,\left(\frac{1}{1\,\text{yr}}\right) + l\,\left(\frac{5.8}{1\,\text{yr}}\right)
+ m\,\left(\frac{0.694}{T}\right)\, \label{eq___peak_freqs}
\end{equation}
for small integer numbers $k$, $l$, and $m$. Here, 
$\nuone \simeq \numthree{31.6880878}$\,nHz corresponds to a 
period of $1\,\mathrm{yr}\equiv3.15576\times 10^7\,\mathrm{s}$, 
$\nusl \simeq \numthree{183.7909093}$\,nHz corresponds to the 
precession period $1\,\mathrm{yr}/5.8\simeq 62.97\,\mathrm{d}$ 
of the scanning law, and 
$\nu_T$ is a frequency depending on the duration $\Tobs$ of the mission,
with $\nu_T\simeq 4.4$\,nHz for $\Tobs=5$\,yr, corresponding to a
period of about 7.2\,yr. The dimensionless constant 0.694 appearing in 
the expression for $\nu_T$ is further discussed below.

Table~\ref{tab__list_peaks} lists the most prominent peaks in the standard 
deviations of the errors in position, parallax, and proper motion, as
measured from the data displayed in Fig.~\ref{fig___freq_overview-source}.
Peaks were detected by fitting a template profile to the data in a sliding 
window of width 15\,nHz, using a cosine-squared profile with a full width 
at half maximum (FWHM) of 7.5\,nHz. This was done separately for the
position, parallax, and proper motion data, but combining the components
in $\alpha$ and $\delta$ for the position and proper motion data for 
improved signal-to-noise ratio. Each peak corresponds to a local maximum
in the amplitude of the fitted profile.

Table~\ref{tab__list_peaks} also shows our interpretation of the peaks 
in terms of $k$, $l$, and $m$ (using $\nu_T=4.4$\,nHz), and the differences 
between the measured peak frequency and $\nu_{k,l,m}$ from 
Eq.~\eqref{eq___peak_freqs}. In two cases the interpretation is ambiguous,
as discussed below. Considering the frequency discretization and the scatter 
in the standard deviations at each frequency point, the agreement between
the measured and calculated frequencies is quite good.

It is noted that for the position and parallax errors, the peaks always have 
$m=0$, while for the proper motion errors the peaks tend to come in pairs 
around the peaks in position error, with $m=\pm 1$. One apparent exception 
is the peak in proper motion errors at 51.1\,nHz, tentatively identified as
$\nu_{-4,1,-1}=52.6$\,nHz: no corresponding peak was detected in position
errors at the expected frequency $\nu_{-4,1,0}=57.0$\,nHz, although there
is one in parallax errors. However, the data do not rule out a small peak 
around 57.0\,nHz in the position errors, blended with the much stronger peak 
at 63.6\,nHz (see the top-left panel of Fig.~\ref{fig___freq_overview-source}).
If this interpretation is correct, there should also be a peak in the proper 
motion errors at $\nu_{-4,1,1}=61.4$\,nHz, which however would be blended
with the strong peak at 59.6\,nHz. The nearly coinciding theoretical frequencies
$\nu_{2,0,-1}=59.0$\,nHz and $\nu_{-4,1,1}=61.4$\,nHz is not the only
example of an ambiguous identification; the peak at 123.9\,nHz, in the table
identified as $\nu_{-2,1,1}=124.8$\,nHz, could just as well be be 
$\nu_{4,0,-1}=122.4$\,nHz. One more potentially ambiguous identification
is the peak at 29.0\,nHz, in the table identified as $\nu_{-5,1,1}$ at 29.8\,nHz. 
This peak could in principle be a blend with $\nu_{1,0,-1}$ at 27.3\,nHz, but 
because neither $\nu_{1,0,0}$ at 31.7\,nHz is seen in the position data, nor 
$\nu_{1,0,1}$ at 36.1\,nHz in the proper motion data, we conclude that the
peak is probably not a blend.
 
\begin{table}[htb]
\caption{List of the most prominent peaks in 
Fig.~\ref{fig___freq_overview-source}.\label{tab__list_peaks}}
\centering
{\small
\begin{tabular}{rrlrrrr}
	\hline\hline\noalign{\smallskip}
 \multicolumn{1}{c}{$\nu$} & \multicolumn{1}{c}{$\Pgw$}  & parameters  & $k$  & $l$ & $m$ & 
 \multicolumn{1}{r}{$\nu-\nu_{k,l,m}$}\\
 \multicolumn{1}{c}{[nHz]} & \multicolumn{1}{c}{[d]} & affected &&&& \multicolumn{1}{r}{[nHz]} \\
 \noalign{\smallskip}\hline\noalign{\smallskip}
  4.3 & 2663.8 & pm & $0$ & $0$ & $1$ & $-0.1$ \\[5pt]
 20.1 &  574.9 & pm & $-5$ & $1$ & $-1$ & $-0.8$ \\
 26.1 &  442.9 & pos & $-5$ & $1$ & $0$ & $+0.7$ \\
 29.0 &  398.1 & pm & $-5$ & $1$ & $1$ & $-0.7$ \\[5pt]  32.3 &  357.9 & plx & $1$ & $0$ & $0$ & $+0.6$ \\[5pt]
 51.1 &  226.0 & pm & $-4$ & $1$ & $-1$ & $-1.5$ \\
 57.2 &  202.2 & plx & $-4$ & $1$ & $0$ & $+0.1$ \\
\multirow[c]{2}[2]{*}[0pt]{59.6} & \multirow[c]{2}[2]{*}[0pt]{194.0} & \multirow{2}{5em}{pm\hfill $\Bigg\{$} & $-4$ & $1$ & $1$ & $-1.8$ \\[5pt]
& & & $2$ & $0$ & $-1$ & $+0.6$ \\
 63.6 &  181.7 & pos & $2$ & $0$ & $0$ & $+0.2$ \\
 68.0 &  170.0 & pm & $2$ & $0$ & $1$ & $+0.2$ \\[5pt]
 84.5 &  136.9 & pm & $-3$ & $1$ & $-1$ & $+0.1$ \\
 88.5 &  130.8 & pos, plx & $-3$ & $1$ & $0$ & $-0.2$ \\
 93.3 &  123.9 & pm & $-3$ & $1$ & $1$ & $+0.2$ \\[5pt]
 96.0 &  120.4 & plx & $3$ & $0$ & $0$ & $+0.9$ \\[5pt]
115.6 &  100.0 & pm & $-2$ & $1$ & $-1$ & $-0.4$ \\
119.8 &   96.6 & pos, plx & $-2$ & $1$ & $0$ & $-0.6$ \\
\multirow[c]{2}[2]{*}[0pt]{123.9} & \multirow[c]{2}[2]{*}[0pt]{93.3} & \multirow{2}{5em}{pm\hfill $\Bigg\{$} & $-2$ & $1$ & $1$ & $-0.9$ \\[5pt]
& & & $4$ & $0$ & $-1$ & $+1.5$ \\
128.3 &   90.1 & pos & $4$ & $0$ & $0$ & $+1.5$ \\
131.4 &   88.0 & pm & $4$ & $0$ & $1$ & $+0.3$ \\[5pt]
147.8 &   78.2 & pm & $-1$ & $1$ & $-1$ & $+0.0$ \\
152.4 &   75.9 & pos, plx & $-1$ & $1$ & $0$ & $+0.3$ \\
156.5 &   73.9 & pm & $-1$ & $1$ & $1$ & $-0.0$ \\[5pt]
183.4 &   63.0 & plx & $0$ & $1$ & $0$ & $-0.4$ \\[5pt]
186.4 &   62.0 & pm & $6$ & $0$ & $-1$ & $+0.7$ \\
190.0 &   60.8 & pos & $6$ & $0$ & $0$ & $-0.1$ \\
195.0 &   59.3 & pm & $6$ & $0$ & $1$ & $+0.5$ \\[5pt]
211.1 &   54.8 & pm & $1$ & $1$ & $-1$ & $+0.0$ \\
215.5 &   53.7 & pos, plx & $1$ & $1$ & $0$ & $+0.0$ \\
219.9 &   52.6 & pm & $1$ & $1$ & $1$ & $-0.0$ \\
\noalign{\smallskip}\hline
\end{tabular}}
\tablefoot{Columns~1--2 give the peak frequency as measured from the simulations 
  and the corresponding period. Column~3 shows the astrometric parameters for which the
  peak was detected in the data at that frequency (pos = $\alpha$ and/or $\delta$;
  plx = $\varpi$; pm = $\mu_{\alpha*}$ and/or $\mu_\delta$). When the peak was
  detected in both position and parallax errors, the mean frequency is given.
  Columns~4--6 give our interpretation of the frequency in terms of the integers
  $k$, $l$, and $m$ in Eq.~\eqref{eq___peak_freqs}. The last column gives the
  difference between the measured and calculated frequency. As discussed in the
  text, the peaks at $\nu=59.6$ and 123.9\,nHz may be blends of $\nu_{k,l,m}$ for
  the given combinations of $k$, $l$, and $m$.
  The GW frequencies and periods given in this Table are shown by red vertical tick marks
  on the left pictures of Fig.~\ref{fig___freq_overview-source}.}
\end{table}

The peaks in position and parallax errors can be qualitatively
understood as an interference phenomenon between the GW signal
and the characteristic frequencies $\nuone$ and $\nusl$ of the
scanning law, or their overtones. When the GW frequency is close
to $k\nuone + l\nusl$ for some $k$ and $l$, the beat frequency
$\Delta\nu=\nu-k\nuone - l\nusl$ will be in the low-frequency
regime ($|\Delta\nu|\lesssim 1/\Tobs$) and large position and/or 
parallax errors may be created by the same mechanism as 
described in Sect.~\ref{sec___attitude_absorption}. We then
expect the peaks in position and parallax errors to have a
FWHM equal to that of the function $f(y)$ (considering both 
positive and negative $y$), or 
$\numfour{1.20670912880323}/T\simeq 7.6$\,nHz
in reasonable agreement with the simulation results.

That the peaks in the proper motion errors are offset by 
$\pm\nu_T\simeq\pm 4.4$\,nHz from the corresponding peak 
in the position errors can also be understood in the framework 
of the theoretical model of Sect.~\ref{sec___attitude_absorption}. 
According to Eqs.~\eqref{lin-pos}--\eqref{lin-y}, the effect in 
proper motion scales as the derivative of the effect in position with 
respect to the GW frequency: $g(y)=-3\text{d}f/\text{d}y$.
Assuming that this relation holds also for the beat signal,\footnote{Two remarks should be made concerning the applicability
of this relation. (i) Although Eqs.~\eqref{lin-pos}--\eqref{lin-y} were 
derived for a continuum of observations uniformly distributed over 
the time span of observations, the model can readily be generalised 
for an arbitrary finite sequence of discrete observations, but the 
effects in proper motion are still proportional to the derivative of the 
effects in position with respect to $\nu$. (ii) The effects in position and 
proper motion depend on different strain parameters, and are therefore 
in principle independent of each other. Statistically speaking, however,
the magnitudes of the effects are still expected to follow this relation.}
we expect the peaks in proper motion errors to occur at the
frequencies where $|g(\Delta\nu)|$ has a maximum, that is 
with an offset of $\pm 0.6626/\Tobs$ from the corresponding
peak in position errors. Theoretically, then, the numerical constant
in Eq.~\eqref{eq___peak_freqs} should be 0.6626 rather than the 
value 0.694 estimated from the simulations. We note that the 
low-frequency peak shown in the second panel of
Fig.~\ref{fig___freq__overviewslow} can formally be identified 
with $\nu_{0,0,1}$, as suggested by the first entry in 
Table~\ref{tab__list_peaks}.
Numerical simulations for a mission duration of $\Tobs=3$\,yr 
confirm that the width of the peaks and the offset of the
peaks in proper motion both scale as $1/\Tobs$.

An in-depth investigation of the mechanisms producing the
increased errors in position and/or parallax around the specific
frequencies listed in Table~\ref{tab__list_peaks} is beyond the
scope of this paper. For example, we have no explanation why
some combinations of $k$ and $l$ affect both positions and 
parallaxes, and others only one of them, or why there are peaks
at $\nu_{k,0,0}$ for $k=1$, 2, 3, 4, and 6, but apparently not 
for $k=5$ (at $\nu_{5,0,0}=158.4$\,nHz). We note that there are 
similarities with the spurious periods that may be detected in 
astrometric and photometric time series of \gaia\ data purely 
as a result of the scanning law. That phenomenon was extensively 
investigated by \citetads{2023A&A...674A..25H}, and indeed their 
Eq.~(3) is equivalent to our Eq.~\eqref{eq___peak_freqs} for $m=0$. 
Clearly, the specific distribution of scanning angles and observation 
times for a given source allows the source model to absorb a larger 
fraction of the GW signal for certain frequencies. As shown in 
Appendix~\ref{section-examples}, the sky distributions of the 
astrometric errors for GW signals close to one of the special 
frequencies $\nu_{k,l,0}$ (e.g.\ $\Pgw=1$\,yr, 96.1\,d, 76.1\,d, 
63.0\,d, and 53.7\,d) display characteristic large-scale patterns, 
not present for other frequencies. Similar patterns were discussed 
by \citetads{2023A&A...674A..25H}.
 
\subsection{Sky distribution of the errors}
\label{sec__spatial_features}

The statistical description of the astrometric errors in previous 
sections does not say anything about how the errors are distributed 
on the sky. Examples of their sky distributions are shown in 
Fig.~\ref{fig___error_sky_maps} of Appendix~\ref{section-examples}. 
One sees that the distributions crucially depend on the GW frequency. 
At some GW frequencies and for some of the parameters, the maps 
display large-scale patterns in which the median astrometric error 
per pixel can be as high as $\simeq 1.2\Delta_{\rm max}$. 
As noted earlier, this happens at the specific frequencies discussed
in Sect.~\ref{sec__freq_bands_high_errors}, while at other frequencies
the spatial features have smaller angular scales and smaller amplitudes.
At all frequencies, however, the spatial distributions are strongly
non-random. 

\begin{figure*}[htb]
\sidecaption
	\begin{minipage}[b]{12.2cm}
    \includegraphics[width=12.2cm]{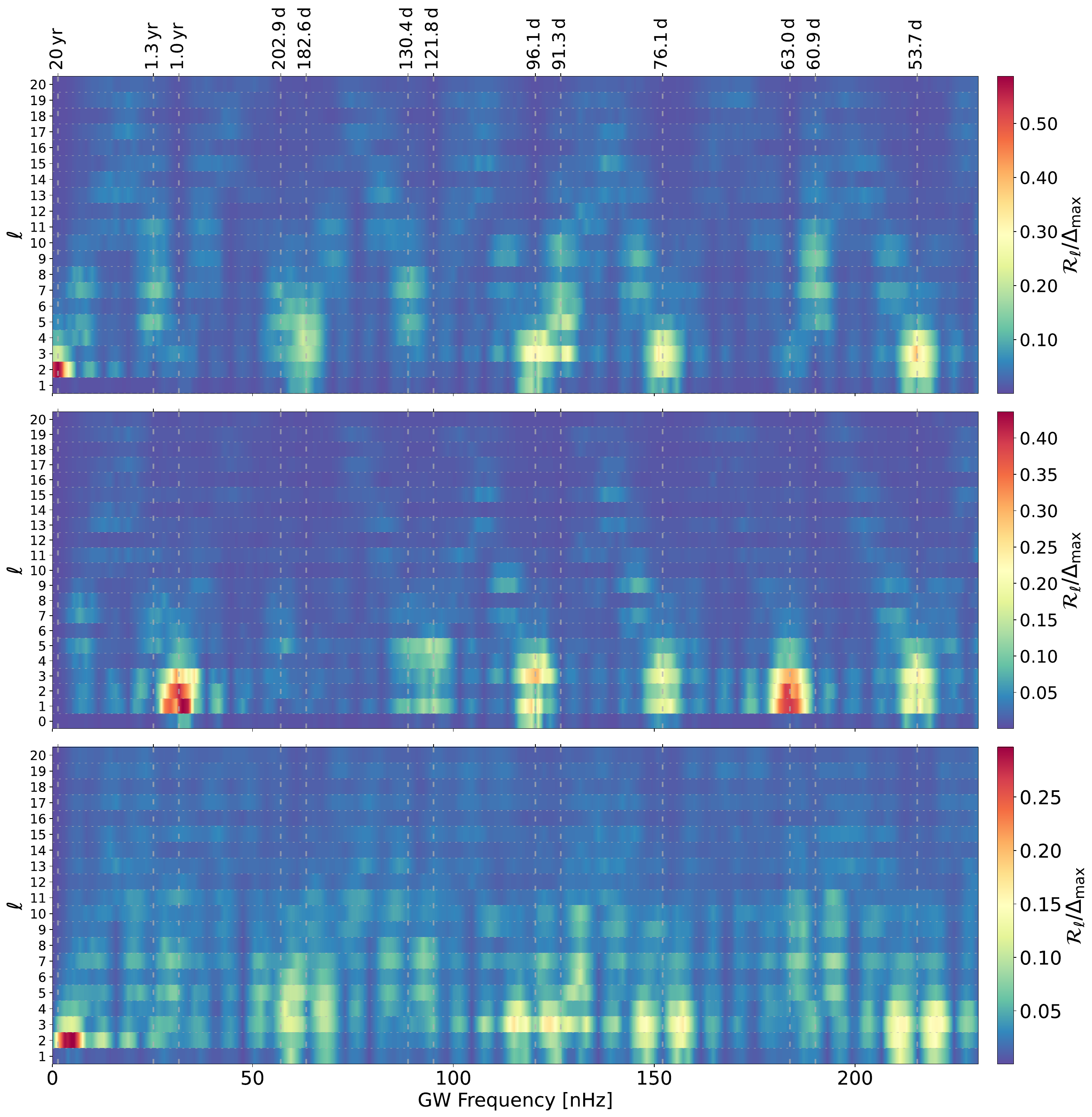}
	\end{minipage}
	\caption{Normalised RMS variation $R_\ell/\Delta_{\rm max}$
          of the SH/VSH expansion of the astrometric errors
          in position (\textit{top}), parallax (\textit{middle}), 
          and proper motion (\textit{bottom}),
          displayed versus GW frequency for degrees $\ell\le 20$. 
          Equivalent data are shown by the solid lines in 
          Figs.~\ref{fig__Rl_pos}--\ref{fig__Rl_pm}
          (see Appendix~\ref{section-shvsh} for details).}
          \label{fig__heatmaps}
\end{figure*}

Although the instantaneous astrometric effect 
$(\delta\alpha^*,\delta\delta)$ of a GW is dominated 
by its quadrupole component
(\citeads{2011PhRvD..83b4024B}; \citeads{2018CQGra..35d5005K}), 
the errors induced in the astrometric parameters rarely have a 
quadrupole spatial distribution. Clearly, the intricate interaction 
between the GW signal, astrometric model, and method of observation 
(where part of the signal is absorbed by the attitude) results in very 
complicated patterns of the astrometric errors on the sky.

In order to provide a more comprehensive description of the 
spatial distributions, we analyse the GW-induced parallax errors
in terms of (scalar) spherical harmonics (SH), and the vector fields
of the errors in position and proper motion in terms of vector 
spherical harmonics (VSH; \citeads{2012A&A...547A..59M}).
Generally speaking, SH and VSH of a given degree $\ell$ quantify 
the strength of the scalar or vector field at a typical angular scale 
of $\simeq 180^\circ/\ell$. 

For each simulation we fit the SH and VSH expansions up to degree 
$\lmax = 20$ separately for the errors in parallax, position, and 
proper motion. For a specific SH or VSH degree $\ell$, the RMS 
value of the error pattern is computed as
\begin{equation}
R_\ell = \sqrt{\frac{1}{4 \pi}\,P_\ell}\, ,\label{RMS_of_VSH}
\end{equation}
\noindent
where $P_\ell$ is the power of the SH components of degree 
$\ell$ for the parallax errors, and the sum of the toroidal $P^t_\ell$ 
and spheroidal $P^s_\ell$ powers of the VSH components for the 
errors in position and proper motion (see Sect.~5.2 of
\citeads{2012A&A...547A..59M}). 
We note that $R_\ell$ is rotationally invariant and therefore independent 
of the coordinate system used (e.g.\ equatorial, ecliptic, or galactic).

Figure~\ref{fig__heatmaps} shows the mean $R_\ell/\Delta_{\rm max}$ 
obtained in the simulations as a function of the GW frequency and $\ell$. 
The peaks at various GW frequencies, already seen in 
Fig.~\ref{fig___freq_overview-att-res}, are easily recognised, but now
we can also see how those errors are distributed in spatial frequency ($\ell$).
These diagrams confirm and further quantify the general conclusions  
drawn from the sky maps, namely that the predominantly quadrupole 
($\ell = 2$) nature of the GW signal does not translate into predominantly 
quadrupole components in the error patterns. Instead, much of the power
is found in higher-degree harmonics, typically $4\lesssim\ell\lesssim 10$,
although lower degrees ($1\le\ell\le3$) dominate in the frequency bands 
with enhanced astrometric errors, where $\ell=3$ is often the strongest
component. It is seen that the parallax errors can have a considerable 
dipole ($\ell = 1$) contribution, and even some global offset ($\ell=0$) 
at certain frequencies. For the proper motion errors, the distribution in 
$\ell$ mirrors that of the position errors, after taking into account the 
frequency splitting by $m=\pm 1$.

In the VSH expansion of the errors in position and proper motion, the 
toroidal components of degree $\ell=1$ represent a global rotation of 
the reference frame, and the spheroidal components of degree $\ell=1$ 
represent a form of distortion known as glide \citepads{2012A&A...547A..59M}.
As implemented in AGISLab, the astrometric solution is set up in such
a way that the resulting positions and proper motions have no net 
rotation with respect to their initial values (in our case the true values). 
Any global rotation in the source errors caused by the GW signal,
if present, is hence removed by the astrometric solution, and the toroidal
components of degree $\ell = 1$ should therefore be zero for position
and proper motion errors. However, technical differences in the way these
computations are made within AGISLab and in our external VSH analysis
result in non-zero rotation components in our VSH coefficients.\footnote{In both AGIS and AGISLab the so-called Frame Rotator is 
configured to fit only toroidal components of order $\ell=1$, whereas
our external VHS analysis fits all VSH harmonics up to $\ell_{\rm max}=20$. 
Moreover, the Frame Rotator takes into account the correlations between 
the $\alpha$ and $\delta$ components of the positions or proper motions 
and uses a special algorithm for outlier rejection, whereas the present VSH 
analysis ignores correlations and assumes no outliers. A detailed 
description of the Frame Rotator algorithm can be found in
Appendix~E of \citeads{2022A&A...667A.148G}.}
Those rotations are small and can be ignored in the discussion below.

\begin{figure}[htb]
 \centering
 \includegraphics[keepaspectratio,width=\hsize]{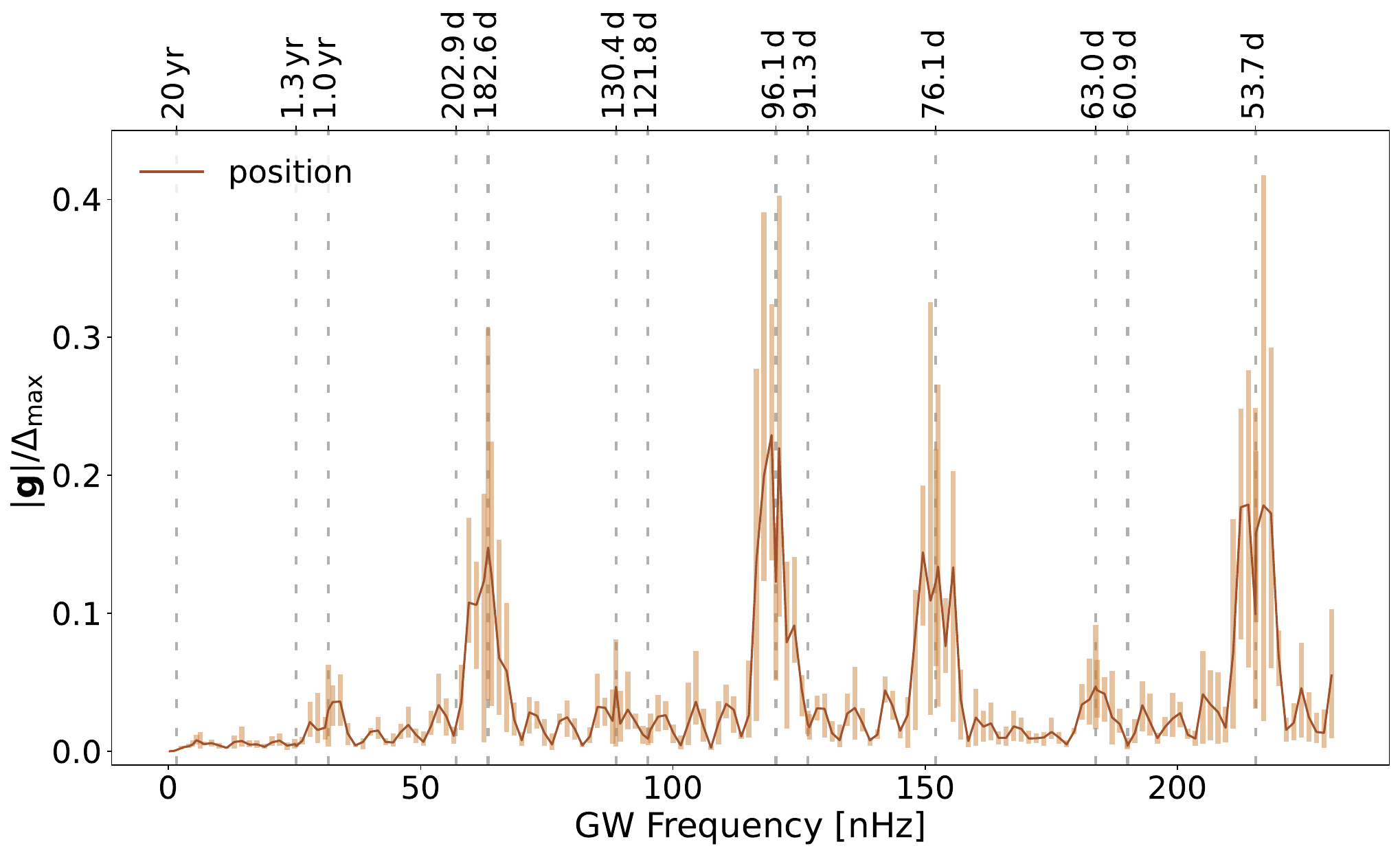}
  \includegraphics[keepaspectratio,width=\hsize]{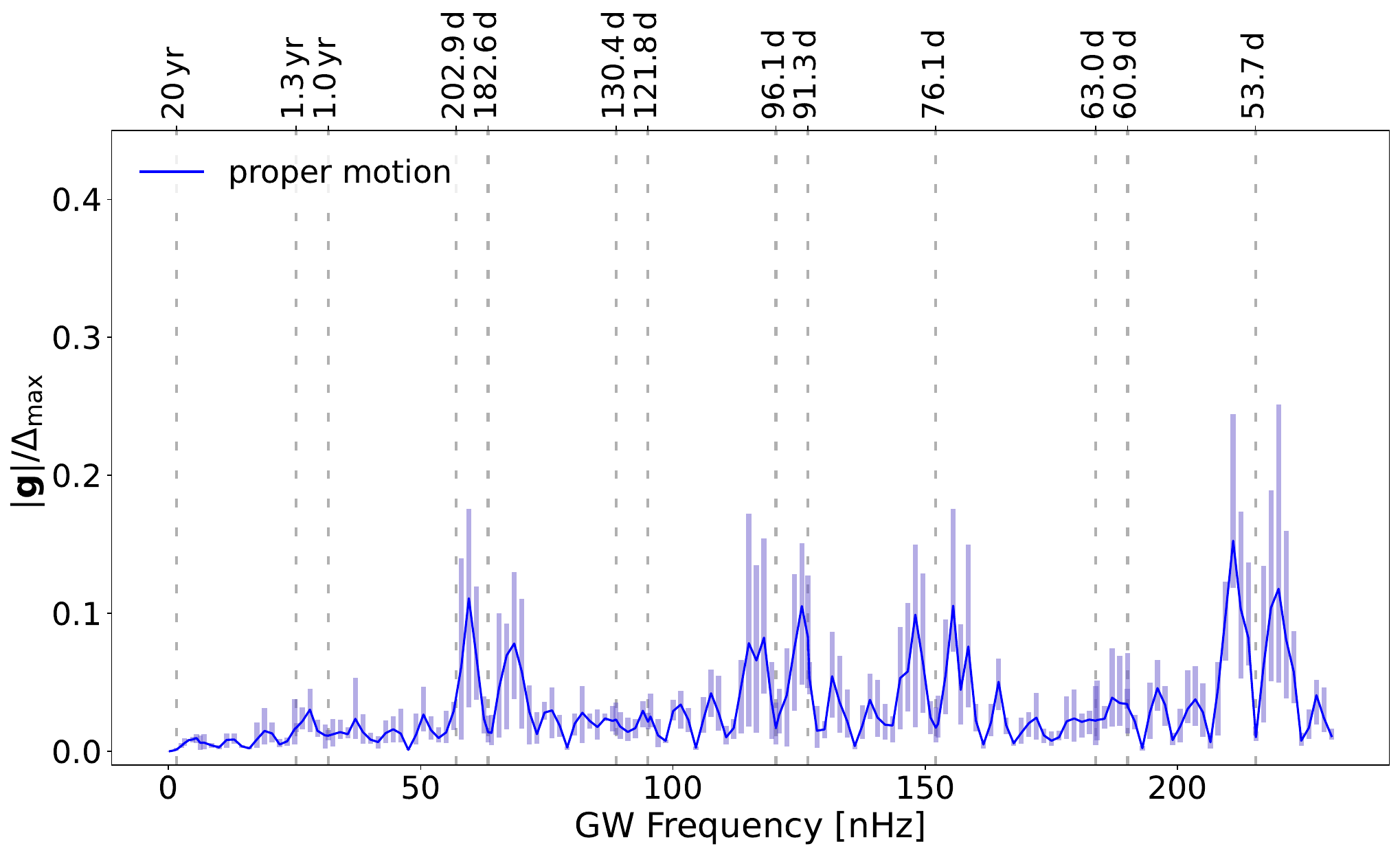}
 \caption{The absolute values of the glide $\vec{g}$ introduced by a GW in
   positions (upper panel) and proper motions (lower panel) of the
   astrometric sources. The coloured bars show the range of values 
  obtained in the five simulations for each frequency (see 
  Sect.~\ref{sec__methodology}); for improved visibility the coloured 
  lines connect the average statistics in the corresponding bars.
  The glide is computed from the spheroidal VSH coefficients of degree
  $\ell=1$ as described in Sect.~6 of \citeads{2021A&A...649A...9G}.
  In our case, with negligible frame rotation in the data, the total glide is
  $|\vec{g}|=(3/2)^{1/2}R_1$.
  \label{fig__glide}}
\end{figure}

The situation is different for the glide, corresponding to the spheroidal 
components of degree $\ell=1$ in the position and proper motion errors:
these components are not removed by the astrometric solution. 
Following the conventions in \citetads{2012A&A...547A..59M}
and \citetads{2021A&A...649A...9G}, the glide is here represented by a
vector $\vec{g}$, given by Eq.~(6) of the latter reference in terms of the 
spheroidal VSH components of degree $\ell=1$. The ICRS components
of $\vec{g}$ are therefore directly obtained from the VHS expansion
of the errors. However, a glide in the proper motion errors caused 
by a GW cannot be distinguished from the glide $\vec{a}/c$
caused by the acceleration $\vec{a}$ of the solar system barycentre. 
Measuring $\vec{a}$ is one of interesting results 
of \gaia-like global astrometry \citepads{2021A&A...649A...9G}, and
a relevant question then is how much glide might be produced by GWs.
Figure~\ref{fig__glide} shows the magnitude of the glide $|\vec{g}|$ 
in position and proper motion, normalised to $\Delta_{\rm max}$. 
For low-frequency GWs ($\Pgw\gtrsim\Tobs$), the glide in position as 
well as in proper motion is small compared to the effect at higher
frequencies. This is expected because the GW model expressed in VSH 
has no coefficients at degree $\ell=1$ \citepads{2018CQGra..35d5005K} 
and the GW signal at these frequencies goes almost unaltered to the 
source errors, as discussed in Sect.~\ref{sec__subsec_interaction_with_sourceparams}. 
For GWs of higher frequency ($\Pgw\lesssim\Tobs$), the glide in both 
position and proper motion is stronger, up to $\sim$0.25$\Delta_{\rm max}$.
But while the acceleration of the solar system barycentre produces a pure 
glide (with no VSH components of degree $\ell>1$), it is evident from 
Fig.~\ref{fig__heatmaps} that the glide component produced by a GW
in the proper motions is only a minor part of the total GW effect at that
frequency. For real \gaia\ measurements, the expected GW amplitudes 
and their astrometric effects are small and, therefore, the glide effect 
generated by them is negligible for the studies of the solar system acceleration.
(The glide in position does not seem to have any physical meaning, 
but is included in Fig.~\ref{fig__glide} for completeness and because 
it is relevant for understanding the peaks in the proper motion diagram.) 
 
\section{Concluding remarks}
\label{sec__conclusion}

In Appendix~\ref{section-GW-model} we present a new theoretical 
formulation of the astrometric effects of a plane GW. 
While mathematically equivalent to previously published models,  
the new formulation permits a simple geometrical interpretation
of the effects: to an astrometric observer, the GW signal appears 
as a synchronized elliptical motion of all sources on the sky. 
The eccentricity of the ellipses is the same for all sources, and 
depends only on the strain parameters, while the size and orientation 
of the ellipses depend also on the celestial position of the source 
relative to the propagation direction of the GW. A useful insight 
from this formulation is that the astrometric effects of a GW resemble, 
to a certain extent, the effects of astrometric binaries, but with the 
important difference that the GW affects all sources on the sky in a 
highly coordinated manner. The resulting pattern of motions for 
many millions of sources potentially makes a global astrometric 
survey mission, such as \gaia, a sensitive detector of GWs.

The detectability of the GW signal depends not only on the properties
of the GW and the accuracy of the astrometric measurements, but also
on the specific conditions under which the measurements are made. 
For example, it is clear from the description above that all astrometric 
sources within a small area of the sky are very similarly affected by 
the GW. Differential measurements in a small field of view are therefore 
not sensitive to the GW effects, whereas global or wide-angle 
astrometry might be. 

In this paper, we analyse the effects of a plane GW with constant 
frequency and strain parameters on a \gaia-like astrometric 
solution. We argue (Sect.~\ref{sec__general_influence}) that the
GW signal, like any global smooth astrometric signal, is 
observationally equivalent to certain time-dependent modifications
of the attitude and basic angle of the instrument. It is only the latter 
(basic-angle like) component of the GW signal that is potentially
observable. However, depending on the frequency of the GW,
that component could also be partly or fully absorbed by the 
astrometric solution in the form of (very small but systematic) errors
in the positions, proper motions, and parallaxes of the astrometric
sources. For the vast majority of sources (ordinary stars), these 
errors are undetectable because the true (unperturbed) values
of the astrometric parameters are not known to sufficient accuracy.
Only for sources at cosmological distances (quasars) might one 
presume to know the true proper motions and parallaxes.
The part of the basic-angle like component that is not absorbed
by the astrometric source parameters creates systematic patterns 
in the residuals that are, at least in principle, possible to detect
also for ordinary stars.

In Sect.~\ref{sec__subsec_interaction_with_sourceparams} we provide 
an analytical model for the astrometric errors in positions and
proper motions, as well as the astrometric residuals, for GWs in the
low frequency regime ($\Pgw\gtrsim\Tobs$). This simple model,
supported by simulations in this frequency range
(Sect.~\ref{sec__errors_slowGWs}), is particularly relevant for the
search of GW signals in the proper motions of quasars.  Previously, 
the general assumption has been that GW signals with such low 
frequencies would manifest themselves solely as systematic 
patterns in the quasar proper motions. However, we find that this 
is not the case. The position parameters can also absorb major parts 
of the signal depending of the phase of the GW. Even more important, 
the magnitude of the GW-induced effects in proper motions (and positions)
depends on the GW frequency even for $\Pgw\gtrsim\Tobs$. The analytical
model defines the sensitivity function for the study of primordial GWs
using quasar astrometry. On the other hand, we find that also in
this regime part of the GW signal remains in the residuals. This
may allow one to use not only the quasar proper motions but also the
astrometric residuals of ordinary stars to search for GWs with 
$\Pgw\gtrsim\Tobs$.

As explained in Sect.~\ref{sec__methodology}, we have performed a large
number of numerical simulations with a realistic \gaia-like
observational setup, using the simulation software AGISLab
\citepads{2012A&A...543A..15H}. The GW periods investigated by
simulations range 
from $\Pgw=100$\,yr to $\simeq\,$50\,d. The reasoning behind this choice
of GW periods is explained in Appendix~\ref{apx__freqRange}. The
simulations demonstrate the complicated character of the GW-induced
errors in a \gaia-like astrometric solution, as discussed in some detail in
Sects.~\ref{sec__source_errors} and \ref{sec__spatial_features}
and in Appendices~\ref{section-examples} and \ref{section-shvsh}.
In particular, our simulations show that a significant part of the
GW signal is absorbed by the astrometric source parameters
(positions, proper motions, and parallaxes) even for GWs with
periods considerably shorter than the time span of observations. 

Although the GW-induced astrometric errors in general cannot be
detected (except possibly for quasars), they are nevertheless interesting
for understanding the fundamental limits of astrometric missions. 
The simulations described here, covering a wide range of GW parameters
and assuming a realistic observational setup, can be used to estimate 
the astrometric noise floor set by the expected GW background (GWB) in the 
relevant frequency ranges.
As demonstrated e.g. in \citetads{2024ApJ...966..105A} and references
therein, the recent Pulsar Timing results show that the GWB likely
exists. This GWB may represent a fundamental limit for future
astrometric projects aiming at much higher accuracy than \gaia.
Both the detectability of the GWB using astrometry and the accuracy limits
imposed by it will be discussed elsewhere.

In this study, we focus on the effects of individual GW
sources that may possibly be detected by astrometry.
Perhaps the most important finding from our simulations is the fact
that, for all frequencies with $\Pgw\lesssim\Tobs$, a significant part
of the GW signal remains in the residuals of the astrometric solution. 
This strengthens our belief that a GW signal, if present 
at a sufficient amplitude, could be detected by a posterior analysis 
of the residuals in the standard astrometric solution. This is a very 
interesting and important task. A separate publication will be devoted 
to the formulation and demonstration of a dedicated GW search 
algorithm in astrometric data, and specifically in \gaia\ data.

Some important aspects of the GW search algorithms can be gleaned
already from the present analysis. 
The residual curves in Fig.~\ref{fig___freq_overview-att-res}
imply that the sensitivity function for astrometric searches of GW 
with $\Pgw\lesssim\Tobs$ is not flat. At frequencies where the
(normalised) astrometric errors are larger and the residuals smaller, 
the possibility to detect the GW signal is correspondingly reduced.
As discussed in Sect.~\ref{sec__freq_bands_high_errors}, this happens 
in particular around the GW frequencies related to the scanning law.
Depending on the position on the sky, the temporal sampling of the 
sources allows the astrometric model to absorb a greater or smaller
part of the GW signal. This could have important implications for
the design of a GW search algorithms based on the residuals: at 
certain GW frequencies, the astrometric residuals from certain parts
of the sky provide little or no information on a possible GW signal.

Considerations in Appendix~\ref{apx__freqRange} allow us to
draw another important conclusion concerning the GW search
algorithm, namely that it should not assume constant GW 
frequency over the whole observational period. That assumption
was expedient and revealing in the present study, but in a search 
algorithm it would effectively preclude the detection of possible 
GW sources with chirp masses above a certain limit, as illustrated 
in Fig.~\ref{fig__maxChirpMassStrain}.
This topic will be further discussed elsewhere.

A theoretical limit for the sensitivity of \gaia\ to a GW signal can be 
derived from the total astrometric weight of the mission, $\omega$, 
calculated by summing up the astrometric weights of the individual 
AL measurements (Eq.~62 in \citeads{2012A&A...538A..78L}). 

In the forthcoming \gaia\ DR4 data we estimate $\omega_\mathrm{GDR4}
\simeq 2\times 10^6\,\muas^{-2}$ if the $\simeq{}10^{12}$ residuals of
all sources with full astrometric solutions are used\footnote{We note
that the astrometric weight of an individual source strongly depends
on the properties of that source (e.g. its magnitude). Therefore, the
total astrometric weight is not a simple function of the number of
sources but crucially depends on various characteristics of the
entirety of observational data.}.  From the bottom left diagram in
Fig.~\ref{fig___freq_overview-att-res} and
Eq.~\eqref{Delta-amplitude-max} we have $\sigma_R\simeq
0.36\Delta_{\rm max}\simeq 0.15h$ (using that $\langle
e^2\rangle\simeq 8/15$ in the simulations).  The minimum detectable
strain at the $1\sigma$ level is therefore $h\gtrsim
\omega^{-1/2}/0.15\simeq 2\times 10^{-14}$.  Considering that the GW
model has seven parameters, and that the GW solution may include
additional nuisance parameters, the actual theoretical limit may
however be higher by a significant factor.

\begin{acknowledgements}
This work is financially supported by ESA grant 4000115263/15/NL/IB
and the German Aerospace Agency (Deutsches Zentrum f\"ur Luft- und
Raumfahrt e.V., DLR) under grants 50QG1402 and 50QG2202. We also
thank the Center for Information Services and High Performance (ZIH)
at TU Dresden for providing a considerable amount of computing time.
The work by LL is supported by the Swedish National Space Agency.
Various software products produced by the \gaia\ DPAC were used in this
work and are gratefully acknowledged. This especially concerns the
AGISLab framework which was developed by Berry Holl, David Hobbs,
Alex Bombrun and other \gaia\ DPAC members.
We also express our gratitude to Hagen Steidelm\"uller, who has maintained
and continued the development of AGISLab for many years and implemented
additional special features used for this project.
Finally, we also thank Enrico Gerlach, Jan Meichsner and Sven Zschocke for
fruitful discussions.
\end{acknowledgements}

\clearpage

\begin{appendix}

\section{Model for the astrometric effects of a plane GW}
\label{section-GW-model}

In this Appendix, we provide a concise description of the model for the 
astrometric effect of a GW and show that it results in an apparent elliptical 
motion of every source on the sky.

\subsection{Basic formulae}
\label{gw-model-general}

A plane continuous GW with constant frequency is completely described by seven
parameters: $\alpha_{\rm gw}$ and $\delta_{\rm gw}$ defining the direction
of propagation, the GW frequency $\nu$, and the four strain parameters
$\hpc$, $\hps$, $\htc$, $\hts$ encoding the strains and phases of the 
two polarisations (we follow here \citeads{2009agwd.book.....J}). 
As presented by \citetads{2018CQGra..35d5005K}, the astrometric effect 
$\delta\vec{u}$ at the point $\vec{u}$ ($|\vec{u}|=1$) with coordinates 
$(\alpha,\delta)$ can be written
\begin{equation}
\label{approximated-model}
\delta u^i=\delta^i_+\left(\hpc\cos\Phi+\hps\sin\Phi\right)
+\delta^i_\times\left(\htc\cos\Phi+\hts\sin\Phi\right)\,,
\end{equation}
where
\begin{eqnarray}
\label{Phase}
\Phi&=&2\pi\nu\left(t-t_{\rm ref}-{1\over c}\vec{p}\cdot\vec{x}_{\rm obs}(t)\right)\,,\\
\label{delta-i-+}
\delta^i_+&=&f^{ijk}\left(\vec{P}\vec{e}^+\vec{P}^{\rm T}\right)_{jk}
\,,\\
\label{delta-i-times}
\delta^i_\times&=&f^{ijk}\left(\vec{P}\vec{e}^\times\vec{P}^{\rm T}\right)_{jk}
\,,\\
\label{fijk}
f^{ijk}&=&{1\over 2}\left({u^i+p^i\over 1+\vec{u}\cdot\vec{p}}u^ju^k-\delta^{ij}u^k\right)\,,\\
\label{vec-u}
\vec{u}&=&
\begin{pmatrix}
~\cos\alpha\cos\delta~\\
\sin\alpha\cos\delta\\
\sin\delta
\end{pmatrix}
\,,\\
\label{vec-p}
\vec{p}&=&
\begin{pmatrix}
~\cos\alpha_{\rm gw}\cos\delta_{\rm gw}~\\
\sin\alpha_{\rm gw}\cos\delta_{\rm gw}\\
\sin\delta_{\rm gw}
\end{pmatrix}
\,,\\
\vec{e}^+&=&
\left(\,\,
\begin{array}{rrr}
1& \phantom{--}0& \phantom{--}0\\[1pt]
0&                    -1&                      0\\[1pt]
0&                      0&                      0
\end{array}\,\,
\right)\,,\\
\vec{e}^\times&=&
\left(\,\,
\begin{array}{rrr}
0& \phantom{--}1& \phantom{--}0\\[1pt]
1&                      0&                      0\\[1pt]
0&                      0&                      0
\end{array}\,\,
\right)\,,\\
\label{matrix-P}
\vec{P}&=&
\left(\hspace*{-4pt}
{\arraycolsep=3pt
\begin{array}{@{\hspace{5pt}}rrr@{\hspace{5pt}}}
-\sin\alpha_{\rm gw} & -\cos\alpha_{\rm gw}\sin\delta_{\rm gw} & \cos\alpha_{\rm gw}\cos\delta_{\rm gw}\\
\cos\alpha_{\rm gw} & -\sin\alpha_{\rm gw}\sin\delta_{\rm gw} & \sin\alpha_{\rm gw}\cos\delta_{\rm gw}\\
0 & \cos\delta_{\rm gw} & \sin\delta_{\rm gw}
\end{array}
}
\hspace*{-4pt}
\right)
\,.
\end{eqnarray} 
Here $t$ is the time of observation, $\vec{x}_{\rm obs}(t)$ the barycentric 
position of the observer at the moment of observation, and the phases of 
the strain parameters are defined with respect to the reference epoch $t_{\rm ref}$. 

The R{\o}mer correction $-c^{-1}\vec{p}\cdot\vec{x}_{\rm obs}(t)$ 
in Eq.~\eqref{Phase} reflects
the fact that the GW phase $\Phi$ at the location of observer is offset
from the GW phase $2\pi\nu(t-t_{\rm ref})$ at the solar system barycentre. 
The R{\o}mer correction is numerically small ($\lesssim 500$\,s at the position of
the Earth of \gaia), and can often be neglected in theoretical considerations.

$\vec{P}$ is the rotation matrix between the reference system in which the 
gravitational wave propagates in the $+z$ direction, and the celestial reference
system in which the propagation direction is $\vec{p}$. 

The dimensionless strain parameters $\hpc$, $\hps$,
$\htc$, and $\hts$ describe the magnitude of the astrometric effects of the GW.
In the astrometric context they are conveniently expressed in angular units 
(e.g.\ mas or \muas).

We remind that the model described here depends only on the gravitational field
of a GW at the observer. As discussed for example in \citetads{2011PhRvD..83b4024B} 
and \citetads{2018CQGra..35d5005K}, one can completely ignore the
`source term', that is the effect depending on the GW at the astrometric sources.

Finally, we note that Eq.~\eqref{approximated-model} describes the
variation of the observable direction towards an astrometric source
as seen by an observer at position $\vec{x}_{\rm obs}(t)$ and having
zero velocity relative to the Barycentric Celestial Reference System
(BCRS; \citeads{2003AJ....126.2687S}). The corresponding direction
observed by an observer moving relative to the BCRS can be computed
using the usual relativistic aberration formulas, for example Eq.~(10)
of \citetads{2003AJ....125.1580K}.

\subsection{Elliptic motion on the sky caused by a GW}
\label{gw-model-ellipses}
The model above can be written in a way that gives an important insight into the 
nature of the effect, namely that the astrometric effect of a GW consists of 
coordinated elliptic motions of all sources on the sky. One can derive the following 
simple representation of the GW effect:
\begin{gather}
  \label{delta=D.b}
  \begin{pmatrix}
 \delta\alpha^*\\
 \delta\delta
 \end{pmatrix}
 =
\begin{pmatrix}
 \delta\vec{u}\cdot\vec{e}_\alpha\\
 \delta\vec{u}\cdot\vec{e}_\delta
 \end{pmatrix}
= \vec{D}\vec{b}\,,\\
\label{e_alpha}
\vec{e}_\alpha={1\over\cos\delta}{\partial\over\partial\alpha}\vec{u}
=
\begin{pmatrix}
  -\sin\alpha\\
  \phantom{-}\cos\alpha\\
  0
\end{pmatrix}\,,\\
\label{e_delta}
\vec{e}_\delta=\vec{u}\times\vec{e}_\alpha={\partial\over\partial\delta}\vec{u}
=\begin{pmatrix}
-\cos\alpha\sin\delta\\
-\sin\alpha\sin\delta\\
\cos\delta
\end{pmatrix}\,,
\end{gather}
where matrix $\vec{D}$ is detailed below, and the vector
\begin{equation}
\label{vector-b}
\vec{b}=
  \begin{pmatrix}
    \hps \sin\Phi + \hpc \cos\Phi\\[3pt]
    \hts \sin\Phi + \htc \cos\Phi
  \end{pmatrix}
\end{equation}
describes an ellipse with semi-major axis $a$,
semi-minor axis $b$, eccentricity $e$ and position
angle $\phi$ (counted from the positive 'horizontal' axis defined by $\vec{e}_\alpha$ to the major axis): 
\begin{eqnarray}
  \label{semi-major-a}
  a&=&{1\over \sqrt{2}}h\sqrt{1+{\cal S}}=h{1\over \sqrt{2-e^2}}\,,\\
  b&\equiv&a \sqrt{1-e^2}={1\over \sqrt{2}}h\sqrt{1-{\cal S}}\,,\\
  \label{eccentricity-e}
  e&=&\sqrt{2{\cal S}\over 1+{\cal S}}\,,\\
  \label{position-angle-phi}
    \phi&=&\begin{cases}
      \displaystyle{\arctan{-h_+^2+h_\times^2+h^2{\cal S}\over 2{\cal B}}}, &{\cal B}\neq 0\\
      0, & {\cal B}= 0,\ h_\times<h_+\\
      \pi/2,& {\cal B}=0,\ h_\times>h_+
\end{cases}\\
  \label{h}
  h^2&=&h_+^2+h_\times^2\,,\\
  h_+^2&=&{\left(\hpc\right)}^2+{\left(\hps\right)}^2\,,\\
  h_\times^2&=&{\left(\htc\right)}^2+{\left(\hts\right)}^2\,,\\
  {\cal D}&=&\htc\hps-\hts\hpc\,,\\
  {\cal S}&=&\sqrt{1-4{\cal D}^2/h^4}\,,\\
  {\cal B}&=&\htc\hpc+\hts\hps\,.   
\end{eqnarray}  
One can show that $0\le{\cal S}\le1$ and, therefore,
$h/\sqrt{2}\le a \le h$ and $0\le b\le h/\sqrt{2}$.
For ${\cal B}=0$ and $h_\times=h_+$ one gets a
circle with $a=b=h/\sqrt{2}$ and undefined $\phi$.

The matrix $\vec{D}$ in Eq.~\eqref{delta=D.b} is obtained as
\begin{equation}
\label{D-split}
\vec{D}={1\over 2}\sin\theta\ \vec{R}\,,\\
\end{equation}
where
\begin{equation}
\label{angular-distance-theta}
\theta=\arccos\left(\sin\delta\sin\delta_{\rm gw}+\cos\delta\cos\delta_{\rm gw}\cos\Delta\alpha\right)
\end{equation}
is the angle ($0\le\theta\le\pi$) between the direction of observation 
$(\alpha,\delta)$ and the direction of GW propagation $(\alpha_{\rm gw},\delta_{\rm gw})$,
and $\vec{R}$ is the rotation matrix
\begin{eqnarray}
\label{matrix-R}
\vec{R}&=&{\cal N}^{-1}\,
  \begin{pmatrix}
   {\cal F} &\ \ -{\cal G} \\
   {\cal G} &\ \ {\cal F}
  \end{pmatrix}\,,\\
{\cal N}&=&4\sin\theta\,(1 + \cos\theta)\,,\\
{\cal F}&=&\cos\delta\,(-3 + \cos2\delta_{\rm gw}-4\sin\delta\sin\delta_{\rm gw})\sin2\Delta\alpha
\nonumber\\
&&
-4\cos\delta_{\rm gw}(\cos2\delta - \sin\delta\sin\delta_{\rm gw})\sin\Delta\alpha\,,\\
{\cal G}&=&
-2\cos\delta\,\bigl(2\sin\delta_{\rm gw}+(1+\sin^2\delta_{\rm gw})\sin\delta\bigr)\cos2\Delta\alpha
\nonumber\\
&&
+4\cos\delta_{\rm gw}(\sin\delta-\cos2\delta\sin\delta_{\rm gw})\cos\Delta\alpha
\nonumber\\
&&
+3\cos^2\delta_{\rm gw}\sin2\delta\,
\end{eqnarray}  
with $\Delta\alpha=\alpha-\alpha_{\rm gw}$. 
The matrix $\vec{R}$ is orthogonal with $\det\vec{R}=1$, so that 
${\cal F}^2+{\cal G}^2={\cal N}^2$. The
signs of ${\cal F}$ and ${\cal G}$ are non-trivial and necessitate the
explicit formulas for both ${\cal F}$ and ${\cal G}$ given above. When
using the formulas for $\vec{R}$, special care must be taken when
$\theta$ is close of 0 or $\pi$. For $\theta=0$ and $\theta=\pi$ we
have $\vec{D}=\vec{0}$, while $\vec{R}$ is undefined.

The astrometric effect of the GW can therefore be described as an
apparent elliptic motion of the source in the plane of the sky, with 
semi-major axis
\begin{equation}
\label{Delta-amplitude}
\Delta={1\over2}a\sin\theta\,,
\end{equation}
eccentricity $e$, and orientation defined by a combination of the angle 
$\phi$ (which depends only on the strain parameters) and the rotation 
matrix $\vec{R}$ (which depends only on the position of the source and
the direction of GW propagation). Here $a$, $e$, $\phi$, $\theta$, and 
$\vec{R}$ are given respectively by Eqs.~\eqref{semi-major-a},
\eqref{eccentricity-e}, \eqref{position-angle-phi}, \eqref{angular-distance-theta}, 
and \eqref{matrix-R}.

The eccentricity of the ellipse is the same for all sources, and the period 
of the effect equals the GW period $\Pgw = \nu^{-1}$. Depending
on the ratio $\Tobs/\Pgw$, the whole ellipse or only part of it is observed
(here $\Tobs$ is the mission duration). 
The unperturbed position of the source is at the centre of the
ellipse and the motion is such that the phase angle is almost linear with time,
$\Phi\simeq 2\pi\nu(t-t_{\rm ref})$. 
The R{\o}mer correction to the phase in Eq.~\eqref{Phase} 
results in a small perturbation of the position of the source along 
the ellipse, depending also on the direction of GW propagation ($\vec{p}$).

\begin{figure}[htb]
 \centering
 \includegraphics[keepaspectratio,width=\hsize]{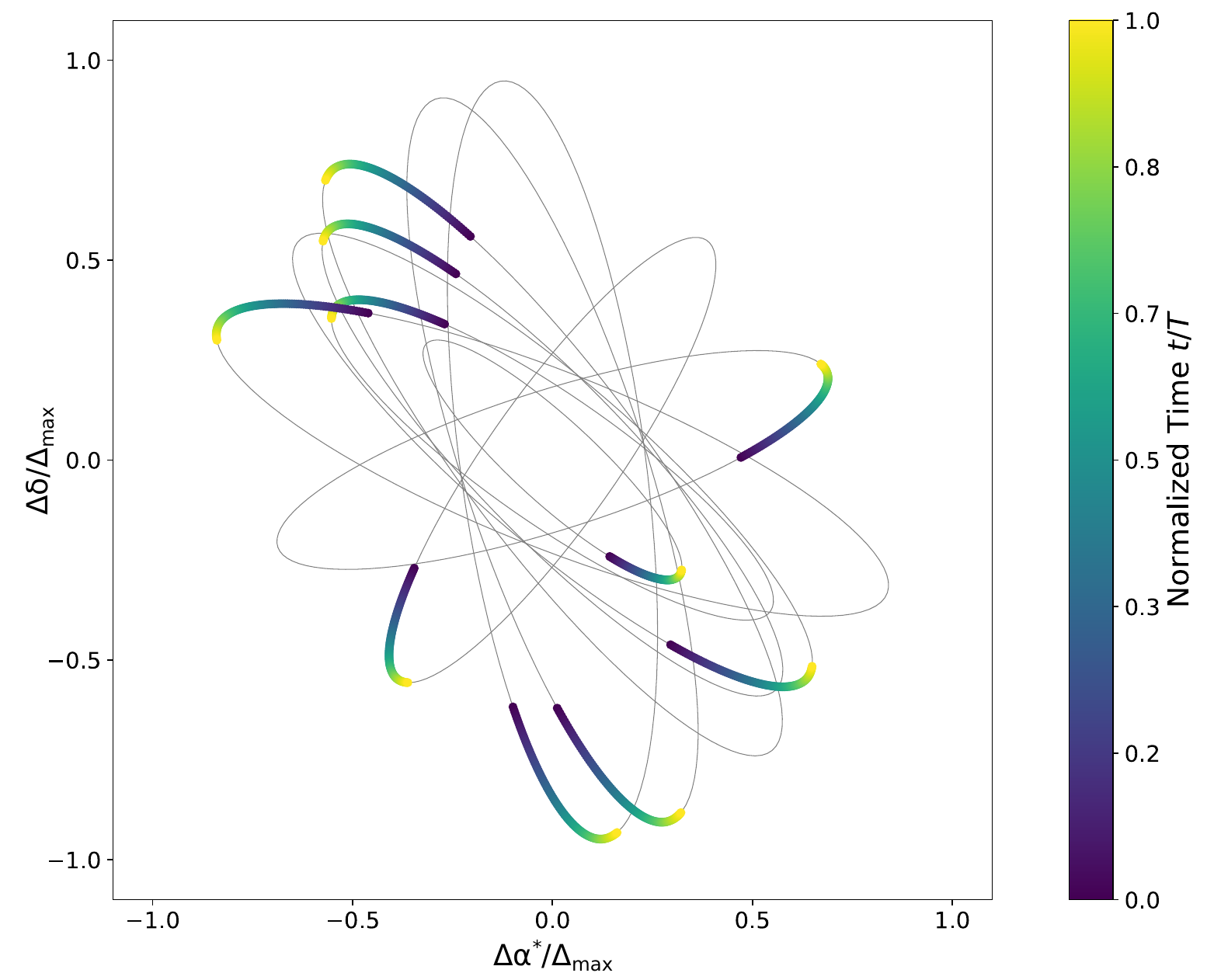}\\
 \includegraphics[keepaspectratio,width=\hsize]{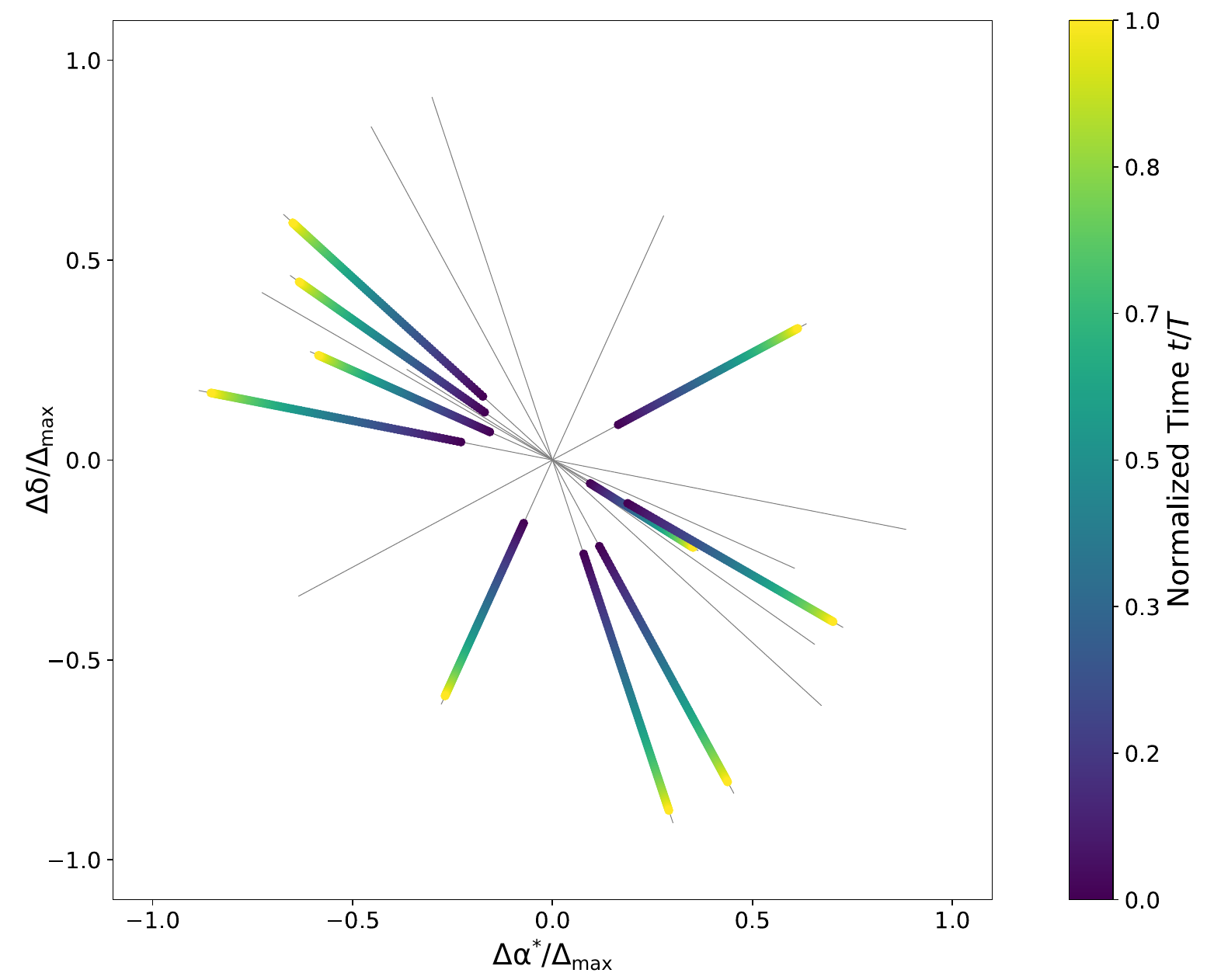}
 \caption{Positional offsets of ten randomly chosen astrometric sources,
   as produced by two GWs propagating in the same direction and having the
   same frequency, but with different strain parameters $\hpc$, $\hps$, $\htc$, and $\hts$. 
   \textit{Top:} for a GW with randomly chosen strain parameters, yielding the 
   (random) axis ratio $\simeq 0.24$ (corresponding to $e\simeq 0.97$). 
   \textit{Bottom:} for a GW with $\hpc=\hps=\htc=\hts$,
   yielding an axis ratio $=0$ (corresponding to $e=1$). The offsets are 
   shown in the local tangential coordinates of respective source, normalised by 
   $\Delta_{\rm max}$, and colour-coded by time normalised to the mission 
   duration $\Tobs$. The GW period is $\Pgw=6T$. The grey lines show the complete
   path that would be traced for $\Tobs\ge\Pgw$.
   None of the randomly selected sources is exactly at $90^\circ$ angle 
   from the GW source, so none of the ellipses attains the maximum semi-major 
   axis $\Delta_{\rm max}$. (The R{\o}mer correction was neglected in these plots, 
   but would hardly be visible if it were included.)
 \label{fig__example_ellipsis}}
\end{figure}

Figure~\ref{fig__example_ellipsis} depicts the positional offsets 
over time for ten randomly selected astrometric sources at different
celestial positions. For each source the effect is plotted in the
respective local plane coordinates $\Delta\alpha*$ (along $\vec{e}_\alpha$) 
and $\Delta\delta$ (along $\vec{e}_\delta$). The rotation and scaling of 
the different ellipses, caused by $\vec{R}$ and ${1\over 2}\sin\theta$, are
clearly seen.

From Eq.~\eqref{Delta-amplitude} it is seen that the maximal amplitude of 
the astrometric effect for a given GW,
\begin{equation}
\label{Delta-amplitude-max}
\Delta_{\rm max}
={1\over 2}a = {h \over 2\sqrt{2-e^2}}\,,
\end{equation}
is reached for sources located normal the direction of GW propagation 
($\sin\theta=1$). From Eq.~\eqref{semi-major-a} we have
$\Delta_{\rm max}\le h/2$.

\clearpage

\section{Statistical properties of the GW-induced signal}
\label{sec__GW-signal}

This Appendix gives some analytical results concerning the statistical 
properties of the astrometric GW signal itself, ignoring its possible 
interaction with the astrometric solution (as discussed in 
Sect.~\ref{sec__general_influence}). We are specifically 
interested in the statistics of $\delta\vec{u}\cdot\vec{s}$, where 
$\delta\vec{u}$ is the instantaneous GW signal for a source at position
$\vec{u}$, and $\vec{s}$ is a unit vector in the tangent plane of the 
sky at $\vec{u}$ (so $\vec{u}\cdot\vec{s}=0$). In the context of \gaia, 
$\vec{s}$ could be the AL or AC direction for a particular observation of 
the source, in which case $\delta\vec{u}\cdot\vec{s}$ is the AL or AC 
component of the GW signal at the source. The 
problem becomes analytically tractable by considering a random time of 
observation (and hence a random phase $\Phi$ of the GW), a random 
propagation direction of the GW, and a randomly selected direction 
$\vec{s}$. By disregarding a number of incidental factors, such as the 
scanning law and the phase of the GW, the resulting statistics are 
conveniently suited for comparison with the error statistics of the 
astrometric solution, and especially its residuals.

In Appendix~\ref{section-GW-model} we demonstrated that
$|\delta\vec{u}|\le\Delta_{\rm max}$. The normalised shift
\begin{equation}
\label{kappa}
\kappa=\delta\vec{u}\cdot\vec{s}/\Delta_{\rm max}
\end{equation}
therefore lies between $-1$ and $+1$.
Using Eqs.~\eqref{delta=D.b}, \eqref{vector-b}, \eqref{D-split}, and
\eqref{Delta-amplitude-max} we find that the normalised shift can be
written $\kappa = r(\Phi)\sin\theta\cos\varphi$, where 
$r(\Phi)=|\vec{b}|/a$, $\theta$ is the angle between the direction 
of GW propagation and the source, and $\varphi$ is the angle 
between the vectors $\vec{s}$ and $\delta\vec{u}$. As before,
$\Phi$ is the GW phase, $\vec{b}$ is the vector given by 
Eq.~\eqref{vector-b} and $a$ is the semi-major axis given by 
Eq.~\eqref{semi-major-a}. 

The distribution of $\kappa = r(\Phi)\sin\theta\cos\varphi$ is derived 
as follows. Using that $\cos\theta$ is uniformly distributed in $[-1,1]$,
and that $\varphi$ is uniformly distributed in $[0,2\pi)$, we see that 
the factor $\sin\theta\cos\varphi$ is uniformly distributed in $[-1,1]$. 
Concerning $r(\Phi)$, we find from Eq.~\eqref{vector-b} that it can be 
written as
\begin{equation}
	\label{r-Phi}
	r(\Phi)=
\sqrt{1-e^2\,\sin^2\left(\Phi-\psi/2\right)}\,,      
\end{equation}
where $e$ is the eccentricity given by Eq.~\eqref{eccentricity-e}, and 
the phase shift $\psi/2$ is defined by 
\begin{eqnarray}
	\label{r-Phi-A}
	\cos\psi&=&{2-e^2\over h^2\,e^2}\,\left({\left(\hpc\right)}^2-{\left(\hps\right)}^2+{\left(\htc\right)}^2-{\left(\hts\right)}^2\right)\,,
	\\
	\label{r-Phi-B}
	\sin\psi&=&2\,{2-e^2\over h^2\,e^2}\,\left(\hpc\,\hps+\htc\,\hts\right)\,.
\end{eqnarray}
We note that $\sqrt{1-e^2}\le r(\Phi)\le 1$ for any strain parameters. 
Because the phase $\Phi$ is uniformly distributed in $[0,2\pi)$, the
constant phase shift $\psi/2$ in the right member of Eq.~\eqref{r-Phi}
obviously plays no role for the distribution of $r(\Phi)$: $\Phi$ and 
$\Phi-\psi/2$ (modulo $2\pi$) are both uniformly distributed 
in $[0,2\pi)$.
Combining the above results, we find that the probability density 
function (PDF) of $\kappa$ for $\left|\kappa\right|\le 1$ reads
\begin{equation}
  {\rm PDF}_\kappa={1\over\pi}\times
  \begin {cases}
      K(e)\,,& \left|\kappa\right|\le\sqrt{1-e^2}\,,\\[6pt]
      F\left(\arcsin\left(e^{-1}\sqrt{1-\kappa^2}\,\right),\,e\right)\,, 
      &\left|\kappa\right|>\sqrt{1-e^2}\,,
      \end{cases}
    \label{eq-PDF}
\end{equation}  
where $F(x,k)=\int_0^x\left(1-k^2\sin^2\phi\right)^{-1/2}\text{d}\phi$ is
the (incomplete) elliptic integral of the first kind and
$K(k)=F(\pi/2,k)$ is the complete elliptic integral of the first kind. 
(For $\left|\kappa\right|>1$ we obviously have ${\rm PDF}_\kappa=0$.)
To derive Eq.~\eqref{eq-PDF} one needs to use several results and
properties of the elliptic integrals, given e.g.\ in
\citetads{1954MitAG...5...99B} and \citetads{1955htf..book.....B}.

Equation~\eqref{eq-PDF} demonstrates one very important result,
namely that under the given assumptions of random sampling, the 
distribution of $\kappa$ only depends on $e$. In particular, it does
not depend on either the GW direction or the GW frequency.
From Eqs.~\eqref{semi-major-a} and \eqref{Delta-amplitude-max}
it is then seen that the distribution of $\delta\vec{u}\cdot\vec{s}$ only
depends on $e$ and the total strain $h$.

Equations~\eqref{r-Phi-A}--\eqref{eq-PDF} become degenerate for
$e=0$; but in that case $r(\Phi)\equiv 1$, so $\kappa$ is uniformly
distributed in $[-1,1]$. More generally, for a given $0\le e\le 1$,
$\kappa$ is uniformly distributed for $\left|\kappa\right|\le \sqrt{1-e^2}$
and falls off quite fast outside this interval. The distribution is clearly
symmetric around $\kappa=0$. Therefore, both the mean value and 
the skewness of $\kappa$ are zero. The standard deviation 
of $\kappa$ is obtained by computing the corresponding integral 
of Eq.~\eqref{eq-PDF}, yielding
\begin{equation}
  \label{sigma-kappa}
\sigma_\kappa={\left(\int_{-\infty}^{\infty}\kappa^2\,{\rm PDF}_\kappa\,
\text{d}\kappa\,\right)}^{1/2}=\sqrt{2-e^2\over6}\,.
\end{equation}
By a similar analytical computation, the excess kurtosis of $\kappa$ is 
found to be $-{3\over 10}(16 - 16 e^2 + e^4)(2-e^2)^{-2}$.
The unnormalised projected signal $\delta\vec{u}\cdot\vec{s}$ 
has zero mean and standard deviation $h/(2\!\sqrt{6})$. 

In the case of \gaia, the above statistics derived for the random sampling
are valid for either the AL and AC components of the GW signal; the only 
difference would be in the interpretation of the angle $\phi$. The standard 
deviation in Eq.~(\ref{sigma-kappa}) nicely agrees with the results 
of Sect.~\ref{sec__errors_slowGWs}, namely that the 
GW signal is distributed between the source parameters, as in 
Eqs.~(\ref{normalized-sigma-pos})--(\ref{normalized-sigma-pm}), 
and the corresponding residuals, as in Eq.~(\ref{normalized-sigma-res}).

\begin{figure*}[htb]
  \centering
  	\includegraphics[keepaspectratio,width=0.49\hsize]{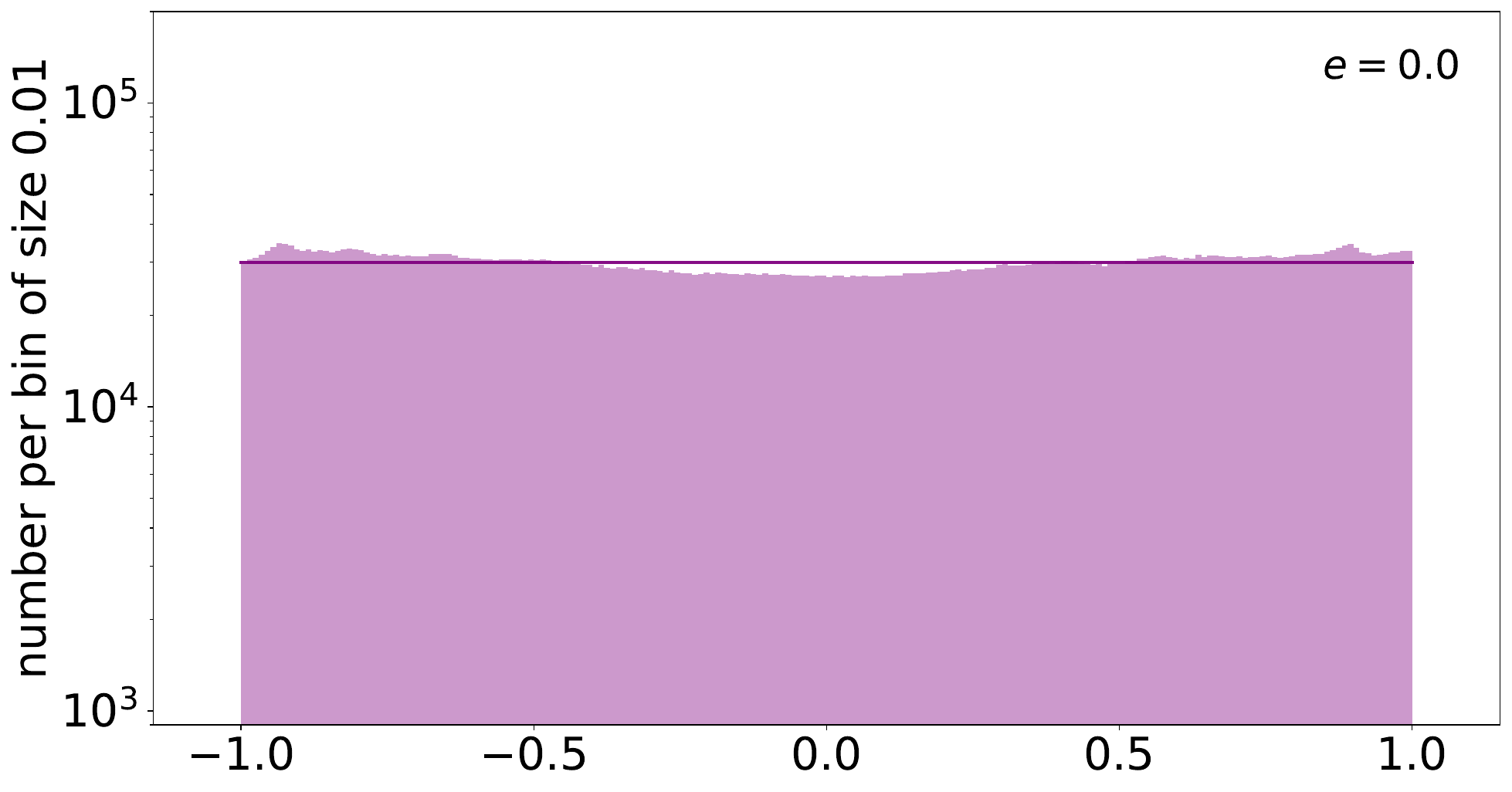}
	\includegraphics[keepaspectratio,width=0.49\hsize]{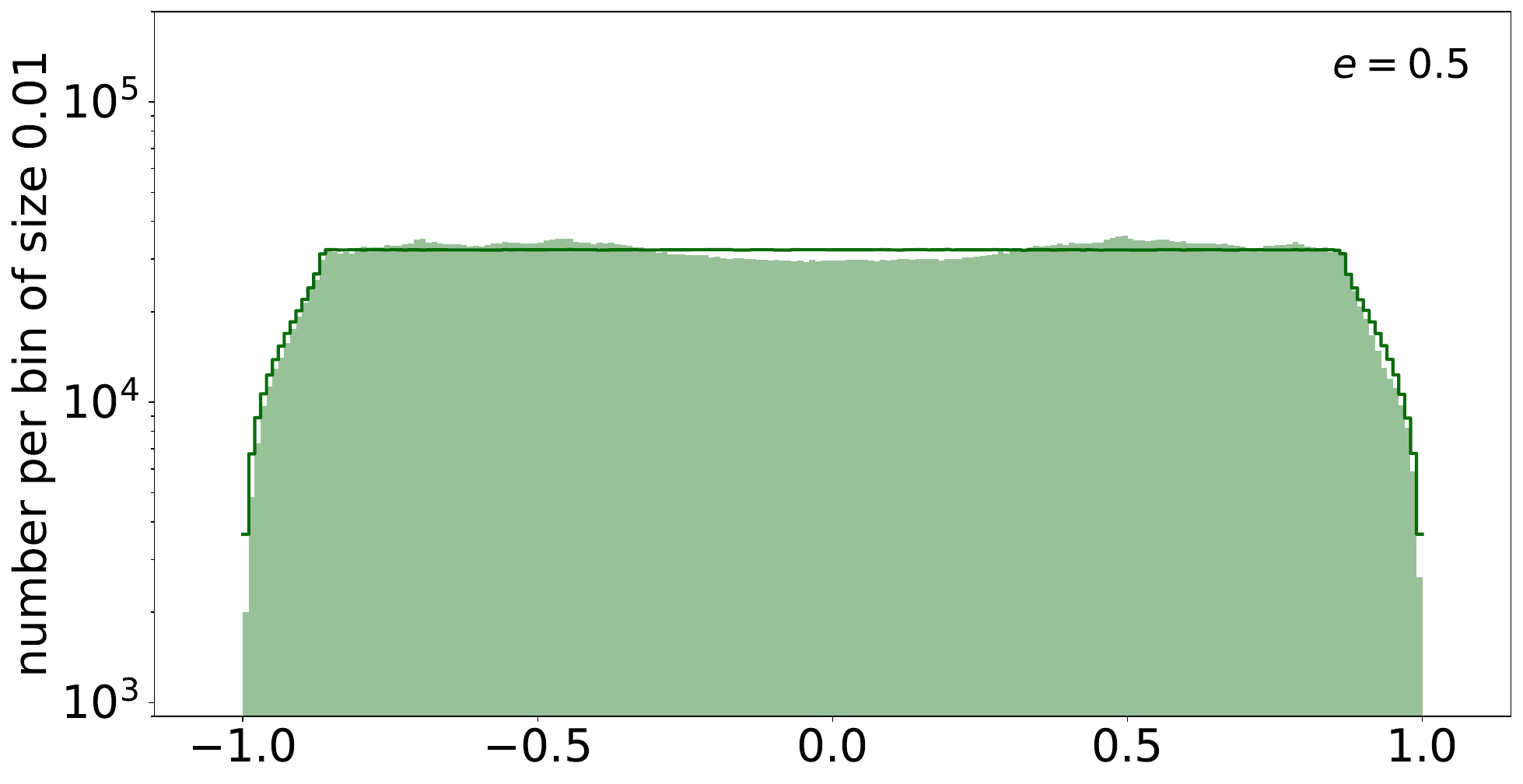}
	\includegraphics[keepaspectratio,width=0.49\hsize]{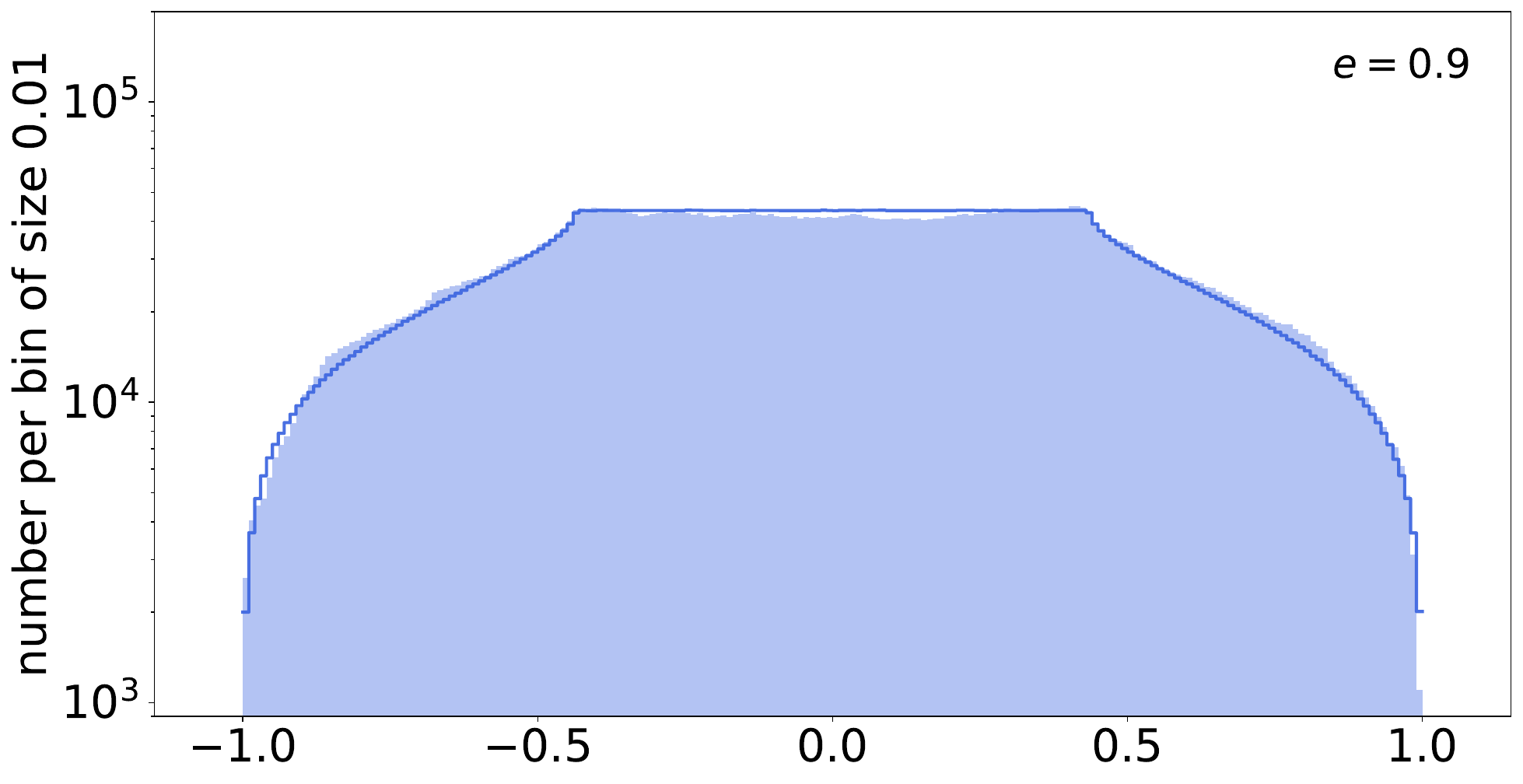}
	\includegraphics[keepaspectratio,width=0.49\hsize]{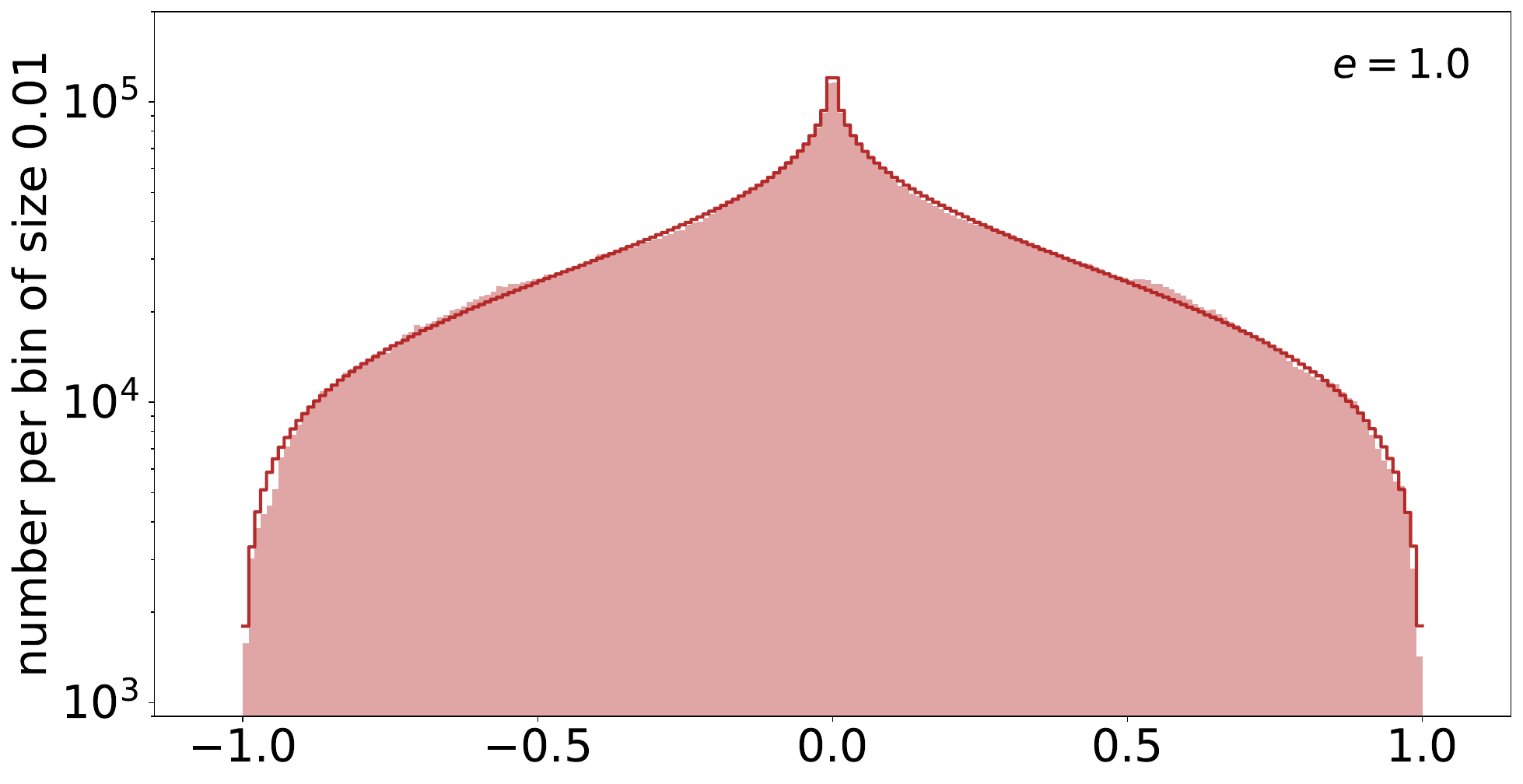}
        \vspace*{1mm}
        \hspace*{8mm}$\delta\vec{u}\cdot\vec{s}\,/\,\Delta_{\rm max}$
	\caption{Histograms of the normalised projections
          $\kappa=\delta\vec{u}\cdot\vec{s}/\Delta_{\rm max}$ of the GW
          signal for different values of the eccentricity $e$. The filled 
          bar charts represent the AL signal components in \gaia\ observations 
          simulated as described in the text. The lines show the
          corresponding distributions from the analytical formula in
          Eq.~\eqref{eq-PDF}.
		\label{fig___mock_AL_Histo}}
\end{figure*}

These theoretical results are confirmed by numerical simulations. We
generated $6\times10^6$ AL signal values. The GW frequency was set
to 200\,nHz, a random GW direction on the sky was taken, and the
nominal \gaia\ scanning law was used with 5\,years of observations. 
The distribution of the simulated values for four values of the eccentricity 
is depicted in Fig.~\ref{fig___mock_AL_Histo} and summary statistics
are given in Table~\ref{tab__gw_sig_stats}. The small differences between
the analytical model and the simulation results can be explained by the
use of a single GW direction in the simulations and the specific 
sampling by the \gaia\ scanning law, which is not completely random.

\begin{table*}[htb]\caption{Statistics of the astrometric GW signal for different
  eccentricities.}
  \label{tab__gw_sig_stats}
	\centering
	\begin{tabular}{lcrrcrrcrrcrr}
	\hline\hline
\noalign{\smallskip}
	&~& \multicolumn{2}{c}{$e=0$}   &~& \multicolumn{2}{c}{$e=0.5$}   &~& \multicolumn{2}{c}{$e=0.9$}   &~& \multicolumn{2}{c}{$e=1$}\\
Statistic	&& Theory                 & Num               && Theory                 & Num                   && Theory                 & Num                    && Theory                 & Num\\
\noalign{\smallskip}\hline\noalign{\smallskip}
	mean                && \numtwo{0.00}          & \numtwo{-0.000607} && \numtwo{0.00}          & \numtwo{0.000758}     && \numtwo{0.00}          & \numtwo{0.000782}      && \numtwo{0.00}          & \numtwo{0.000127}\\
	median              && \numtwo{0.000598275}   & \numtwo{0.000598} && \numtwo{0.000189}      & \numtwo{0.000539}     && \numtwo{0.000130}      & \numtwo{-0.00012}      && \numtwo{0.000024}      & \numtwo{-0.00015}\\
	standard deviation  && \numtwo{0.577350}      & \numtwo{0.592594}  && \numtwo{0.540062}      & \numtwo{0.542721}     && \numtwo{0.445346}      & \numtwo{0.453999}      && \numtwo{0.408248}      & \numtwo{0.409938}\\
	minimum/maximum     && $\mp\numtwo{0.999999}$ & $\mp\numtwo{1.0}$  && $\mp\numtwo{0.999974}$ & $\mp\numtwo{0.99989}$ && $\mp\numtwo{0.999923}$ & $\mp\numtwo{0.999959}$ && $\mp\numtwo{0.999923}$ & $\mp\numtwo{0.99995}$\\
	excess kurtosis     && \numtwo{-1.20}         & \numtwo{-1.268023} && \numtwo{-1.18163}      & \numtwo{-1.23092}    && \numtwo{-0.783017}     & \numtwo{-0.84798}      && \numtwo{-0.30}         & \numtwo{-0.351089}\\
	skewness            && \numtwo{0.00}          & \numtwo{0.000038}  && \numtwo{0.00}          & \numtwo{-0.00116}     && \numtwo{0.00}          & \numtwo{0.002223}      && \numtwo{0.00}          & \numtwo{0.001121}\\
	$Q_{0.01}$          && \numtwo{-0.98}         & \numtwo{-0.980053} && \numtwo{-0.937381}      & \numtwo{-0.93150}     && \numtwo{-0.905347}     & \numtwo{-0.90116}      && \numtwo{-0.898253}     & \numtwo{-0.891415}\\
	$Q_{0.05}$          && \numtwo{-0.90}         & \numtwo{-0.907626} && \numtwo{-0.838628}      & \numtwo{-0.83326}     && \numtwo{-0.734333}     & \numtwo{-0.74297}      && \numtwo{-0.712924}     & \numtwo{-0.710598}\\
	$Q_{0.10}$          && \numtwo{-0.80}         & \numtwo{-0.815765} && \numtwo{-0.745447}      & \numtwo{-0.74107}     && \numtwo{-0.595442}     & \numtwo{-0.60892}      && \numtwo{-0.559660}     & \numtwo{-0.564561}\\
	$Q_{0.25}$          && \numtwo{-0.50}         & \numtwo{-0.529484} && \numtwo{-0.465904}      & \numtwo{-0.47520}     && \numtwo{-0.344390}     & \numtwo{-0.35652}      && \numtwo{-0.258237}     & \numtwo{-0.262389}\\
	$Q_{0.75}$          && \numtwo{0.50}          & \numtwo{0.526712}  && \numtwo{0.465904}      & \numtwo{0.478900}     && \numtwo{0.344390}      & \numtwo{0.357627}      && \numtwo{0.258237}      & \numtwo{0.264814}\\
	$Q_{0.90}$          && \numtwo{0.80}          & \numtwo{0.815291}  && \numtwo{0.745447}      & \numtwo{0.743328}     && \numtwo{0.595442}      & \numtwo{0.610960}      && \numtwo{0.559660}      & \numtwo{0.563313}\\
	$Q_{0.95}$          && \numtwo{0.90}          & \numtwo{0.906873}  && \numtwo{0.838628}      & \numtwo{0.833398}     && \numtwo{0.734333}      & \numtwo{0.746429}      && \numtwo{0.712924}      & \numtwo{0.710106}\\
	$Q_{0.99}$          && \numtwo{0.98}          & \numtwo{0.981652}  && \numtwo{0.937381}      & \numtwo{0.931941}     && \numtwo{0.905347}      & \numtwo{0.906497}      && \numtwo{0.898253}      & \numtwo{0.894598}\\
\noalign{\smallskip}\hline
\end{tabular}
\tablefoot{The table gives the statistics listed in the first column of the quantity
$\kappa$ defined by Eq.~\eqref{kappa}. Four different eccentricities $e$ of the
GW signal are considered, as indicated in the top row. Values in the columns 
marked Theory are computed using the model distribution in Eq.~\eqref{eq-PDF}. 
Values in the columns marked Num are computed from numerical simulations 
as described in the text. $Q_p$ ($p=0.01,~0.05,~\dots,~0.99$) denotes the 
$p$-th quantile. All statistics are rounded to two decimal places. }
\end{table*}
 
\clearpage

\clearpage

\section{Relevant GW frequency range}
\label{apx__freqRange}

In this study we primarily consider a GW frequency range from
approximately 0.317\,nHz to ${\simeq}231$\,nHz, corresponding to 
periods between 100\,yr and 50.2\,d. Here we present and discuss
the rationale for this choice.

The lower frequency bound, corresponding to a maximum GW period 
of 100\,yr, is related to the mission duration, taken to be $\Tobs=5$\,yr. 
From the discussions in \citetads{2018CQGra..35d5005K} and in 
Sects.~\ref{sec__subsec_interaction_with_sourceparams} and
\ref{sec__errors_slowGWs} of this paper, it is clear that an increasing
fraction of the GW signal is absorbed by the astrometric parameters
for $\Pgw\gtrsim\Tobs$. According to Eqs.~\eqref{normalized-sigma-res}
and \eqref{sigma-kappa}
that fraction, in terms of signal variance, exceeds 0.997 for 
$\Pgw=25\Tobs$. Although that limit is somewhat arbitrary, there is 
obviously not much point in considering even longer periods.

The choice of upper frequency limit is also somewhat arbitrary, but
based on several practical and theoretical considerations. One purely
practical thing is that the amount of computations required for the
numerical simulations increases in direct proportion to the maximal
GW frequency. Thus, it is desirable not to choose an upper limit 
higher than motivated by other requirements. On the other hand, 
it is desirable to include, with some margin, the special frequency
$\nusl \simeq \numthree{184}$\,nHz related to the \gaia\ scanning
law, discussed in Sect.~\ref{sec__freq_bands_high_errors}. A third
requirement follows from the expected properties of supermassive 
black hole binaries that could generate GWs of sufficiently stable 
frequencies, and which are strong enough to be potentially 
detectable with \gaia. The rest of this Appendix is an attempt 
to quantify this requirement. 

One of the limitations of the present study is that the GW frequency 
is assumed to be constant throughout the length of the mission. 
We consequently want to avoid considering cases where the GW 
from an inspiralling binary system demonstrates considerable 
frequency drift during the period of observation $\Tobs=5$\,yr.
Section~8 of \citetads{2018CQGra..35d5005K} contains a brief 
discussion of the GW parameters expected from such binary 
systems, based on \citetads{2007arXiv0709.4682B} and 
\citetads{2009agwd.book.....J}. 

One of the limitations of the present study is that the GW frequency 
is assumed to be constant throughout the length of the mission. 
Since we are thus only interested in the regime where the GW frequency 
is changing very slowly, it is sufficient to consider the post-Newtonian 
theory. We also ignore alternative theories to General Relativity. 
Equation~(59) of \citetads{2018CQGra..35d5005K} then gives a
differential equation for the evolution of the GW frequency from a 
binary on a circular orbit. Integrating this formula from the initial
frequency $\nu_0$ at time $t_0$ gives
\begin{equation}
  \label{nu(t)-explicit}
  \nu(t)=\left[\nu_0^{-8/3}-{2^8\pi^{8/3}\over 5c^5}\,
  \bigl(G{\cal M}\,\bigr)^{5/3}(t-t_0)\right]^{-3/8}
  =\nu_0(1-S\!x)^{-3/8}\,,
\end{equation}
where ${\cal M}$ is the chirp mass of the binary and
\begin{align}
  S&={2^8\pi^{8/3}\over 5c^5}\,\bigl(G{\cal M}\,\bigr)^{5/3}\nu_0^{8/3}T\,,\\
  x&=(t-t_0)/T\,.
\end{align}
Requiring that the relative change of the GW frequency over the mission,
$(\nu(t_0+\Tobs)-\nu_0)/\nu_0$, does not exceed $\epsilon$ then leads
to an upper limit on the chirp mass,
\begin{equation}
  \label{eq-chirpM-max}
  \left({{\cal M}\over 10^9\,M_\odot}\right)^{5/3}
  <2.030\,\left(1-(1+\epsilon)^{-8/3}\right)\,\left({P_{\rm gw0}\over 50\,{\rm d}}\right)^{8/3}
  \left({\Tobs\over 5\,{\rm yr}}\right)^{-1},
\end{equation}
where $M_\odot$ is the mass of the Sun and $P_{\rm gw0}=1/\nu_0$. Using Eq.~(57) of \citetads{2018CQGra..35d5005K} now gives the maximal possible 
strain $h$ created by a GW source at the (luminosity) distance $r$: 
\begin{equation}
  \label{eq-strain-max}
  h<9.09\times 10^{-14}\,\left(1-(1+\epsilon)^{-8/3}\right)\,
  \left({P_{\rm gw0}\over 50\,{\rm d}}\right)^2\,
  \left({\Tobs\over 5\,{\rm yr}}\right)^{-1}\,
  \left({r\over 100\,{\rm Mpc}}\right)^{-1}\,.
\end{equation}
For a given maximal relative change $\epsilon$ of the GW frequency, 
Eqs.~\eqref{eq-chirpM-max} and \eqref{eq-strain-max} give the 
maximal allowed chirp mass ${\cal M}$ and the maximal expected strain as
functions of the initial GW period $P_{\rm gw0}$ and other parameters.
From Eq.~(60) of \citetads{2018CQGra..35d5005K} we note that
\begin{equation}
  \label{S-epsilon}
S=1-(1+\epsilon)^{-8/3}\,
\end{equation}
equals $\Tobs/\tau$, where $\tau$ is the time to coalescence (at $t_0+\tau$).
In the regime of interest here we have $S\ll 1$ or $\Tobs\ll\tau$.

From the point of view of data processing the relative frequency change 
$\epsilon$ may not be the most interesting quantity. In this study we 
have assumed a GW of constant frequency, and a more relevant quantity 
is then the coherence of the phase over the mission; in other words, how 
much the actual phase $\Phi(t)$ deviates from the best-fitting linearly 
varying phase $\overline\Phi(t)$. The true phase is defined by the differential
equation ${\rm d}\Phi/{\rm d}t=2\pi\nu(t)$ and the initial condition 
$\Phi(t_0)=\Phi_0$. Integrating Eq.~\eqref{nu(t)-explicit} gives
\begin{equation}
\Phi(t)=\Phi_0 + 2\pi\nu_0(t-t_0)\,\left({8\over 5}\,{1-(1-S\!x)^{5/8}\over S\!x}\right)\,.
\end{equation}
where, for small $S\!x$, the phase dilation factor in large brackets equals 
$1+{3\over 16}S\!x+O(S^2\!x^2)$.  
In the linear approximation it can be shown that the condition
\begin{equation}  
\left|\Phi-\overline\Phi\right|\le\varrho \,,
\end{equation}
where $\varrho$ is the maximal phase error in radians, 
is equivalent to a maximal relative change in frequency \begin{equation}
  \label{epsilon-varrho}
  \epsilon={24\over 7\pi}\,{P_{gw0}\over\Tobs}\,\varrho
\end{equation}  
or
\begin{equation}
  \label{epsilon-varrho-lqs-num}
  \epsilon=0.0697\,\varrho\,\left({P_{\rm gw0}\over 50\,{\rm d}}\right)\,
  \left({\Tobs\over 5\,{\rm yr}}\right)^{-1}\,.
\end{equation}
This $\epsilon$ can be used in Eqs.~\eqref{eq-chirpM-max} and
\eqref{eq-strain-max} to derive the maximal chirp mass and maximal
expected strain for a given maximal phase error $\varrho$ in the
constant frequency model.

\begin{figure}[htb]
	\centering
	\includegraphics[keepaspectratio,width=\hsize]{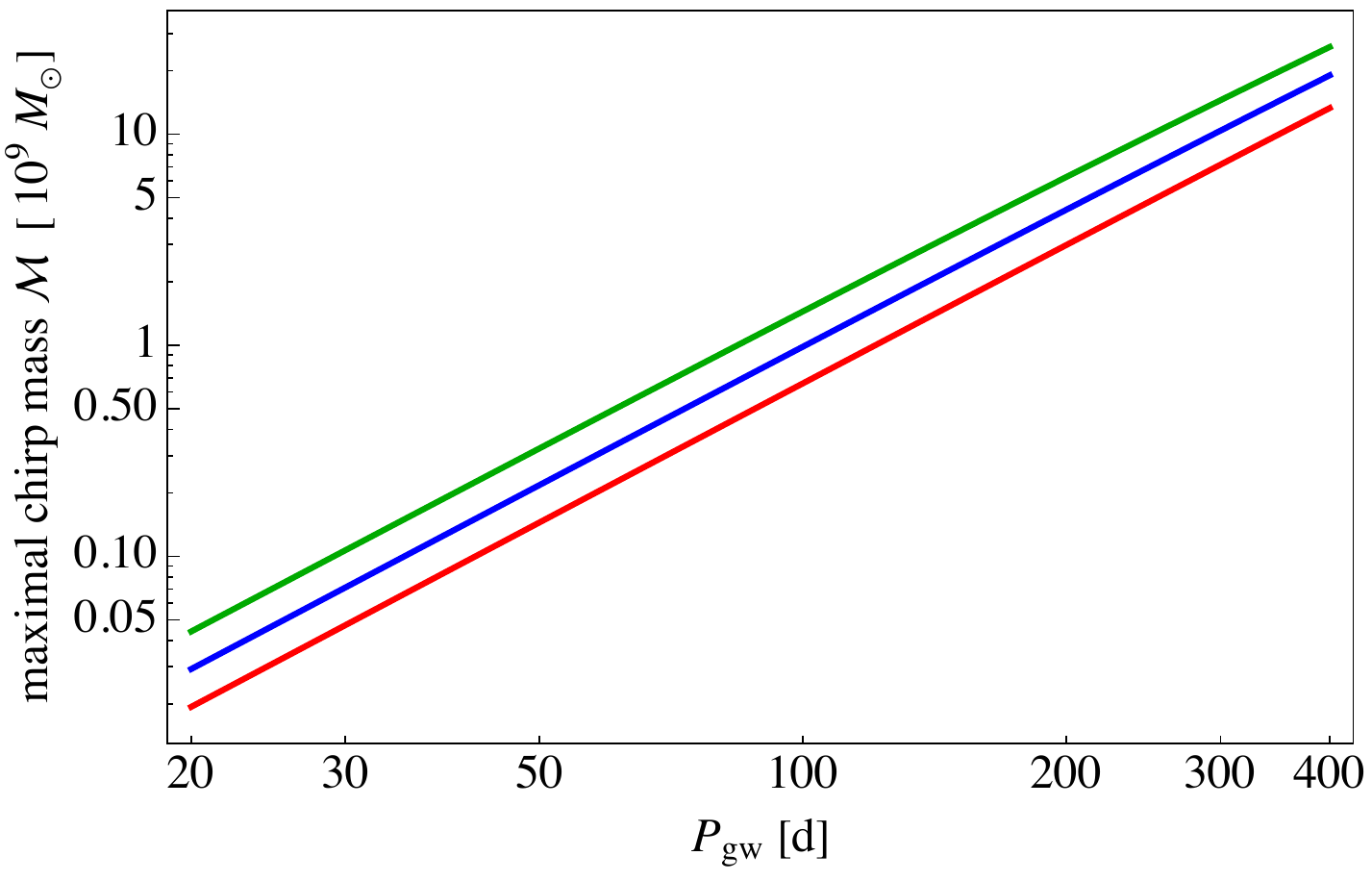}
\includegraphics[keepaspectratio,width=\hsize]{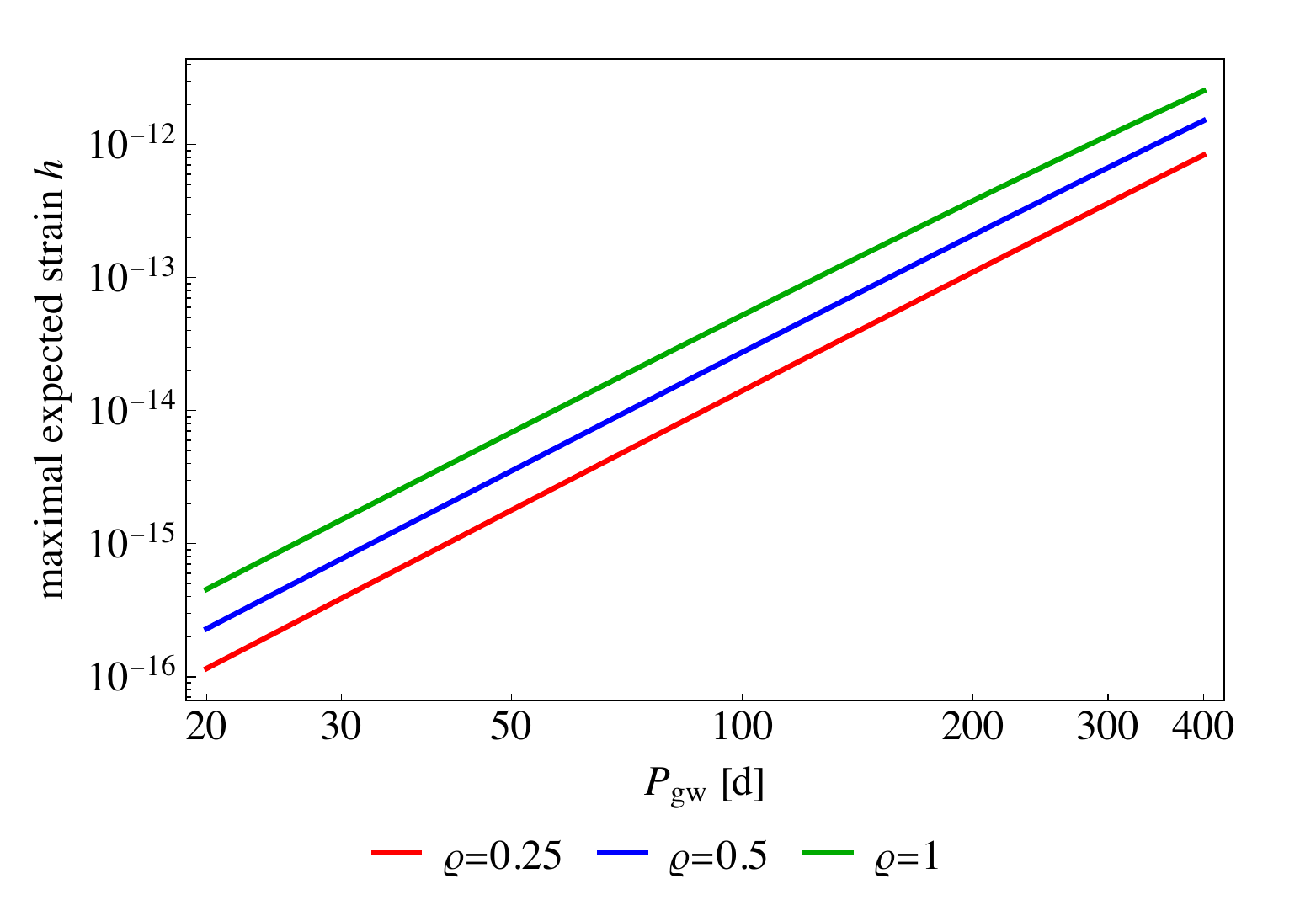}
	\caption{\textit{Top:} Maximal chirp mass $\mathcal{M}$ 
          of a binary system for which the assumption of a constant GW frequency
          $\Pgw$ gives a maximal phase error of $\varrho=0.25$, 0.5, and 1\,rad 
          over the duration of observations $\Tobs=5$\,yr (circular orbit assumed for the binary). 
          \textit{Bottom:} Maximal strain $h$ under the same conditions, 
          assuming a luminosity distance of $r=100$\,Mpc. 
	\label{fig__maxChirpMassStrain}}
\end{figure}

Figure~\ref{fig__maxChirpMassStrain} shows the maximal chirp mass and
the maximal expected strain as functions of the GW period for three
different values of $\varrho$. The duration of the observations is
taken to be $\Tobs=5$\,yr and a minimal plausible distance
$r=100$\,Mpc is assumed. For a maximal phase error of
$\varrho=1$\,rad, we find that $\Pgw=50$\,d gives a maximal strain
slightly below $10^{-14}$, roughly compatible with the theoretical
sensitivity of \gaia\ DR4 as discussed in Sect.~\ref{sec__conclusion}
(see also Sect.~7 of \citeads{2018CQGra..35d5005K} for a related
earlier discussion).  This justifies not considering shorter periods
than $\Pgw\simeq 50$\,d in the present study, where the GW is assumed
to have negligible frequency drift over 5\,yr.

\section{Sky distributions of the astrometric errors}
\label{section-examples}

Figure~\ref{fig___error_sky_maps} shows the sky distributions of the
astrometric errors obtained in noise-free astrometric solutions perturbed
by GWs of selected frequencies. The plots illustrate the complexity of the 
astrometric errors generated by GWs in the high-frequency regime 
($\Pgw\lesssim\Tobs$) compared to the much simpler patterns obtained 
when the GW period $\Pgw$ significantly exceeds the duration of the 
observations $\Tobs$, as exemplified in Fig.~\ref{fig_simulated20yr}. 
At the special frequencies discussed in Sect.~\ref{sec__freq_bands_high_errors}, 
however, the error patterns become relatively simple if the corresponding 
diagram in Fig.~\ref{fig___freq_overview-source} has a strong peak at that 
frequency. Additional statistics are given in Table~\ref{tab_statstable2}. 
All simulations in this Appendix have $\aGW = 0$, $\dGW = 0$ and the 
four strain parameters are set to equal values ($e=1$).  

\begin{landscape}
	\def\mygridwidth{0.24\textwidth}
	\def\mylegwidth{0.1\textwidth}
	\begin{figure}[htb]
		\centering
		\setlength\tabcolsep{2pt}
		\begin{tabular}{>{\centering\arraybackslash}m{0.02\textwidth}>{\centering\arraybackslash}m{\mygridwidth}>{\centering\arraybackslash}m{\mygridwidth}>{\centering\arraybackslash}m{\mygridwidth}>{\centering\arraybackslash}m{\mygridwidth}>{\centering\arraybackslash}m{\mygridwidth}>{\centering\arraybackslash}m{\mylegwidth}}
			~     & \large${\Delta\alpha^*}$     & \large${\Delta\delta}$ & \large${\Delta\varpi}$  & \large${\Delta\mu_{\alpha*}}$ & \large${\Delta\mu_{\delta}}$ &\\
			\rotatebox{90}{\large\textbf{1\,yr}} &
			\includegraphics[keepaspectratio,width=\mygridwidth]{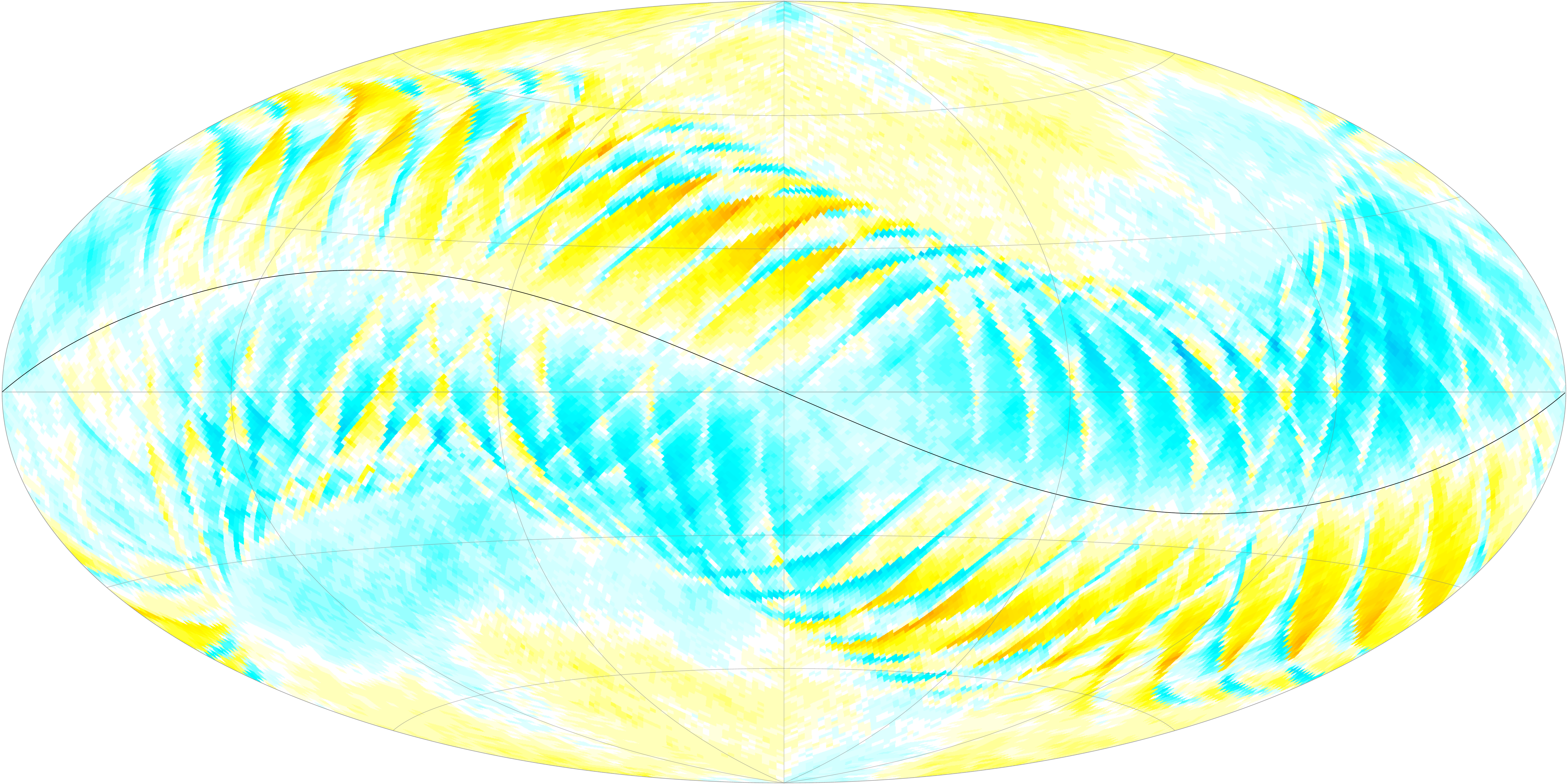}&
			\includegraphics[keepaspectratio,width=\mygridwidth]{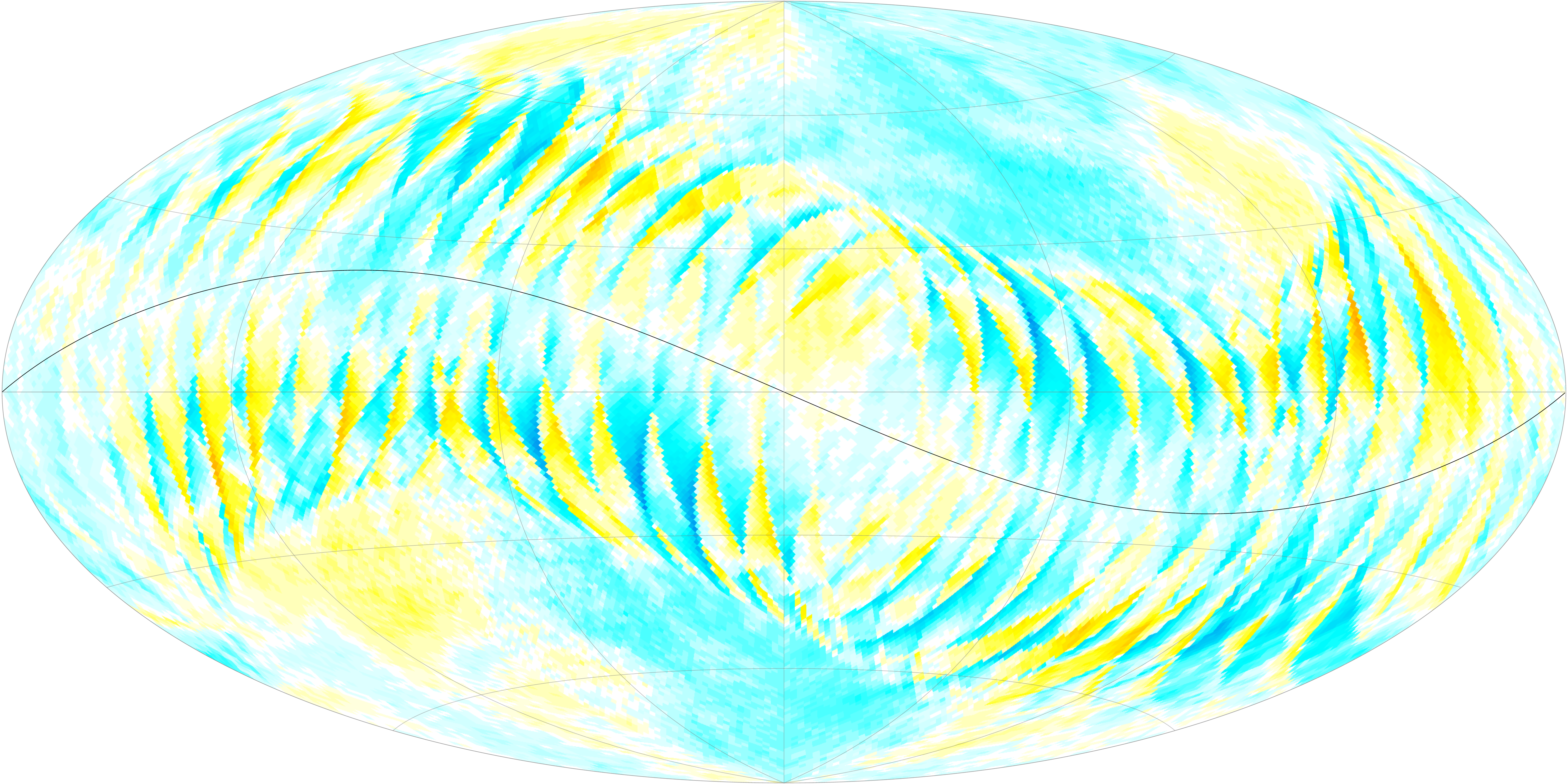}&
			\includegraphics[keepaspectratio,width=\mygridwidth]{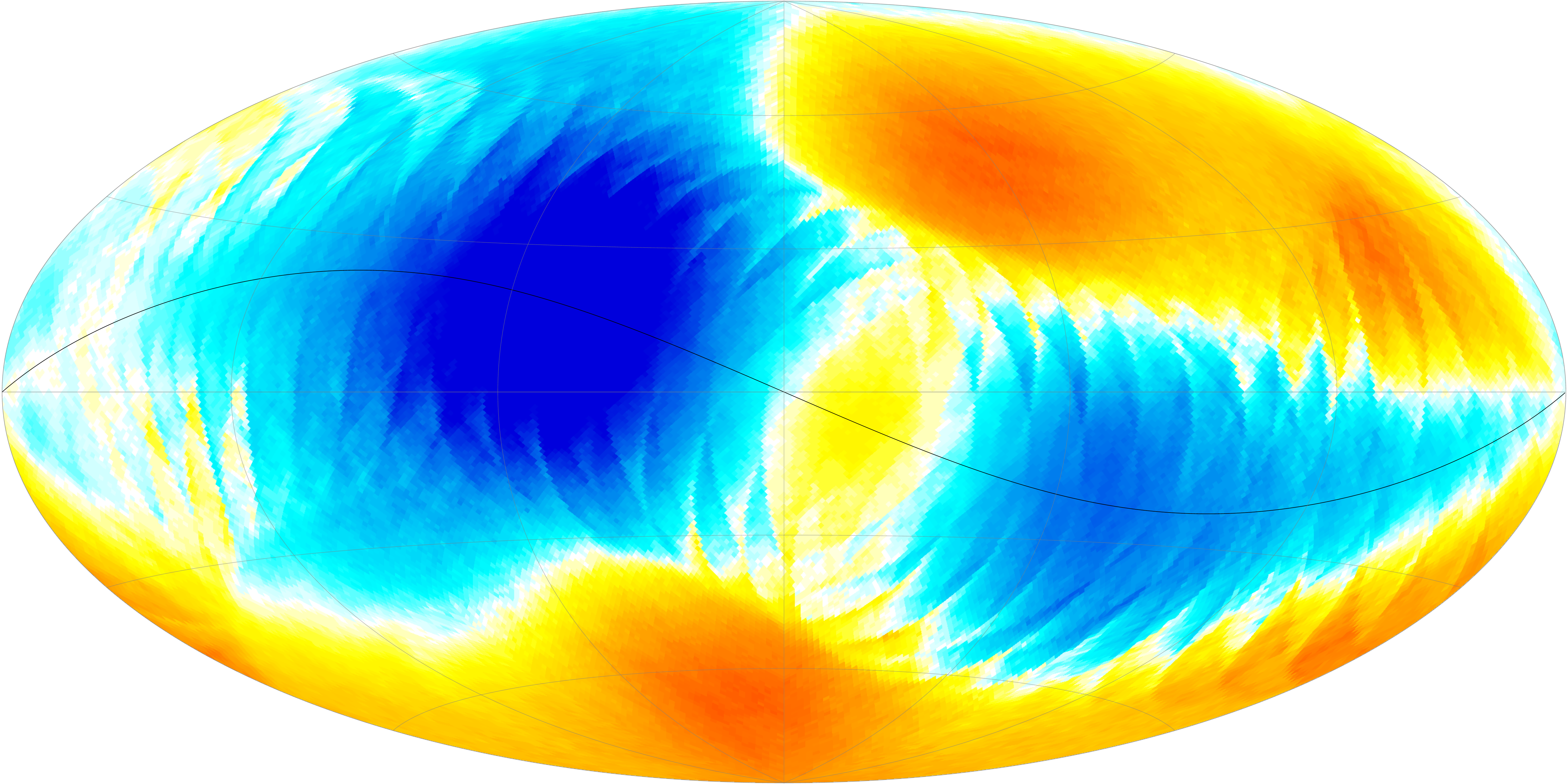}&
			\includegraphics[keepaspectratio,width=\mygridwidth]{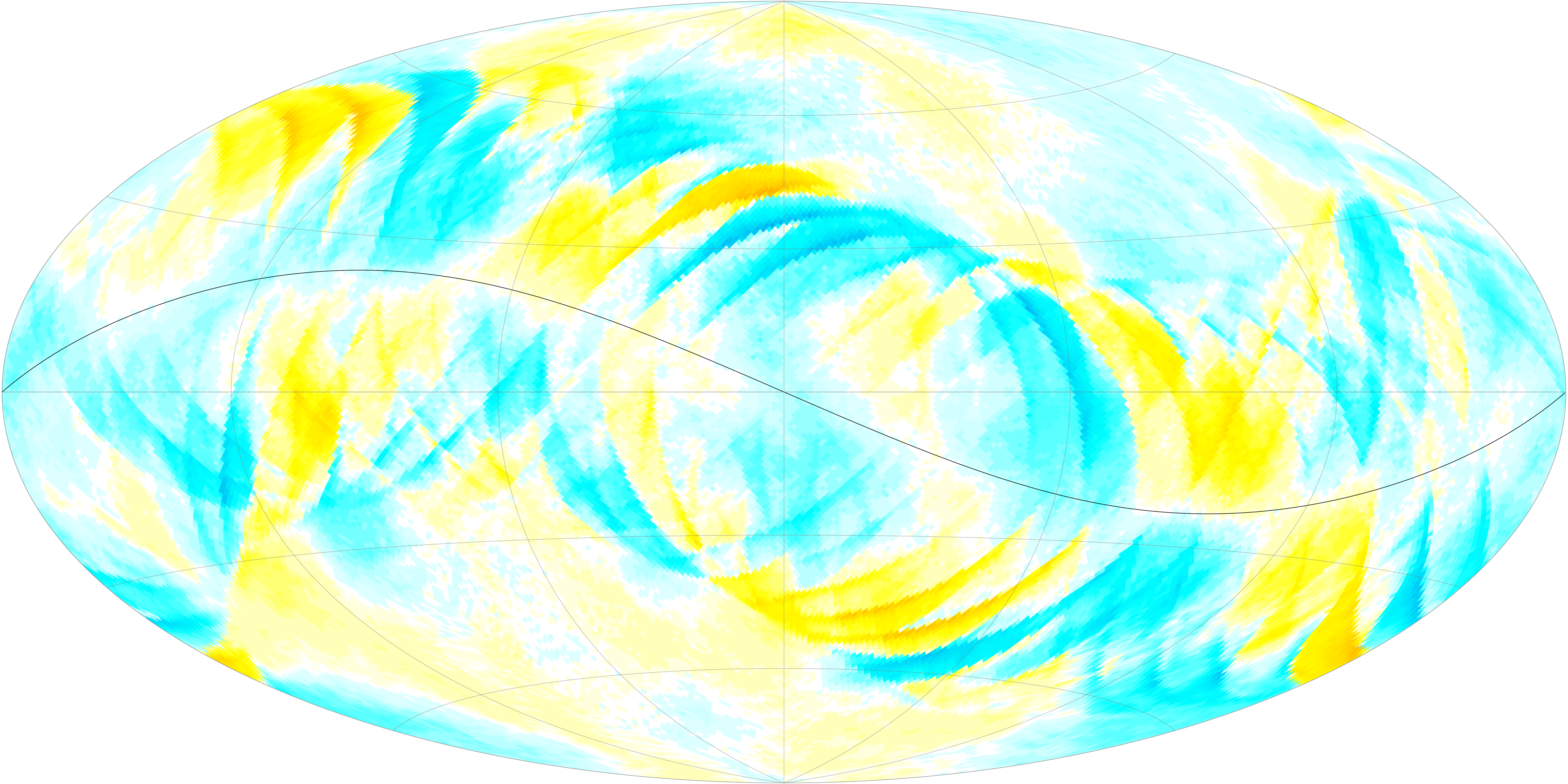}&
			\includegraphics[keepaspectratio,width=\mygridwidth]{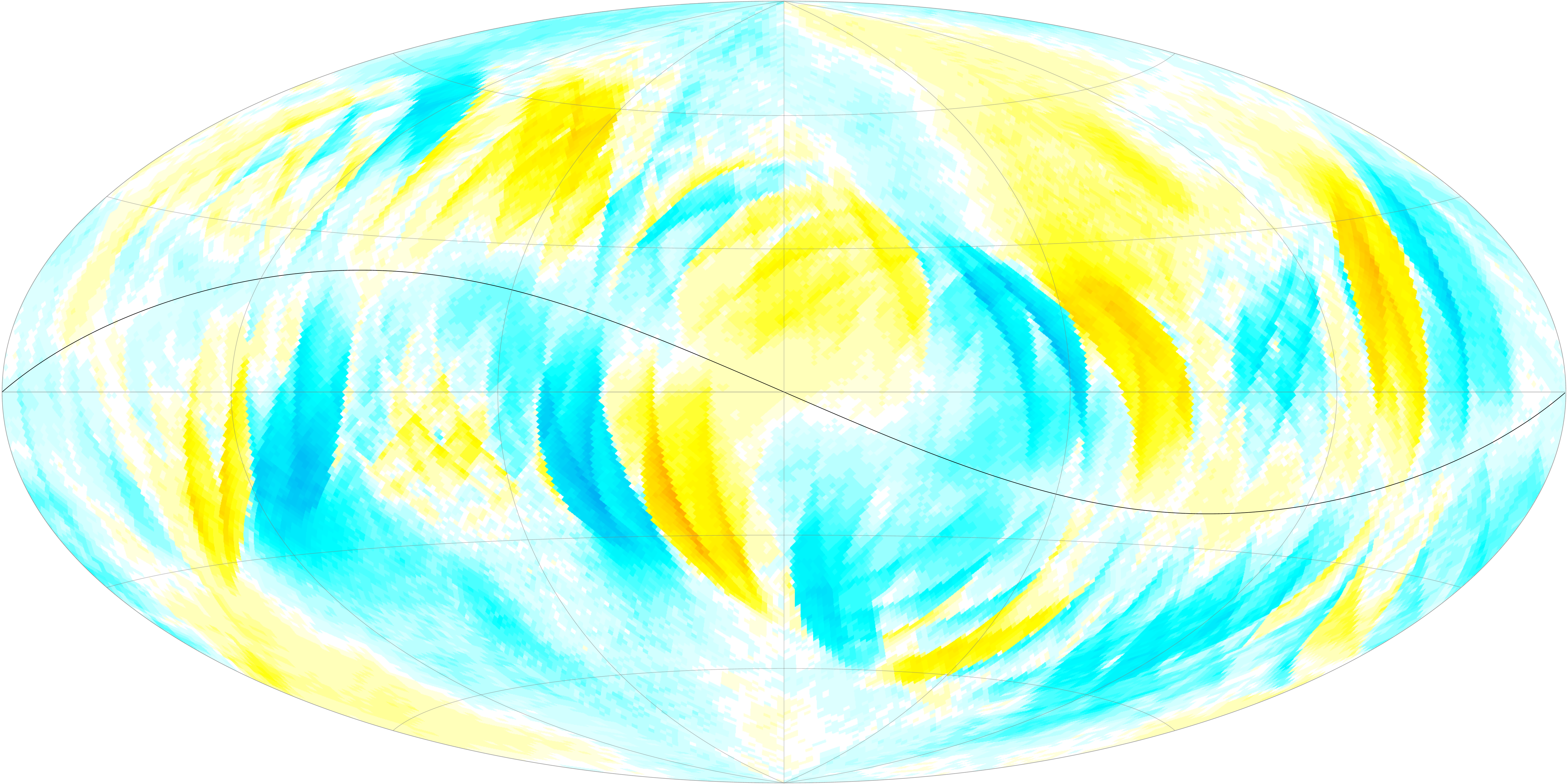}&
			\multirow{7}{*}{\includegraphics[keepaspectratio,height=0.77\textheight]{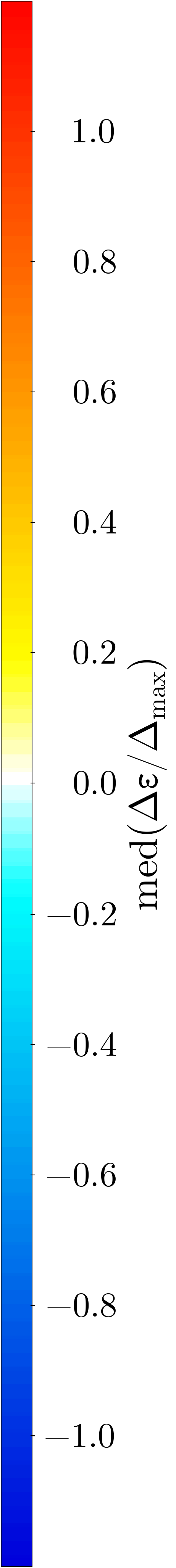}}\\
			\rotatebox{90}{\large\textbf{279\,d}} &
			\includegraphics[keepaspectratio,width=\mygridwidth]{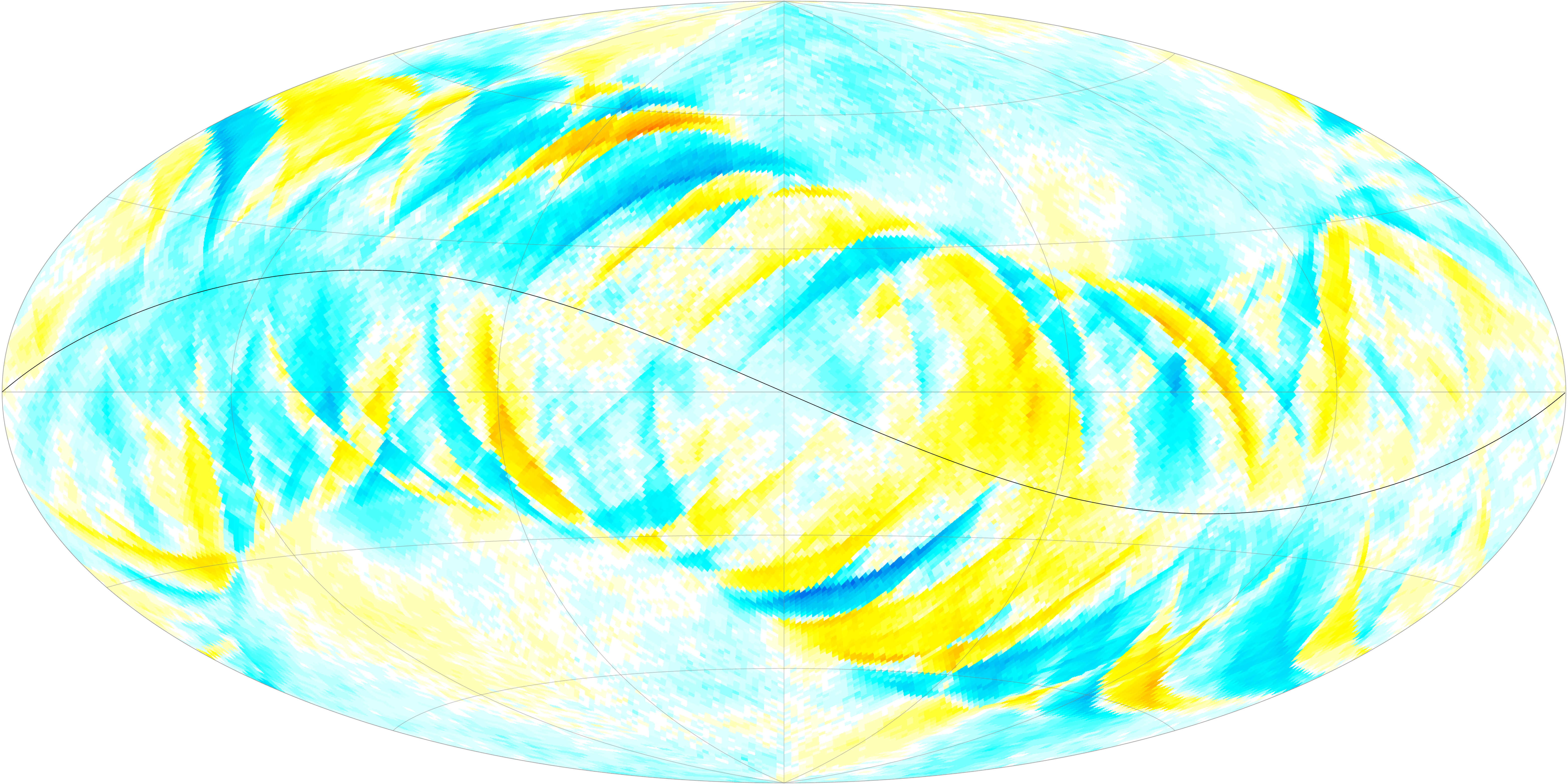}&
			\includegraphics[keepaspectratio,width=\mygridwidth]{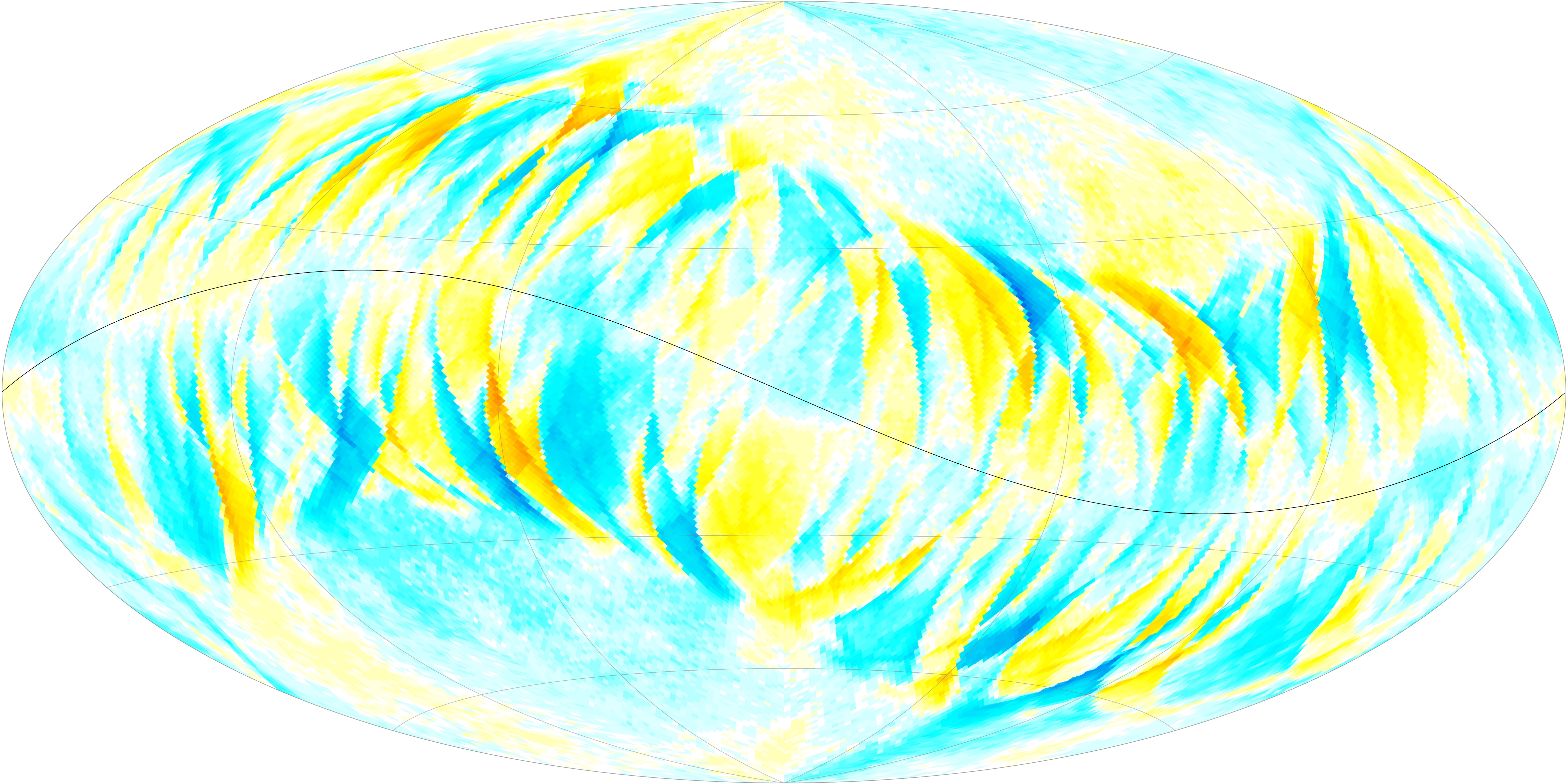}&
			\includegraphics[keepaspectratio,width=\mygridwidth]{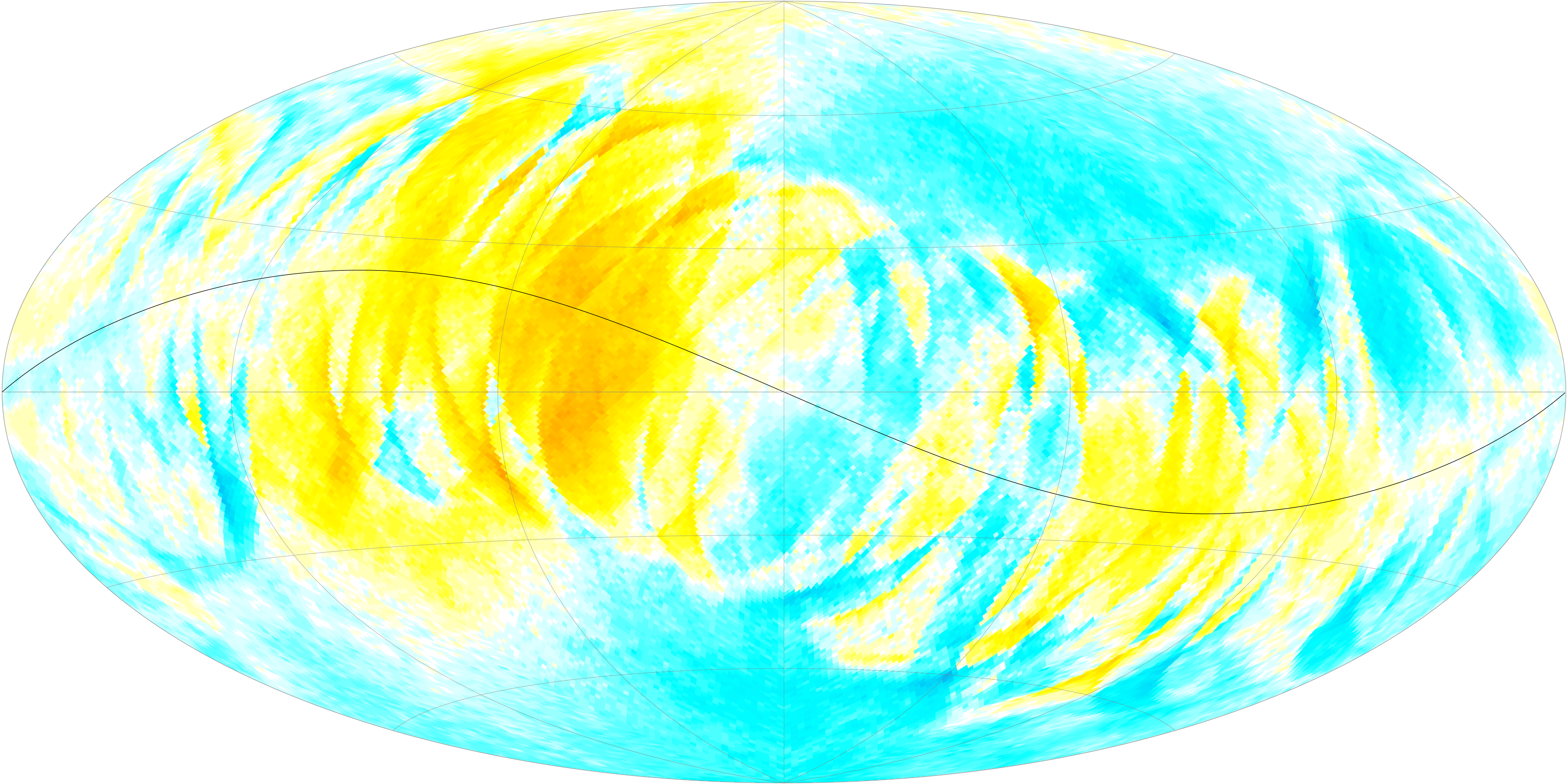}&
			\includegraphics[keepaspectratio,width=\mygridwidth]{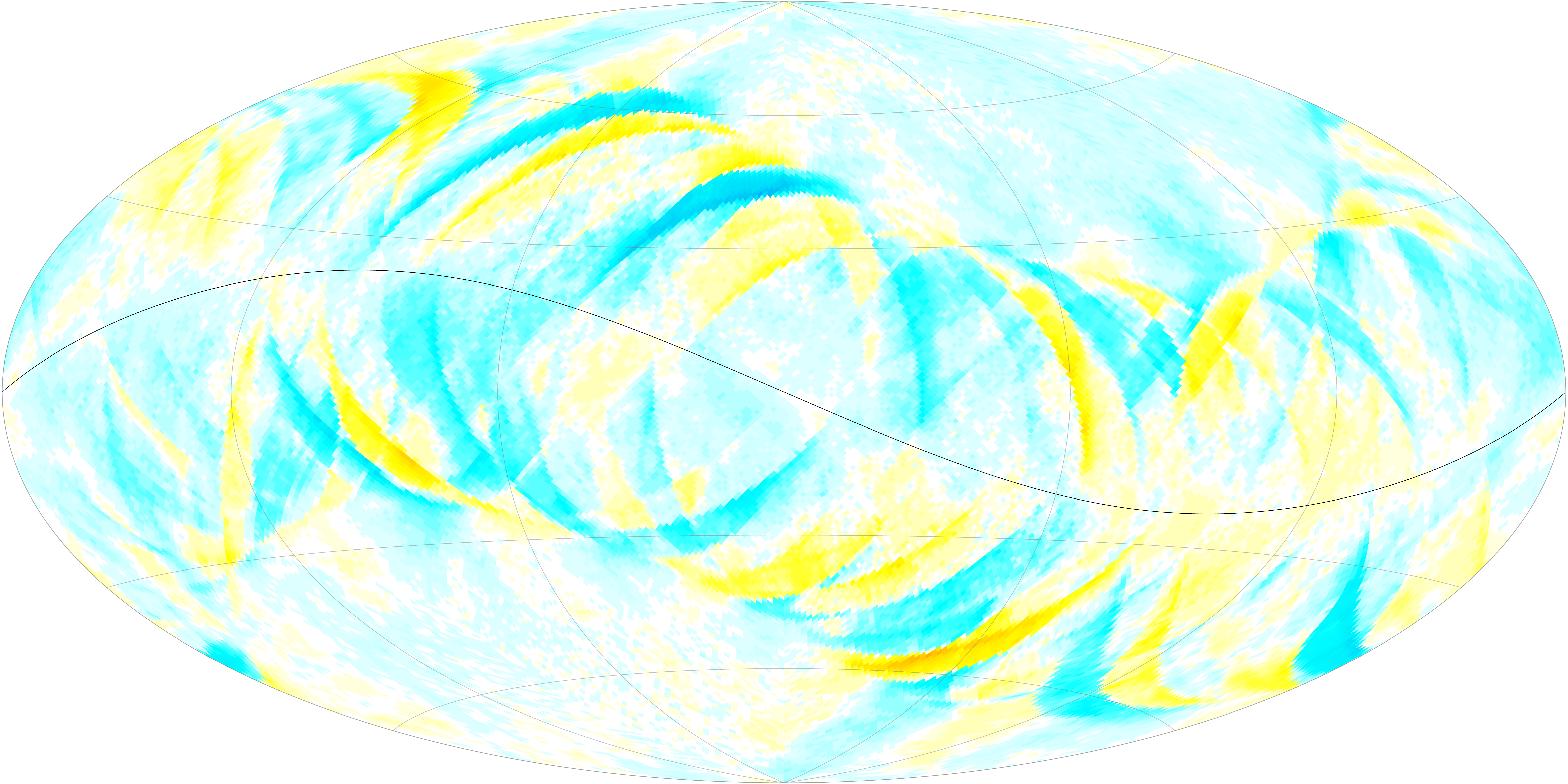}&
			\includegraphics[keepaspectratio,width=\mygridwidth]{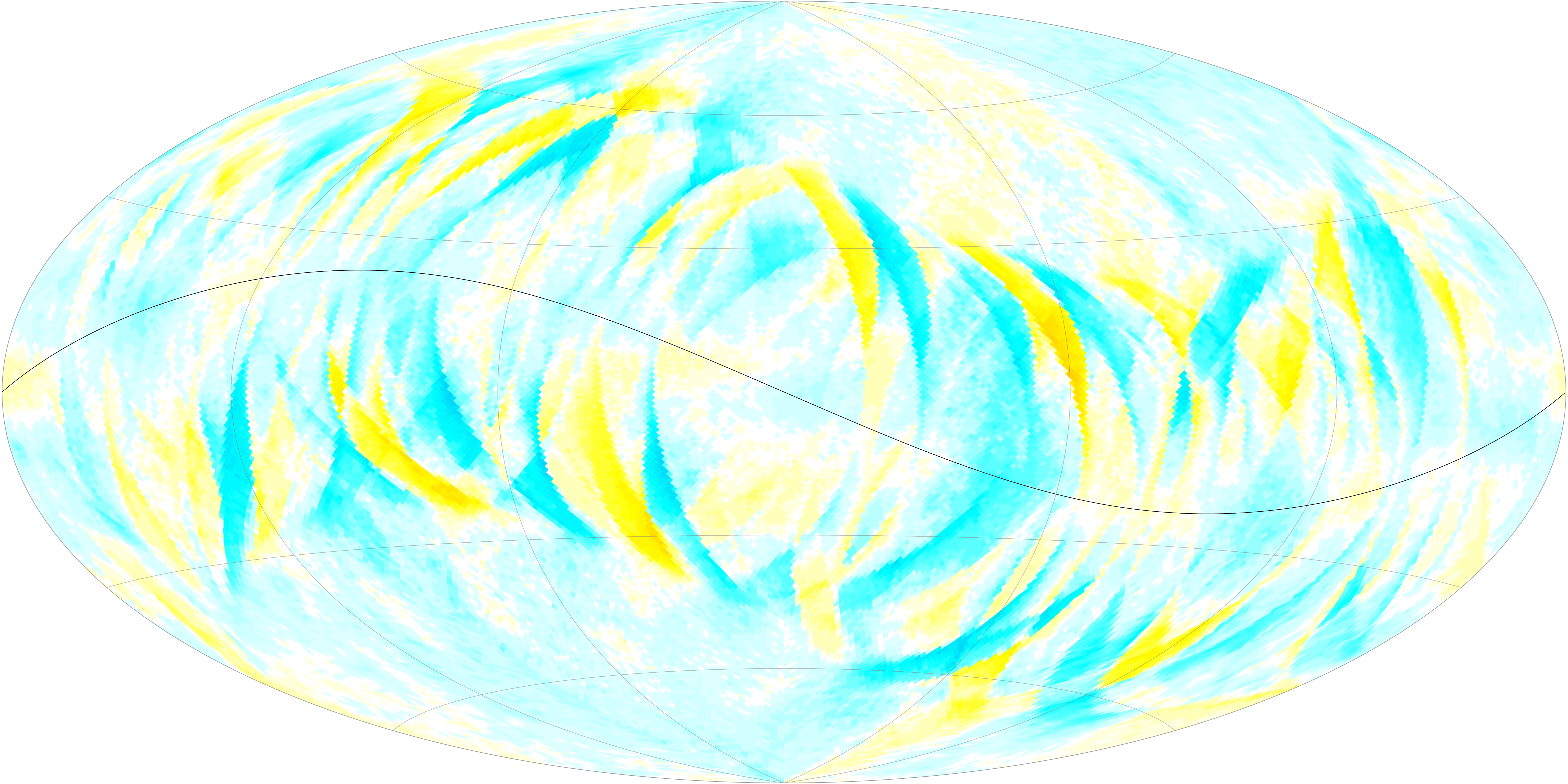}& \\
			\rotatebox{90}{\large\textbf{96.1\,d}} &
			\includegraphics[keepaspectratio,width=\mygridwidth]{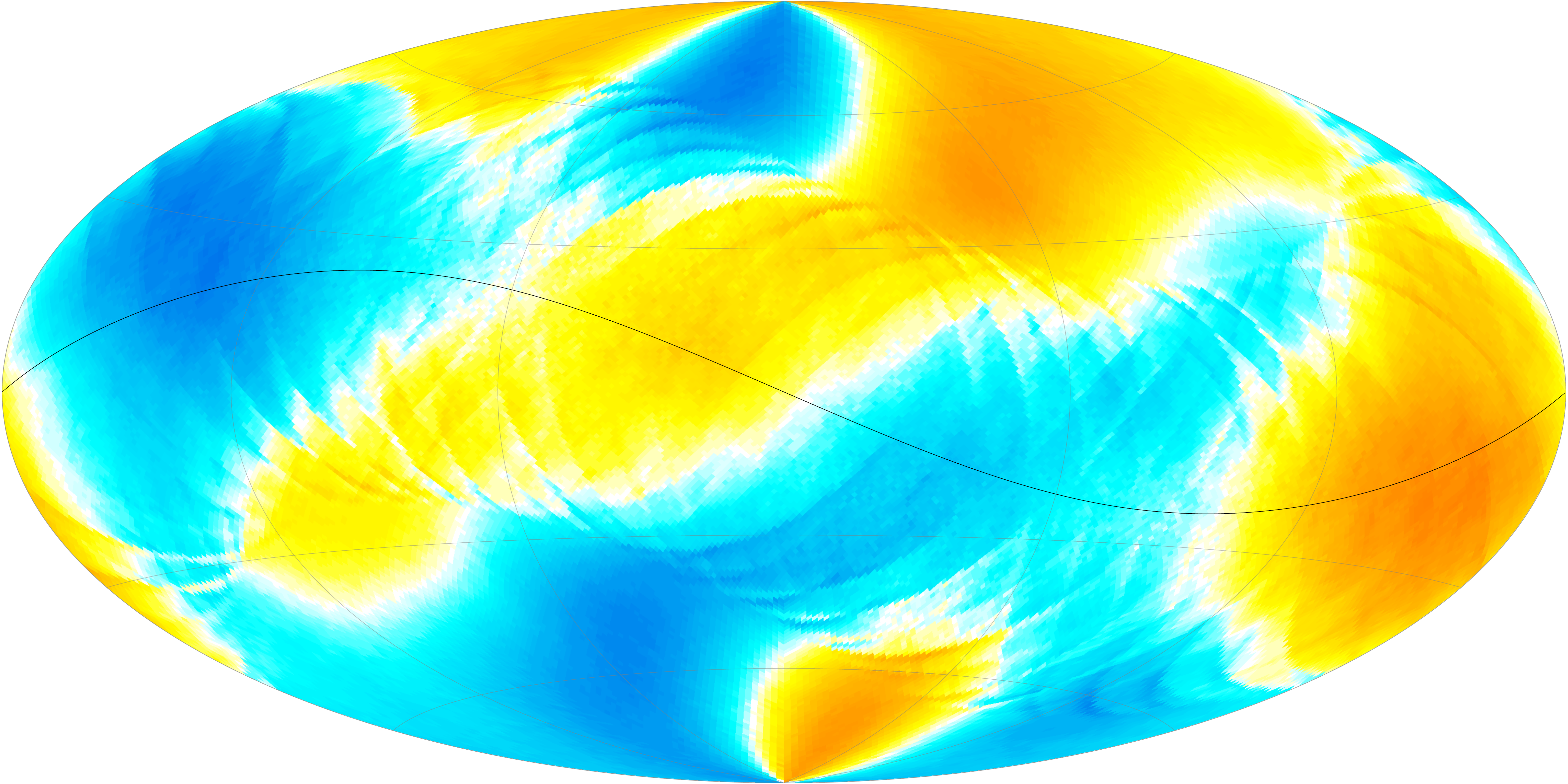}&
			\includegraphics[keepaspectratio,width=\mygridwidth]{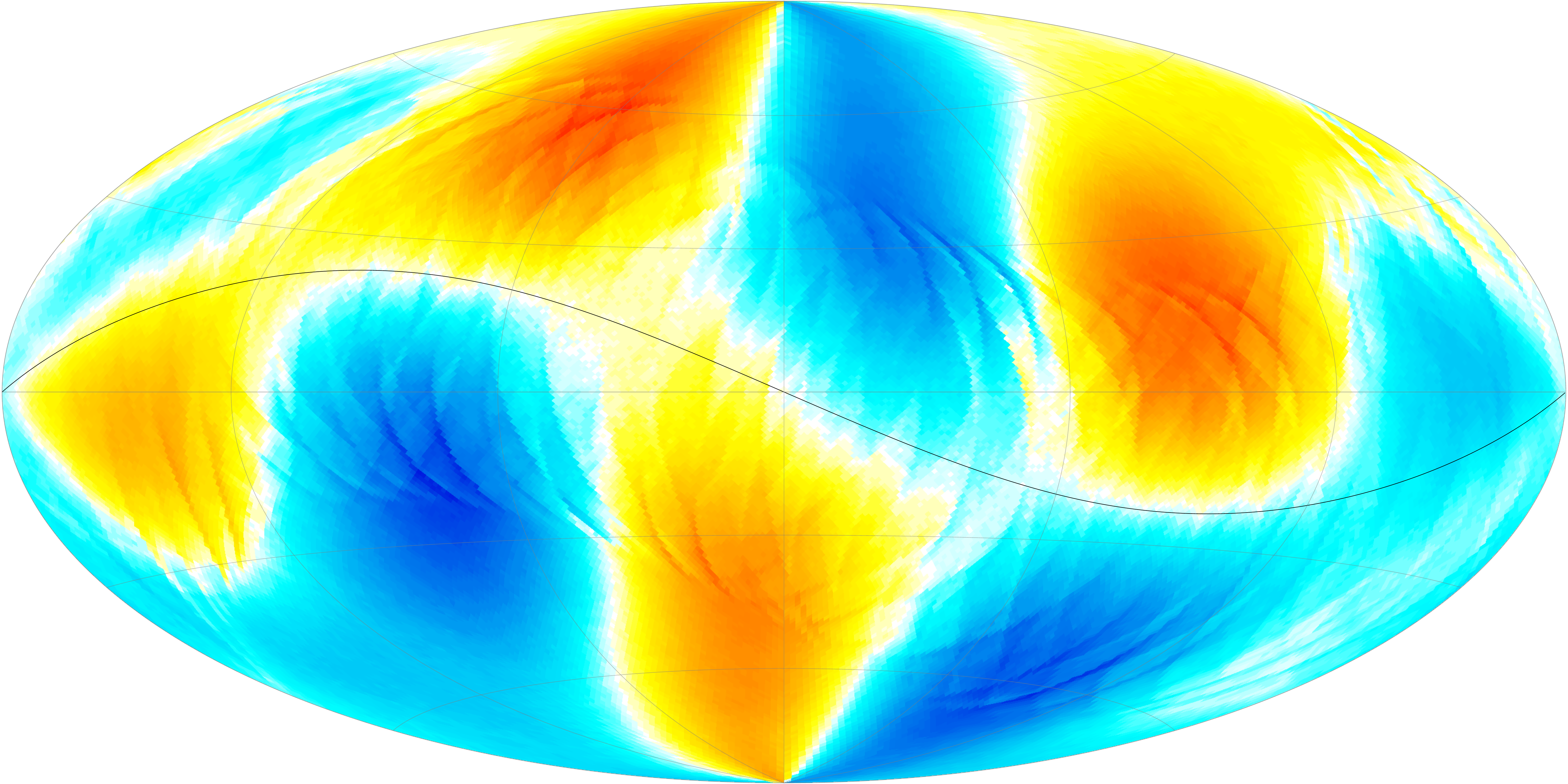}&
			\includegraphics[keepaspectratio,width=\mygridwidth]{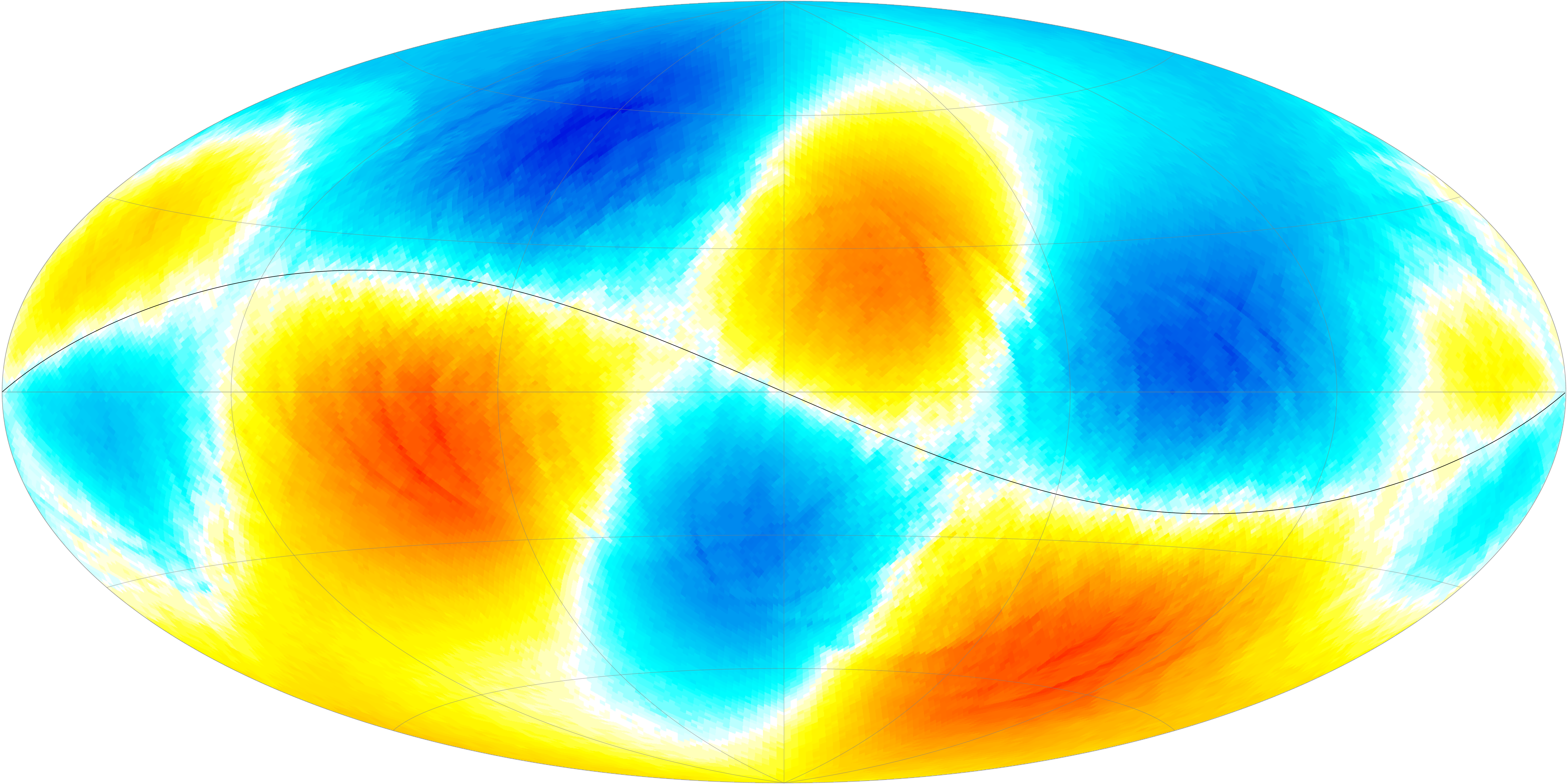}&
			\includegraphics[keepaspectratio,width=\mygridwidth]{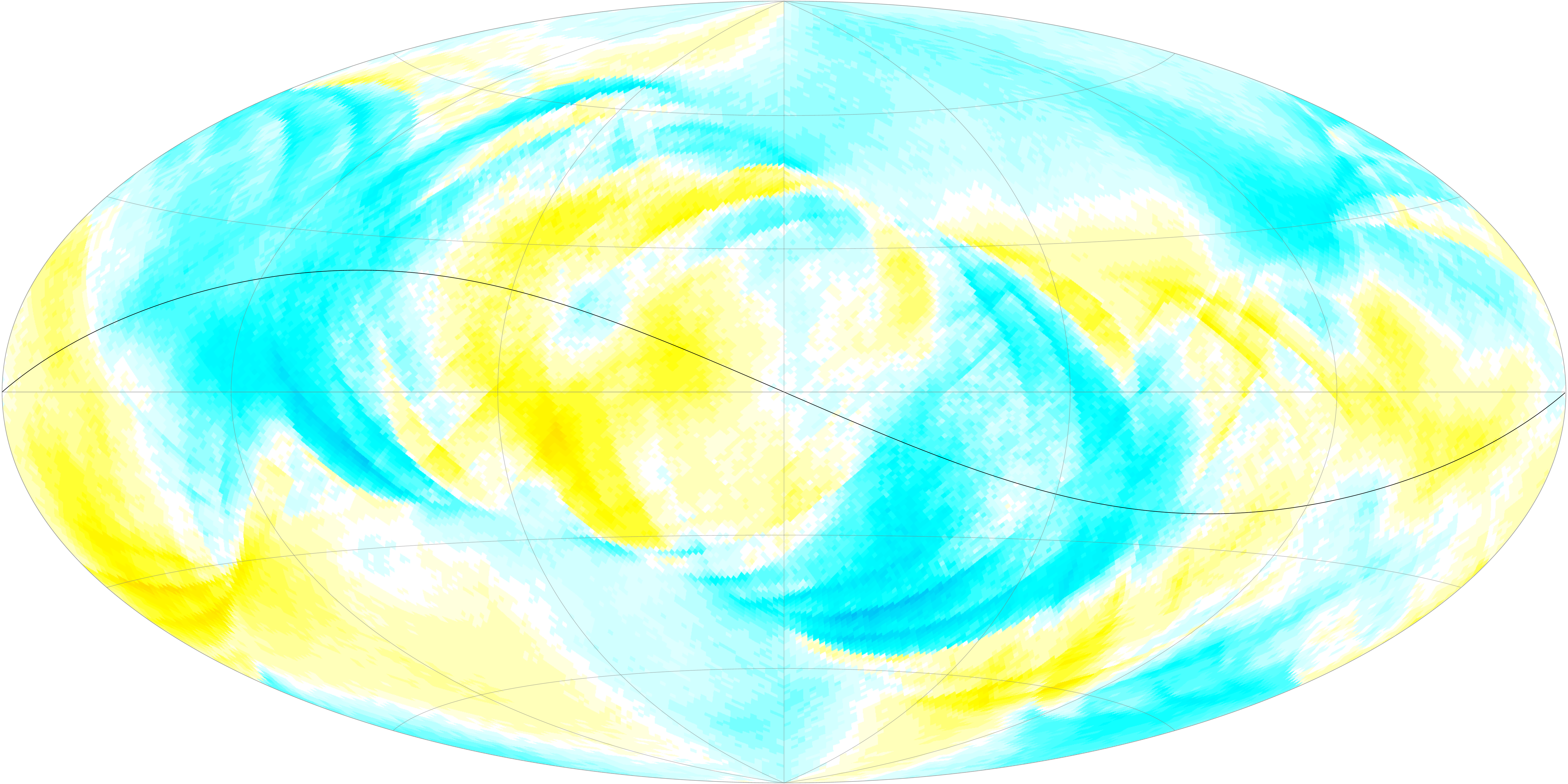}&
			\includegraphics[keepaspectratio,width=\mygridwidth]{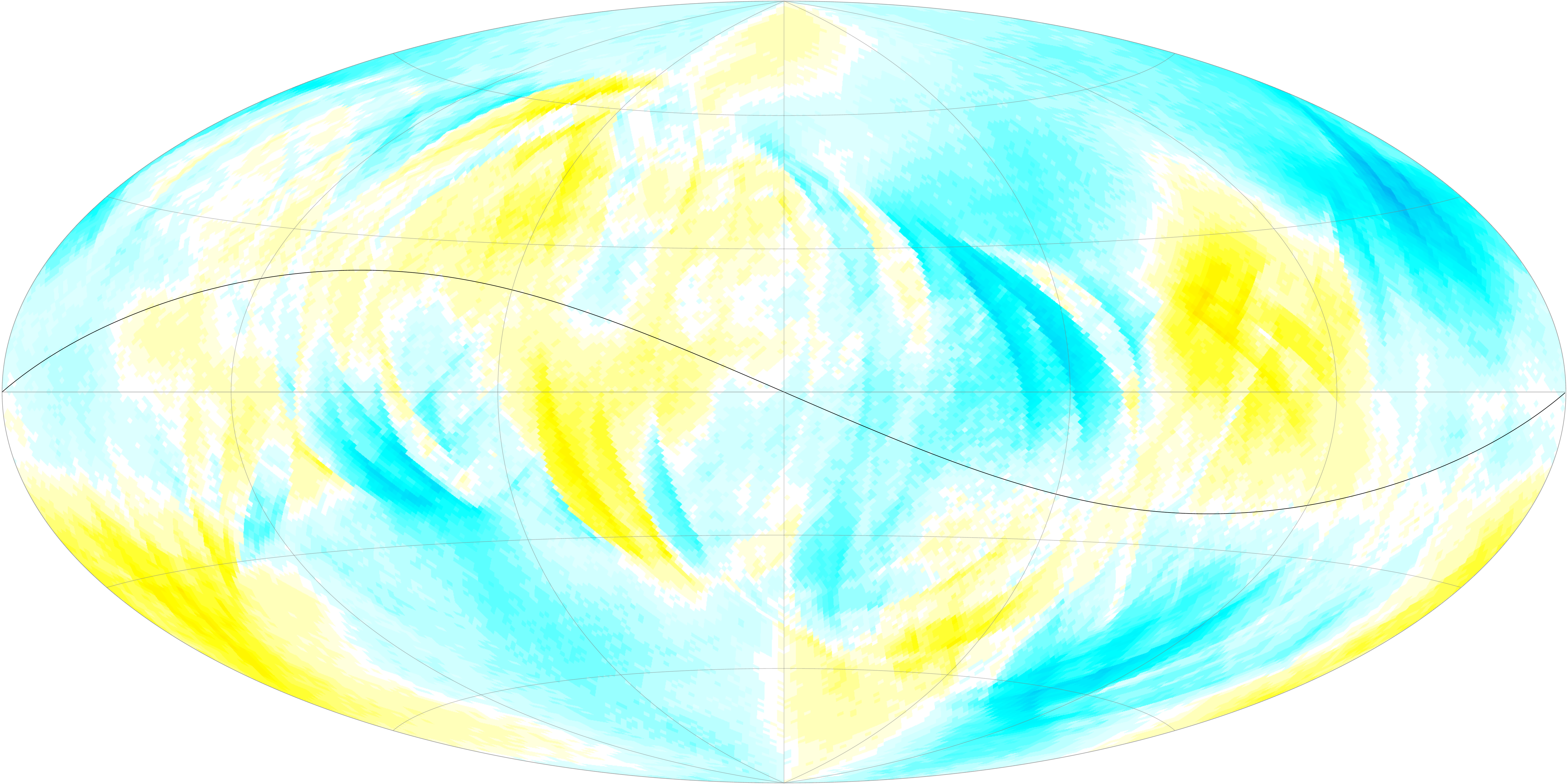}& \\
			\rotatebox{90}{\large\textbf{76.1\,d}} &
			\includegraphics[keepaspectratio,width=\mygridwidth]{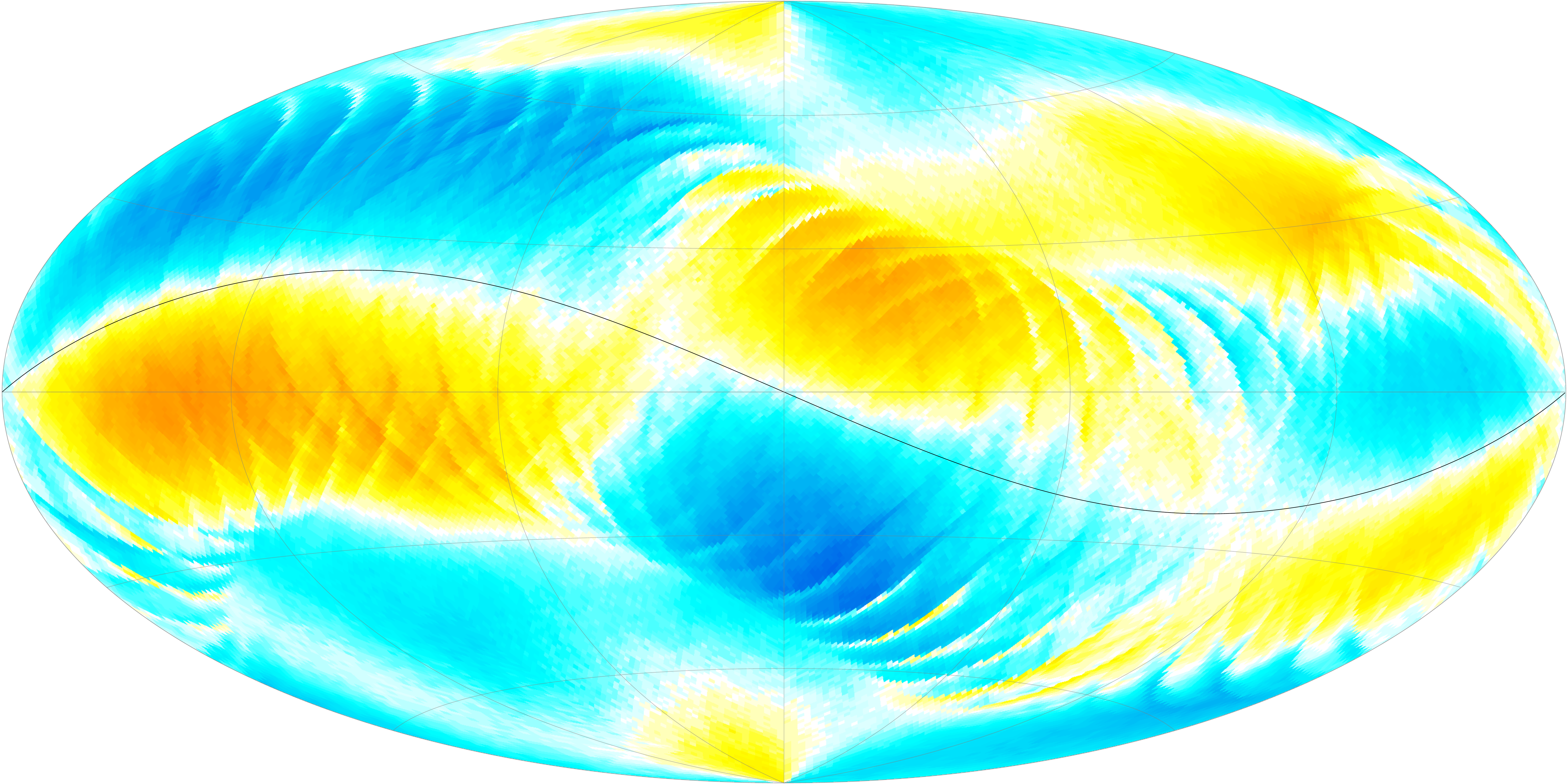}&
			\includegraphics[keepaspectratio,width=\mygridwidth]{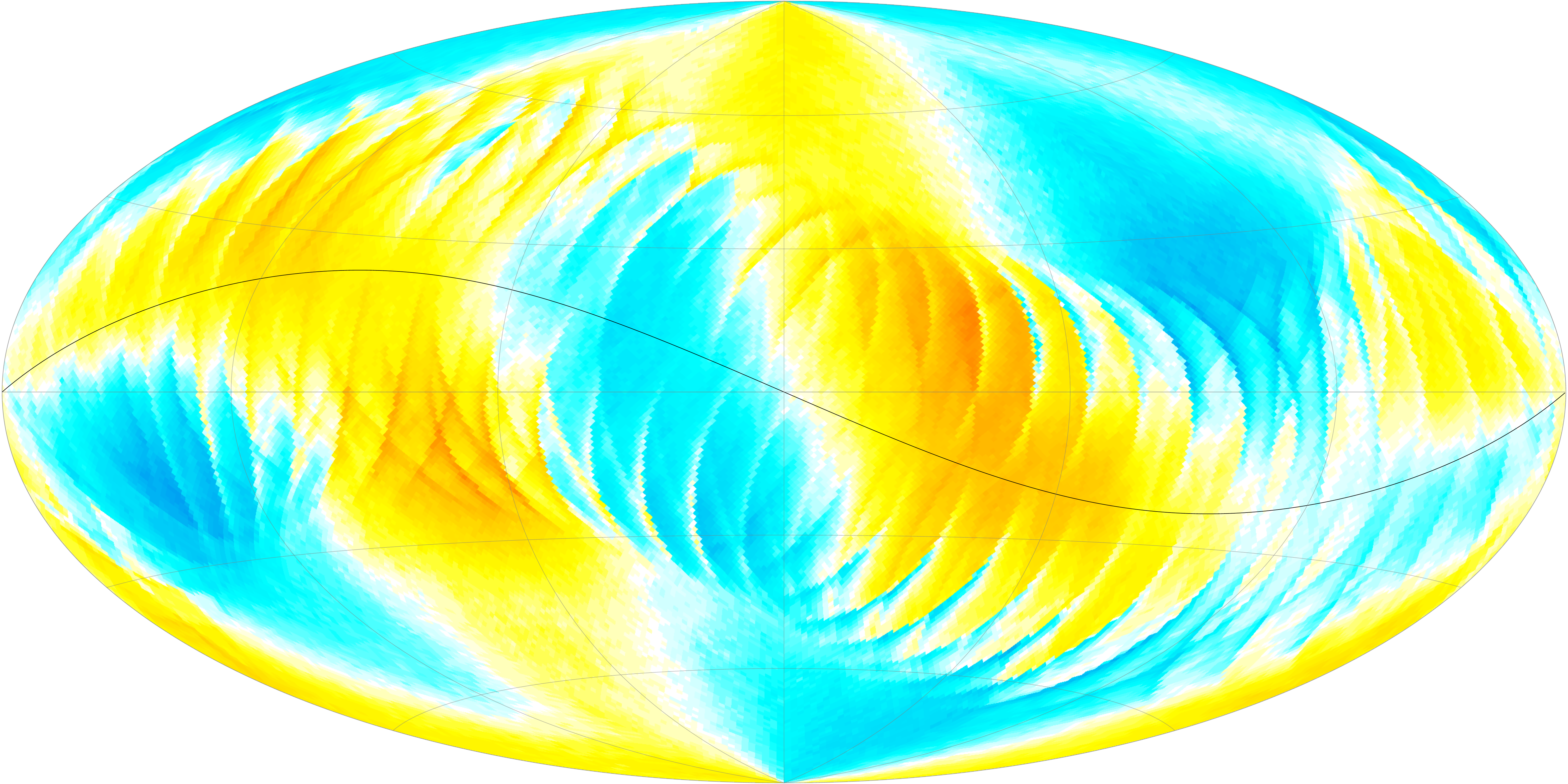}&
			\includegraphics[keepaspectratio,width=\mygridwidth]{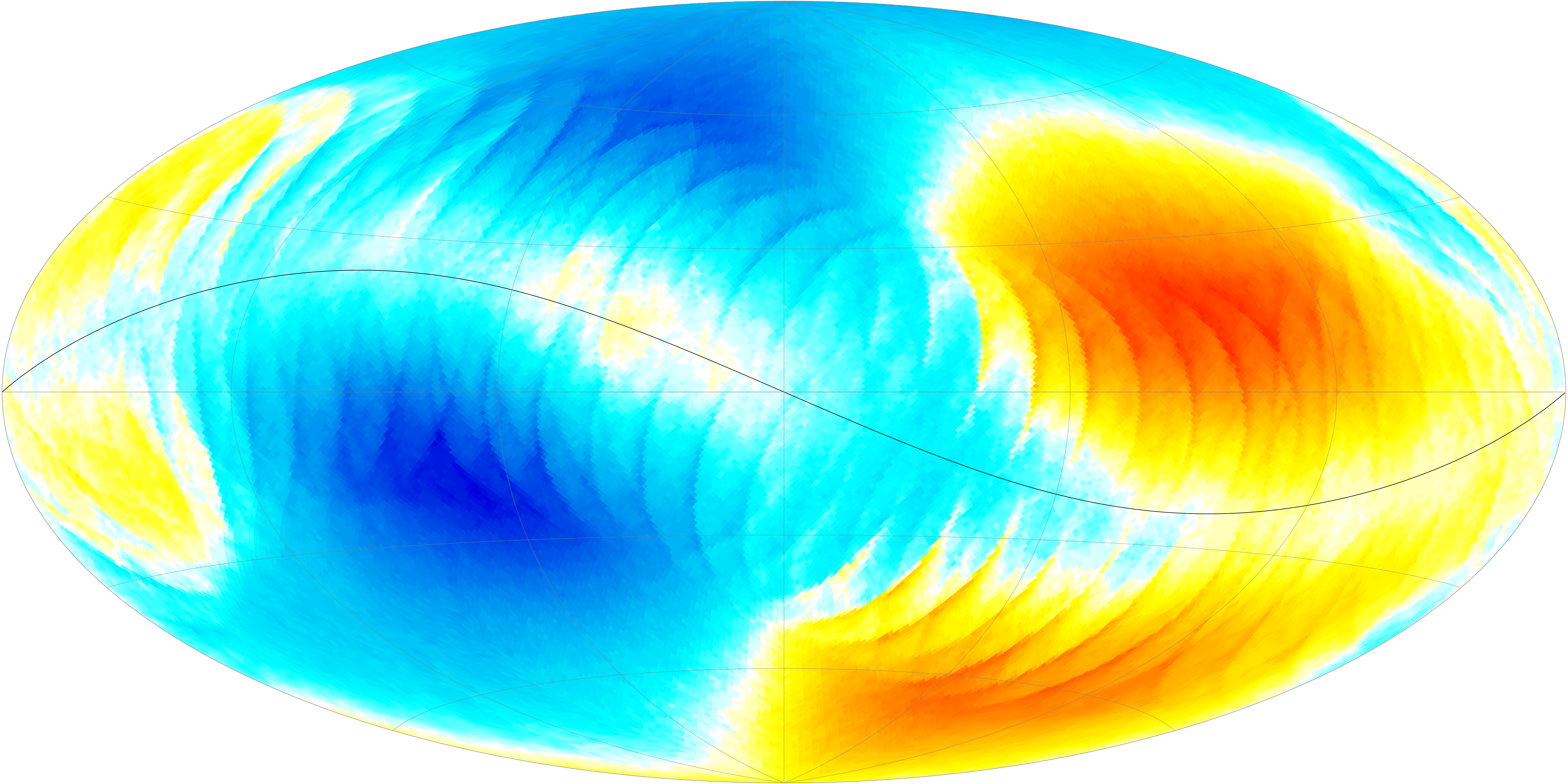}&
			\includegraphics[keepaspectratio,width=\mygridwidth]{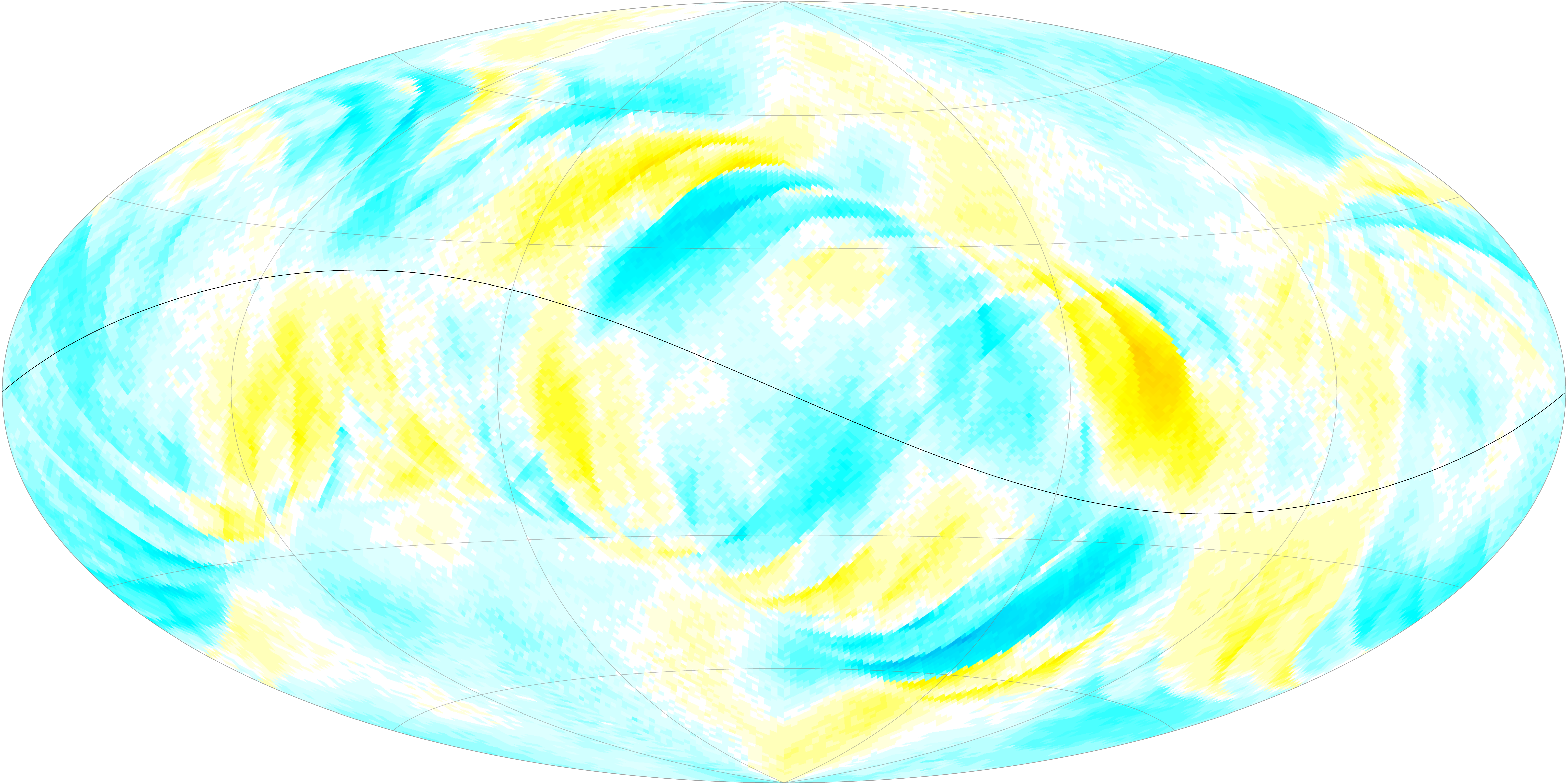}&
			\includegraphics[keepaspectratio,width=\mygridwidth]{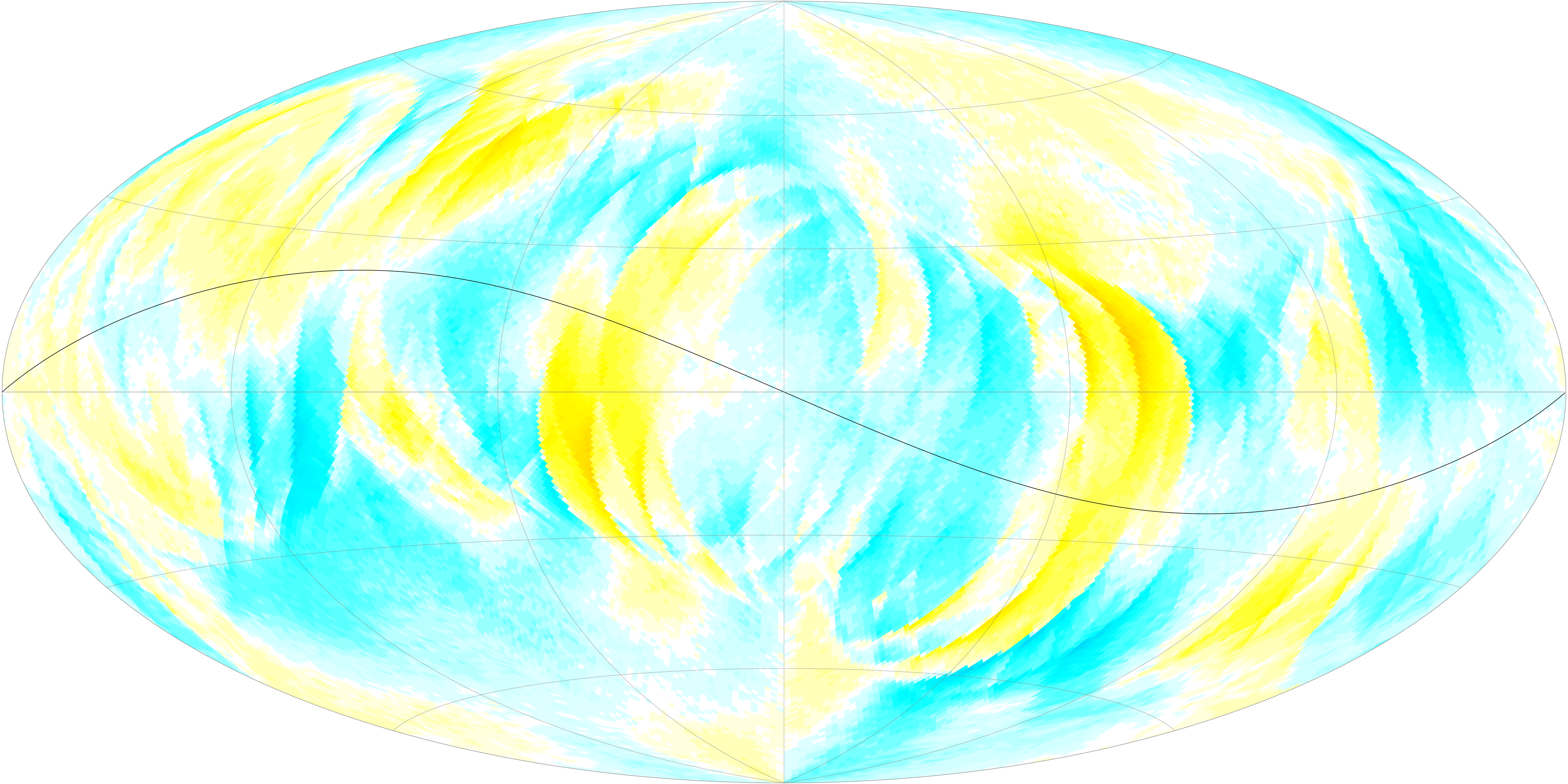}& \\
			\rotatebox{90}{\large\textbf{67.9\,d}} &
			\includegraphics[keepaspectratio,width=\mygridwidth]{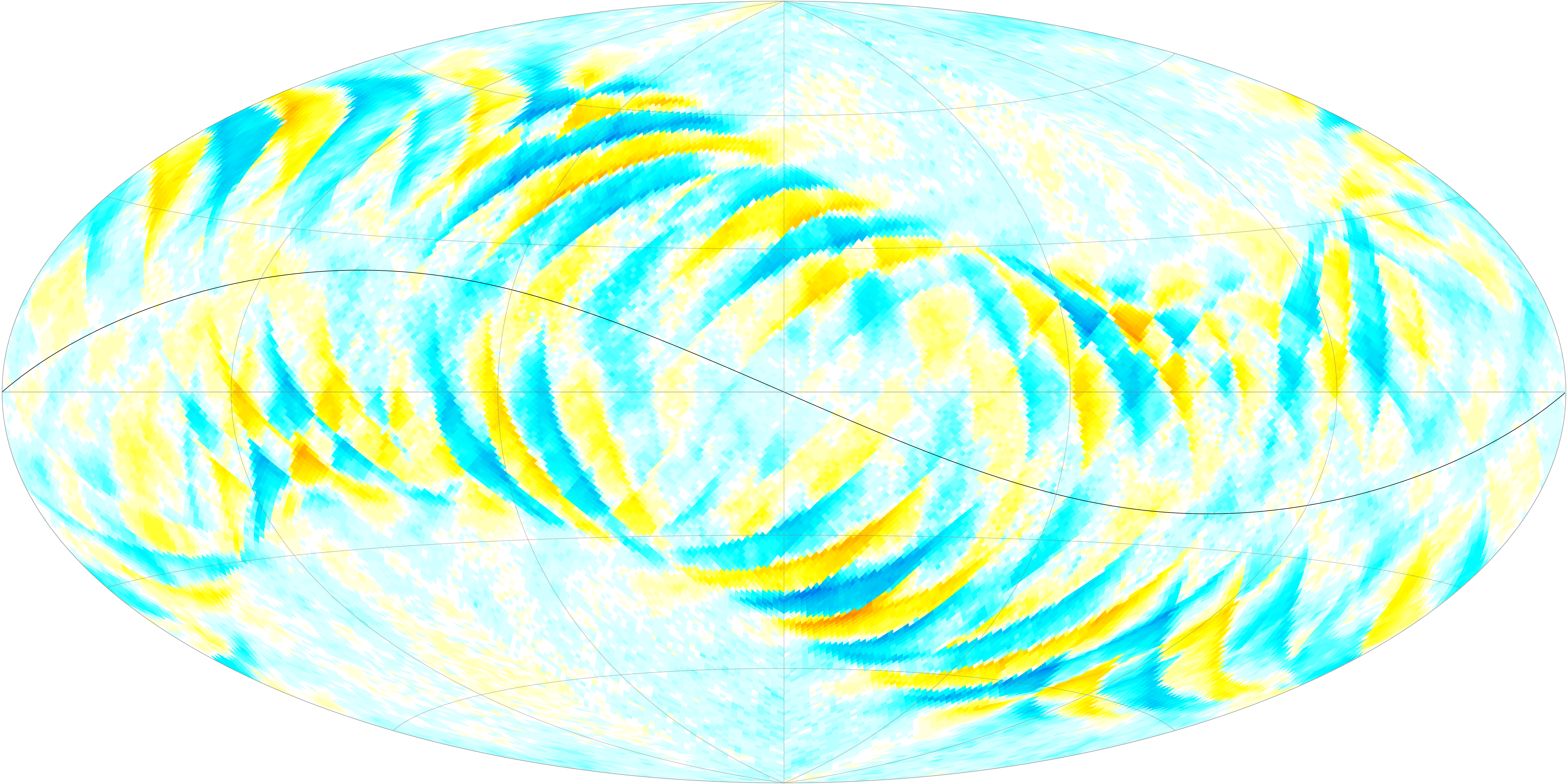}&
			\includegraphics[keepaspectratio,width=\mygridwidth]{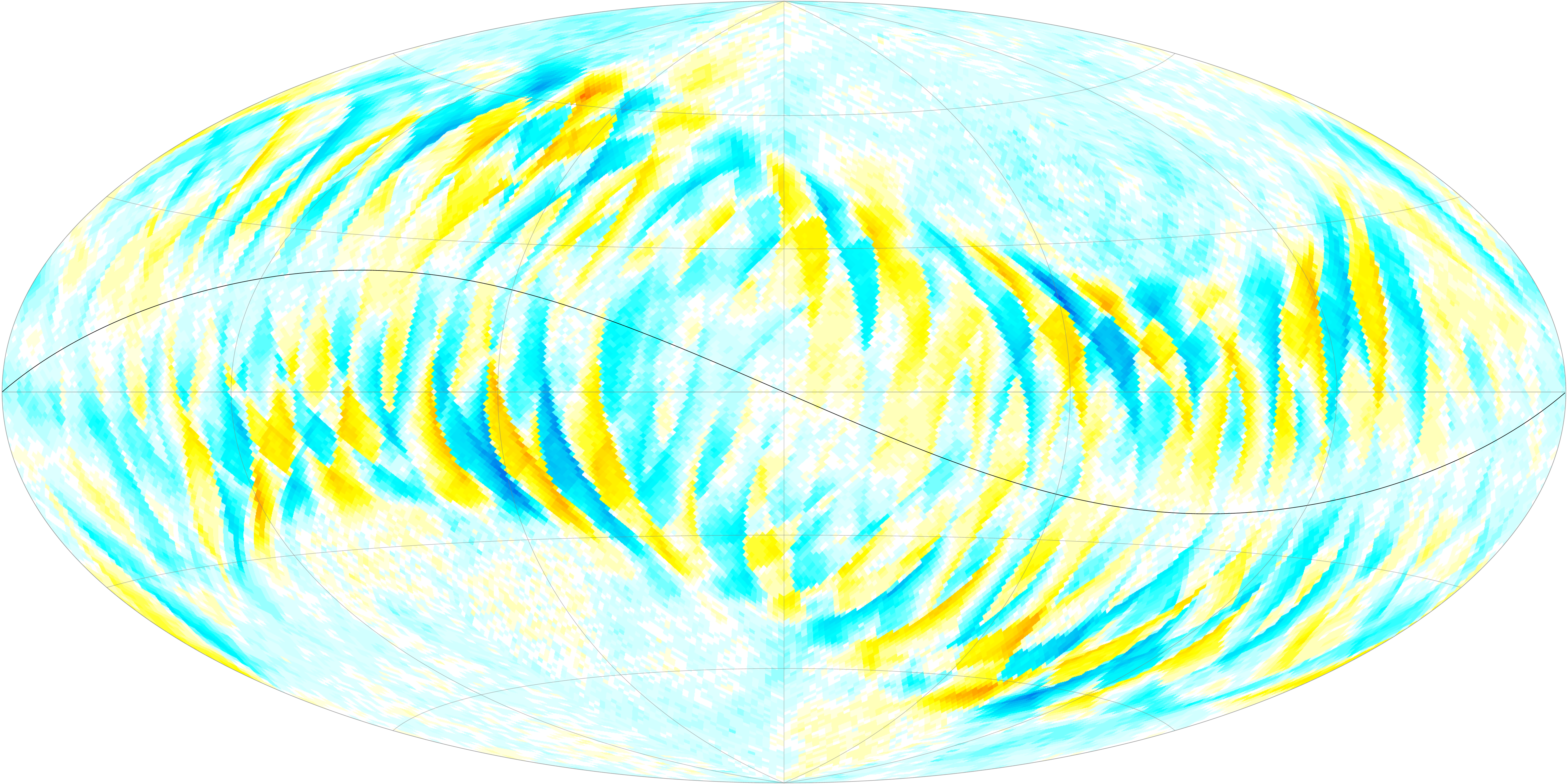}&
			\includegraphics[keepaspectratio,width=\mygridwidth]{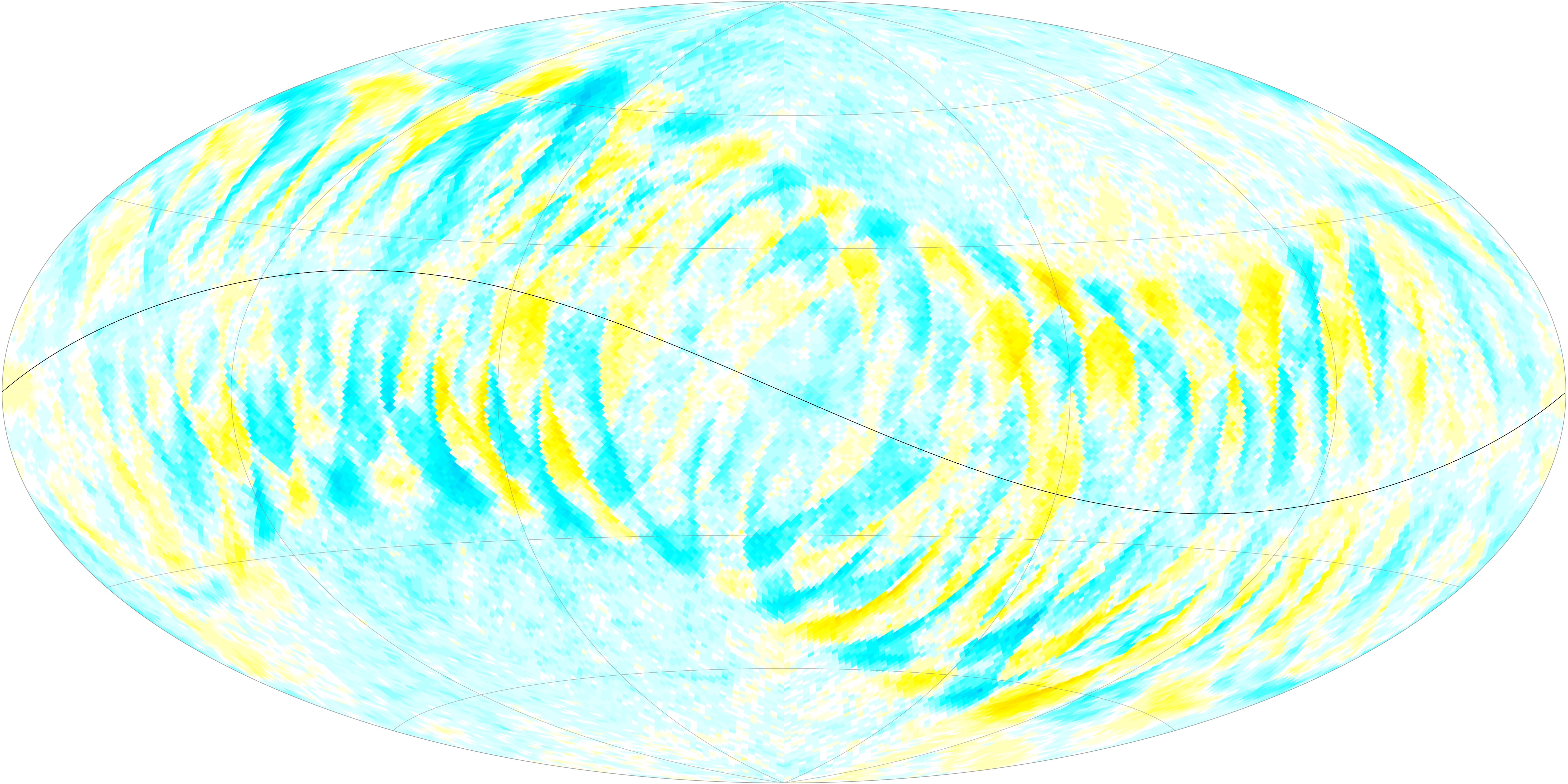}&
			\includegraphics[keepaspectratio,width=\mygridwidth]{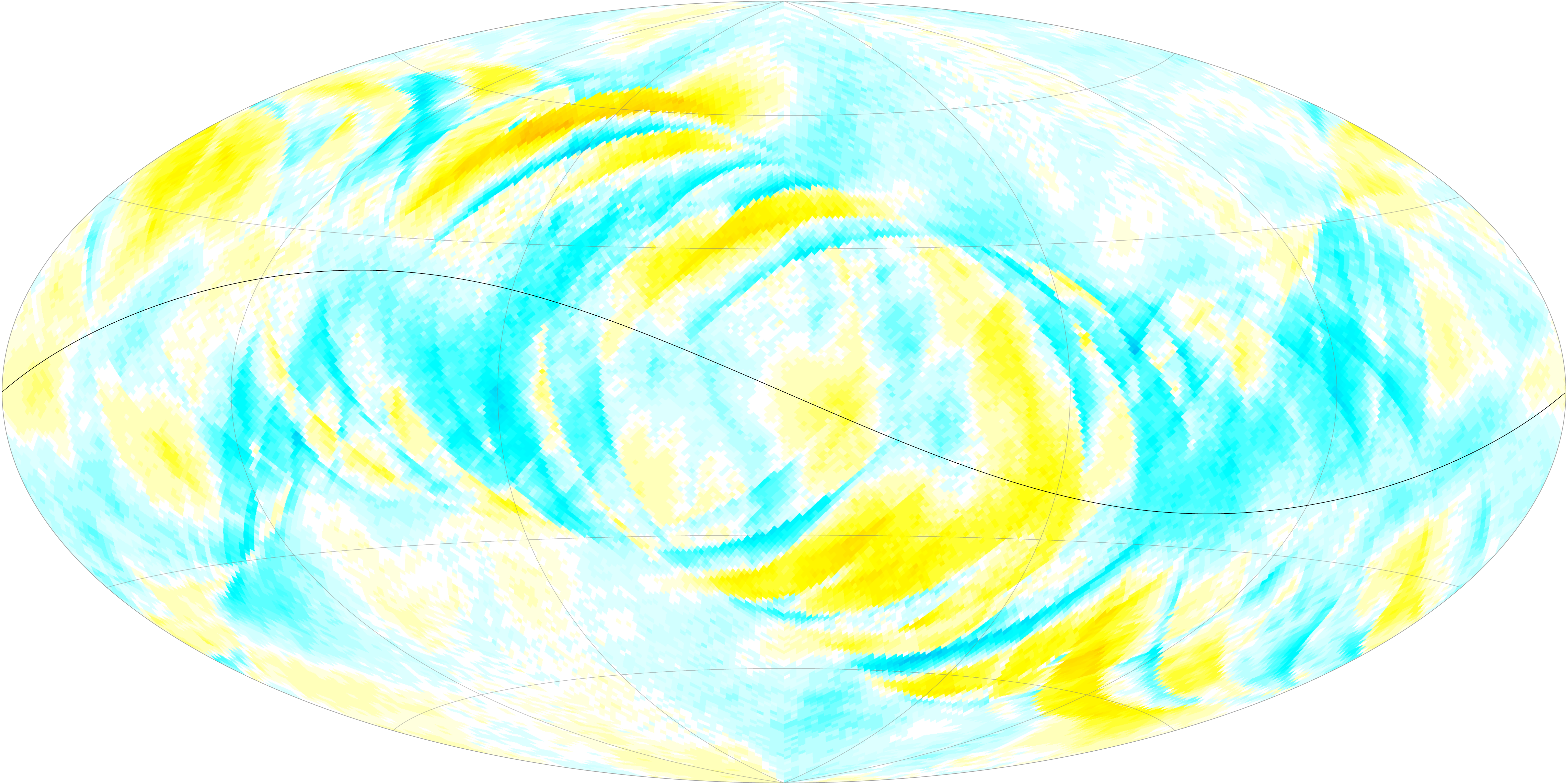}&
			\includegraphics[keepaspectratio,width=\mygridwidth]{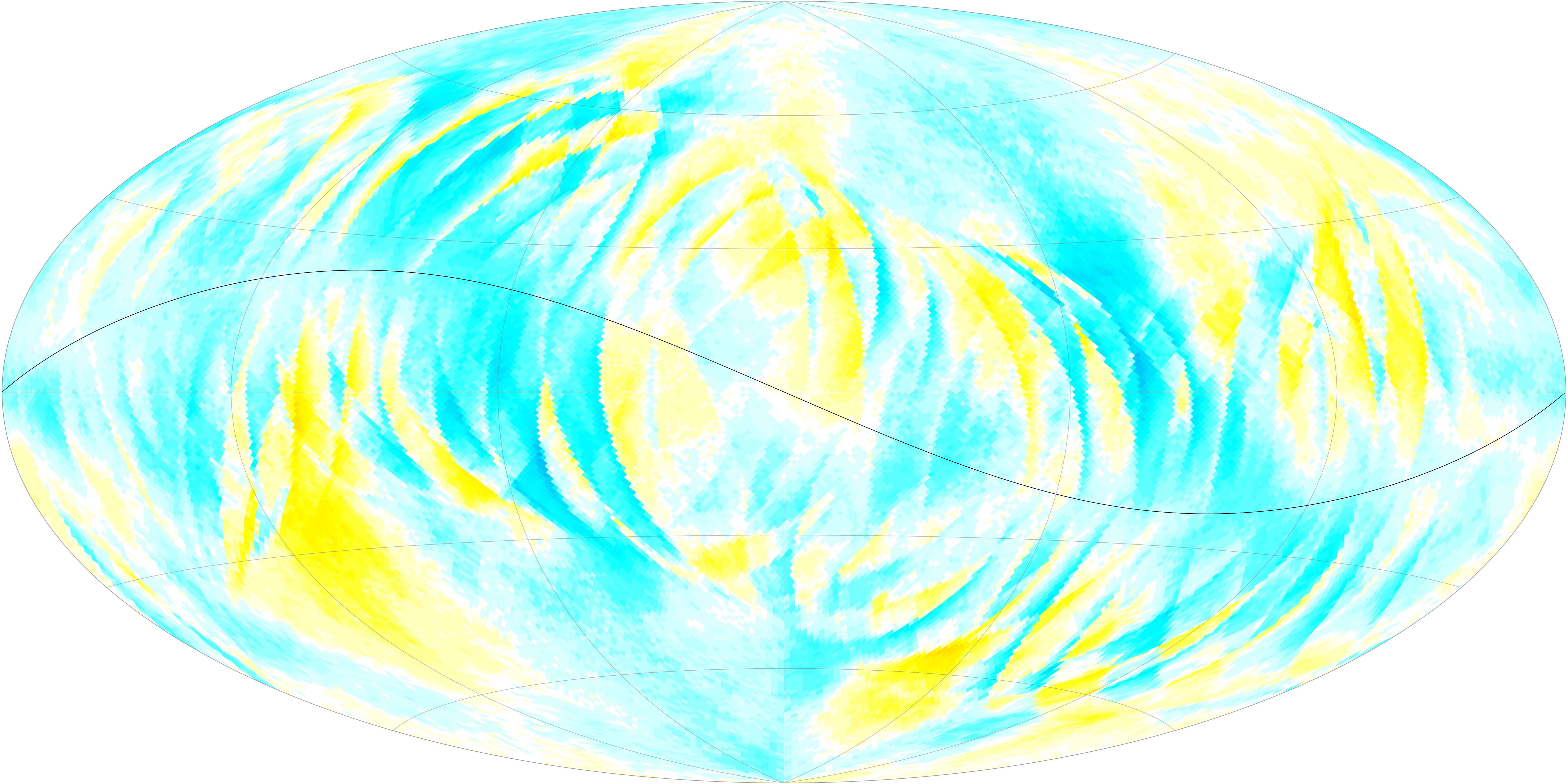}& \\
			\rotatebox{90}{\large\textbf{63\,d}} &
			\includegraphics[keepaspectratio,width=\mygridwidth]{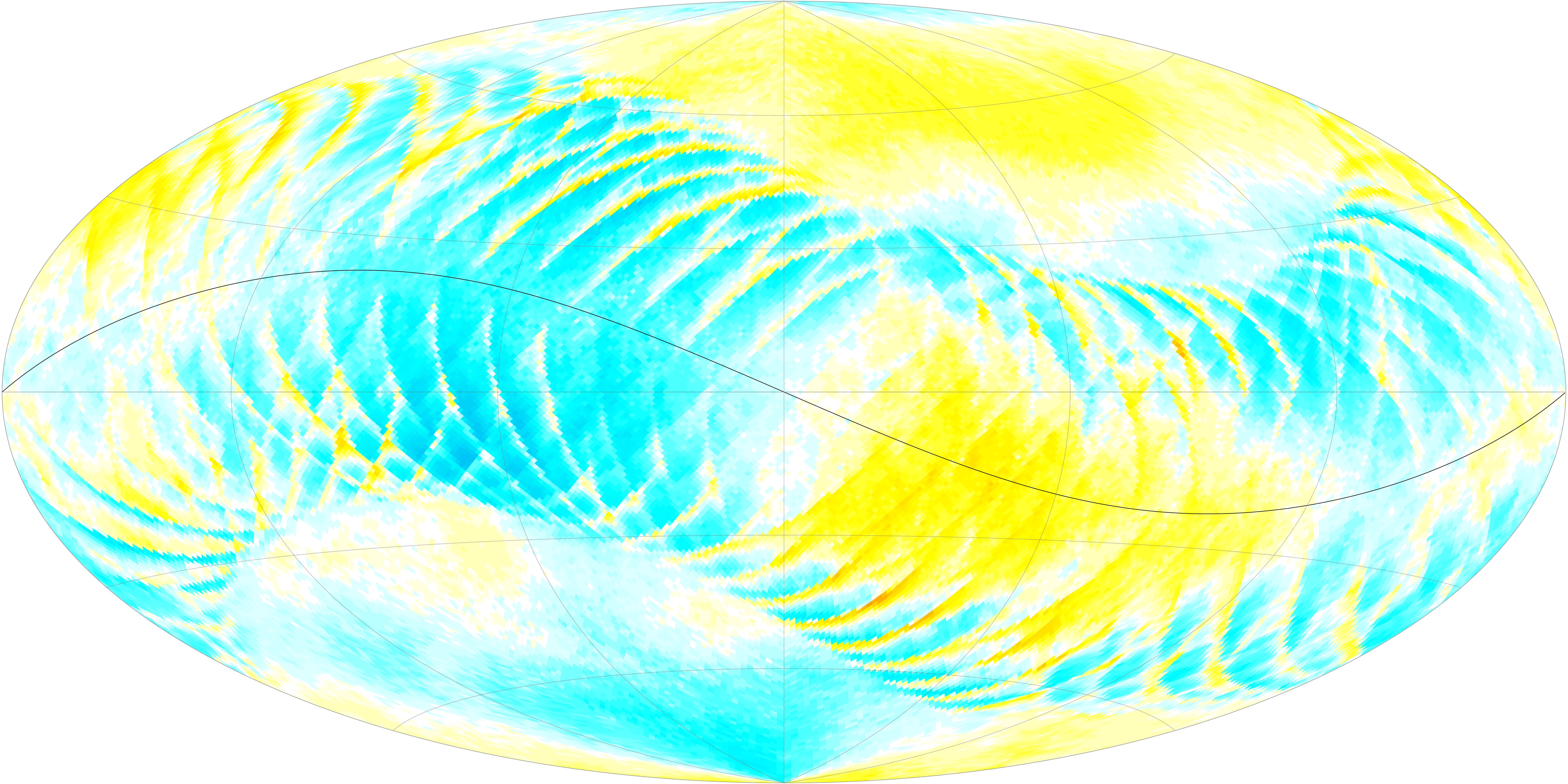}&
			\includegraphics[keepaspectratio,width=\mygridwidth]{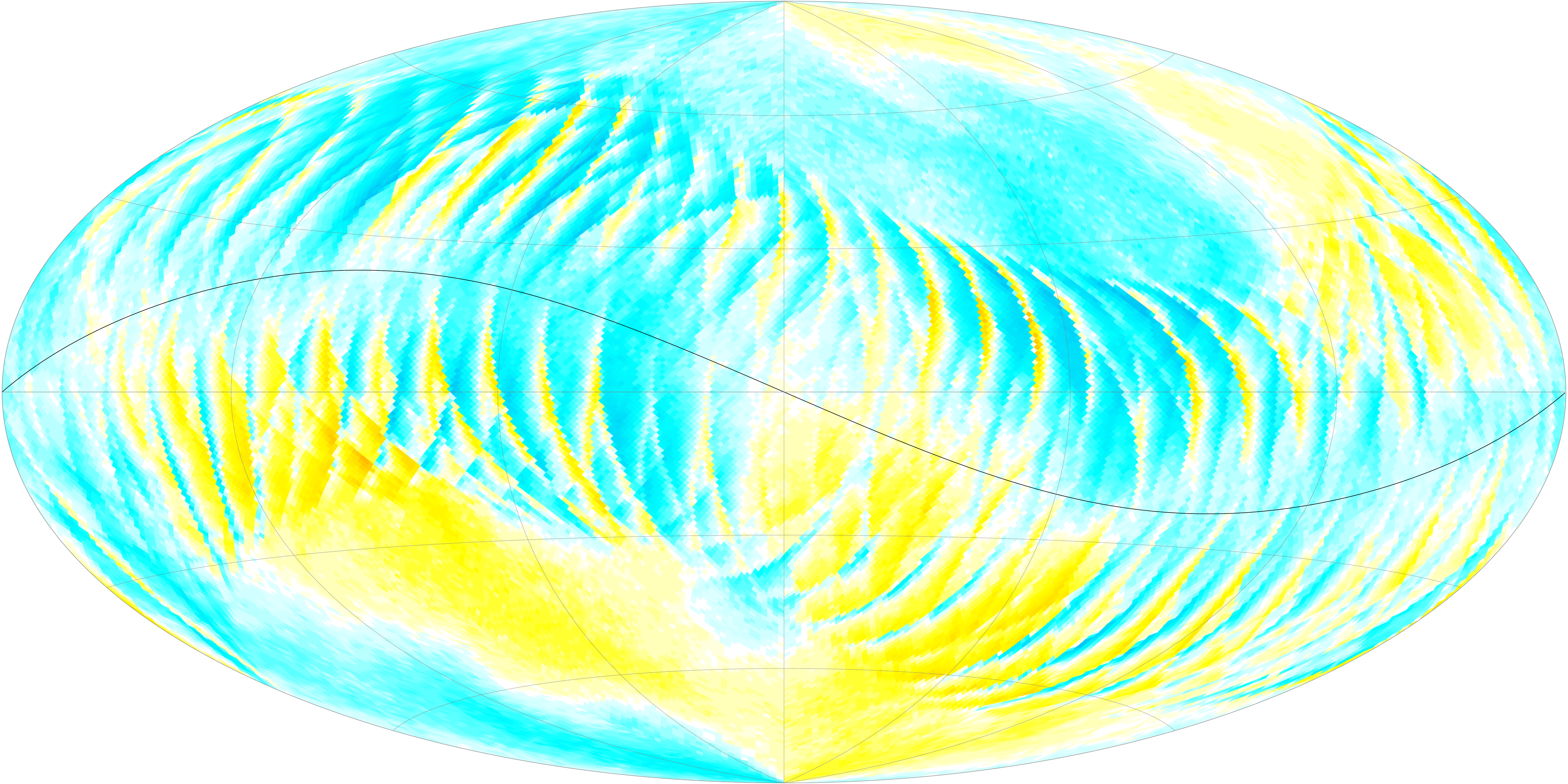}&
			\includegraphics[keepaspectratio,width=\mygridwidth]{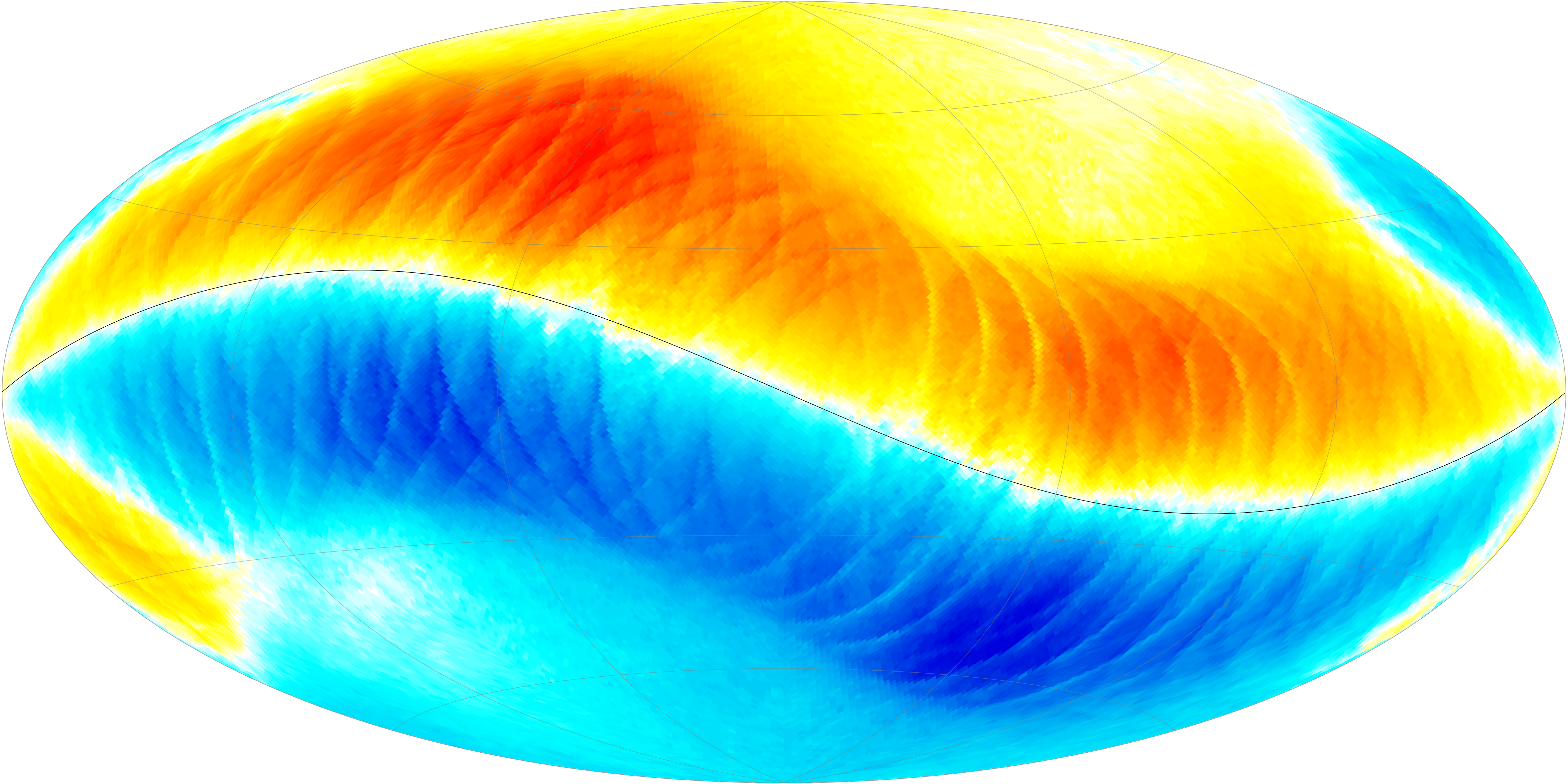}&
			\includegraphics[keepaspectratio,width=\mygridwidth]{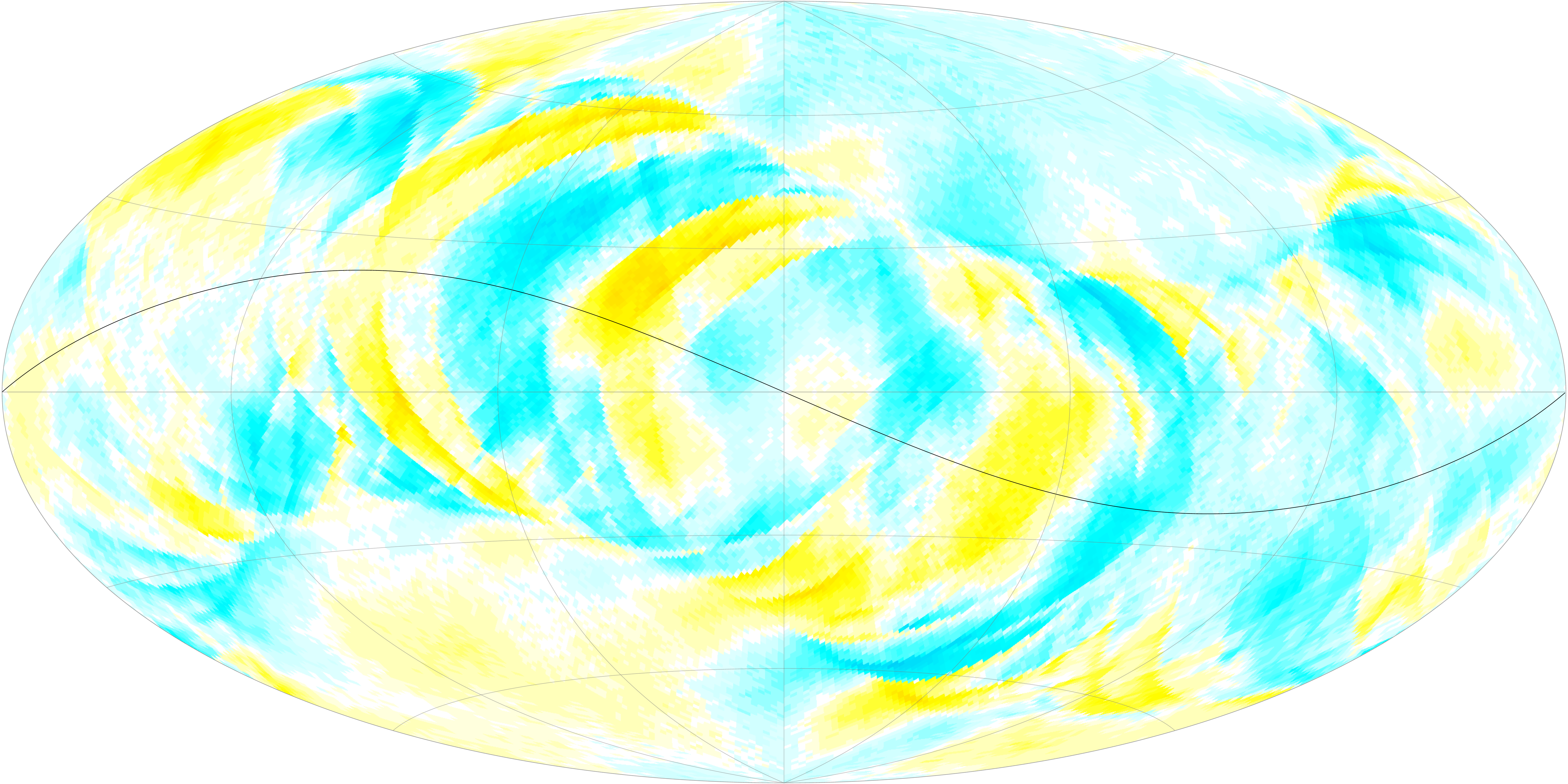}&
			\includegraphics[keepaspectratio,width=\mygridwidth]{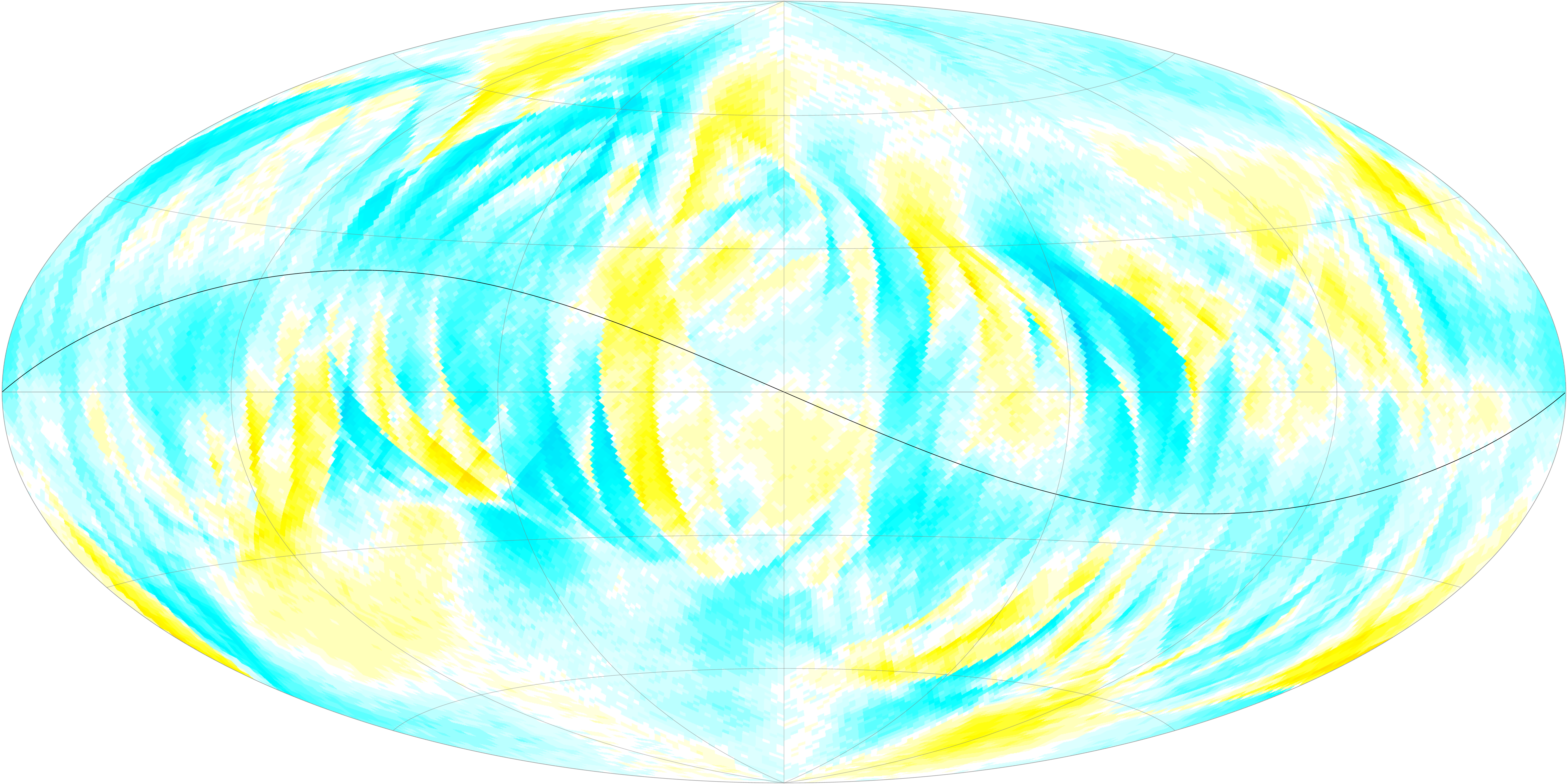}&\\
			\rotatebox{90}{\large\textbf{53.7\,d}} &
			\includegraphics[keepaspectratio,width=\mygridwidth]{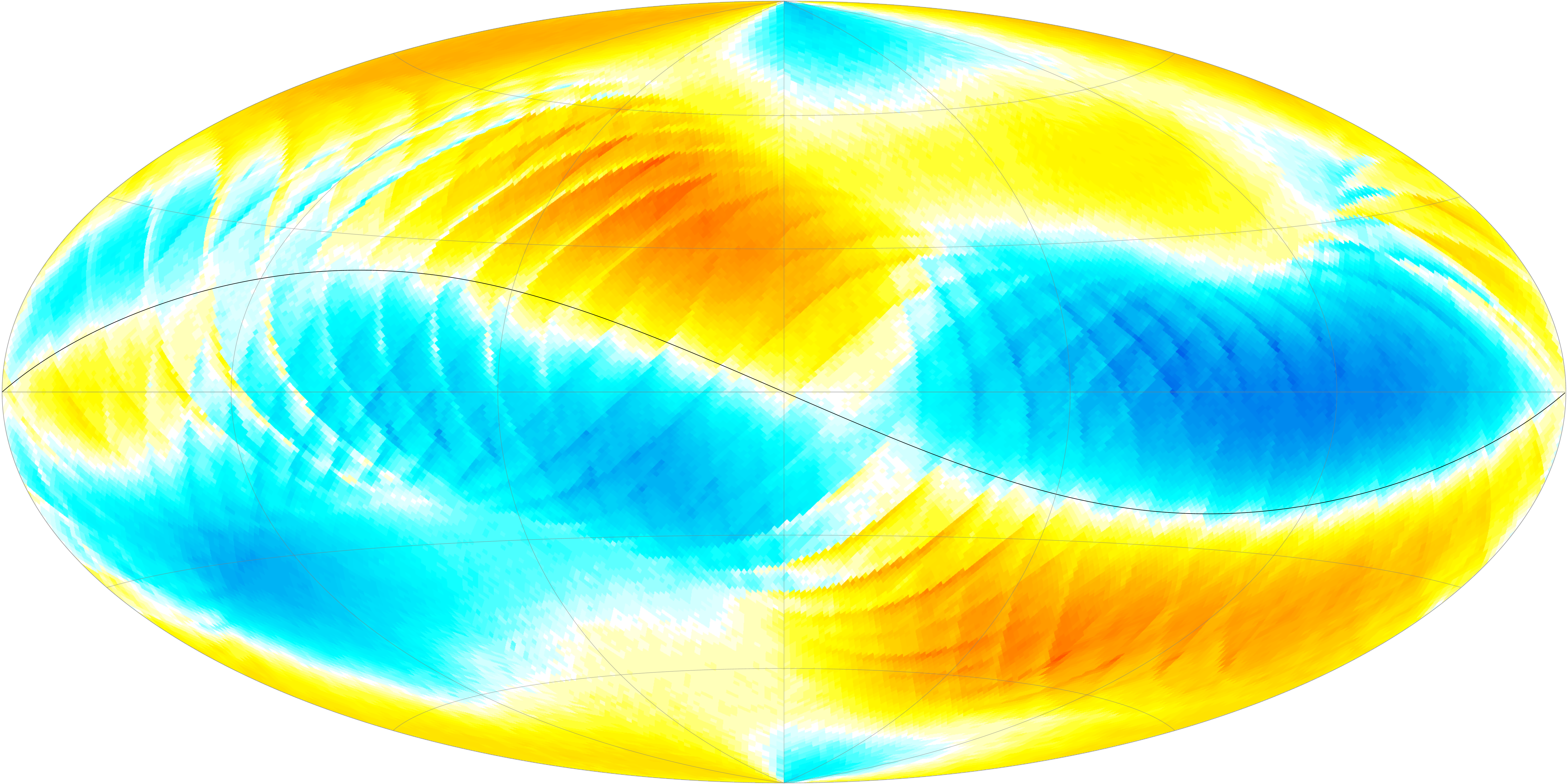}&
			\includegraphics[keepaspectratio,width=\mygridwidth]{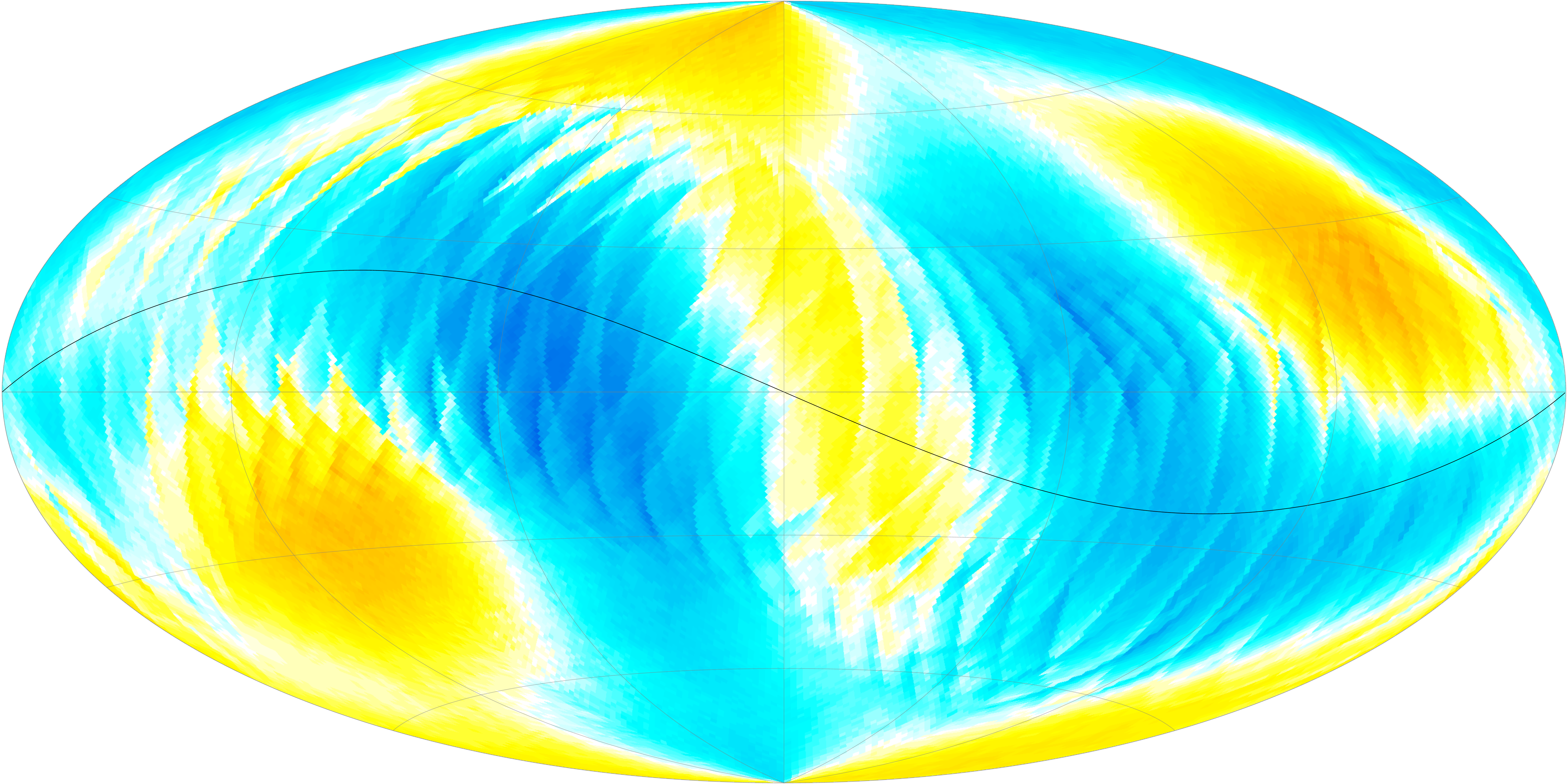}&
			\includegraphics[keepaspectratio,width=\mygridwidth]{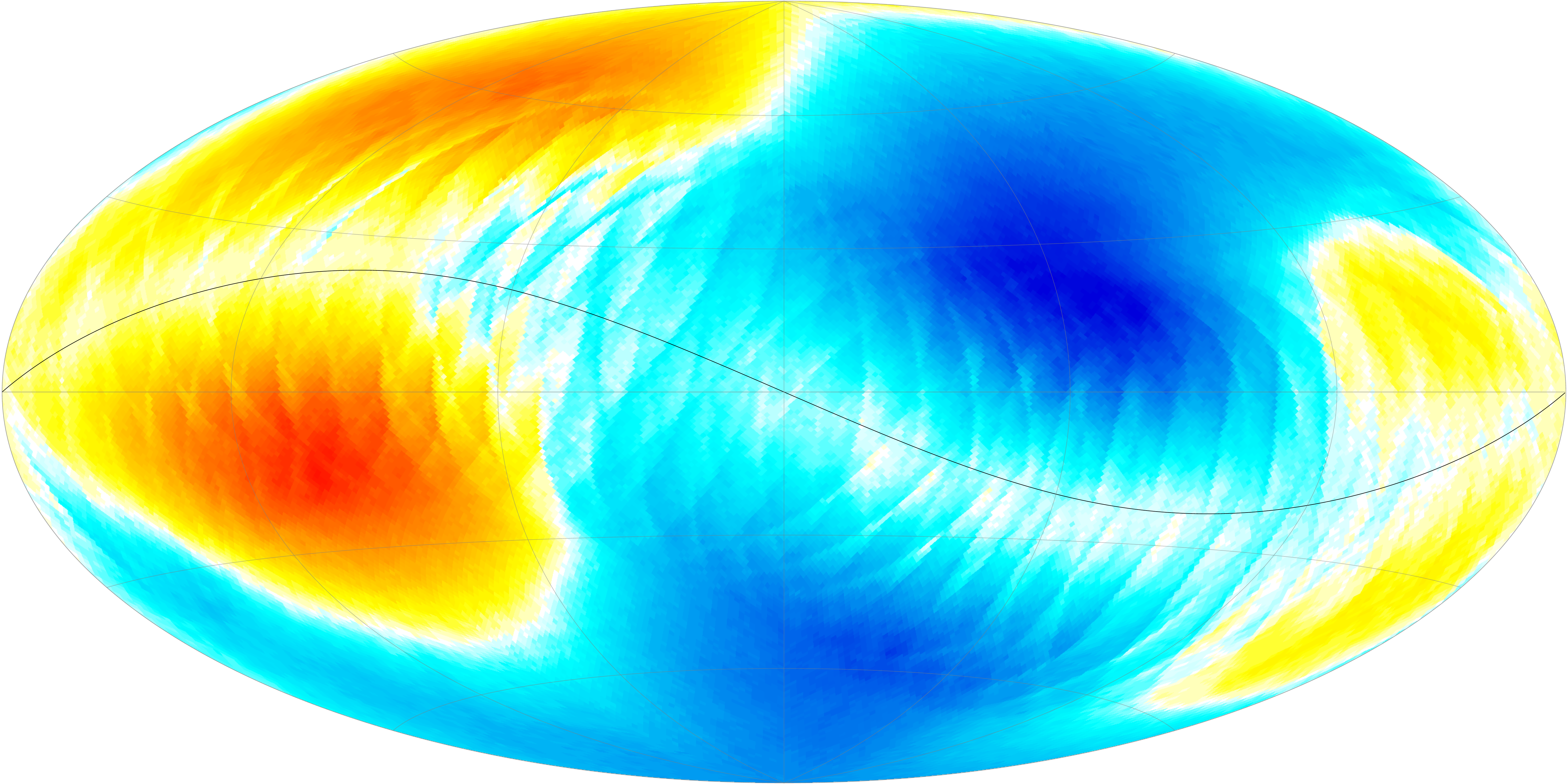}&
			\includegraphics[keepaspectratio,width=\mygridwidth]{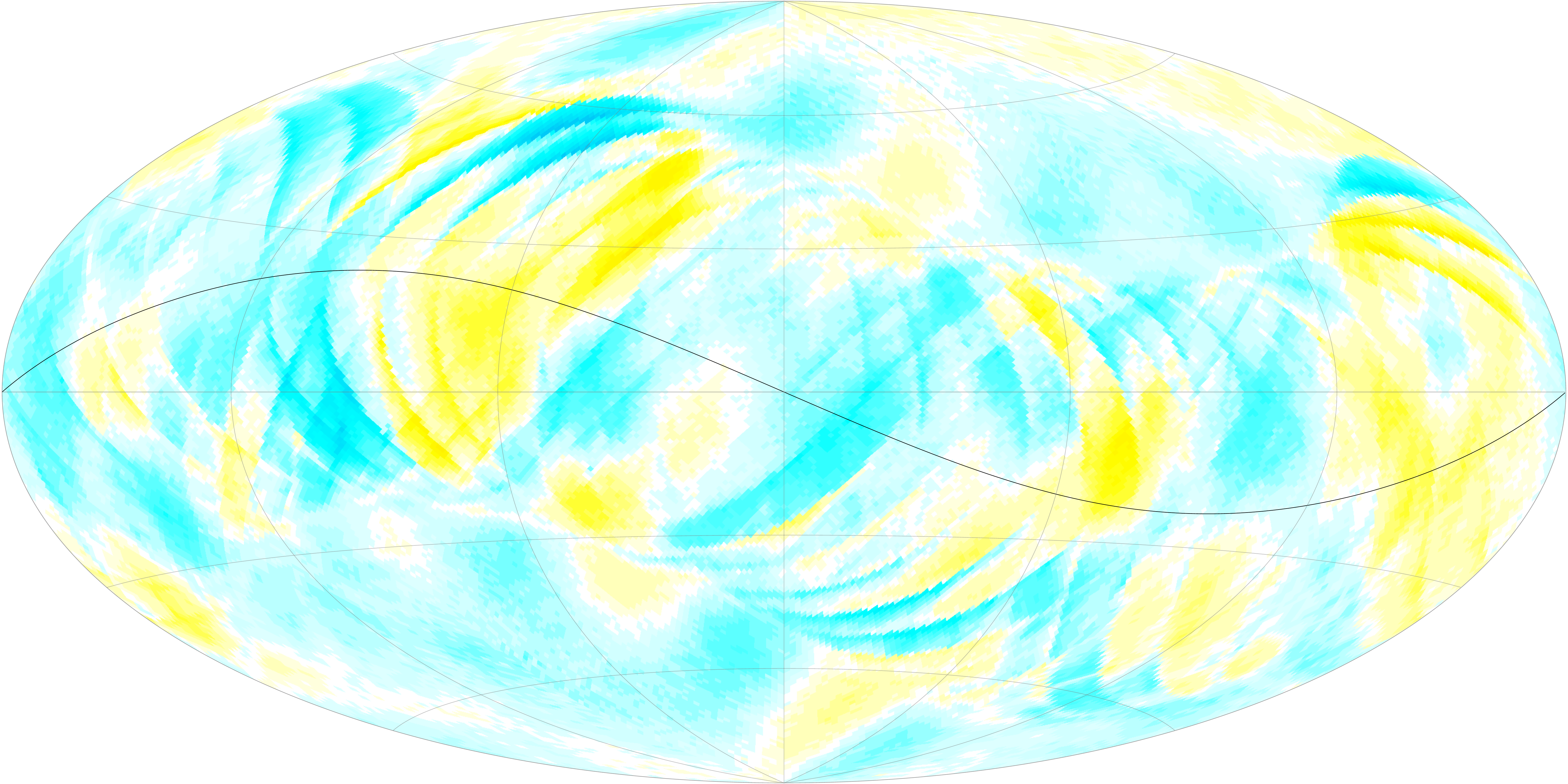}&
			\includegraphics[keepaspectratio,width=\mygridwidth]{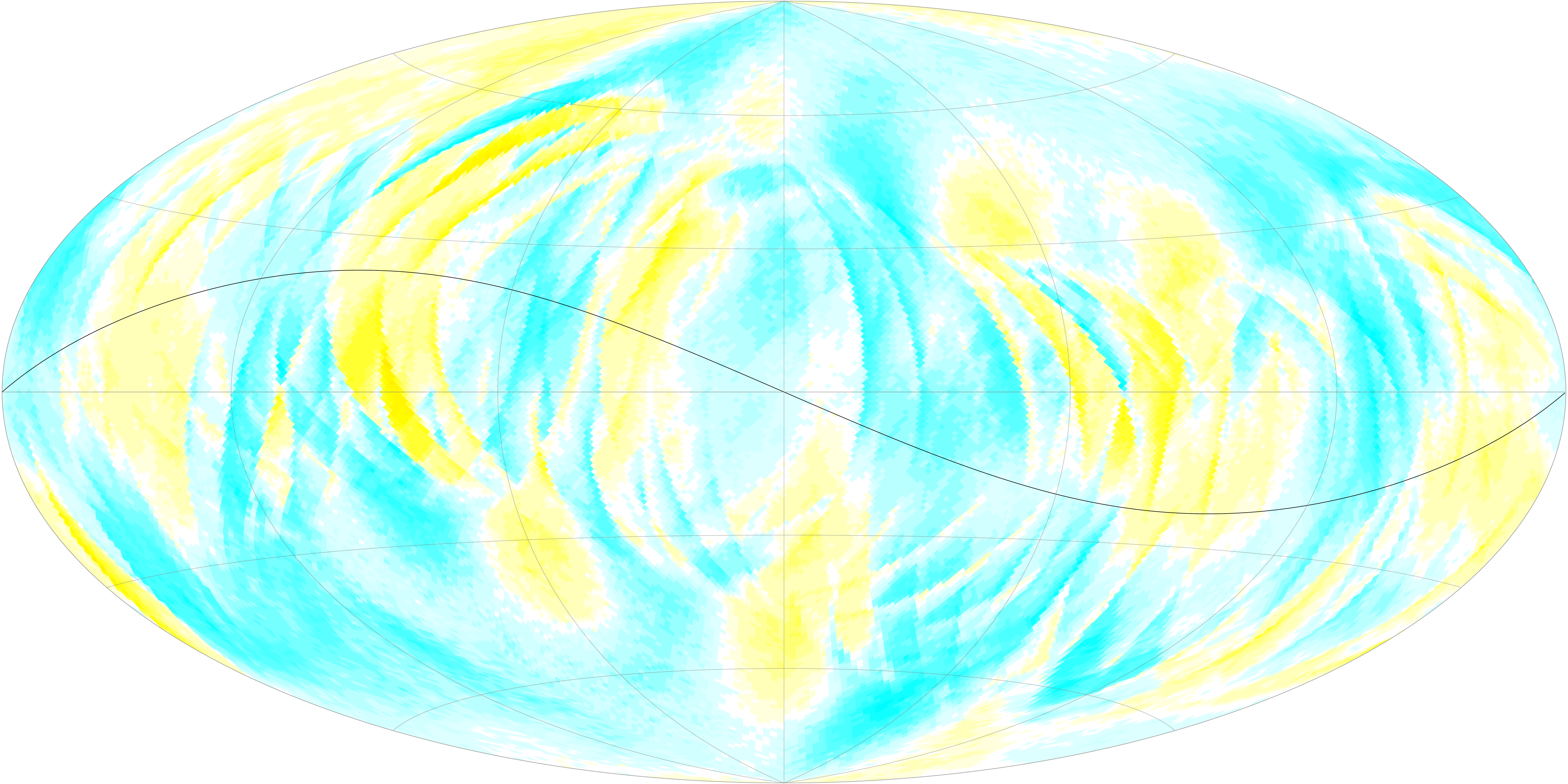}&\\
		\end{tabular}
		\caption{Errors of the five astrometric parameters
			(top label) depending on the GW period $\Pgw$ (left label).
			The propagation direction of the GW is towards the
			centre of the map ($\aGW=\dGW=0$). The
			strain parameters are all equal giving eccentricity
			$e=1$. The plots show the median error per HEALPix
			at a HEALPix level of 6. The black line represents
			the ecliptic. Additional maps for $\Pgw=20$\,yr are 
			shown in Fig.~\ref{fig_simulated20yr}.
			\label{fig___error_sky_maps}}
	\end{figure}
\end{landscape}

\clearpage

\begin{table*}[h!]
	\caption{Statistics of the normalized astrometric errors $\Delta\epsilon/\Delta_{\rm max}$ produced by GWs of selected frequencies.\label{tab_statstable2}}
	\begin{minipage}{1\columnwidth}
		{\setlength{\extrarowheight}{1.5pt}% top pading of cells
			\resizebox{1\columnwidth}{!}{%
				\begin{tabular}{llSSSSSS[table-format=2.3,table-number-alignment=left]}
					\hline\hline
					$P_{gw}$   & $\Delta\epsilon$     & $\overline{x}$  & $\sigma$ & $s$    & $k$    & min    & max\\\hline
					20.00\,yr     & $\Delta\alpha^*$  & 0.00            & 0.37     & 0.00   & -1.24  & -0.65  & 0.66\\
					& $\Delta\delta$                  & 0.00            & 0.37     & 0.00   & -1.23  & -0.69  & 0.67\\
					& $\Delta\varpi$                  & 0.00            & 0.01     & 0.14   & 2.17   & -0.06  & 0.07\\
					& $\Delta\mu_{\alpha *}$          & -0.00           & 0.12     & -0.03  & -1.19  & -0.23  & 0.22\\
					& $\Delta\mu_\delta$              & 0.00            & 0.12     & -0.02  & -1.23  & -0.25  & 0.24\\\hline
					5.00\,yr     & $\Delta\alpha^*$   & -0.00           & 0.12     & 0.34   & 3.78   & -0.88  & 0.79\\
					& $\Delta\delta$                  & 0.00            & 0.16     & -0.05  & 2.39   & -0.93  & 0.80\\
					& $\Delta\varpi$                  & 0.00            & 0.10     & 0.57   & 0.97   & -0.65  & 0.62\\
					& $\Delta\mu_{\alpha *}$          & -0.00           & 0.18     & -0.16  & -0.73  & -0.78  & 0.64\\
					& $\Delta\mu_\delta$              & 0.00            & 0.18     & -0.09  & -0.54  & -0.67  & 0.62\\\hline
					1.58\,yr   & $\Delta\alpha^*$     & -0.00           & 0.12     & -0.06  & 6.10   & -1.09  & 0.91\\
					& $\Delta\delta$                  & 0.00            & 0.13     & 0.07   & 3.02   & -0.73  & 0.79\\
					& $\Delta\varpi$                  & 0.01            & 0.11     & 0.64   & 1.29   & -0.56  & 0.65\\
					& $\Delta\mu_{\alpha *}$          & -0.01           & 0.11     & -0.23  & 1.12   & -0.56  & 0.51\\
					& $\Delta\mu_\delta$              & 0.01            & 0.11     & -0.15  & 2.03   & -0.69  & 0.56\\\hline
					1.00\,yr      & $\Delta\alpha^*$  & 0.01            & 0.10     & 0.46   & 2.62   & -0.55  & 0.72\\
					& $\Delta\delta$                  & -0.00           & 0.10     & -0.12  & 3.50   & -0.73  & 0.66\\
					& $\Delta\varpi$                  & -0.09           & 0.52     & -0.76  & 0.48   & -1.83  & 0.96\\
					& $\Delta\mu_{\alpha *}$          & 0.00            & 0.09     & 0.21   & 2.29   & -0.53  & 0.57\\
					& $\Delta\mu_\delta$              & 0.00            & 0.11     & 0.06   & 2.10   & -0.51  & 0.61\\\hline
					279.00\,d     & $\Delta\alpha^*$  & -0.00           & 0.12     & -0.14  & 3.08   & -0.84  & 0.80\\
					& $\Delta\delta$                  & 0.00            & 0.12     & -0.08  & 3.66   & -0.84  & 0.74\\
					& $\Delta\varpi$                  & 0.02            & 0.14     & 0.76   & 0.83   & -0.55  & 0.70\\
					& $\Delta\mu_{\alpha *}$          & 0.00            & 0.07     & 0.00   & 4.21   & -0.51  & 0.55\\
					& $\Delta\mu_\delta$              & -0.00           & 0.07     & 0.10   & 3.33   & -0.38  & 0.45\\\hline
					230.94\,d  & $\Delta\alpha^*$     &-0.00            & 0.10     & 0.02   & 2.75   & -0.66  & 0.78\\
					& $\Delta\delta$                  &-0.00            & 0.10     & 0.05   & 4.48   & -0.71  & 0.75\\
					& $\Delta\varpi$                  &-0.01            & 0.08     & -0.36  & 1.74   & -0.53  & 0.50\\
					& $\Delta\mu_{\alpha *}$          &0.00             & 0.10     & 0.10   & 1.45   & -0.50  & 0.50\\
					& $\Delta\mu_\delta$              &-0.00            & 0.09     & 0.07   & 1.48   & -0.49  & 0.51\\\hline
					182.75\,d  & $\Delta\alpha^*$     & 0.00            & 0.28     & -0.67  & 0.68   & -0.96  & 1.08\\
					& $\Delta\delta$                  & -0.04           & 0.16     & -0.00  & 0.56   & -0.74  & 0.88\\
					& $\Delta\varpi$                  & 0.01            & 0.08     & 0.02   & 2.63   & -0.52  & 0.63\\
					& $\Delta\mu_{\alpha *}$          & -0.00           & 0.10     & -0.73  & 2.27   & -0.59  & 0.53\\
					& $\Delta\mu_\delta$              & -0.00           & 0.08     & -0.13  & 1.33   & -0.44  & 0.38\\\hline
					144.62\,d  & $\Delta\alpha^*$     & 0.00            & 0.12     & -0.21  & 2.49   & -0.76  & 0.76\\
					& $\Delta\delta$                  & -0.01           & 0.12     & -0.13  & 3.49   & -0.77  & 0.81\\
					& $\Delta\varpi$                  & 0.01            & 0.11     & 0.02   & 1.41   & -0.63  & 0.59\\
					& $\Delta\mu_{\alpha *}$          & 0.00            & 0.07     & 0.38   & 2.55   & -0.43  & 0.47\\
					& $\Delta\mu_\delta$              & 0.00            & 0.07     & 0.05   & 3.84   & -0.45  & 0.50\\\hline
					132.17\,d  & $\Delta\alpha^*$     & 0.00            & 0.18     & 0.16   & 1.76   & -0.89  & 0.99\\
					& $\Delta\delta$                  & 0.00            & 0.15     & -0.03  & 1.37   & -0.97  & 0.80\\
					& $\Delta\varpi$                  & -0.01           & 0.18     & -0.13  & 0.95   & -0.78  & 0.69\\
					& $\Delta\mu_{\alpha *}$          & -0.00           & 0.09     & 0.13   & 1.74   & -0.52  & 0.52\\
					& $\Delta\mu_\delta$              & -0.01           & 0.08     & -0.13  & 2.65   & -0.54  & 0.40\\\hline
					127.50\,d  & $\Delta\alpha^*$     & -0.00           & 0.17     & -0.12  & 2.10   & -1.07  & 1.07\\
					& $\Delta\delta$                  & -0.01           & 0.15     & 0.04   & 1.09   & -0.73  & 0.80\\
					& $\Delta\varpi$                  & 0.01            & 0.18     & -0.42  & 0.48   & -0.85  & 0.72\\
					& $\Delta\mu_{\alpha *}$          & 0.00            & 0.10     & -0.38  & 1.75   & -0.58  & 0.54\\
					& $\Delta\mu_\delta$              & -0.00           & 0.08     & 0.01   & 0.58   & -0.42  & 0.49\\\hline
					121.89\,d    & $\Delta\alpha^*$   & -0.00           & 0.09     & -0.42  & 4.09   & -0.84  & 0.69\\
					& $\Delta\delta$                  & 0.00            & 0.09     & 0.08   & 3.65   & -0.57  & 0.70\\
					& $\Delta\varpi$                  & 0.01            & 0.25     & -0.72  & 0.28   & -0.97  & 0.76\\
					& $\Delta\mu_{\alpha *}$          & 0.00            & 0.11     & -0.13  & 2.48   & -0.74  & 0.61\\
					& $\Delta\mu_\delta$              & 0.01            & 0.11     & 0.17   & 1.46   & -0.46  & 0.60\\\hline
				\end{tabular}
		}}
	\end{minipage}
	\hfill
	\begin{minipage}{1\columnwidth}
		%	\vspace*{-2ex}
		{\setlength{\extrarowheight}{1.5pt}% top pading of cells
			\resizebox{1\columnwidth}{!}{%
				\begin{tabular}{llSSSSSS[table-format=2.3,table-number-alignment=left]}
					\hline\hline
					$P_{gw}$   & $\Delta\epsilon$     & $\overline{x}$ & $\sigma$ & $s$    & $k$    & min    & max\\\hline
					105.54\,d  & $\Delta\alpha^*$     & -0.00          & 0.13     & -0.10  & 2.91   & -0.88  & 0.82\\
					& $\Delta\delta$                  & 0.01           & 0.14     & -0.03  & 2.10   & -0.82  & 0.87\\
					& $\Delta\varpi$                  & -0.00          & 0.12     & -0.09  & 1.04   & -0.71  & 0.61\\
					& $\Delta\mu_{\alpha *}$          & 0.00           & 0.09     & 0.44   & 0.97   & -0.45  & 0.56\\
					& $\Delta\mu_\delta$              & 0.01           & 0.09     & -0.11  & 1.14   & -0.43  & 0.49\\\hline
					96.10\,d    & $\Delta\alpha^*$    & -0.01          & 0.32     & 0.07   & -0.92  & -0.78  & 0.74\\
					& $\Delta\delta$                  & -0.01          & 0.36     & 0.05   & -0.31  & -1.09  & 1.09\\
					& $\Delta\varpi$                  & -0.01          & 0.39     & 0.04   & -0.41  & -1.13  & 1.07\\
					& $\Delta\mu_{\alpha *}$          & 0.00           & 0.12     & 0.18   & -0.29  & -0.46  & 0.42\\
					& $\Delta\mu_\delta$              & 0.01           & 0.11     & -0.06  & 0.95   & -0.52  & 0.48\\\hline
					91.38\,d   & $\Delta\alpha^*$     & -0.01          & 0.28     & -0.47  & 0.69   & -1.50  & 1.02\\
					& $\Delta\delta$                  & 0.00           & 0.29     & -0.11  & 0.91   & -1.26  & 0.96\\
					& $\Delta\varpi$                  & 0.00           & 0.07     & 0.04   & 1.72   & -0.43  & 0.47\\
					& $\Delta\mu_{\alpha *}$          & 0.01           & 0.14     & 0.10   & -0.77  & -0.42  & 0.49\\
					& $\Delta\mu_\delta$              & 0.01           & 0.13     & 0.07   & -0.07  & -0.42  & 0.58\\\hline
					82.77\,d   & $\Delta\alpha^*$     & 0.00           & 0.10     & 0.07   & 4.62   & -0.98  & 0.82\\
					& $\Delta\delta$                  & 0.00           & 0.12     & 0.08   & 3.11   & -0.74  & 0.67\\
					& $\Delta\varpi$                  & 0.00           & 0.10     & -0.29  & 2.60   & -0.71  & 0.64\\
					& $\Delta\mu_{\alpha *}$          & 0.02           & 0.13     & 0.54   & 1.24   & -0.55  & 0.63\\
					& $\Delta\mu_\delta$              & -0.01          & 0.10     & -0.29  & 1.77   & -0.54  & 0.52\\\hline
					76.10\,d    & $\Delta\alpha^*$    & -0.02          & 0.24     & -0.02  & 0.18   & -0.82  & 0.69\\
					& $\Delta\delta$                  & 0.03           & 0.20     & 0.11   & -0.34  & -0.58  & 0.72\\
					& $\Delta\varpi$                  & -0.06          & 0.37     & 0.10   & 0.34   & -1.30  & 1.16\\
					& $\Delta\mu_{\alpha *}$          & -0.00          & 0.07     & 0.04   & 2.67   & -0.52  & 0.43\\
					& $\Delta\mu_\delta$              & 0.00           & 0.08     & 0.33   & 1.27   & -0.38  & 0.42\\\hline
					69.83\,d    & $\Delta\alpha^*$    & -0.00          & 0.12     & -0.30  & 5.08   & -0.99  & 0.80\\
					& $\Delta\delta$                  & 0.00           & 0.10     & -0.01  & 4.88   & -0.74  & 0.75\\
					& $\Delta\varpi$                  & 0.00           & 0.10     & 0.02   & 2.42   & -0.67  & 0.57\\
					& $\Delta\mu_{\alpha *}$          & -0.01          & 0.09     & -0.32  & 1.86   & -0.59  & 0.41\\
					& $\Delta\mu_\delta$              & 0.01           & 0.08     & 0.06   & 1.50   & -0.54  & 0.47\\\hline
					67.90\,d    & $\Delta\alpha^*$    & -0.00          & 0.12     & -0.35  & 3.89   & -0.77  & 0.77\\
					& $\Delta\delta$                  & 0.00           & 0.11     & -0.13  & 4.21   & -0.86  & 0.81\\
					& $\Delta\varpi$                  & -0.00          & 0.08     & 0.06   & 1.97   & -0.50  & 0.56\\
					& $\Delta\mu_{\alpha *}$          & 0.01           & 0.08     & 0.50   & 2.05   & -0.48  & 0.51\\
					& $\Delta\mu_\delta$              & -0.01          & 0.08     & -0.07  & 1.34   & -0.46  & 0.39\\\hline
					63.00\,d      & $\Delta\alpha^*$  & 0.00           & 0.10     & 0.02   & 0.54   & -0.47  & 0.67\\
					& $\Delta\delta$                  & -0.01          & 0.11     & 0.04   & 0.53   & -0.50  & 0.51\\
					& $\Delta\varpi$                  & -0.01          & 0.47     & 0.01   & -0.63  & -1.33  & 1.38\\
					& $\Delta\mu_{\alpha *}$          & 0.00           & 0.09     & 0.18   & 1.28   & -0.38  & 0.45\\
					& $\Delta\mu_\delta$              & -0.01          & 0.08     & -0.03  & 0.92   & -0.43  & 0.39\\\hline
					60.95\,d   & $\Delta\alpha^*$     & 0.00           & 0.28     & -0.03  & 0.50   & -1.17  & 1.17\\
					& $\Delta\delta$                  & 0.00           & 0.17     & -0.02  & 1.75   & -0.79  & 0.84\\
					& $\Delta\varpi$                  & 0.00           & 0.09     & 0.18   & 2.32   & -0.57  & 0.60\\
					& $\Delta\mu_{\alpha *}$          & -0.01          & 0.08     & 0.26   & 0.41   & -0.41  & 0.40\\
					& $\Delta\mu_\delta$              & 0.02           & 0.09     & 0.26   & 0.90   & -0.43  & 0.46\\\hline
					53.70\,d    & $\Delta\alpha^*$    & 0.03           & 0.28     & -0.03  & -0.25  & -0.84  & 0.91\\
					& $\Delta\delta$                  & -0.07          & 0.24     & -0.04  & -0.51  & -0.85  & 0.58\\
					& $\Delta\varpi$                  & -0.11          & 0.41     & 0.08   & 0.08   & -1.31  & 1.20\\
					& $\Delta\mu_{\alpha *}$          & -0.00          & 0.07     & 0.03   & 1.70   & -0.44  & 0.34\\
					& $\Delta\mu_\delta$              & -0.00          & 0.06     & 0.28   & 0.14   & -0.25  & 0.31\\\hline
					50.29\,d    & $\Delta\alpha^*$    & 0.00           & 0.13     & -0.03  & 2.49   & -0.81  & 0.87\\
					& $\Delta\delta$                  & -0.01          & 0.14     & 0.05   & 1.77   & -0.72  & 0.69\\
					& $\Delta\varpi$                  & -0.01          & 0.12     & 0.14   & 1.65   & -0.84  & 0.72\\
					& $\Delta\mu_{\alpha *}$          & 0.00           & 0.08     & 0.14   & 1.01   & -0.46  & 0.55\\
					& $\Delta\mu_\delta$              & -0.01          & 0.08     & 0.21   & 2.96   & -0.58  & 0.58\\\hline
				\end{tabular}
		}}
	\end{minipage}
	\tablefoot{Statistics are grouped by the GW period $\Pgw$ and computed over 
		all sources in the simulations. The following statistics 
		are given for each astrometric parameter ($\Delta\epsilon$) considered: 
		mean $\overline{x}$, standard deviation $\sigma$, skewness $s$,
		excess kurtosis $k$ ($=0$ for a normal distribution), minimal and maximal values.
		All astrometric errors were normalised by $\Delta_{\rm max}$ before computing the
		statistics. All statistics are rounded to two decimal places.
		Sky distributions for some of the simulations are shown in
		Figs.~\ref{fig___error_sky_maps} and \ref{fig_simulated20yr}.}
\end{table*}

\section{SH/VSH expansions of astrometric error patterns}
\label{section-shvsh}

\begin{figure*}[htb]
	\centering
	\includegraphics[keepaspectratio,width=0.76\textwidth]{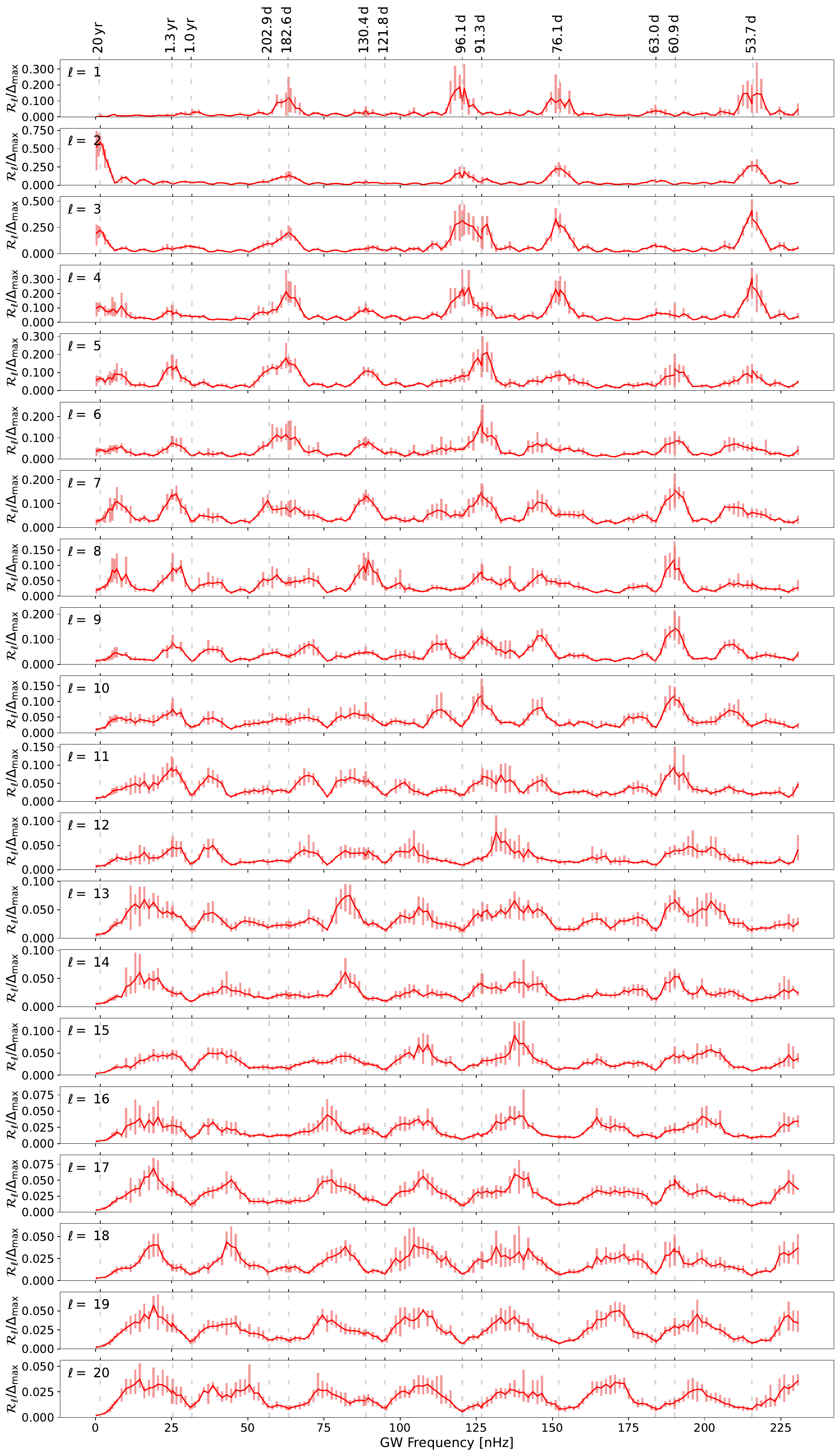}
	\caption{Normalised RMS variations $R_\ell/\Delta_{\rm max}$ of the 
		vector field of positional errors caused by GWs of different frequencies.
		\label{fig__Rl_pos}}
\end{figure*}

\begin{figure*}[htb]
	\centering
	\includegraphics[keepaspectratio,width=0.76\textwidth]{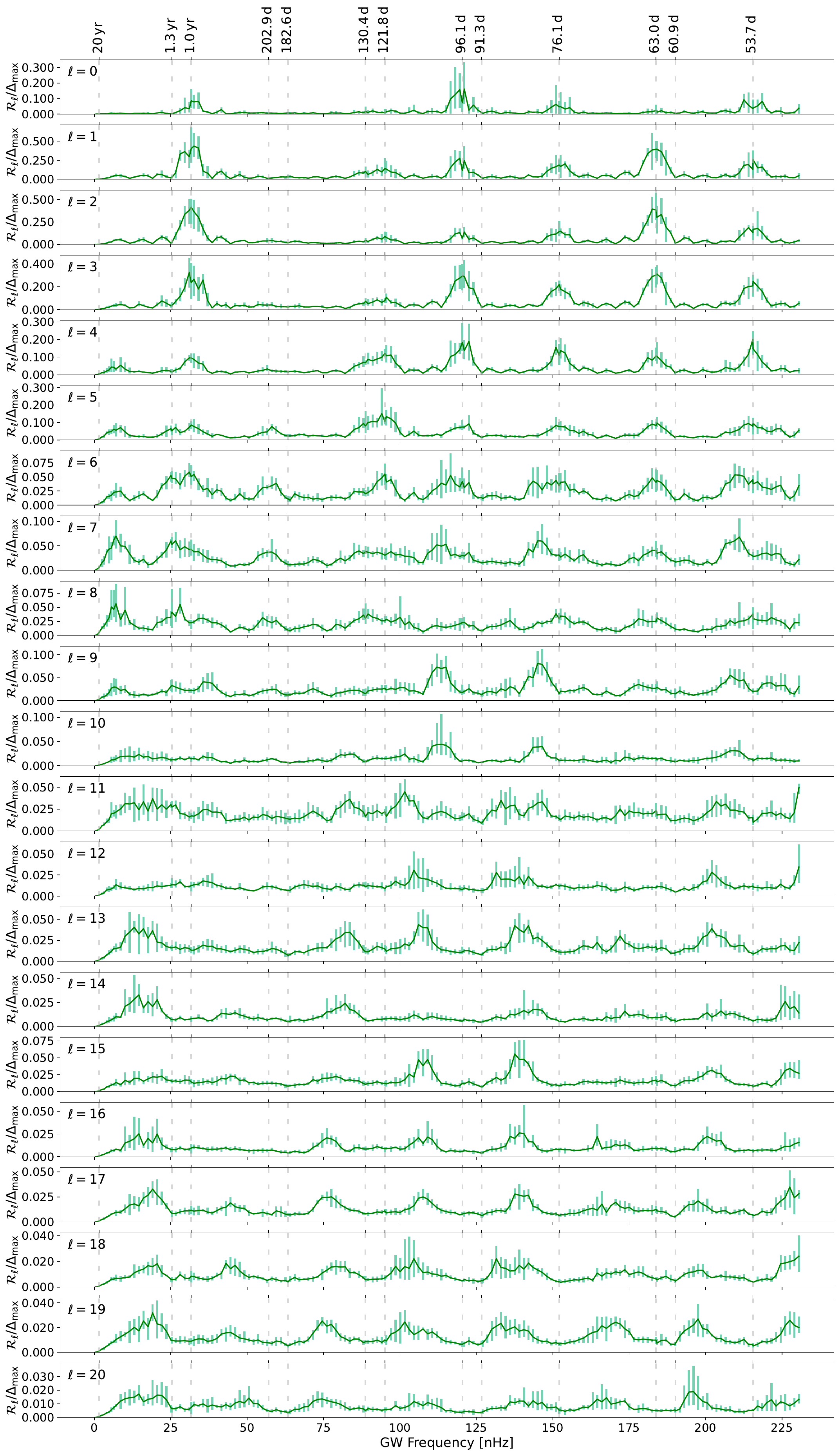}
	\caption{Normalised RMS variations $R_\ell/\Delta_{\rm max}$ of the 
		scalar field of parallax errors caused by GWs of different frequencies.
		\label{fig__Rl_para}}
\end{figure*}

\begin{figure*}[htb]
	\centering
	\includegraphics[keepaspectratio,width=0.76\textwidth]{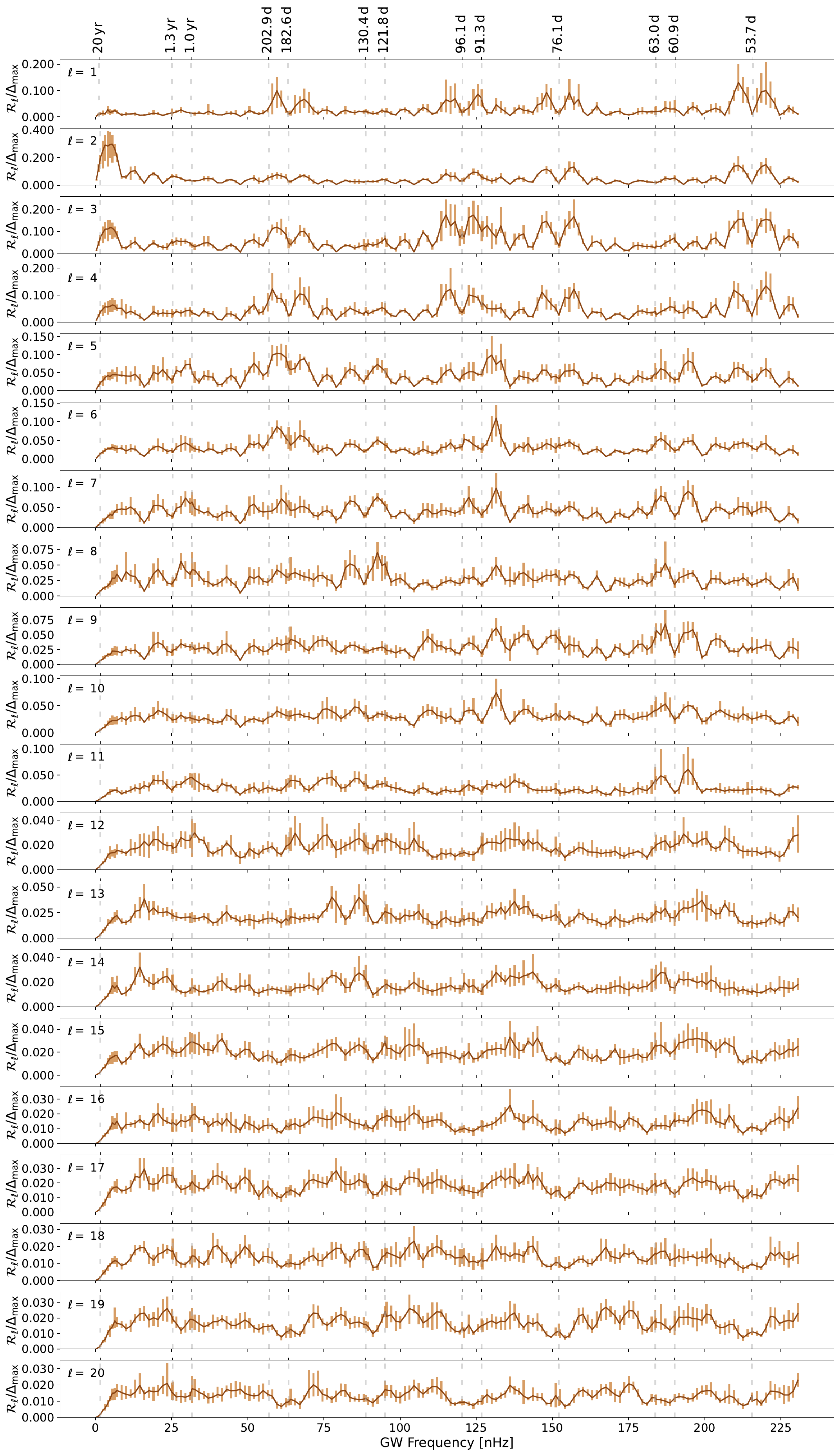}
	\caption{Normalised RMS variations $R_\ell/\Delta_{\rm max}$ of the 
		vector field of proper motion errors caused by GWs of different frequencies.
		\label{fig__Rl_pm}}
\end{figure*}

Figures~\ref{fig__Rl_pos}--\ref{fig__Rl_pm} show the
RMS values $R_\ell$ computed by 
Eq.~\eqref{RMS_of_VSH} from the SH or VSH expansions of the 
astrometric error patterns up to degree $\ell=20$.
$R_\ell$ characterises the components of the errors that have 
a typical spatial scale of $180^\circ/\ell$.
The RMS values $R_\ell$ are normalised by $\Delta_{\rm max}$, the
maximum astrometric effect of the GW at any point on the sky.
See Sect.~\ref{sec__spatial_features} for further explanation.

Similar to other plots in this paper, the coloured bars show the 
range of values obtained in the five simulations for each frequency 
(see Sect.~\ref{sec__methodology}). The solid
lines show the mean values computed over all simulations with a
given GW frequency. These values can be considered as typical values
and are used for the maps on Fig.~\ref{fig__heatmaps}. It is however
clear that the scatter of $R_\ell$ among different simulations is not
negligible. In particular, the scatter mainly shows the dependence of the
error patterns on the direction of the GW. Although the $R_\ell$ itself
is rotationally invariant, the scanning law used in the simulations is
not, which means that the error patterns and $R_\ell$ will depend on 
the GW direction as well as on the strain parameters. 

The plots characterise a complicated structure of the error
fields as a function of GW frequency. As already discussed in 
connection with Fig.~\ref{fig__heatmaps}, the errors increase for the
special frequencies $\nu_{k,l,m}$ in Eq.~\eqref{eq___peak_freqs} and
Table~\ref{tab__list_peaks}. At some of these 
frequencies the errors have significant harmonics starting from the 
lowest degree $\ell$ (0 or 1), at others the significant 
harmonics start only from $\ell=5$.
It should be noted that the vertical scale was adjusted individually 
for each $\ell$ and varies bu more than a factor ten. Generally
speaking, the RMS values decrease strongly with increasing $\ell$.
The maximal degree $\ell_{\rm max}=20$ has no special significance
and was chosen only for computational convenience. There may well be
significant power in the harmonics of even higher degree. 

%Very few rules can be formulated here. For example, one can
%claim that for $\nu=\nu_{2k,0,0}$ the significant VSH harmonics in the
%errors of positions begin with $\ell=2k-1$. The errors pattern for
%some GW frequencies are relatively simple and higher harmonics (say,
%$\ell>10$ become insignificant), while for other frequencies the error
%fields have a finer structure and the harmonics even with $\ell=20$
%remain important. 

\end{appendix}

\begin{thebibliography}{33}
\expandafter\ifx\csname natexlab\endcsname\relax\def\natexlab#1{#1}\fi

\bibitem[{{Agazie} {et~al.}(2024){Agazie}, {Antoniadis}, {Anumarlapudi},
  {Archibald}, {Arumugam}, {Arumugam}, {Arzoumanian}, {Askew}, {Babak},
  {Bagchi}, {Bailes}, {Bak Nielsen}, {Baker}, {Bassa}, {Bathula}, {B{\'e}csy},
  {Berthereau}, {Bhat}, {Blecha}, {Bonetti}, {Bortolas}, {Brazier}, {Brook},
  {Burgay}, {Burke-Spolaor}, {Burnette}, {Caballero}, {Cameron}, {Case},
  {Chalumeau}, {Champion}, {Chanlaridis}, {Charisi}, {Chatterjee},
  {Chatziioannou}, {Cheeseboro}, {Chen}, {Chen}, {Cognard}, {Cohen}, {Coles},
  {Cordes}, {Cornish}, {Crawford}, {Cromartie}, {Crowter}, {Cury{\l}o},
  {Cutler}, {Dai}, {Dandapat}, {Deb}, {DeCesar}, {DeGan}, {Demorest}, {Deng},
  {Desai}, {Desvignes}, {Dey}, {Dhanda-Batra}, {Di Marco}, {Dolch}, {Drachler},
  {Dwivedi}, {Ellis}, {Falxa}, {Feng}, {Ferdman}, {Ferrara}, {Fiore},
  {Fonseca}, {Franchini}, {Freedman}, {Gair}, {Garver-Daniels}, {Gentile},
  {Gersbach}, {Glaser}, {Good}, {Goncharov}, {Gopakumar}, {Graikou},
  {Griessmeier}, {Guillemot}, {G{\"u}ltekin}, {Guo}, {Gupta}, {Grunthal},
  {Hazboun}, {Hisano}, {Hobbs}, {Hourihane}, {Hu}, {Iraci}, {Islo},
  {Izquierdo-Villalba}, {Jang}, {Jawor}, {Janssen}, {Jennings}, {Jessner},
  {Johnson}, {Jones}, {Joshi}, {Kaiser}, {Kaplan}, {Kapur}, {Kareem},
  {Karuppusamy}, {Keane}, {Keith}, {Kelley}, {Kerr}, {Key}, {Kharbanda},
  {Kikunaga}, {Klein}, {Kolhe}, {Kramer}, {Krishnakumar}, {Kulkarni}, {Laal},
  {Lackeos}, {Lam}, {Lamb}, {Larsen}, {Lazio}, {Lee}, {Levin}, {Lewandowska},
  {Littenberg}, {Liu}, {Liu}, {Liu}, {Lommen}, {Lorimer}, {Lower}, {Luo},
  {Luo}, {Lynch}, {Lyne}, {Ma}, {Maan}, {Madison}, {Main}, {Manchester},
  {Mandow}, {Mattson}, {McEwen}, {McKee}, {McLaughlin}, {McMann}, {Meyers},
  {Meyers}, {Mickaliger}, {Miles}, {Mingarelli}, {Mitridate}, {Natarajan},
  {Nathan}, {Ng}, {Nice}, {Ni{\c{t}}u}, {Nobleson}, {Ocker}, {Olum},
  {Os{\l}owski}, {Paladi}, {Parthasarathy}, {Pennucci}, {Perera}, {Perrodin},
  {Petiteau}, {Petrov}, {Pol}, {Porayko}, {Possenti}, {Prabu}, {Quelquejay
  Leclere}, {Radovan}, {Rana}, {Ransom}, {Ray}, {Reardon}, {Rogers}, {Romano},
  {Russell}, {Samajdar}, {Sanidas}, {Sardesai}, {Schmiedekamp}, {Schmiedekamp},
  {Schmitz}, {Schult}, {Sesana}, {Shaifullah}, {Shannon}, {Shapiro-Albert},
  {Siemens}, {Simon}, \& {Singha}}]{2024ApJ...966..105A}
{Agazie}, G., {Antoniadis}, J., {Anumarlapudi}, A., {et~al.} 2024, \apj, 966,
  105

\bibitem[{{Bini} \& {Geralico}(2018)}]{2018PhRvD..98l4036B}
{Bini}, D. \& {Geralico}, A. 2018, \prd, 98, 124036

\bibitem[{{Book} \& {Flanagan}(2011)}]{2011PhRvD..83b4024B}
{Book}, L.~G. \& {Flanagan}, {\'E}.~{\'E}. 2011, \prd, 83, 024024

\bibitem[{{Braginsky} {et~al.}(1990){Braginsky}, {Kardashev}, {Polnarev}, \&
  {Novikov}}]{1990NCimB.105.1141B}
{Braginsky}, V.~B., {Kardashev}, N.~S., {Polnarev}, A.~G., \& {Novikov}, I.~D.
  1990, Nuovo Cimento B Serie, 105, 1141

\bibitem[{{Buonanno}(2007)}]{2007arXiv0709.4682B}
{Buonanno}, A. 2007, arXiv e-prints, arXiv:0709.4682

\bibitem[{{Butkevich} {et~al.}(2017){Butkevich}, {Klioner}, {Lindegren},
  {Hobbs}, \& {van Leeuwen}}]{2017A&A...603A..45B}
{Butkevich}, A.~G., {Klioner}, S.~A., {Lindegren}, L., {Hobbs}, D., \& {van
  Leeuwen}, F. 2017, \aap, 603, A45

\bibitem[{{Byrd} \& {Friedman}(1971)}]{1954MitAG...5...99B}
{Byrd}, P.~F. \& {Friedman}, M. 1971, {Handbook of Elliptic Integrals for
  Engineers and Physicists}, 2nd edn. (Berlin--Heidelberg--New York:
  Springer-Verlag)

\bibitem[{{{\c{C}}al{\i}{\c{s}}kan} {et~al.}(2024){{\c{C}}al{\i}{\c{s}}kan},
  {Chen}, {Dai}, {Kumar}, {Stomberg}, \& {Xue}}]{2024JCAP...05..030C}
{{\c{C}}al{\i}{\c{s}}kan}, M., {Chen}, Y., {Dai}, L., {et~al.} 2024, \jcap,
  2024, 030

\bibitem[{{Darling} {et~al.}(2018){Darling}, {Truebenbach}, \&
  {Paine}}]{2018ApJ...861..113D}
{Darling}, J., {Truebenbach}, A.~E., \& {Paine}, J. 2018, \apj, 861, 113

\bibitem[{{Erd\'elyi} {et~al.}(1953){Erd\'elyi}, {Magnus}, {Oberhettinger}, \&
  {Tricomi}}]{1955htf..book.....B}
{Erd\'elyi}, A., {Magnus}, W., {Oberhettinger}, F., \& {Tricomi}, F. 1953,
  {Higher transcendental functions, Volume II} (New York-Toronto-London:
  McGraw-Hill Book Company, Inc.)

\bibitem[{{Gaia Collaboration} {et~al.}(2022){Gaia Collaboration}, {Klioner},
  {Lindegren}, {Mignard}, {Hern{\'a}ndez}, {Ramos-Lerate}, {Bastian},
  {Biermann}, {Bombrun}, {de Torres}, {Gerlach}, {Geyer}, {Hilger}, {Hobbs},
  {Lammers}, {McMillan}, {Steidelm{\"u}ller}, {Teyssier}, {Raiteri},
  {Bartolom{\'e}}, {Bernet}, {Casta{\~n}eda}, {Clotet}, {Davidson},
  {Fabricius}, {Garralda Torres}, {Gonz{\'a}lez-Vidal}, {Portell}, {Rowell},
  {Torra}, {Torra}, {Brown}, {Vallenari}, {Prusti}, {de Bruijne}, {Arenou},
  {Babusiaux}, {Creevey}, {Ducourant}, {Evans}, {Eyer}, {Guerra}, {Hutton},
  {Jordi}, {Luri}, {Panem}, {Pourbaix}, {Randich}, {Sartoretti}, {Soubiran},
  {Tanga}, {Walton}, {Bailer-Jones}, {Drimmel}, {Jansen}, {Katz}, {Lattanzi},
  {van Leeuwen}, {Bakker}, {Cacciari}, {De Angeli}, {Fouesneau}, {Fr{\'e}mat},
  {Galluccio}, {Guerrier}, {Heiter}, {Masana}, {Messineo}, {Mowlavi},
  {Nicolas}, {Nienartowicz}, {Pailler}, {Panuzzo}, {Riclet}, {Roux},
  {Seabroke}, {Sordo}, {Th{\'e}venin}, {Gracia-Abril}, {Altmann}, {Andrae},
  {Audard}, {Bellas-Velidis}, {Benson}, {Berthier}, {Blomme}, {Burgess},
  {Busonero}, {Busso}, {C{\'a}novas}, {Carry}, {Cellino}, {Cheek},
  {Clementini}, {Damerdji}, {de Teodoro}, {Nu{\~n}ez Campos}, {Delchambre},
  {Dell'Oro}, {Esquej}, {Fern{\'a}ndez-Hern{\'a}ndez}, {Fraile}, {Garabato},
  {Garc{\'\i}a-Lario}, {Gosset}, {Haigron}, {Halbwachs}, {Hambly}, {Harrison},
  {Hestroffer}, {Hodgkin}, {Holl}, {Jan{\ss}en}, {Jevardat de Fombelle},
  {Jordan}, {Krone-Martins}, {Lanzafame}, {L{\"o}ffler}, {Marchal}, {Marrese},
  {Moitinho}, {Muinonen}, {Osborne}, {Pancino}, {Pauwels}, {Recio-Blanco},
  {Reyl{\'e}}, {Riello}, {Rimoldini}, {Roegiers}, {Rybizki}, {Sarro}, {Siopis},
  {Smith}, {Sozzetti}, {Utrilla}, {van Leeuwen}, {Abbas}, {{\'A}brah{\'a}m},
  {Abreu Aramburu}, {Aerts}, {Aguado}, {Ajaj}, {Aldea-Montero}, {Altavilla},
  {{\'A}lvarez}, {Alves}, {Anderson}, {Anglada Varela}, {Antoja}, {Baines},
  {Baker}, {Balaguer-N{\'u}{\~n}ez}, {Balbinot}, {Balog}, {Barache}, {Barbato},
  {Barros}, {Barstow}, {Bassilana}, {Bauchet}, {Becciani}, {Bellazzini},
  {Berihuete}, {Bertone}, {Bianchi}, {Binnenfeld}, {Blanco-Cuaresma}, {Boch},
  {Bossini}, {Bouquillon}, {Bragaglia}, {Bramante}, {Breedt}, {Bressan},
  {Brouillet}, {Brugaletta}, {Bucciarelli}, {Burlacu}, {Butkevich}, {Buzzi},
  {Caffau}, {Cancelliere}, {Cantat-Gaudin}, {Carballo}, {Carlucci},
  {Carnerero}, {Carrasco}, {Casamiquela}, {Castellani}, {Castro-Ginard},
  {Chaoul}, {Charlot}, {Chemin}, {Chiaramida}, {Chiavassa}, {Chornay},
  {Comoretto}, {Contursi}, {Cooper}, {Cornez}, {Cowell}, {Crifo}, {Cropper},
  {Crosta}, {Crowley}, {Dafonte}, {Dapergolas}, {David}, {de Laverny}, {De
  Luise}, {De March}, {De Ridder}, {de Souza}, {del Peloso}, {del Pozo},
  {Delbo}, {Delgado}, {Delisle}, {Demouchy}, {Dharmawardena}, {Diakite},
  {Diener}, {Distefano}, {Dolding}, {Enke}, {Fabre}, {Fabrizio}, {Faigler},
  {Fedorets}, {Fernique}, {Fienga}, {Figueras}, {Fournier}, {Fouron},
  {Fragkoudi}, {Gai}, {Garcia-Gutierrez}, {Garcia-Reinaldos},
  {Garc{\'\i}a-Torres}, {Garofalo}, {Gavel}, {Gavras}, {Giacobbe}, {Gilmore},
  {Girona}, {Giuffrida}, {Gomel}, {Gomez}, {Gonz{\'a}lez-N{\'u}{\~n}ez},
  {Gonz{\'a}lez-Santamar{\'\i}a}, {Granvik}, {Guillout}, {Guiraud},
  {Guti{\'e}rrez-S{\'a}nchez}, {Guy}, {Hatzidimitriou}, {Hauser}, {Haywood},
  {Helmer}, {Helmi}, {Sarmiento}, {Hidalgo}, {H{\l}adczuk}, {Holland},
  {Huckle}, {Jardine}, {Jasniewicz}, {Jean-Antoine Piccolo},
  {Jim{\'e}nez-Arranz}, {Juaristi Campillo}, {Julbe}, {Karbevska}, {Kervella},
  {Khanna}, {Kordopatis}, {Korn}, {K{\'o}sp{\'a}l}, {Kostrzewa-Rutkowska},
  {Kruszy{\'n}ska}, {Kun}, {Laizeau}, {Lambert}, {Lanza}, {Lasne}, {Le
  Campion}, {Lebreton}, {Lebzelter}, {Leccia}, {Leclerc}, {Lecoeur-Taibi},
  {Liao}, {Licata}, {Lindstr{\o}m}, {Lister}, {Livanou}, {Lobel}, {Lorca},
  {Loup}, {Madrero Pardo}, {Magdaleno Romeo}, {Managau}, {Mann}, {Manteiga},
  {Marchant}, {Marconi}, {Marcos}, {Santos}, {Mar{\'\i}n Pina}, {Marinoni},
  {Marocco}, {Marshall}, {Polo}, {Mart{\'\i}n-Fleitas}, {Marton}, {Mary},
  {Masip}, {Massari}, {Mastrobuono-Battisti}, {Mazeh}, {Messina}, {Michalik},
  {Millar}, {Mints}, {Molina}, {Molinaro}, {Moln{\'a}r}, {Monari},
  {Mongui{\'o}}, {Montegriffo}, {Montero}, {Mor}, {Mora}, {Morbidelli},
  {Morel}, {Morris}, {Muraveva}, {Murphy}, {Musella}, {Nagy}, {Noval},
  {Oca{\~n}a}, {Ogden}, {Ordenovic}, {Osinde}, {Pagani}, {Pagano}, {Palaversa},
  {Palicio}, {Pallas-Quintela}, {Panahi}, {Payne-Wardenaar}, {Pe{\~n}alosa
  Esteller}, {Penttil{\"a}}, {Pichon}, {Piersimoni}, {Pineau}, {Plachy},
  {Plum}, {Poggio}, {Pr{\v{s}}a}, {Pulone}, {Racero}, {Ragaini}, {Rainer},
  {Rambaux}, {Ramos}, {Re Fiorentin}, {Regibo}, {Richards}, {Diaz}, {Ripepi},
  {Riva}, {Rix}, {Rixon}, {Robichon}, {Robin}, {Robin}, {Roelens}, {Rogues},
  {Rohrbasser}, {Romero-G{\'o}mez}, {Royer}, {Ruz Mieres}, {Rybicki},
  {Sadowski}, {S{\'a}ez N{\'u}{\~n}ez}, {Sagrist{\`a} Sell{\'e}s}, {Sahlmann},
  {Salguero}, {Samaras}, {Sanchez Gimenez}, {Sanna}, {Santove{\~n}a},
  {Sarasso}, {Schultheis}, {Sciacca}, {Segol}, {Segovia}, {S{\'e}gransan},
  {Semeux}, {Shahaf}, {Siddiqui}, {Siebert}, {Siltala}, {Silvelo}, {Slezak},
  {Slezak}, {Smart}, {Snaith}, {Solano}, {Solitro}, {Souami}, {Souchay},
  {Spagna}, {Spina}, {Spoto}, {Steele}, {Stephenson}, {S{\"u}veges}, {Surdej},
  {Szabados}, {Szegedi-Elek}, {Taris}, {Taylor}, {Teixeira}, {Tolomei},
  {Tonello}, {Torralba Elipe}, {Trabucchi}, {Tsounis}, {Turon}, {Ulla},
  {Unger}, {Vaillant}, {van Dillen}, {van Reeven}, {Vanel}, {Vecchiato},
  {Viala}, {Vicente}, {Voutsinas}, {Weiler}, {Wevers}, {Wyrzykowski}, {Yoldas},
  {Yvard}, {Zhao}, {Zorec}, {Zucker}, \& {Zwitter}}]{2022A&A...667A.148G}
{Gaia Collaboration}, {Klioner}, S.~A., {Lindegren}, L., {et~al.} 2022, \aap,
  667, A148

\bibitem[{{Gaia Collaboration} {et~al.}(2021){Gaia Collaboration}, {Klioner},
  {Mignard}, {Lindegren}, {Bastian}, {McMillan}, {Hern{\'a}ndez}, {Hobbs},
  {Ramos-Lerate}, {Biermann}, {Bombrun}, {de Torres}, {Gerlach}, {Geyer},
  {Hilger}, {Lammers}, {Steidelm{\"u}ller}, {Stephenson}, {Brown}, {Vallenari},
  {Prusti}, {de Bruijne}, {Babusiaux}, {Creevey}, {Evans}, {Eyer}, {Hutton},
  {Jansen}, {Jordi}, {Luri}, {Panem}, {Pourbaix}, {Randich}, {Sartoretti},
  {Soubiran}, {Walton}, {Arenou}, {Bailer-Jones}, {Cropper}, {Drimmel}, {Katz},
  {Lattanzi}, {van Leeuwen}, {Bakker}, {Casta{\~n}eda}, {De Angeli},
  {Ducourant}, {Fabricius}, {Fouesneau}, {Fr{\'e}mat}, {Guerra}, {Guerrier},
  {Guiraud}, {Jean-Antoine Piccolo}, {Masana}, {Messineo}, {Mowlavi},
  {Nicolas}, {Nienartowicz}, {Pailler}, {Panuzzo}, {Riclet}, {Roux},
  {Seabroke}, {Sordo}, {Tanga}, {Th{\'e}venin}, {Gracia-Abril}, {Portell},
  {Teyssier}, {Altmann}, {Andrae}, {Bellas-Velidis}, {Benson}, {Berthier},
  {Blomme}, {Brugaletta}, {Burgess}, {Busso}, {Carry}, {Cellino}, {Cheek},
  {Clementini}, {Damerdji}, {Davidson}, {Delchambre}, {Dell'Oro},
  {Fern{\'a}ndez-Hern{\'a}ndez}, {Galluccio}, {Garc{\'\i}a-Lario},
  {Garcia-Reinaldos}, {Gonz{\'a}lez-N{\'u}{\~n}ez}, {Gosset}, {Haigron},
  {Halbwachs}, {Hambly}, {Harrison}, {Hatzidimitriou}, {Heiter}, {Hestroffer},
  {Hodgkin}, {Holl}, {Jan{\ss}en}, {Jevardat de Fombelle}, {Jordan},
  {Krone-Martins}, {Lanzafame}, {L{\"o}ffler}, {Lorca}, {Manteiga}, {Marchal},
  {Marrese}, {Moitinho}, {Mora}, {Muinonen}, {Osborne}, {Pancino}, {Pauwels},
  {Recio-Blanco}, {Richards}, {Riello}, {Rimoldini}, {Robin}, {Roegiers},
  {Rybizki}, {Sarro}, {Siopis}, {Smith}, {Sozzetti}, {Ulla}, {Utrilla}, {van
  Leeuwen}, {van Reeven}, {Abbas}, {Abreu Aramburu}, {Accart}, {Aerts},
  {Aguado}, {Ajaj}, {Altavilla}, {{\'A}lvarez}, {{\'A}lvarez Cid-Fuentes},
  {Alves}, {Anderson}, {Anglada Varela}, {Antoja}, {Audard}, {Baines}, {Baker},
  {Balaguer-N{\'u}{\~n}ez}, {Balbinot}, {Balog}, {Barache}, {Barbato},
  {Barros}, {Barstow}, {Bartolom{\'e}}, {Bassilana}, {Bauchet},
  {Baudesson-Stella}, {Becciani}, {Bellazzini}, {Bernet}, {Bertone}, {Bianchi},
  {Blanco-Cuaresma}, {Boch}, {Bossini}, {Bouquillon}, {Bramante}, {Breedt},
  {Bressan}, {Brouillet}, {Bucciarelli}, {Burlacu}, {Busonero}, {Butkevich},
  {Buzzi}, {Caffau}, {Cancelliere}, {C{\'a}novas}, {Cantat-Gaudin}, {Carballo},
  {Carlucci}, {Carnerero}, {Carrasco}, {Casamiquela}, {Castellani},
  {Castro-Ginard}, {Castro Sampol}, {Chaoul}, {Charlot}, {Chemin}, {Chiavassa},
  {Comoretto}, {Cooper}, {Cornez}, {Cowell}, {Crifo}, {Crosta}, {Crowley},
  {Dafonte}, {Dapergolas}, {David}, {David}, {de Laverny}, {De Luise}, {De
  March}, {De Ridder}, {de Souza}, {de Teodoro}, {del Peloso}, {del Pozo},
  {Delgado}, {Delgado}, {Delisle}, {Di Matteo}, {Diakite}, {Diener},
  {Distefano}, {Dolding}, {Eappachen}, {Enke}, {Esquej}, {Fabre}, {Fabrizio},
  {Faigler}, {Fedorets}, {Fernique}, {Fienga}, {Figueras}, {Fouron},
  {Fragkoudi}, {Fraile}, {Franke}, {Gai}, {Garabato}, {Garcia-Gutierrez},
  {Garc{\'\i}a-Torres}, {Garofalo}, {Gavras}, {Giacobbe}, {Gilmore}, {Girona},
  {Giuffrida}, {Gomez}, {Gonzalez-Santamaria}, {Gonz{\'a}lez-Vidal}, {Granvik},
  {Guti{\'e}rrez-S{\'a}nchez}, {Guy}, {Hauser}, {Haywood}, {Helmi}, {Hidalgo},
  {H{\l}adczuk}, {Holland}, {Huckle}, {Jasniewicz}, {Jonker}, {Juaristi
  Campillo}, {Julbe}, {Karbevska}, {Kervella}, {Khanna}, {Kochoska},
  {Kordopatis}, {Korn}, {Kostrzewa-Rutkowska}, {Kruszy{\'n}ska}, {Lambert},
  {Lanza}, {Lasne}, {Le Campion}, {Le Fustec}, {Lebreton}, {Lebzelter},
  {Leccia}, {Leclerc}, {Lecoeur-Taibi}, {Liao}, {Licata}, {Lindstr{\o}m},
  {Lister}, {Livanou}, {Lobel}, {Madrero Pardo}, {Managau}, {Mann}, {Marchant},
  {Marconi}, {Marcos Santos}, {Marinoni}, {Marocco}, {Marshall}, {Martin Polo},
  {Mart{\'\i}n-Fleitas}, {Masip}, {Massari}, {Mastrobuono-Battisti}, {Mazeh},
  {Messina}, {Michalik}, {Millar}, {Mints}, {Molina}, {Molinaro}, {Moln{\'a}r},
  {Montegriffo}, {Mor}, {Morbidelli}, {Morel}, {Morris}, {Mulone}, {Munoz},
  {Muraveva}, {Murphy}, {Musella}, {Noval}, {Ord{\'e}novic}, {Orr{\`u}},
  {Osinde}, {Pagani}, {Pagano}, {Palaversa}, {Palicio}, {Panahi}, {Pawlak},
  {Pe{\~n}alosa Esteller}, {Penttil{\"a}}, {Piersimoni}, {Pineau}, {Plachy},
  {Plum}, {Poggio}, {Poretti}, {Poujoulet}, {Pr{\v{s}}a}, {Pulone}, {Racero},
  {Ragaini}, {Rainer}, {Raiteri}, {Rambaux}, {Ramos}, {Re Fiorentin}, {Regibo},
  {Reyl{\'e}}, {Ripepi}, {Riva}, {Rixon}, {Robichon}, {Robin}, {Roelens},
  {Rohrbasser}, {Romero-G{\'o}mez}, {Rowell}, {Royer}, {Rybicki}, {Sadowski},
  {Sagrist{\`a} Sell{\'e}s}, {Sahlmann}, {Salgado}, {Salguero}, {Samaras},
  {Sanchez Gimenez}, {Sanna}, {Santove{\~n}a}, {Sarasso}, {Schultheis},
  {Sciacca}, {Segol}, {Segovia}, {S{\'e}gransan}, {Semeux}, {Siddiqui},
  {Siebert}, {Siltala}, {Slezak}, {Smart}, {Solano}, {Solitro}, {Souami},
  {Souchay}, {Spagna}, {Spoto}, {Steele}, {S{\"u}veges}, {Szabados},
  {Szegedi-Elek}, {Taris}, {Tauran}, {Taylor}, {Teixeira}, {Thuillot},
  {Tonello}, {Torra}, {Torra}, {Turon}, {Unger}, {Vaillant}, {van Dillen},
  {Vanel}, {Vecchiato}, {Viala}, {Vicente}, {Voutsinas}, {Weiler}, {Wevers},
  {Wyrzykowski}, {Yoldas}, {Yvard}, {Zhao}, {Zorec}, {Zucker}, {Zurbach}, \&
  {Zwitter}}]{2021A&A...649A...9G}
{Gaia Collaboration}, {Klioner}, S.~A., {Mignard}, F., {et~al.} 2021, \aap,
  649, A9

\bibitem[{{Gaia Collaboration} {et~al.}(2016){Gaia Collaboration}, {Prusti},
  {de Bruijne}, {Brown}, {Vallenari}, {Babusiaux}, {Bailer-Jones}, {Bastian},
  {Biermann}, {Evans}, {Eyer}, {Jansen}, {Jordi}, {Klioner}, {Lammers},
  {Lindegren}, {Luri}, {Mignard}, {Milligan}, {Panem}, {Poinsignon},
  {Pourbaix}, {Randich}, {Sarri}, {Sartoretti}, {Siddiqui}, {Soubiran},
  {Valette}, {van Leeuwen}, {Walton}, {Aerts}, {Arenou}, {Cropper}, {Drimmel},
  {H{\o}g}, {Katz}, {Lattanzi}, {O'Mullane}, {Grebel}, {Holland}, {Huc},
  {Passot}, {Bramante}, {Cacciari}, {Casta{\~n}eda}, {Chaoul}, {Cheek}, {De
  Angeli}, {Fabricius}, {Guerra}, {Hern{\'a}ndez}, {Jean-Antoine-Piccolo},
  {Masana}, {Messineo}, {Mowlavi}, {Nienartowicz}, {Ord{\'o}{\~n}ez-Blanco},
  {Panuzzo}, {Portell}, {Richards}, {Riello}, {Seabroke}, {Tanga},
  {Th{\'e}venin}, {Torra}, {Els}, {Gracia-Abril}, {Comoretto},
  {Garcia-Reinaldos}, {Lock}, {Mercier}, {Altmann}, {Andrae}, {Astraatmadja},
  {Bellas-Velidis}, {Benson}, {Berthier}, {Blomme}, {Busso}, {Carry},
  {Cellino}, {Clementini}, {Cowell}, {Creevey}, {Cuypers}, {Davidson}, {De
  Ridder}, {de Torres}, {Delchambre}, {Dell'Oro}, {Ducourant}, {Fr{\'e}mat},
  {Garc{\'\i}a-Torres}, {Gosset}, {Halbwachs}, {Hambly}, {Harrison}, {Hauser},
  {Hestroffer}, {Hodgkin}, {Huckle}, {Hutton}, {Jasniewicz}, {Jordan},
  {Kontizas}, {Korn}, {Lanzafame}, {Manteiga}, {Moitinho}, {Muinonen},
  {Osinde}, {Pancino}, {Pauwels}, {Petit}, {Recio-Blanco}, {Robin}, {Sarro},
  {Siopis}, {Smith}, {Smith}, {Sozzetti}, {Thuillot}, {van Reeven}, {Viala},
  {Abbas}, {Abreu Aramburu}, {Accart}, {Aguado}, {Allan}, {Allasia},
  {Altavilla}, {{\'A}lvarez}, {Alves}, {Anderson}, {Andrei}, {Anglada Varela},
  {Antiche}, {Antoja}, {Ant{\'o}n}, {Arcay}, {Atzei}, {Ayache}, {Bach},
  {Baker}, {Balaguer-N{\'u}{\~n}ez}, {Barache}, {Barata}, {Barbier}, {Barblan},
  {Baroni}, {Barrado y Navascu{\'e}s}, {Barros}, {Barstow}, {Becciani},
  {Bellazzini}, {Bellei}, {Bello Garc{\'\i}a}, {Belokurov}, {Bendjoya},
  {Berihuete}, {Bianchi}, {Bienaym{\'e}}, {Billebaud}, {Blagorodnova},
  {Blanco-Cuaresma}, {Boch}, {Bombrun}, {Borrachero}, {Bouquillon}, {Bourda},
  {Bouy}, {Bragaglia}, {Breddels}, {Brouillet}, {Br{\"u}semeister},
  {Bucciarelli}, {Budnik}, {Burgess}, {Burgon}, {Burlacu}, {Busonero}, {Buzzi},
  {Caffau}, {Cambras}, {Campbell}, {Cancelliere}, {Cantat-Gaudin}, {Carlucci},
  {Carrasco}, {Castellani}, {Charlot}, {Charnas}, {Charvet}, {Chassat},
  {Chiavassa}, {Clotet}, {Cocozza}, {Collins}, {Collins}, {Costigan}, {Crifo},
  {Cross}, {Crosta}, {Crowley}, {Dafonte}, {Damerdji}, {Dapergolas}, {David},
  {David}, {De Cat}, {de Felice}, {de Laverny}, {De Luise}, {De March}, {de
  Martino}, {de Souza}, {Debosscher}, {del Pozo}, {Delbo}, {Delgado},
  {Delgado}, {di Marco}, {Di Matteo}, {Diakite}, {Distefano}, {Dolding}, {Dos
  Anjos}, {Drazinos}, {Dur{\'a}n}, {Dzigan}, {Ecale}, {Edvardsson}, {Enke},
  {Erdmann}, {Escolar}, {Espina}, {Evans}, {Eynard Bontemps}, {Fabre},
  {Fabrizio}, {Faigler}, {Falc{\~a}o}, {Farr{\`a}s Casas}, {Faye}, {Federici},
  {Fedorets}, {Fern{\'a}ndez-Hern{\'a}ndez}, {Fernique}, {Fienga}, {Figueras},
  {Filippi}, {Findeisen}, {Fonti}, {Fouesneau}, {Fraile}, {Fraser}, {Fuchs},
  {Furnell}, {Gai}, {Galleti}, {Galluccio}, {Garabato}, {Garc{\'\i}a-Sedano},
  {Gar{\'e}}, {Garofalo}, {Garralda}, {Gavras}, {Gerssen}, {Geyer}, {Gilmore},
  {Girona}, {Giuffrida}, {Gomes}, {Gonz{\'a}lez-Marcos},
  {Gonz{\'a}lez-N{\'u}{\~n}ez}, {Gonz{\'a}lez-Vidal}, {Granvik}, {Guerrier},
  {Guillout}, {Guiraud}, {G{\'u}rpide}, {Guti{\'e}rrez-S{\'a}nchez}, {Guy},
  {Haigron}, {Hatzidimitriou}, {Haywood}, {Heiter}, {Helmi}, {Hobbs},
  {Hofmann}, {Holl}, {Holland}, {Hunt}, {Hypki}, {Icardi}, {Irwin}, {Jevardat
  de Fombelle}, {Jofr{\'e}}, {Jonker}, {Jorissen}, {Julbe}, {Karampelas},
  {Kochoska}, {Kohley}, {Kolenberg}, {Kontizas}, {Koposov}, {Kordopatis},
  {Koubsky}, {Kowalczyk}, {Krone-Martins}, {Kudryashova}, {Kull}, {Bachchan},
  {Lacoste-Seris}, {Lanza}, {Lavigne}, {Le Poncin-Lafitte}, {Lebreton},
  {Lebzelter}, {Leccia}, {Leclerc}, {Lecoeur-Taibi}, {Lemaitre}, {Lenhardt},
  {Leroux}, {Liao}, {Licata}, {Lindstr{\o}m}, {Lister}, {Livanou}, {Lobel},
  {L{\"o}ffler}, {L{\'o}pez}, {Lopez-Lozano}, {Lorenz}, {Loureiro},
  {MacDonald}, {Magalh{\~a}es Fernandes}, {Managau}, {Mann}, {Mantelet},
  {Marchal}, {Marchant}, {Marconi}, {Marie}, {Marinoni}, {Marrese},
  {Marschalk{\'o}}, {Marshall}, {Mart{\'\i}n-Fleitas}, {Martino}, {Mary},
  {Matijevi{\v{c}}}, {Mazeh}, {McMillan}, {Messina}, {Mestre}, {Michalik},
  {Millar}, {Miranda}, {Molina}, {Molinaro}, {Molinaro}, {Moln{\'a}r},
  {Moniez}, {Montegriffo}, {Monteiro}, {Mor}, {Mora}, {Morbidelli}, {Morel},
  {Morgenthaler}, {Morley}, {Morris}, {Mulone}, {Muraveva}, {Musella},
  {Narbonne}, {Nelemans}, {Nicastro}, {Noval}, {Ord{\'e}novic},
  {Ordieres-Mer{\'e}}, {Osborne}, {Pagani}, {Pagano}, {Pailler}, {Palacin},
  {Palaversa}, {Parsons}, {Paulsen}, {Pecoraro}, {Pedrosa}, {Pentik{\"a}inen},
  {Pereira}, {Pichon}, {Piersimoni}, {Pineau}, {Plachy}, {Plum}, {Poujoulet},
  {Pr{\v{s}}a}, {Pulone}, {Ragaini}, {Rago}, {Rambaux}, {Ramos-Lerate},
  {Ranalli}, {Rauw}, {Read}, {Regibo}, {Renk}, {Reyl{\'e}}, {Ribeiro},
  {Rimoldini}, {Ripepi}, {Riva}, {Rixon}, {Roelens}, {Romero-G{\'o}mez},
  {Rowell}, {Royer}, {Rudolph}, {Ruiz-Dern}, {Sadowski}, {Sagrist{\`a}
  Sell{\'e}s}, {Sahlmann}, {Salgado}, {Salguero}, {Sarasso}, {Savietto},
  {Schnorhk}, {Schultheis}, {Sciacca}, {Segol}, {Segovia}, {Segransan},
  {Serpell}, {Shih}, {Smareglia}, {Smart}, {Smith}, {Solano}, {Solitro},
  {Sordo}, {Soria Nieto}, {Souchay}, {Spagna}, {Spoto}, {Stampa}, {Steele},
  {Steidelm{\"u}ller}, {Stephenson}, {Stoev}, {Suess}, {S{\"u}veges}, {Surdej},
  {Szabados}, {Szegedi-Elek}, {Tapiador}, {Taris}, {Tauran}, {Taylor},
  {Teixeira}, {Terrett}, {Tingley}, {Trager}, {Turon}, {Ulla}, {Utrilla},
  {Valentini}, {van Elteren}, {Van Hemelryck}, {van Leeuwen}, {Varadi},
  {Vecchiato}, {Veljanoski}, {Via}, {Vicente}, {Vogt}, {Voss}, {Votruba},
  {Voutsinas}, {Walmsley}, {Weiler}, {Weingrill}, {Werner}, {Wevers},
  {Whitehead}, {Wyrzykowski}, {Yoldas}, {{\v{Z}}erjal}, {Zucker}, {Zurbach},
  {Zwitter}, {Alecu}, {Allen}, {Allende Prieto}, {Amorim},
  {Anglada-Escud{\'e}}, {Arsenijevic}, {Azaz}, {Balm}, {Beck}, {Bernstein},
  {Bigot}, {Bijaoui}, {Blasco}, {Bonfigli}, {Bono}, {Boudreault}, {Bressan},
  {Brown}, {Brunet}, {Bunclark}, {Buonanno}, {Butkevich}, {Carret}, {Carrion},
  {Chemin}, {Ch{\'e}reau}, {Corcione}, {Darmigny}, {de Boer}, {de Teodoro}, {de
  Zeeuw}, {Delle Luche}, {Domingues}, {Dubath}, {Fodor}, {Fr{\'e}zouls},
  {Fries}, {Fustes}, {Fyfe}, {Gallardo}, {Gallegos}, {Gardiol}, {Gebran},
  {Gomboc}, {G{\'o}mez}, {Grux}, {Gueguen}, {Heyrovsky}, {Hoar}, {Iannicola},
  {Isasi Parache}, {Janotto}, {Joliet}, {Jonckheere}, {Keil}, {Kim},
  {Klagyivik}, {Klar}, {Knude}, {Kochukhov}, {Kolka}, {Kos}, {Kutka}, {Lainey},
  {LeBouquin}, {Liu}, {Loreggia}, {Makarov}, {Marseille}, {Martayan},
  {Martinez-Rubi}, {Massart}, {Meynadier}, {Mignot}, {Munari}, {Nguyen},
  {Nordlander}, {Ocvirk}, {O'Flaherty}, {Olias Sanz}, {Ortiz}, {Osorio},
  {Oszkiewicz}, {Ouzounis}, {Palmer}, {Park}, {Pasquato}, {Peltzer}, {Peralta},
  {P{\'e}turaud}, {Pieniluoma}, {Pigozzi}, {Poels}, {Prat}, {Prod'homme},
  {Raison}, {Rebordao}, {Risquez}, {Rocca-Volmerange}, {Rosen}, {Ruiz-Fuertes},
  {Russo}, {Sembay}, {Serraller Vizcaino}, {Short}, {Siebert}, {Silva},
  {Sinachopoulos}, {Slezak}, {Soffel}, {Sosnowska}, {Strai{\v{z}}ys}, {ter
  Linden}, {Terrell}, {Theil}, {Tiede}, {Troisi}, {Tsalmantza}, {Tur},
  {Vaccari}, {Vachier}, {Valles}, {Van Hamme}, {Veltz}, {Virtanen}, {Wallut},
  {Wichmann}, {Wilkinson}, {Ziaeepour}, \& {Zschocke}}]{2016A&A...595A...1G}
{Gaia Collaboration}, {Prusti}, T., {de Bruijne}, J.~H.~J., {et~al.} 2016,
  \aap, 595, A1

\bibitem[{{Gaia Collaboration} {et~al.}(2023){Gaia Collaboration}, {Vallenari},
  {Brown}, {Prusti}, {de Bruijne}, {Arenou}, {Babusiaux}, {Biermann},
  {Creevey}, {Ducourant}, {Evans}, {Eyer}, {Guerra}, {Hutton}, {Jordi},
  {Klioner}, {Lammers}, {Lindegren}, {Luri}, {Mignard}, {Panem}, {Pourbaix},
  {Randich}, {Sartoretti}, {Soubiran}, {Tanga}, {Walton}, {Bailer-Jones},
  {Bastian}, {Drimmel}, {Jansen}, {Katz}, {Lattanzi}, {van Leeuwen}, {Bakker},
  {Cacciari}, {Casta{\~n}eda}, {De Angeli}, {Fabricius}, {Fouesneau},
  {Fr{\'e}mat}, {Galluccio}, {Guerrier}, {Heiter}, {Masana}, {Messineo},
  {Mowlavi}, {Nicolas}, {Nienartowicz}, {Pailler}, {Panuzzo}, {Riclet}, {Roux},
  {Seabroke}, {Sordo}, {Th{\'e}venin}, {Gracia-Abril}, {Portell}, {Teyssier},
  {Altmann}, {Andrae}, {Audard}, {Bellas-Velidis}, {Benson}, {Berthier},
  {Blomme}, {Burgess}, {Busonero}, {Busso}, {C{\'a}novas}, {Carry}, {Cellino},
  {Cheek}, {Clementini}, {Damerdji}, {Davidson}, {de Teodoro}, {Nu{\~n}ez
  Campos}, {Delchambre}, {Dell'Oro}, {Esquej}, {Fern{\'a}ndez-Hern{\'a}ndez},
  {Fraile}, {Garabato}, {Garc{\'\i}a-Lario}, {Gosset}, {Haigron}, {Halbwachs},
  {Hambly}, {Harrison}, {Hern{\'a}ndez}, {Hestroffer}, {Hodgkin}, {Holl},
  {Jan{\ss}en}, {Jevardat de Fombelle}, {Jordan}, {Krone-Martins}, {Lanzafame},
  {L{\"o}ffler}, {Marchal}, {Marrese}, {Moitinho}, {Muinonen}, {Osborne},
  {Pancino}, {Pauwels}, {Recio-Blanco}, {Reyl{\'e}}, {Riello}, {Rimoldini},
  {Roegiers}, {Rybizki}, {Sarro}, {Siopis}, {Smith}, {Sozzetti}, {Utrilla},
  {van Leeuwen}, {Abbas}, {{\'A}brah{\'a}m}, {Abreu Aramburu}, {Aerts},
  {Aguado}, {Ajaj}, {Aldea-Montero}, {Altavilla}, {{\'A}lvarez}, {Alves},
  {Anders}, {Anderson}, {Anglada Varela}, {Antoja}, {Baines}, {Baker},
  {Balaguer-N{\'u}{\~n}ez}, {Balbinot}, {Balog}, {Barache}, {Barbato},
  {Barros}, {Barstow}, {Bartolom{\'e}}, {Bassilana}, {Bauchet}, {Becciani},
  {Bellazzini}, {Berihuete}, {Bernet}, {Bertone}, {Bianchi}, {Binnenfeld},
  {Blanco-Cuaresma}, {Blazere}, {Boch}, {Bombrun}, {Bossini}, {Bouquillon},
  {Bragaglia}, {Bramante}, {Breedt}, {Bressan}, {Brouillet}, {Brugaletta},
  {Bucciarelli}, {Burlacu}, {Butkevich}, {Buzzi}, {Caffau}, {Cancelliere},
  {Cantat-Gaudin}, {Carballo}, {Carlucci}, {Carnerero}, {Carrasco},
  {Casamiquela}, {Castellani}, {Castro-Ginard}, {Chaoul}, {Charlot}, {Chemin},
  {Chiaramida}, {Chiavassa}, {Chornay}, {Comoretto}, {Contursi}, {Cooper},
  {Cornez}, {Cowell}, {Crifo}, {Cropper}, {Crosta}, {Crowley}, {Dafonte},
  {Dapergolas}, {David}, {David}, {de Laverny}, {De Luise}, {De March}, {De
  Ridder}, {de Souza}, {de Torres}, {del Peloso}, {del Pozo}, {Delbo},
  {Delgado}, {Delisle}, {Demouchy}, {Dharmawardena}, {Di Matteo}, {Diakite},
  {Diener}, {Distefano}, {Dolding}, {Edvardsson}, {Enke}, {Fabre}, {Fabrizio},
  {Faigler}, {Fedorets}, {Fernique}, {Fienga}, {Figueras}, {Fournier},
  {Fouron}, {Fragkoudi}, {Gai}, {Garcia-Gutierrez}, {Garcia-Reinaldos},
  {Garc{\'\i}a-Torres}, {Garofalo}, {Gavel}, {Gavras}, {Gerlach}, {Geyer},
  {Giacobbe}, {Gilmore}, {Girona}, {Giuffrida}, {Gomel}, {Gomez},
  {Gonz{\'a}lez-N{\'u}{\~n}ez}, {Gonz{\'a}lez-Santamar{\'\i}a},
  {Gonz{\'a}lez-Vidal}, {Granvik}, {Guillout}, {Guiraud},
  {Guti{\'e}rrez-S{\'a}nchez}, {Guy}, {Hatzidimitriou}, {Hauser}, {Haywood},
  {Helmer}, {Helmi}, {Sarmiento}, {Hidalgo}, {Hilger}, {H{\l}adczuk}, {Hobbs},
  {Holland}, {Huckle}, {Jardine}, {Jasniewicz}, {Jean-Antoine Piccolo},
  {Jim{\'e}nez-Arranz}, {Jorissen}, {Juaristi Campillo}, {Julbe}, {Karbevska},
  {Kervella}, {Khanna}, {Kontizas}, {Kordopatis}, {Korn}, {K{\'o}sp{\'a}l},
  {Kostrzewa-Rutkowska}, {Kruszy{\'n}ska}, {Kun}, {Laizeau}, {Lambert},
  {Lanza}, {Lasne}, {Le Campion}, {Lebreton}, {Lebzelter}, {Leccia}, {Leclerc},
  {Lecoeur-Taibi}, {Liao}, {Licata}, {Lindstr{\o}m}, {Lister}, {Livanou},
  {Lobel}, {Lorca}, {Loup}, {Madrero Pardo}, {Magdaleno Romeo}, {Managau},
  {Mann}, {Manteiga}, {Marchant}, {Marconi}, {Marcos}, {Marcos Santos},
  {Mar{\'\i}n Pina}, {Marinoni}, {Marocco}, {Marshall}, {Martin Polo},
  {Mart{\'\i}n-Fleitas}, {Marton}, {Mary}, {Masip}, {Massari},
  {Mastrobuono-Battisti}, {Mazeh}, {McMillan}, {Messina}, {Michalik}, {Millar},
  {Mints}, {Molina}, {Molinaro}, {Moln{\'a}r}, {Monari}, {Mongui{\'o}},
  {Montegriffo}, {Montero}, {Mor}, {Mora}, {Morbidelli}, {Morel}, {Morris},
  {Muraveva}, {Murphy}, {Musella}, {Nagy}, {Noval}, {Oca{\~n}a}, {Ogden},
  {Ordenovic}, {Osinde}, {Pagani}, {Pagano}, {Palaversa}, {Palicio},
  {Pallas-Quintela}, {Panahi}, {Payne-Wardenaar}, {Pe{\~n}alosa Esteller},
  {Penttil{\"a}}, {Pichon}, {Piersimoni}, {Pineau}, {Plachy}, {Plum}, {Poggio},
  {Pr{\v{s}}a}, {Pulone}, {Racero}, {Ragaini}, {Rainer}, {Raiteri}, {Rambaux},
  {Ramos}, {Ramos-Lerate}, {Re Fiorentin}, {Regibo}, {Richards}, {Rios Diaz},
  {Ripepi}, {Riva}, {Rix}, {Rixon}, {Robichon}, {Robin}, {Robin}, {Roelens},
  {Rogues}, {Rohrbasser}, {Romero-G{\'o}mez}, {Rowell}, {Royer}, {Ruz Mieres},
  {Rybicki}, {Sadowski}, {S{\'a}ez N{\'u}{\~n}ez}, {Sagrist{\`a} Sell{\'e}s},
  {Sahlmann}, {Salguero}, {Samaras}, {Sanchez Gimenez}, {Sanna},
  {Santove{\~n}a}, {Sarasso}, {Schultheis}, {Sciacca}, {Segol}, {Segovia},
  {S{\'e}gransan}, {Semeux}, {Shahaf}, {Siddiqui}, {Siebert}, {Siltala},
  {Silvelo}, {Slezak}, {Slezak}, {Smart}, {Snaith}, {Solano}, {Solitro},
  {Souami}, {Souchay}, {Spagna}, {Spina}, {Spoto}, {Steele},
  {Steidelm{\"u}ller}, {Stephenson}, {S{\"u}veges}, {Surdej}, {Szabados},
  {Szegedi-Elek}, {Taris}, {Taylor}, {Teixeira}, {Tolomei}, {Tonello}, {Torra},
  {Torra}, {Torralba Elipe}, {Trabucchi}, {Tsounis}, {Turon}, {Ulla}, {Unger},
  {Vaillant}, {van Dillen}, {van Reeven}, {Vanel}, {Vecchiato}, {Viala},
  {Vicente}, {Voutsinas}, {Weiler}, {Wevers}, {Wyrzykowski}, {Yoldas}, {Yvard},
  {Zhao}, {Zorec}, {Zucker}, \& {Zwitter}}]{2023A&A...674A...1G}
{Gaia Collaboration}, {Vallenari}, A., {Brown}, A.~G.~A., {et~al.} 2023, \aap,
  674, A1

\bibitem[{{Gwinn} {et~al.}(1997){Gwinn}, {Eubanks}, {Pyne}, {Birkinshaw}, \&
  {Matsakis}}]{1997ApJ...485...87G}
{Gwinn}, C.~R., {Eubanks}, T.~M., {Pyne}, T., {Birkinshaw}, M., \& {Matsakis},
  D.~N. 1997, \apj, 485, 87

\bibitem[{{Holl} {et~al.}(2023){Holl}, {Fabricius}, {Portell}, {Lindegren},
  {Panuzzo}, {Bernet}, {Casta{\~n}eda}, {Jevardat de Fombelle}, {Audard},
  {Ducourant}, {Harrison}, {Evans}, {Busso}, {Sozzetti}, {Gosset}, {Arenou},
  {De Angeli}, {Riello}, {Eyer}, {Rimoldini}, {Gavras}, {Mowlavi},
  {Nienartowicz}, {Lecoeur-Ta{\"\i}bi}, {Garc{\'\i}a-Lario}, \&
  {Pourbaix}}]{2023A&A...674A..25H}
{Holl}, B., {Fabricius}, C., {Portell}, J., {et~al.} 2023, \aap, 674, A25

\bibitem[{{Holl} {et~al.}(2012){Holl}, {Lindegren}, \&
  {Hobbs}}]{2012A&A...543A..15H}
{Holl}, B., {Lindegren}, L., \& {Hobbs}, D. 2012, \aap, 543, A15

\bibitem[{{Jaranowski} \& {Krolak}(2009)}]{2009agwd.book.....J}
{Jaranowski}, P. \& {Krolak}, A. 2009, {Analysis of Gravitational-Wave Data}
  (Cambridge University Press)

\bibitem[{{Kaiser} \& {Jaffe}(1997)}]{1997ApJ...484..545K}
{Kaiser}, N. \& {Jaffe}, A. 1997, \apj, 484, 545

\bibitem[{{Klioner}(2003)}]{2003AJ....125.1580K}
{Klioner}, S.~A. 2003, \aj, 125, 1580

\bibitem[{{Klioner}(2004)}]{2004PhRvD..69l4001K}
{Klioner}, S.~A. 2004, \prd, 69, 124001

\bibitem[{{Klioner}(2014)}]{Klioner2014}
{Klioner}, S.~A. 2014, Velocity error and effective basic angle calibration
  (VBAC): basic principles and possible applications, Tech. Rep.
  GAIA-C3-TN-LO-SK-020, available from the Gaia document archive:
  \url{https://dms.cosmos.esa.int/cs/livelink?objId=3268461}

\bibitem[{{Klioner}(2018)}]{2018CQGra..35d5005K}
{Klioner}, S.~A. 2018, Classical and Quantum Gravity, 35, 045005

\bibitem[{{Kopeikin} {et~al.}(1999){Kopeikin}, {Sch{\"a}fer}, {Gwinn}, \&
  {Eubanks}}]{1999PhRvD..59h4023K}
{Kopeikin}, S.~M., {Sch{\"a}fer}, G., {Gwinn}, C.~R., \& {Eubanks}, T.~M. 1999,
  \prd, 59, 084023

\bibitem[{Lindegren(1977)}]{lindegren77}
Lindegren, L. 1977, {Thermal stability and the determination of parallaxes},
  Tech. rep., Lund Observatory, available from
  \url{https://doi.org/10.5281/zenodo.7642744}

\bibitem[{{Lindegren} \& {Bastian}(2010)}]{2010EAS....45..109L}
{Lindegren}, L. \& {Bastian}, U. 2010, in EAS Publications Series, Vol.~45, EAS
  Publications Series, ed. C.~{Turon}, F.~{Meynadier}, \& F.~{Arenou}, 109--114

\bibitem[{{Lindegren} {et~al.}(2012){Lindegren}, {Lammers}, {Hobbs},
  {O'Mullane}, {Bastian}, \& {Hern{\'a}ndez}}]{2012A&A...538A..78L}
{Lindegren}, L., {Lammers}, U., {Hobbs}, D., {et~al.} 2012, \aap, 538, A78

\bibitem[{{Mignard} \& {Klioner}(2012)}]{2012A&A...547A..59M}
{Mignard}, F. \& {Klioner}, S. 2012, \aap, 547, A59

\bibitem[{{Mihaylov} {et~al.}(2018){Mihaylov}, {Moore}, {Gair}, {Lasenby}, \&
  {Gilmore}}]{2018PhRvD..97l4058M}
{Mihaylov}, D.~P., {Moore}, C.~J., {Gair}, J.~R., {Lasenby}, A., \& {Gilmore},
  G. 2018, \prd, 97, 124058

\bibitem[{{Moore} {et~al.}(2017){Moore}, {Mihaylov}, {Lasenby}, \&
  {Gilmore}}]{2017PhRvL.119z1102M}
{Moore}, C.~J., {Mihaylov}, D.~P., {Lasenby}, A., \& {Gilmore}, G. 2017, \prl,
  119, 261102

\bibitem[{{O'Beirne} \& {Cornish}(2018)}]{2018PhRvD..98b4020O}
{O'Beirne}, L. \& {Cornish}, N.~J. 2018, \prd, 98, 024020

\bibitem[{{Pyne} {et~al.}(1996){Pyne}, {Gwinn}, {Birkinshaw}, {Eubanks}, \&
  {Matsakis}}]{1996ApJ...465..566P}
{Pyne}, T., {Gwinn}, C.~R., {Birkinshaw}, M., {Eubanks}, T.~M., \& {Matsakis},
  D.~N. 1996, \apj, 465, 566

\bibitem[{{Soffel} {et~al.}(2003){Soffel}, {Klioner}, {Petit}, {Wolf},
  {Kopeikin}, {Bretagnon}, {Brumberg}, {Capitaine}, {Damour}, {Fukushima},
  {Guinot}, {Huang}, {Lindegren}, {Ma}, {Nordtvedt}, {Ries}, {Seidelmann},
  {Vokrouhlick{\'y}}, {Will}, \& {Xu}}]{2003AJ....126.2687S}
{Soffel}, M., {Klioner}, S.~A., {Petit}, G., {et~al.} 2003, \aj, 126, 2687

\end{thebibliography}
\end{document}